%% file: EQMCCRfinal.tex
\documentclass{article}
\usepackage{graphicx}
\usepackage{amsmath}
\usepackage{amsfonts}
\usepackage{amssymb}

\numberwithin{equation}{section}

\newtheorem{theorem}{Theorem}

\newtheorem{example}[theorem]{Example}

\newtheorem{proposition}[theorem]{Proposition}
\newtheorem{remark}[theorem]{Remark}

\begin{document}

\title{FROM THE EQUATIONS OF MOTION TO THE CANONICAL COMMUTATION RELATIONS}
\maketitle

\begin{center}
E.Ercolessi\\ Dipartimento di Fisica and INFN. Universita' di Bologna,\\
46 v. Irnerio, I-40126 Bologna. Italy. \textit{e-mail: ercolessi@bo.infn.it}\\
G.Marmo\\ Dipartimento di Scienze Fisiche and INFN. Universita' di Napoli "Federico II", \\
v.Cinthia, I-80125 Napoli. Italy. \textit{e-mail: marmo@na.infn.it}\\
G.Morandi\\ Dipartimento di Fisica and INFN. Universita' di Bologna, \\
6/2 v.le B. Pichat, I-40127 Bologna. Italy.
\textit{e-mail: morandi@bo.infn.it}
\end{center}

\begin{abstract}
The problem of whether or not the equations of motion of a quantum system
determine the commutation relations was posed by E.P.Wigner in $1950$. A
similar problem (known as \textit{"The Inverse Problem in the Calculus of
Variations"}) was posed in a classical setting as back as in $1887$ by
H.Helmoltz and has received great attention also in recent times. The aim of
this paper is to discuss how these two apparently unrelated problems can
actually be discussed in a somewhat unified framework. After reviewing briefly
the Inverse Problem and the existence of alternative structures for classical
systems, we discuss the geometric structures that are intrinsically present in
Quantum Mechanics, starting from finite-level systems and then moving to a
more general setting by using the Weyl-Wigner approach, showing how this
approach can accomodate in an almost natural way the existence of alternative
structures in Quantum Mechanics as well. 

\end{abstract}

\vskip1cm



\newpage
\tableofcontents
\newpage

\input Chapt1rev.tex
\input Chapt2rev.tex

\input Chapt3rev.tex
\input Chapt4rev.tex

\input Chapt5rev.tex
\input Chapt6rev.tex

\input Chapt7rev.tex

\appendix
\input AppArev.tex
\input AppBrev.tex
\input AppCrev.tex

\end{document}

%% file: Chapt1rev.tex
\section{Introduction and Motivations}\label{introduc}

\bigskip
\subsection{Introductory Considerations \label{introd}}
\bigskip

Back in $1950$, E.P.Wigner \cite{Wig3} (see also
Refs.\cite{BDe,MMSZ,OW}) raised the problem of whether the
equations of motion determine or not the quantum commutation
relations. A few papers \cite{Pu,Ya} followed immediately, and the
same problem was considered by S.Schweber \cite{S1} in the
framework of Quantum Field Theory. It also originated the interest
for parastatistics \cite{Ga,Gre,GreMe}. Physicists were apparently
motivated in this research by the search of a way out of the
apparently uncontrollable divergences that were plaguing
Relativistic Quantum Field Theory.

As reported by F.Dyson \cite{Dys}, also Feynman addressed the same
problem, looking for commutation relations not associated with
Lagrangian descriptions. One would have also avoided in this way
\cite{CIMS} the introduction of gauge potentials. In the classical
setting the problem, known as the \textit{"Inverse Problem in the
Calculus of Variations" }\cite{MFVMR}, was stated and clearly
formulated already by H.Helmoltz \cite{Hel}. An example of a system
admitting of two alternative Hamiltonian descriptions had already
been given by J.L.Lagrange \cite{Lag} when dealing with linear
problems.

With the advent of Relativity. T.Levi-Civita \cite{TLC} considered a
similar problem when looking for a Lagrangian description of
massless particles in General Relativity. P.Bergmann also noticed,
in his famous book on Relativity \cite{Ber}, that, when the
Lagrangian function is itself a constant of the motion, as it
happens, e.g., for geodesic motions in General Relativity, then
any function of the Lagrangian can be shown to provide, under very
mild assumptions, a possible alternative Lagrangian description of
the same dynamical system.

Other motivations for interest in the same problem arose from the
so-called "no-interaction theorem" \cite{BMSL,CJS,MMSL} concerning the
covariant canonical description of relativistic interacting
particles \cite{BMMNSSZ}. Here too alternative Lagrangian
descriptions were sought that could allow to evade the theorem
\cite{CS}. The so-called "quadratic Hamiltonian theorem"
\cite{CS2} was also considered in the same spirit.

A complete mathematical investigation of the inverse problem was
initiated by J.Douglas \cite{Doug} (who was also one of the first
Field medalists) back \ in 1941. Many investigators considered in
particular the problem with reference to the N\"{o}ther theorem
\cite{AM} connecting symmetries and constants of the motion
\cite{MFVMR}.

A first differential-geometric formulation of the problem appeared
in the mid-Seventies \cite{MarG}. A few years later, R.M.Santilli \cite{Sant} initiated a
systematic presentation of the problem for both particles and
fields. 

The Inverse Problem arises quite naturally if one starts from the
"experimentalist's" point of view \cite{Mar10} that the trajectories
(think of the observations in a bubble-chamber experiment) are the
first raw data that are provided by the direct observation of a
dynamical evolution. It is therefore natural to start from the
trajectories to build up a vector field and, afterwards, to look for
Lagrangian and/or Hamiltonian descriptions. A first attempt in this direction
had been made by E.K.Kasner \cite{Ka} already in 1913.

As the "raw data" are usually given on some configuration space, the
first problem one is faced with are the ambiguities that are present
when trying to go from a second-order differential equation on a
configuration space to a first-order one (i.e. a vector field) on a
larger carrier space. This problem was analyzed in detail in
Ref.\cite{Mar10}.

To clearly identify and formulate the problem, it is very useful to
consider linear dynamical systems first, and to investigate the
existence of Hamiltonian descriptions from the point of view of
Poisson brackets.

In this context, writing the equations of motion in Hamiltonian form, i.e.:%

\begin{equation}
\left\vert
\begin{array}
[c]{c}%
\frac{dq^{i}}{dt}\\
\frac{dp_{i}}{dt}%
\end{array}
\right\vert =\left\vert
\begin{array}
[c]{cc}%
\mathbf{0}_{n\times n} & \mathbf{1}_{n\times n}\\
\mathbf{-1}_{n\times n} & \mathbf{0}_{n\times n}%
\end{array}
\right\vert \left\vert
\begin{array}
[c]{c}%
\frac{\partial H}{\partial q^{i}}\\
\frac{\partial H}{\partial p_{i}}%
\end{array}
\right\vert
\end{equation}
or, in collective coordinates:%
\begin{equation}
\overset{\mathbf{.}}{\xi}^{i}=\Lambda^{ij}\frac{\partial H}{\partial\xi^{j}%
}=A^{i}\text{ }_{j}\xi^{j}%
\end{equation}
amounts to looking for a decomposition \cite{GMR} of the matrix
representing the (linear) dynamics, say $A$, into the product of a
skew-symmetric matrix $\Lambda$, which stays for the Poisson tensor and
defines the Poisson brackets and, if it is non-singular, the
symplectic structure, and of a symmetric
matrix $H$ which represents the Hamiltonian, i.e.:%
\begin{equation}
A=\Lambda\cdot H
\end{equation}
Out of all possible such decompositions we obtain all the
alternative quadratic Hamiltonian descriptions for a given dynamical system.
It is easy to realize (see below, Chapt.\ref{ch:3}) that all
symmetries for $A$, once applied to the factorization, will take
from one factorization to another one unless they correspond to
canonical transformations.

When going from a linear vector space to a generic differentiable
manifold, matrices should be replaced by tensor fields and, when
"moving from a point" to a neighboring one, we will have to take
into account also differential relations (partial differential
equations will arise in addition to algebraic relations).

One may trace the existence of alternative Lagrangian and/or
Hamiltonian descriptions to the existence of a large group of
symmetries for the dynamics, some of them being non-canonical
symmetries.

The most obvious transformation taking one Lagrangian into another
one is a scale transformation. For instance, we might scale the mass
in a Lagrangian containing only a kinetic term, or we could do that,
thanks to the equivalence principle \cite{Ber}, for a massive
particle moving in a gravitational field.

When moving to the quantum descriptions, it becomes already clear
that the scaling of the Lagrangian will give rise to a selection of
the "allowed" periodic motion within a Bohr-Sommerfeld quantization
scheme which will depend on the scale. This is not surprising, as
the Lagrangian times the period is measured in units of Planck's
constant.

This observation shows that we should not expect that the quantum
description of a dynamical evolution would trivially exhibit
properties similar to the classical ones.

On the other hand, there is a strong belief that Classical Mechanics
should be a suitable limit of Quantum Mechanics. To quote from
Dirac's book \cite{Dir3}:

\textit{\textquotedblleft Classical mechanics must be a limiting
case of quantum mechanics. We should thus expect to find that
important concepts in classical mechanics correspond to important
concepts in quantum mechanics and, from an understanding of the
general nature of the analogy between classical and quantum
mechanics, we may hope to get laws and theorems in quantum mechanics
appearing as simple generalizations of well known results in
classical mechanics.\textquotedblright}

This, along with the existence of alternative Hamiltonian
descriptions for solitonic equations \cite{Mar18}, strongly
suggests that a proper formulation of bi-Hamiltonian descriptions
should exist for quantum dynamical systems as well.

Here one can be more or less demanding. For instance, one may
require that known situations of bi-Hamiltonian descriptions of
specific classical dynamical systems be fully recovered in the
quantum framework. As we shall see, these requirements may have
far-reaching consequences in the acceptable formulations of Quantum
Mechanics.

For instance, one of the fundamental principles of Quantum Mechanics
as formulated by Dirac \cite{Dir3} is the existence of a
superposition rule for wave functions in order to deal with
interference phenomena. This is usually translated into the
requirement \cite{Dir3} that the carrier space should be a vector
space.

On the other hand, the approach in terms of $C^{\ast}$-algebras
shows clearly that the Hilbert space we arrive at with the $GNS$
construction \cite{Ha} depends on the initial state we choose, which
is obviously "prepared", so-to-speak, "in the laboratory".

A spin-off of this construction is also the need for a clear
distinction between the "abstract" $C^{\ast}$-algebra and its
specific realizations in terms of operators acting on the Hilbert
space that results from the $GNS$ construction.

Considering next more closely the Dirac prescription of  replacing
Poisson brackets with commutator brackets, one finds that, while in
the classical case all possible Poisson brackets generate
derivations for the pointwise product of functions on the carrier
space (i.e. the classical observables), in the quantum setting
another result \ by Dirac (see Chapt. $IV$ of Ref.\cite{Dir3}) shows
that the associative product of operators identifies completely (up
to a scale factor) the associated Lie algebra structure (the
commutator brackets). In some sense, therefore, the associative
product and the Lie product strongly determine each other in the
quantum case.

Many of these issues will be closely scrutinized in the present
Report, which has been organized in the following way.

The remainder of this Chapter and Chapt.$2$ \ serve to, so-to-speak,
"set the stage" for the analysis of the following Chapters,
discussing, to begin with, how the Schr\"{o}dinger equation can be
recast in the form of a Hamiltonian system, both in the finite and
the infinite-dimensional case, and how alternative Hamiltonian
descriptions of the same quantum system can be generated. As
bi-Hamiltonian systems are usually associated with complete
integrability \cite{Das,DSVM2,Ma}, Chapt.$2$ reviews some general
problems concerning complete (Liouville) integrability and related
invariant structures. In Chapt.$3$ we discuss the existence of
alternative structures at the classical level starting, as
anticipated in these \ introductory notes, with a discussion of the
case of linear vector fields. Chapt.$4$ moves to the quantum
setting. Also in order to set the problem within a framework similar
to that of the classical case, and to take into account the fact
that pure states in Quantum Mechanics are a manifold rather than a
vector space,  we begin with a discussion of how geometric
(tensorial) structures that are somehow hidden by the linear vector
space structure of the Hilbert space emerge nonetheless as
fundamental structures. We emphasize there how the proper carrier
space for quantum dynamical system is instead the (no more linear)
complex projective space associated with the Hilbert space. We
conclude by discussing here too possible bi-Hamiltonian descriptions
of quantum systems and with a brief account of the extensions of the
concepts developed along the Chapter to the infinite-dimensional
case. In Chapt.$5$ we discuss the Wigner-Weyl approach to Quantum
Mechanics, beginning with a review of the Weyl map, illustrated also
with a good number of examples, we continue with the Wigner map, the
Moyal product, Quantum Mechanics in phase space and we discuss also
the quantum-classical transition. In the following Chapt.$6$
 we discuss how one can induce either on the same space or on
spaces that are diffeomorphically related alternative linear
structures, i.e. linear structures on the same carrier space that
are however not linearly related. We discuss how alternative linear
structures can offer a way of "reformulating", in a sense explained
in the text, the von Neumann uniqueness theorem \cite{Neu1}, as well
as their r\^{o}le in Statistical Mechanics.  Chapt.$7$ contains some further generalizations and our concluding remarks.

In order to make the paper more readable, some technical matters
have been discussed in details in the Appendices, that expert
readers can of course skip reading.

\subsection{The Schr\"{o}dinger Equation as a (Classical) Dynamical System}\label{SE}

\subsubsection{The Finite-Dimensional case}\label{SE1}
\bigskip

We begin by considering the Schr\"{o}dinger equation:%
\begin{equation}
\frac{d}{dt}\psi\left(  t\right)  =-\frac{i}{\hbar}H\psi\left(  t\right)
;\text{ \ }\psi\left(  0\right)  =\psi\label{Schroedinger}%
\end{equation}
on a \textit{finite-dimensional} (complex) Hilbert space $\mathcal{H}$,
deferring the discussion of some infinite-dimensional examples to the end of
this Chapter. Hence, for the time being: $\mathcal{H}\approx\mathbb{C}^{n}$
for some $n$, As $\mathcal{H}$ is a vector \ space, there is a natural
identification of the tangent space at any point $\psi\in\mathcal{H}$ with
$\mathcal{H}$ \ itself: $T_{\psi}\mathcal{H\approx H}$ . In other words,
\textit{vectors in a Hilbert space play\footnote{As in any linear vector
space.} a double r\^{o}le}, as "points" in the space and as tangent vectors at
a given point. Which r\^{o}le they play should be (hopefully) clear from the
context. More generally, we have the identification: $T\mathcal{H}%
\approx\mathcal{H}\times\mathcal{H}$, with $T\mathcal{H}$ the tangent bundle
of $\mathcal{H}$.

As in the case of differentiable manifolds, $\psi=\psi\left(  t\right)
,\psi\left(  0\right)  =\psi$ will define a curve in $\mathcal{H}$, and hence
the quantity $\left(  d\psi\left(  t\right)  /dt\right)  |_{t=0}$ will define
the tangent vector at the curve at $\psi\in\mathcal{H}$. \ A smooth assignment
of tangent vectors at every point $\psi\in\mathcal{H}$ will define then a
\textit{vector field}, i.e. a smooth (and
global) section \ of \ $T\mathcal{H}$:%
\begin{equation}
\Gamma:\mathcal{H}\rightarrow T\mathcal{H};\text{ }\psi\mapsto\left(
\psi,\phi\right)  ,\psi\in\mathcal{H},\text{ }\phi\in T_{\psi}\mathcal{H}%
\approx\mathcal{H} \label{gamma1}%
\end{equation}
where the second argument may depend in a smooth way on $\psi$ and with the
tangent bundle projection:
\begin{equation}
\pi:\left(  \psi,\phi\right)  \mapsto\psi
\end{equation}
such that: $\pi\circ\Gamma=Id_{\mathcal{H}}$.\ We will employ \ the notation:
$\Gamma\left(  \psi\right)  $ for the
vector field evaluated at the point
$\psi$ with tangent vector at $\psi$ given by Eqn.(\ref{gamma1}). The latter
defines \ a flow on $\mathcal{H}$ determined by the differential equation:%
\begin{equation}
\frac{d}{dt}\psi\left(  t\right)  =\phi\left(  \psi\left(  t\right)  \right)
,\text{ }\psi\left(  0\right)  =\psi\label{flow1}%
\end{equation}

Every vector field will define a derivation on the algebra of functions just
as in the case of real manifolds. Specifically, if: $\phi=\left(  d\psi\left(
t\right)  /dt\right)  |_{t=0}$, $\psi\left(  0\right)  =\psi$ and:
$f:\mathcal{H}\mapsto\mathbb{R}$ is a function, then, in intrinsic terms:%
\begin{equation}
(\mathcal{L}_{\Gamma}\left(  f\right)  )\left(  \psi\right)  =\frac{d}%
{dt}f\left(  \psi\left(  t\right)  \right)  \left\vert _{t=0}\right.
\end{equation}
will define the Lie derivative along ${\Gamma}$ on the algebra of functions.

In local coordinates, choosing, e.g., an orthonormal ($O.N.$ from
now on) basis $\left\{ e_{i}\right\} _{i}^{n}$ $\left(
n=\dim\mathcal{H}\right) $, vectors (and tangent vectors) will be
represented by $n$-tuples of complex numbers ($\psi=\left(  \psi
^{1},...,\psi^{n}\right),\psi^{j}=:\left\langle e_{j}|\psi\right\rangle $ and so on), and\footnote{As $\psi_{j}$ is complex: $\psi_{j}=q_{j}+ip_{j}$, $q_{j},p_{j}%
\in\mathbb{R},$ the derivative here has to be understood simply as:
$\partial/\partial\psi_{j}=\partial/\partial q_{j}-i\partial/\partial p_{j}$.}:

\begin{equation}
(\mathcal{L}_{\Gamma}\left(  f\right)  )\left(  \psi\right)  =\phi^{i}\left(
\psi\right)  \frac{\partial f}{\partial\psi^{i}}\left(  \psi\right)
\end{equation}

Notice that, in the infinite-dimensional case (for a separable and
infinite-dimensional Hilbert space), "functions" will become
\textit{functionals}, and ordinary derivatives will have to be replaced by
properly defined functional derivatives.

\textit{Constant} as well as \textit{linear} (with respect to the
linear structure identified by the vector space) vector fields will
play a role in what follows. The former are characterized by:
$\phi=const.$ in the second
argument of Eqn.(\ref{gamma1}), and give rise to the one-parameter group:%
\begin{equation}
\mathbb{R}\ni t\mapsto\psi\left(  t\right)  =\psi+t\phi
\end{equation}
The latter are characterized instead by $\phi\left(  \psi\right)  $ being a
linear and homogeneous function of $\psi$, i.e.: $\phi=A\psi$ for some linear
operator $A$. Eqn.(\ref{flow1}) integrates in this case to\footnote{in the
finite-dimensional case there are of course no problems in exponentiating a
linear operator.}:%
\begin{equation}
\psi\left(  t\right)  =\exp\left\{  tA\right\}  \psi\label{flow2} %
\end{equation}
Of particular interest is the \textit{dilation vector field} $\Delta$:%
\begin{equation}
\Delta:\psi\mapsto\left(  \psi,\psi\right)  \label{delta}%
\end{equation}
which corresponds to: $A=Id_{\mathcal{H}}$. In this case Eqns.(\ref{flow1})
and (\ref{flow2}) become:%
\begin{equation}
\frac{d}{dt}\psi\left(  t\right)  =\psi\Rightarrow\psi\left(  t\right)
=e^{t}\psi
\end{equation}
Eqn.(\ref{delta}) exhibits clearly the fact that the dilation field
leads to an identification of $\mathcal{H}$ with the fiber
$T_{\psi}\mathcal{H}$. The latter carrying a natural linear
structure, Eqn.(\ref{delta}) provides a tensorial characterization
of the linear structure of the base space $\mathcal{H}$ by means of
the vector field $\Delta$. For more details, see, e.g.,
Ref.\cite{DLMV}.
\bigskip

With  every linear operator\footnote{Not considering questions of domain, which are
of no relevance in the finite-dimensional case.} \ $\mathbb{A}$ there is therefore associated the
linear vector field:%
\begin{equation}
\mathbb{X}_{\mathbb{A}}:\mathcal{H}\rightarrow T\mathcal{H};\text{ }
\psi  \rightarrow\left( \psi,\mathbb{A}\psi\right)
\end{equation}
In local coordinates, this vector field can be written as:%
\begin{equation}
\mathbb{X}_{\mathbb{A}}=:A^{i}\text{ }_{j}\psi^{j}\frac{\partial}{\partial
\psi^{i}}%
\end{equation}
and is of course entirely defined by the representative matrix: $\mathbb{A=}%
\left\Vert A^{i}\text{ }_{j}\right\Vert $ of the linear operator. In
particular, then:%
\begin{equation}
\Delta=\psi^{i}\frac{\partial}{\partial\psi^{i}}%
\end{equation}

Notice however that, while linear operators form an associative algebra,
vector fields do not : they form instead only a Lie algebra. An associative algebra
can be recovered by using the same matrix $\mathbb{A}$ to define instead the
$\left(  1,1\right)  $ tensor\footnote{Notice that, while $\mathbb{X}%
_{\mathbb{A}}$ depends on the choice of the origin of the coordinates,
$\mathbb{T}_{\mathbb{A}}$ does not, i.e. it has an affine character.}:%
\begin{equation}
\mathbb{T}_{\mathbb{A}}=:\text{ }A^{i}\text{ }_{j}d\psi^{j}\otimes
\frac{\partial}{\partial\psi^{i}}%
\end{equation}
Then it is easy to check that the vector field $\mathbb{X}_{\mathbb{A}}$ is
recovered from $\mathbb{T}_{\mathbb{A}}$ and the dilation field as:%
\begin{equation}
\mathbb{X}_{\mathbb{A}}=\mathbb{T}_{\mathbb{A}}\left(  \Delta\right)
\end{equation}

\bigskip

Coming back to the Schr\"{o}dinger equation, the linear operator
$H$ \ will
define a linear vector field that we will denote\footnote{We use here the notation $\Gamma_{H}$ instead of $\mathbb{X}_{H}$ as a reminder of the fact that we had to include the "extra" factor $(-i/\hbar)$ in its definition.} for short as $\Gamma_{H}$:%
\begin{equation}
\Gamma_{H}:\mathcal{H}\rightarrow T\mathcal{H};\text{ \ }\Gamma_{H}%
:\psi\mapsto\left(  \psi,-\left(  i/\hbar\right)  H\psi\right)
\label{vectorfield1}%
\end{equation}
 and then:%
\begin{equation}
\mathcal{L}_{\Gamma_{H}}\psi\equiv\frac{d}{dt}\psi=-\frac{i}{\hbar}H\psi
\end{equation}
In this sense, the Schr\"{o}dinger equation (\ref{Schroedinger}) can
be viewed as a classical evolution equation on a complex vector
space.

At variance with the infinite-dimensional case, every linear vector
field is complete in finite dimensions. Then, if in addition we
require conservation of probability, Wigner's theorem \cite{Wig} \
states that the associated one-parameter group has to be
unitary\footnote{To be a bit more precise, \textit{pure states} in
Quantum Mechanics are described by elements of the projective
Hilbert space $P\mathcal{H}$ (for instance, one-dimensional
projectors of the form:
$P_{\psi}=|\psi\rangle\langle\psi|/\left\langle \psi|\psi
\right\rangle $, $|\psi\rangle\in\mathcal{H}$. The Hermitian
structure on $\mathcal{H}$ induces a binary product: $\left\langle
.,.\right\rangle $ on $P\mathcal{H}$ via: $\left\langle
P_{\psi},P_{\phi}\right\rangle =:Tr\left\{ P_{\psi}P_{\phi}\right\}
=\left\vert \left\langle \phi|\psi\right\rangle \right\vert
^{2}/\left(  \left\langle \phi|\phi\right\rangle \left\langle
\psi|\psi\right\rangle \right) $ and yields a transition
probability. Wigner's theorem states then that any bijective map on
$P\mathcal{H}$ preserving transition probabilities can be realized
as a unitary or anti-unitary transformation on the original Hilbert
space.} and, by Stone-von Neumann's theorem \cite{RS}, $H$ has to be
essentially self-adjoint, i.e. it will be symmetric with a unique
self-adjoint extension. In the sequel we will refer always to the
latter, and will simply say that $H$ is self-adjoint. In the
finite-dimensional case no distinctions between Hermitian, symmetric
and self-adjoint operators \cite{RS} need to be made, of course.

Let now:%
\begin{equation}
h:\mathcal{H\times H\rightarrow}\mathbb{C} \label{scalar1}%
\end{equation}
be a Hermitian structure on $\mathcal{H}$, i.e. let:%
\begin{equation}
h\left(  \phi,\psi\right)  =:\left\langle \phi|\psi\right\rangle
\label{scalar2}%
\end{equation}
define an Hermitian scalar product on $\mathcal{H}$ with the usual properties, namely;

\begin{itemize}
\item $h\left(  \phi,\psi\right)  =\overline{h\left(  \psi,\phi\right)  }$

\item $h\left(  \phi,\phi\right)  \geq0,$ \ $h\left(  \phi,\phi\right)
=0\leftrightarrow\phi=0$

\item $h\left(  \lambda\phi,\psi\right)  =\overline{\lambda}h\left(  \phi
,\psi\right)  ,$ \ $h\left(  \phi,\lambda\psi\right)  =\lambda h\left(
\phi,\psi\right)  $
\end{itemize}

\bigskip

\begin{remark}
If \ $h$ \ is viewed more properly as a $\left(  0,2\right)  $ tensor
field, then $\phi$ and $\psi$ in Eqn.(\ref{scalar2}) have to be viewed as
tangent vectors at a point in $\mathcal{H}$, and a more complete (albeit a bit
more cumbersome) notation should be:%
\begin{equation}
h\left(  \varphi\right)  \left(  \Gamma_{\phi}\left(  \varphi\right)
,\Gamma_{\psi}\left(  \varphi\right)  \right)  =\left\langle \phi
|\psi\right\rangle \label{herm0}
\end{equation}
where $h\left(  \varphi\right)  $ stands for $h$ evaluated at point $\varphi
\in\mathcal{H}$. As the r.h.s. of this equation does not depend on $\varphi$,
this implies : $\mathcal{L}_{\Gamma_{H}}\left\langle \phi|\psi\right\rangle
\equiv\mathcal{L}_{\Gamma_{H}}(h\left(  \phi,\psi\right)  )=0$ and, using
Eqn.(\ref{Schroedinger}):%
\begin{eqnarray}
0&=&\mathcal{L}_{\Gamma_{H}}(h\left(  \phi,\psi\right)  )=\left(  \mathcal{L}%
_{\Gamma_{H}}h\right)  \left(  \phi,\psi\right)  +h\left(  \mathcal{L}%
_{\Gamma_{H}}\phi,\psi\right)  +h\left(  \phi,\mathcal{L}_{\Gamma_{H}}%
\psi\right)  =\nonumber\\
&=&\left(  \mathcal{L}_{\Gamma_{H}}h\right)  \left(  \phi,\psi\right)  +\frac
{i}{\hbar}\left\{  \left\langle H\phi|\psi\right\rangle -\left\langle
\phi|H\psi\right\rangle \right\}
\end{eqnarray}
which implies in turn, as $H$ is self-adjoint, that:%
\begin{equation}
\mathcal{L}_{\Gamma_{H}}h=0 \label{invariance}%
\end{equation}
i.e. that the Hermitian structure be \textit{invariant} under the (unitary)
flow of $\Gamma_{H}$ (and viceversa), or, stated equivalently, that
$\Gamma_{H}$ be a \textit{Killing vector field} for the Hermitian structure.
If \ instead the Hermitian structure is not invariant, then $H$ will fail to
be self-adjoint \ w.r.t. the given Hermitian structure.
\end{remark}
\begin{remark}
A family of privileged (actually global) charts for $\mathcal{H}$, all
unitarily related to each other, is provided by the choice of any $O.N.$ basis
$\left\{  \left.  |k\right\rangle \right\}  _{i}^{n},\left\langle
h|k\right\rangle =\delta_{hk}$. In any such basis: $h\left(  \phi,\psi\right)
=:\left\langle \phi|\psi\right\rangle =h_{ij}\overline{\phi^{i}}\psi^{j}$
with: $h_{ij}=\delta_{ij}$, and all the above statements (in particular
Eqn.(\ref{invariance})) are self-evident. However, the statements of the
previous Remark have a tensorial meaning. As such, they will remain true also
under (possible) non-linear changes of coordinates.
\end{remark}
\begin{remark}
We can decompose the Hermitian structure into real and imaginary parts
as:%
\begin{equation}
h\left(  .,.\right)  =g\left(  .,.\right)  +i\omega\left(  .,.\right)
\end{equation}
where:%
\begin{equation}
g\left(  \phi,\psi\right)  =\frac{1}{2}\left[  \left\langle \phi
|\psi\right\rangle +\left\langle \psi|\phi\right\rangle \right]
\label{metr}%
\end{equation}
and:%
\begin{equation}
\omega\left(  \phi,\psi\right)  =\frac{1}{2i}\left[  \left\langle \phi
|\psi\right\rangle -\left\langle \psi|\phi\right\rangle \right]
\label{symplectic}%
\end{equation}
According to Eqn.(\ref{herm0}) we may consider $h$ as an Hermitian
tensor. It is clear that \ both $g$ and $\omega$ are $\left(
0,2\right) $ tensors, and that $g$ is symmetric, while $\omega$ is
skew-symmetric, hence a two-form. \ Eqn.(\ref{invariance}) implies
then that both tensors are (separately) invariant under
$\Gamma_{H}$. Notice that: $\omega\left(  \phi,i\psi\right) =g\left(
\phi,\psi\right)  $. Hence, non-degeneracy of $h$ entails separately
that of $\omega$ and of $g$. \
\end{remark}
\begin{remark}
The non-degenerate two-form $\omega$ will be represented, in any one of
the privileged charts, by a constant (and unitarily invariant) matrix. Hence
it will be closed:
\begin{equation}
d\omega=0 \label{symplectic2}%
\end{equation}
But, again, we stress that an equation like Eqn.(\ref{symplectic2}) has a
tensorial meaning. Hence, $\omega$ will be a \textit{symplectic form}, while
$g$ will be a ( non-degenerate and constant in any privileged chart)
\textit{metric tensor.}
\end{remark}

\bigskip Let now $\Gamma_{H}$ be a vector field of the form
(\ref{vectorfield1}). Then, a little algebra shows that:%
\begin{equation}
\left(  i_{\Gamma_{H}}\omega\right)  \left(  \psi\right)  =\omega\left(
-\frac{i}{\hbar}H\phi,\psi\right)  =\frac{1}{2\hbar}\left[  \left\langle
H\phi|\psi\right\rangle +\left\langle \psi|H\phi\right\rangle \right]
\end{equation}
On the other hand, if we define the quadratic function:
\begin{equation}
f_{H}\left(  \phi\right)  =\frac{1}{2\hbar}\left\langle \phi|H\phi
\right\rangle
\end{equation}
we can define its differential as the one-form:%
\begin{equation}
df_{H}\left(  \phi\right)  =\frac{1}{2}\left[  \left\langle .|H\phi
\right\rangle +\left\langle \phi|H.\right\rangle \right]  =\frac{1}{2}\left[
\left\langle .|H\phi\right\rangle +\left\langle H\phi|.\right\rangle \right]
\end{equation}
the last passage following from $H$ being self-adjoint. Therefore: $\left(
i_{\Gamma_{H}}\omega\right)  \left(  \psi\right)  =df_{H}\left(  \phi\right)
\left(  \psi\right)  \forall\psi$, and hence:%
\begin{equation}
i_{\Gamma_{H}}\omega=df_{H}%
\end{equation}
i.e. $\Gamma_{H}$ is \textit{Hamiltonian} w.r.t. the symplectic structure with
the quadratic Hamiltonian $f_{H}$.

As a further remark, we recall that $\mathcal{H}$ \ is endowed with a natural
complex structure $J$ \ defined simply by :%
\begin{equation}
J:\phi\rightarrow i\phi
\end{equation}
Then: $J^{2}=-\mathbb{I}$ \ (the identity on $\mathcal{H}$) and:%
\begin{equation}
\omega\left(  \phi,J\psi\right)  =g\left(  \phi,\psi\right)  \label{compl1}%
\end{equation}
Therefore the complex structure $J$ is
\textit{compatible}\cite{Mar7} with the pair
$\left(  g,\omega\right)  $ and we can reconstruct the Hermitian structure as:%
\begin{equation}
h\left(  \phi,\psi\right)  =\omega\left(  \phi,J\psi\right)  +i\omega\left(
\phi,\psi\right)\label{compl2}
\end{equation}
or equivalently, as:%
\begin{equation}
h\left(  \phi,\psi\right)  =g\left(  \phi,\psi\right)  -ig\left(  \phi
,J\psi\right)
\end{equation}

Notice also that:%
\begin{equation}
\omega\left(  J\phi,J\psi\right)  =\omega\left(  \phi,\psi\right)
\end{equation}
as well as:%
\begin{equation}
g\left(  J\phi,J\psi\right)  =g\left(  \phi,\psi\right)
\end{equation}

We can summarize what has been proved up to now by saying that
$\mathcal{H}$ \ is a K\"{a}hler manifold\cite{Ch,CM,Weil}, and that
$h$ is the associated Hermitian metric, while $g$ is the Riemannian
metric and $\omega$ the fundamental two-form. As $\omega$ is closed,
$g$ is also \cite{Weil} a K\"{a}hler metric.

Choosing\footnote{Of course the best choice would be a basis in which the
Hamiltonian is diagonal.} an $O.N.$ basis $\left\{  |k\rangle\right\}
_{1}^{n}$, $\left\langle h|k\right\rangle =\delta_{hk}$, the Hermitian product
can be written as:%
\begin{equation}
h\left(  \phi,\psi\right)  =\delta_{ij}\overline{\phi}^{i}\psi^{j}
\label{scalar3}%
\end{equation}
where: $|\phi\rangle=\phi^{k}|k\rangle$, and similarly for $\psi$.

Writing: $\phi=\phi_{1}+i\phi_{2},$ $\phi_{1,2}\in\mathbb{R}^{n}$, we can
\textit{realify} \cite{Ar,Ger} $\ \mathbb{C}^{n}$ to $\mathbb{R}^{2n}$ via:%
\begin{equation}
\mathbb{C}^{n}\ni\phi\rightarrow\left\vert
\begin{array}
[c]{c}%
\phi_{1}\\
\phi_{2}%
\end{array}
\right\vert
\end{equation}
In this way:
\begin{equation}
g\left(  \phi,\psi\right)  =\operatorname{Re}\left\{  \delta_{ij}%
\overline{\phi}^{i}\psi^{j}\right\}  =\left\vert
\begin{array}
[c]{cc}%
\phi_{1} & \phi_{2}%
\end{array}
\right\vert G\left\vert
\begin{array}
[c]{c}%
\psi_{1}\\
\psi_{2}%
\end{array}
\right\vert
\end{equation}
where $G$ is the matrix:%
\begin{equation}
G=\mathbb{I}_{2n}\equiv\left\vert
\begin{array}
[c]{cc}%
\mathbb{I}_{n} & \mathbf{0}_{n}\\
\mathbf{0}_{n} & \mathbb{I}_{n}%
\end{array}
\right\vert
\end{equation}
the $\mathbb{I}$'s being the identity matrices. Quite similarly, we find that
$\omega$ has the representative matrix $\Omega$ given by:%
\begin{equation}
\Omega=\left\vert
\begin{array}
[c]{cc}%
\mathbf{0}_{n} & \mathbb{I}_{n}\\
\mathbf{-}\mathbb{I}_{n} & \mathbf{0}_{n}%
\end{array}
\right\vert
\end{equation}
in $\mathbb{R}^{2n}$, and $J$ is represented by the matrix:%
\begin{equation}
J=\left\vert
\begin{array}
[c]{cc}%
\mathbf{0}_{n} & -\mathbb{I}_{n}\\
\mathbb{I}_{n} & \mathbf{0}_{n}%
\end{array}
\right\vert =-\Omega=\Omega^{-1}%
\end{equation}
consistently with Eqn.(\ref{compl1}) which implies,in terms of the
representative matrices:%
\begin{equation}
J=\Omega^{-1}G \label{complex2}%
\end{equation}
Notice, however, that while $G$ and $\Omega$ are representatives of $\left(
0,2\right)  $ tensors, $J$ \textit{is the representative of a} $\left(
1,1\right)  $ \textit{tensor. }Explicitly, denoting with $\left\Vert
\Omega^{ij}\right\Vert $ the inverse of $\Omega$ (i.e. a $\left(  2,0\right)
$ tensor):%
\begin{equation}
\Omega^{ij}\Omega_{jk}=\delta^{i}\text{ }_{k}%
\end{equation}
then:%
\begin{equation}
J^{i}\text{ }_{j}=\Omega^{ik}G_{kj}%
\end{equation}

\bigskip

Let us turn now to the Schr\"{o}dinger equation
(\ref{Schroedinger}). Written in components, it reads\footnote{It
is clear that the matrix elements of the Hamiltonian have
to be viewed as those of a $\left(  1,1\right)  $ tensor.}:%
\begin{equation}
\frac{d}{dt}\psi^{h}=-\frac{i}{\hbar}\left\langle h|H|k\right\rangle \psi^{k}%
\end{equation}

Writing then, as before, $\psi=\psi_{1}+i\psi_{2},\psi_{1,2}\in\mathbb{R}^{n}$
and introducing the real column vector:%
\begin{equation}
\left\vert
\begin{array}
[c]{c}%
\psi_{1}\\
\psi_{2}%
\end{array}
\right\vert \in\mathbb{R}^{2n}%
\end{equation}
we find (separating real and imaginary parts) the equation:
\begin{equation}
\frac{d}{dt}\left\vert
\begin{array}
[c]{c}%
\psi_{1}\\
\psi_{2}%
\end{array}
\right\vert =A\left\vert
\begin{array}
[c]{c}%
\psi_{1}\\
\psi_{2}%
\end{array}
\right\vert
\end{equation}
where $A$ is the skew-symmetric matrix:%
\begin{equation}
A=:\frac{1}{\hbar}\left\vert
\begin{array}
[c]{cc}%
\operatorname{Im}H & \operatorname{Re}H\\
-\operatorname{Re}H & \operatorname{Im}H
\end{array}
\right\vert
\end{equation}
and $\operatorname{Im}H$ and $\operatorname{Re}H$ are the $n\times n$
matrices:%
\begin{equation}
(\operatorname{Im}H)^{h}\text{ }_{k}=\operatorname{Im}\left\langle
h|H|k\right\rangle ,\text{ }\left(  \operatorname{Re}H\right)  ^{h}\text{
}_{k}=\operatorname{Re}\left\langle h|H|k\right\rangle
\end{equation}
Just as before, $\operatorname{Im}H$ will be skew-symmetric and
$\operatorname{Re}H$ symmetric.

\bigskip

\begin{remark}
If we write the representative matrix of the Hamiltonian
as: $H=\operatorname{Re}H+i\operatorname{Im}H$, then the "realified" version
of it is \cite{Ar} the symmetric matrix:
\begin{equation}
^{R}H=\left\vert
\begin{array}
[c]{cc}%
\operatorname{Re}H & -\operatorname{Im}H\\
\operatorname{Im}H & \operatorname{Re}H
\end{array}
\right\vert
\end{equation}
Then it is easy to check that:
\begin{equation}
A=-J\circ(^{R}H/\hbar)
\end{equation}
This completes the identification of the Schr\"{o}dinger equation as a real
dynamical system on a real space of dimension $2n$.
\end{remark}

\bigskip

Taking a further time derivative, we obtain:%
\begin{equation}
\frac{d^{2}}{dt^{2}}\left\vert
\begin{array}
[c]{c}%
\psi_{1}\\
\psi_{2}%
\end{array}
\right\vert =A^{2}\left\vert
\begin{array}
[c]{c}%
\psi_{1}\\
\psi_{2}%
\end{array}
\right\vert
\end{equation}
and a simple calculation shows that:%
\begin{equation}
A^{2}=-\left(  \frac{^{R}H}{\hbar}\right)  ^{2}%
\end{equation}

\bigskip

Actually this result follows simply from the fact that the complex structure
and the realified form of $H$ commute, i.e.:

\bigskip%

\begin{equation}
J\circ^{R}H=^{R}H\circ J
\end{equation}
and from: $J^{2}=-\mathbb{I}$.

As already remarked, things simplify if the basis in $\mathbb{C}^{n}$ is
chosen as the basis of the eigenvectors of $H$ itself: $H|k\rangle
=E_{k}|k\rangle$. Then it is immediate to see that:%
\begin{equation}
A=\frac{1}{\hbar}\left\vert
\begin{array}
[c]{cc}%
\mathbf{0} & H\\
-H & \mathbf{0}%
\end{array}
\right\vert
\end{equation}
where $H$ is now the diagonal \ $n\times n$ matrix:
\begin{equation}
H=diag\left\{  E_{1},...,E_{n}\right\}
\end{equation}
Then we obtain the equations of motion:%
\begin{equation}
\frac{d}{dt}\psi_{1}=H\psi_{2}\text{ },\text{ \ }\frac{d}{dt}\psi_{2}%
=-H\psi_{1}%
\end{equation}
or:%
\begin{equation}
\frac{d^{2}}{dt^{2}}\psi_{i}+\left(  \frac{H}{\hbar}\right)  ^{2}\psi
_{i}=0\text{ },\text{ }i=1,2
\end{equation}
Explicitly:%
\begin{equation}
\frac{d^{2}}{dt^{2}}\psi_{i}^{k}+\left(  \frac{E_{k}}{\hbar}\right)  ^{2}%
\psi_{i}^{k}=0\text{ },k=1,...,n,\text{ }i=1,2
\end{equation}
i.e. in this basis each one of the components of the real vectors
$\psi_{1}$ and $\psi_{2}$ behaves as a simple harmonic oscillator
with frequency $\nu_{k}=E_{k}/\hbar$.

\subsubsection{Alternative Schrodinger and Heisenberg
descriptions via modified Hermitian structures}\label{Alt1}
\bigskip

Let now $K$ \ be a (strictly) \textit{positive }\ linear operator on $\mathcal{H}%
$, and consider the bilinear (sesquilinear) functional:%
\begin{equation}
\left\langle \phi|K\psi\right\rangle \equiv h\left(
\phi,K\psi\right)
,\text{ }\phi,\psi\in T\mathcal{H} \label{newscalar}%
\end{equation}
It is immediate to check that \ this functional enjoys all the
three properties listed after Eqn.(\ref{scalar2}). Hence it
defines a \textit{new} Hermitian structure that we will denote as
$h_{K}\left(  .,.\right)  $ or as:
$\left\langle .|.\right\rangle _{K}$:%
\begin{equation}
h\left(  \phi,K\psi\right)  =:h_{K}\left(  \phi,\psi\right)
=:\left\langle \phi|\psi\right\rangle _{K}\label{sesqui}%
\end{equation}

\bigskip

It is easy to show now that, as a consequence of the Hermiticity of $H$:%
\begin{equation}
\mathcal{L}_{\Gamma_{H}}\left(  h_{K}\left(  \phi,\psi\right)
\right) =\frac{i}{\hbar}h\left(  \phi,\left[  H,K\right]
\psi\right)
\end{equation}

Invariance of the new Hermitian structure w.r.t. the dynamics
requires then
that $K$ be a "constant of the motion" for $H$:%
\begin{equation}
\left[  H,K\right]  =0\label{comhk}
\end{equation}
$h_{K}$ will now be given explicitly as: $h_{K}\left(
\phi,\psi\right) =(h_{K})_{ij}\overline{\phi}^{i}\psi^{j},$
$(h_{K})_{ij}=\left\langle i|K|j\right\rangle =S_{ij}+iA_{ij}$, with
$S,A$ $n\times n$ real matrices. Hermiticity implies then:
$\widetilde{S}=S$ and $\widetilde{A}=-A$, i.e. that $S$ be
symmetric and $A$ skew-symmetric. Proceeding as before, it is not
difficult to see that the new metric tensor, symplectic form and
complex structure $g_{k}$ ,$\omega_{k}$ and $J_{K}$ would be
represented in the
previous basis by the matrices:%
\begin{equation}
G_{K}=\left\vert
\begin{array}
[c]{cc}%
S & A\\
-A & S
\end{array}
\right\vert ,\text{ \ }\Omega_{K}=\left\vert
\begin{array}
[c]{cc}%
A & S\\
-S & A
\end{array}
\right\vert
\end{equation}
with $J_{K}$ being given again by Eqn.(\ref{complex2}).

The above results have been derived by considering "time" (i.e.
Hamiltonian) evolution of vectors in the Hilbert space, i.e. in
the framework of the Schr\"{o}dinger picture.

It is not hard to show that similar results can be achieved in the
context of the Heisenberg picture. Indeed, the new scalar product
(\ref{sesqui}) induces a new associative product among linear
operators, namely\footnote{See also Ref. \cite{Ru}
for the Abelian case.}:%
\begin{equation}
A,B\rightarrow A\underset{\left(  K\right)  }{\cdot}B=:AKB\label{prod2}%
\end{equation}
and a new commutator:%
\begin{equation}
\left[  A,B\right]  _{\left(  K\right)  }=:A\underset{\left(
K\right) }{\cdot}B-B\underset{\left(  K\right)  }{\cdot}A=AKB-BKA
\end{equation}
that will fulfill the Jacobi identity in view of the associativity
of the product (\ref{prod2}).

Now, if we want to represent the same dynamics in terms of the
new commutator bracket, we will have to define a new
Hamiltonian $H^{\prime}$ such that:%
\begin{equation}\label{Heis}
i\hbar\frac{dA}{dt}=\left[  H^{\prime},A\right]  _{\left(
K\right)  }=\left[ H,A\right]
\end{equation}
As $A$ is generic,this requires: $H^{\prime}K=KH^{\prime}=H$, and hence:%
\begin{equation}
H^{\prime}=HK^{-1}%
\end{equation}
as well as:%
\begin{equation}
\left[  H,K\right]  =0
\end{equation}
as before. Notice that this will ensure that "time" evolution will
be a
derivation on the new product algebra, i.e. that:%
\begin{equation}\label{Kder}
\frac{d}{dt}\left(  A\underset{\left(  K\right)  }{\cdot}B\right)
=\frac {dA}{dt}\underset{\left(  K\right)
}{\cdot}B+A\underset{\left(  K\right)
}{\cdot}\frac{dB}{dt}%
\end{equation}
for all $A,B$.

 Let us summarize at this point what \ we have found
starting from the Schr\"{o}dinger equation (\ref{Schroedinger}):

\begin{itemize}
\item Eqn.(\ref{Schroedinger}) defines a real, linear Hamiltonian vector field
on the realification of the complex (and finite-dimensional, for the time
being) Hilbert space $\mathcal{H}$.

\item On this space, Eqn.(\ref{Schroedinger}) defines a Killing vector field
for the Euclidean metric tensor associated with the real part of the Hermitian
scalar product.

\item Eqn.(\ref{Schroedinger}) decomposes into $n$ non-interacting
harmonic oscillators with proper frequencies $E_{k}/\hbar$ and is
therefore \cite{DSVM2, Ma} (see also next Chapter) a completely
integrable system. Finally:

\item Eqn.(\ref{Schroedinger}) preserves alternative Hermitian structures
associated with positive linear operators $K$ which commute with $H$.
Therefore, $\Gamma_{H}$ is also Killing for the new metric tensor and
Hamiltonian for the new symplectic structure.
\end{itemize}

\subsubsection{From Finite to Infinite Dimensions}\label{SE2}
\bigskip

We turn now to the infinite-dimensional case, concentrating on a quantum
system described, in the Schr\"{o}dinger picture, on the Hilbert space
$\mathcal{L}_{2}\left(  \mathbb{R}^{d},\mathbb{C}\right)  $, $d\geq1$, of
complex, square-integrable\footnote{With respect to the Lebesgue measure.}
functions. Defining real variables $q$ and $p$ via:%
\begin{equation}
\mathcal{L}_{2}\left(  \mathbb{R}^{d},\mathbb{C}\right)  \ni\psi\left(
\mathbf{r},t\right)  =:q\left(  \mathbf{r},t\right)  +ip\left(  \mathbf{r}%
,t\right)  ,\text{ }\mathbf{r\in}\mathbb{R}^{d}%
\end{equation}
$q$ and $p$ will be functions in $\mathcal{L}_{2}\left(  \mathbb{R}%
^{d},\mathbb{R}\right)  $\footnote{One can also identify
\cite{Mar18} $\mathcal{L}_{2}(\mathbb{R}^{d},\mathbb{C)}$ with the
cotangent bundle of $\mathcal{L}_{2}(\mathbb{R}^{d},\mathbb{R)}$.}.

With a Schr\"{o}dinger operator of the form:%
\begin{equation}
\mathcal{H=}-\frac{\hbar^{2}}{2m}\nabla^{2}+U\left(  \mathbf{r}\right)
\label{Schro1}%
\end{equation}
(with $U\left(  \mathbf{r}\right)  $ a potential), the (time-dependent)
Schr\"{o}dinger equation will be:%
\begin{equation}
i\hbar\frac{d\psi}{dt}=\mathcal{H}\psi\label{Schro2}%
\end{equation}
In a natural way, we will have to deal here with (real) functionals instead of
functions. We will consider functionals such that the functional differential
$\delta F$ of any one of them, $F=F\left[  q,p\right]  $ ($\int
d\mathbf{r...=:}\int d^{d}r...$):%
\begin{equation}
\delta F=\int d\mathbf{r}\left\{  \frac{\delta F}{\delta q\left(
\mathbf{r}\right)  }\delta q\left(  \mathbf{r}\right)  +\frac{\delta F}{\delta
p\left(  \mathbf{r}\right)  }\delta p\left(  \mathbf{r}\right)  \right\}
\label{diff1}%
\end{equation}
is well defined, and this will require both the "differentials" (i.e. the
variations) $\delta q$ and $\delta p$ and the functional derivatives $\delta
F/\delta q$ and $\delta F/\delta p$ to be (real) square-integrable functions.

Defining \ a Hamiltonian functional $H_{1}\left[  q,p\right]  $ as:%
\begin{equation}
H_{1}[q,p]=\frac{1}{2}\int d\mathbf{r}\left\{  \frac{\hbar^{2}}{2m}\left[
\left(  \nabla q\right)  ^{2}+\left(  \nabla p\right)  ^{2}\right]  +U\left(
\mathbf{r}\right)  \left(  q^{2}+p^{2}\right)  \right\}  \label{ham1}%
\end{equation}
or (integrating by parts):
\begin{equation}
H_{1}[q,p]=\frac{1}{2}\left\{  \left\langle q,\mathcal{H}q\right\rangle
+\left\langle p,\mathcal{H}p\right\rangle \right\}
\end{equation}%
with $\left\langle .,.\right\rangle $ denoting the (real) scalar product in
$\mathcal{L}_{2}\left(  \mathbb{R}^{d},\mathbb{R}\right)  $, we have, taking
functional derivatives:%
\begin{equation}
\frac{\delta H_{1}}{\delta q\left(  \mathbf{r}\right)  }=\mathcal{H}q\left(
\mathbf{r}\right)  ,\text{ \ }\frac{\delta H_{1}}{\delta p\left(
\mathbf{r}\right)  }=\mathcal{H}p\left(  \mathbf{r}\right)
\end{equation}
and the Schr\"{o}dinger equation (\ref{Schro2}) can be rewritten as the
(infinite-dimensional) Hamiltonian system:
\begin{equation}
\frac{d}{dt}\left\vert
\begin{array}
[c]{c}%
p\\
q
\end{array}
\right\vert =\frac{1}{\hbar}J\left\vert
\begin{array}
[c]{c}%
\frac{\delta H_{1}}{\delta p}\\
\frac{\delta H_{1}}{\delta q}%
\end{array}
\right\vert \label{Schro3}%
\end{equation}
where:%
\begin{equation}
J=\left\vert
\begin{array}
[c]{cc}%
0 & -1\\
1 & 0
\end{array}
\right\vert
\end{equation}

As:%
\begin{equation}
J\left\vert
\begin{array}
[c]{c}%
p\\
q
\end{array}
\right\vert =\left\vert
\begin{array}
[c]{c}%
-q\\
p
\end{array}
\right\vert
\end{equation}
the tensor $J$ is the realified \cite{Ar} version of the standard complex
structure $J_{0}$ on $\mathcal{L}_{2}\left(  \mathbb{R}^{d},\mathbb{C}\right)
$ defined by:%
\begin{equation}
J_{0}:\psi\rightarrow i\psi
\end{equation}
Explicitly:%
\begin{equation}
J=\int d\mathbf{r}\left(  \delta p\left(  \mathbf{r}\right)
\otimes \frac{\delta}{\delta q\left(  \mathbf{r}\right)  }-\delta
q\left( \mathbf{r}\right)  \otimes\frac{\delta}{\delta p\left(
\mathbf{r}\right) }\right)
\end{equation}

The Schr\"{o}dinger equation (\ref{Schro3}) can be rewritten as:%
\begin{equation}
\frac{d}{dt}\left\vert
\begin{array}
[c]{c}%
p\\
q
\end{array}
\right\vert =\left\vert
\begin{array}
[c]{c}%
\left\{  p,H\right\}  _{1}\\
\left\{  q,H\right\}  _{1}%
\end{array}
\right\vert
\end{equation}
where the Poisson bracket $\left\{  .,.\right\}  _{1}$ and the associated
Poisson tensor $\Lambda_{1}\left(  .,.\right)  $ are defined, for any two
functionals $F\left[  q,p\right]  $ and $G\left[  q,p\right]  $, as:%
\begin{equation}
\Lambda_{1}\left(  \delta F,\delta G\right)  =:\left\{  F,G\right\}
_{1}=\frac{1}{\hbar}\int d\mathbf{r}\left\{  \frac{\delta F}{\delta q\left(
\mathbf{r}\right)  }\frac{\delta G}{\delta p\left(  \mathbf{r}\right)  }%
-\frac{\delta F}{\delta p\left(  \mathbf{r}\right)  }\frac{\delta G}{\delta
q\left(  \mathbf{r}\right)  }\right\}  %
\end{equation}
or:
\begin{equation}
\left\{  F,G\right\}  _{1}=\frac{1}{\hbar}\int d\mathbf{r}\left\{  \left\vert
\begin{array}
[c]{cc}%
\delta F/\delta p & \delta F/\delta q
\end{array}
\right\vert J\left\vert
\begin{array}
[c]{c}%
\delta G/\delta p\\
\delta G/\delta q
\end{array}
\right\vert \right\}\label{1:PB1}
\end{equation}

The corresponding symplectic structure\footnote{A non-degenerate two-form
which is closed, being constant in the (global) $\left(  q,p\right)  $ chart.}
$\omega_{1}$ is given by:%
\begin{equation}
\omega_{1}=\hbar\int d\mathbf{r}\left(  \delta q\wedge\delta p\right)
\label{symp1}%
\end{equation}
or:%
\begin{equation}
\omega_{1}=\hbar\int d\mathbf{r}\left\vert
\begin{array}
[c]{cc}%
\delta p & \delta q
\end{array}
\right\vert \otimes J\left\vert
\begin{array}
[c]{c}%
\delta p\\
\delta q
\end{array}
\right\vert
\end{equation}
and the composition of the symplectic and the complex structures
gives rise
\cite{Mar7,Mar18} to the metric tensor:%
\begin{equation}
g=:J\circ\omega_{1}=\hbar\int d\mathbf{r}\left(  \delta p\left(
\mathbf{r}\right)  \otimes\delta p\left(  \mathbf{r}\right)
+\delta q\left( \mathbf{r}\right)  \otimes\delta q\left(
\mathbf{r}\right)  \right)
\end{equation}

 Given any functional $F=F\left[  q,p\right]  $, the Hamiltonian vector field
$X_{F}$ associated with $F$ via:%
\begin{equation}
i_{X_{F}}\omega_{1}=\delta F
\end{equation}
is easily seen to be:%
\begin{equation}
X_{F}=\frac{1}{\hbar}\int d\mathbf{r}\left\{  \frac{\delta F}{\delta p\left(
\mathbf{r}\right)  }\frac{\delta}{\delta q\left(  \mathbf{r}\right)  }%
-\frac{\delta F}{\delta q\left(  \mathbf{r}\right)  }\frac{\delta}{\delta
p\left(  \mathbf{r}\right)  }\right\}  \label{hamfield1}%
\end{equation}
In particular:%
\begin{equation}
X_{H_{1}}=\frac{1}{\hbar}\int d\mathbf{r}\left\{  \mathcal{H}p\left(
\mathbf{r}\right)  \frac{\delta}{\delta q\left(  \mathbf{r}\right)
}-\mathcal{H}q\left(  \mathbf{r}\right)  \frac{\delta}{\delta p\left(
\mathbf{r}\right)  }\right\}\label{hamfield2}
\end{equation}
The Poisson bracket (\ref{1:PB1}) can then be written also as:%
\begin{equation}
\left\{  F,G\right\}  _{1}=\omega_{1}\left(  X_{G},X_{F}\right)
\end{equation}

\bigskip

\textbf{Digression.}

Things acquire a more familiar (and manageable) form if we introduce a (real)
complete orthonormal set of functions\footnote{They could be, e.g., the
eigenfunctions of a $d$-dimensional isotropic harmonic oscillator.}:%
\begin{equation}
\left\{  \psi_{n}\left(  \mathbf{r}\right)  \right\}  _{1}^{\infty
};\left\langle \psi_{n},\psi_{m}\right\rangle =\delta_{nm};%
{\displaystyle\sum\limits_{n}}
\int d\mathbf{r}\psi_{n}\left(  \mathbf{r}\right)  \psi_{n}\left(
\mathbf{r}^{\prime}\right)  =\delta\left(  \mathbf{r}-\mathbf{r}^{\prime
}\right)  \label{basis}%
\end{equation}
in $\mathcal{L}_{2}\left(  \mathbb{R}^{d},\mathbb{R}\right)  $. Then,
defining:%
\begin{equation}
\delta q\left(  \mathbf{r}\right)  =%
{\displaystyle\sum\limits_{n}}
\psi_{n}\left(  \mathbf{r}\right)  dq_{n},\text{ \ }dq_{n}=:\left\langle
\psi_{n},\delta q\right\rangle
\end{equation}
and similarly for $\delta p$, the functional differential (\ref{diff1})
becomes:%
\begin{equation}
\delta F=%
{\displaystyle\sum\limits_{n}}
\left\{  \frac{\partial F}{\partial q_{n}}dq_{n}+\frac{\partial F}{\partial
p_{n}}dp_{n}\right\}  \label{diff2}%
\end{equation}
where:%
\begin{equation}
\frac{\partial F}{\partial q_{n}}=:\left\langle \psi_{n},\frac{\delta F}{\delta
q}\right\rangle
\end{equation}
(and similarly for $\partial F/\partial p_{n}$)\footnote{Note that, under the
stated assumptions, the series on the r.h.s. of Eqn.(\ref{diff2}) will be
convergent.}. In other words:
\begin{equation}
\delta F=\left\langle \frac{\delta F}{\delta q},\delta q\right\rangle
+\left\langle \frac{\delta F}{\delta p},\delta p\right\rangle
\end{equation}

Proceeding in a similar way, it is easy to check that the Poisson tensor
(\ref{PB1}), the symplectic form (\ref{symp1}) and the Hamiltonian vector
field (\ref{hamfield1}) associated with $F$ can be written in this basis as:%
\begin{eqnarray}
&&\Lambda_{1}=\frac{1}{\hbar}%
{\displaystyle\sum\limits_{n}}
\frac{\partial}{\partial p_{n}}\wedge\frac{\partial}{\partial q_{n}}\\
&& \omega_{1}=\hbar%
{\displaystyle\sum\limits_{n}}
dq_{n}\wedge dp_{n}%
\end{eqnarray}
and:%
\begin{equation}
X_{F}=\frac{1}{\hbar}%
{\displaystyle\sum\limits_{n}}
\left\{  \frac{\partial F}{\partial p_{n}}\frac{\partial}{\partial q_{n}%
}-\frac{\partial F}{\partial q_{n}}\frac{\partial}{\partial p_{n}}\right\}
\end{equation}

\subsubsection{Alternative Hamiltonian Descriptions}
\bigskip

Let's assume now the Schr\"{o}dinger operator \ (\ref{Schro1}) to be
\textit{positive}\footnote{It could be, e.g., the Schr\"{o}dinger
operator for the isotropic harmonic oscillator:
$\mathcal{H}=-\frac{\hslash^{2}}{2m}\nabla^{2}+U(\mathbf{r})$
with: $U\left( \mathbf{r}\right)  =m\omega
^{2}\mathbf{r}^{2}/2$.}or, more generally, invertible, and let,
for simplicity, the $\psi_{n}$'s be the
associated eigenfunctions:%
\begin{equation}
\mathcal{H}\psi_{n}=E_{n}\psi_{n},\ E_{n}>0\forall n
\end{equation}
Then, defining \cite{Mar18} a new Poisson tensor and Poisson bracket as:%
\begin{equation}
\Lambda_{0}\left(  \delta F,\delta G\right)  =:\left\{  F,G\right\}
_{0}=\frac{1}{\hbar}\int d\mathbf{r}\left\{  \frac{\delta F}{\delta q\left(
\mathbf{r}\right)  }\mathcal{H}\frac{\delta G}{\delta p\left(  \mathbf{r}%
\right)  }-\frac{\delta F}{\delta p\left(  \mathbf{r}\right)  }\mathcal{H}%
\frac{\delta G}{\delta q\left(  \mathbf{r}\right)  }\right\}  \label{PBrac2}%
\end{equation}
the same Schr\"{o}dinger equation can be written also as:%
\begin{equation}
\frac{d}{dt}\left\vert
\begin{array}
[c]{c}%
p\\
q
\end{array}
\right\vert =\frac{1}{\hbar}\left\vert
\begin{array}
[c]{cc}%
0 & -\mathcal{H}\\
\mathcal{H} & 0
\end{array}
\right\vert \left\vert
\begin{array}
[c]{c}%
\frac{\delta H_{0}}{\delta p}\\
\frac{\delta H_{0}}{\delta q}%
\end{array}
\right\vert
\end{equation}
or:%
\begin{equation}
\frac{d}{dt}\left\vert
\begin{array}
[c]{c}%
p\left(  \mathbf{r}\right)  \\
q\left(  \mathbf{r}\right)
\end{array}
\right\vert =\left\vert
\begin{array}
[c]{c}%
\left\{  p\left(  \mathbf{r}\right)  ,H_{0}\right\}  _{0}\\
\left\{  q\left(  \mathbf{r}\right)  ,H_{0}\right\}  _{0}%
\end{array}
\right\vert
\end{equation}

where:%
\begin{equation}
H_{0}\left[  q,p\right]  =\frac{1}{2}\int d\mathbf{r}\left(  q^{2}%
+p^{2}\right)
\end{equation}
is a sort of "universal" Hamiltonian functional.

In the basis of the eigenfunctions of $\mathcal{H}$ the Poisson
bracket (\ref{PBrac2}) can be written as:
\begin{equation}
\left\{  F,G\right\}  _{0}=\frac{1}{\hbar}%
{\displaystyle\sum\limits_{n}}
E_{n}\left\{  \frac{\partial F}{\partial q_{n}}\frac{\partial
G}{\partial p_{n}}-\frac{\partial F}{\partial p_{n}}\frac{\partial
G}{\partial q_{n}}\right\}
\end{equation}
and the associated symplectic form will be given by:%
\begin{equation}
\omega_{0}=\hbar%
{\displaystyle\sum\limits_{n}}
E_{n}^{-1}dq_{n}\wedge dp_{n} \label{symp}%
\end{equation}
or, in a basis-free notation:%

\begin{equation}
\omega_{0}=\hbar\int d\mathbf{r}\left(  \mathcal{H}^{-1}\delta q\wedge\delta
p\right)  \label{symp3}%
\end{equation}
Moreover, the Hamiltonian vector field associated, via $\omega_{0}$ now, with
the functional $F=F\left[  q,p\right]  $ is given by:%
\begin{equation}
X_{F}=\frac{1}{\hbar}%
{\displaystyle\sum\limits_{n}}
\epsilon_{n}\left\{  \frac{\partial F}{\partial p_{n}}\frac{\partial}{\partial
q_{n}}-\frac{\partial F}{\partial q_{n}}\frac{\partial}{\partial p_{n}%
}\right\}
\end{equation}
or, in \ basis-independent form:%
\begin{equation}
X_{F}=\frac{1}{\hbar}\int d\mathbf{r}\left\{  \mathcal{H}\frac{\delta
F}{\delta p\left(  \mathbf{r}\right)  }\frac{\delta}{\delta q\left(
\mathbf{r}\right)  }-\mathcal{H}\frac{\delta F}{\delta q\left(  \mathbf{r}%
\right)  }\frac{\delta}{\delta p\left(  \mathbf{r}\right)  }\right\}
\end{equation}
In particular:%
\begin{equation}
X_{H_{0}}=\frac{1}{\hbar}\int d\mathbf{r}\left\{  \mathcal{H}p\left(
\mathbf{r}\right)  \frac{\delta}{\delta q\left(  \mathbf{r}\right)
}-\mathcal{H}q\left(  \mathbf{r}\right)  \frac{\delta}{\delta p\left(
\mathbf{r}\right)  }\right\}
\end{equation}
which coincides with the Hamiltonian vector field (\ref{hamfield2}).

\bigskip

\begin{remark}
One could have also rewritten $\omega_{0}$ as:%
\begin{equation}
\omega_{0}=\hbar\int d\mathbf{r}\left(  \delta q\wedge\mathcal{H}^{-1}\delta
p\right)
\end{equation}
but the two forms of course coincide, in view of the fact that
$\mathcal{H}$ is self-adjoint.
\end{remark}

What has been proved up to here is that the \textit{same} vector field,
namely:%
\begin{equation}
\Gamma=\frac{1}{\hbar}\int d\mathbf{r}\left\{  \mathcal{H}p\left(
\mathbf{r}\right)  \frac{\delta}{\delta q\left(  \mathbf{r}\right)
}-\mathcal{H}q\left(  \mathbf{r}\right)  \frac{\delta}{\delta p\left(
\mathbf{r}\right)  }\right\}  \label{vecfield3}%
\end{equation}
is Hamiltonian w.r.t. two different \ Poisson brackets\footnote{I.e.: $\Gamma=\left\{  \mathcal{H}_{1},.\right\}  _{1}=\left\{  \mathcal{H}%
_{0},.\right\}  _{0}$.} $\left(  \left\{  .,.\right\}  _{1}\text{ and
}\left\{  .,.\right\}  _{0}\right)  $ and Hamiltonian functionals \ $\left(
H_{1}\text{ and }H_{0}\right)  $, i.e. that it is \textit{bi-Hamiltonian}. It turns out \cite{DSVM2,Ma} that this, together with the compatibility condition, can lead to complete integrability.

The procedure can actually be iterated, leading to the conclusion
\cite{Mar18} that the Schr\"{o}dinger equation admits of infinitely many
alternative Hamiltonian descriptions, with Hamiltonians:%
\begin{equation}
H_{n}\left[  q,p\right] =\frac{1}{2}\left\{  \left\langle
q,\mathcal{H}^{n}q\right\rangle +\left\langle p,\mathcal{H}^{n}p\right\rangle
\right\}  ,\text{ \ }n\geq1  ;  H_{1}[q,p]=H[q,p]
\end{equation}
with associated symplectic forms:%
\begin{equation}
\omega_{n}=\hbar\int d\mathbf{r}\left(  \mathcal{H}^{n-1}\delta q\wedge\delta
p\right)
\end{equation}
and Poisson tensors:%
\begin{equation}
\Lambda_{n}\left(  \delta F,\delta G\right)  =\left\{  F,G\right\}  _{n}%
=\frac{1}{\hbar}\int d\mathbf{r}\left\{  \frac{\delta F}{\delta q\left(
\mathbf{r}\right)  }\mathcal{H}^{1-n}\frac{\delta G}{\delta p\left(
\mathbf{r}\right)  }-\frac{\delta F}{\delta p\left(  \mathbf{r}\right)
}\mathcal{H}^{1-n}\frac{\delta G}{\delta q\left(  \mathbf{r}\right)  }\right\}
\end{equation}
such that:%
\begin{equation}
i_{\Gamma}\omega_{n}=\delta H_{n}\text{ }\forall n
\end{equation}
where $\Gamma$ is the vector field (\ref{vecfield3}) and that the Hamiltonian functionals $H_{n}$ are pairwise in involution w.r.t.
\textit{all} the Poisson brackets, i.e.:%
\begin{equation}
\left\{  H_{n},H_{m}\right\}  _{k}=0\text{ \ }\forall n,m,k
\end{equation}
In other words, the Schr\"{o}dinger equation admits of infinitely many
constants of the motion pairwise in involution, which is another hallmark \cite{DSVM2,Ma} of
complete integrability. Having established this, as well as the fact that the Schr\"{o}dinger equation admits of infinitely many Hamiltonian descriptions, and that it can be considered as an infinite-dimensional Hamiltonian system on some infinite-dimensional space, it will be appropriate to devote the next Chapter to the study of completely-integrable dynamical systems and of their alternative Hamiltonian descriptions.
\newpage

%% file: Chapt2rev.tex
\section{Completely Integrable Systems and Bi-Hamiltonian Descriptions}

\subsection{Liouville Integrability and Linearization}
\bigskip
In order to avoid reducing the generality of our treatment, and for future
reference, when the carrier space of a quantum system may be a manifold (like
the complex projective Hilbert space (see below Sect.\ref{GQM})) instead of a
vector space, we will work here in the framework of symplectic manifolds and
Hamiltonian systems. So, let $\left(  \mathcal{M},\omega\right)  $ be a
symplectic manifold ($\dim\mathcal{M}\mathbf{=}2n$ for some $n$ and $\omega$ a
symplectic form). A dynamical system, i.e. a vector field \ $\Gamma\in
T\mathcal{M}$ \ is \textit{$\omega$-Hamiltonian} or, for
short,\textit{Hamiltonian} iff:%
\begin{equation}
i_{\Gamma}\omega=d\mathcal{H} \label{hamiltonian}%
\end{equation}
for some $\mathcal{H}\in\mathcal{F}\left(  \mathcal{M}\right)  $. \ A
Hamiltonian dynamical system is said to be \textit{completely integrable} if
it has $n$ constants of the motion $f_{1},...,f_{n}$ that are:

$i)$ functionally independent:%
\begin{equation}
df_{1}\wedge...\wedge df_{n}\neq0
\end{equation}
and:

$ii)$ pairwise in involution, i.e.:%
\begin{equation}
\left\{  f_{i},f_{j}\right\}  =0\text{ \ }\forall i,j \label{poisson}%
\end{equation}
where $\left\{  .,.\right\}  $ is the Poisson bracket associated with the
symplectic form $\omega$. The \textit{Arnold-Liouville theorem}\cite{Ar}
states then that the level sets:%
\begin{equation}
\mathbb{M}_{\mathbf{c}}=f^{-1}\left(  \mathbf{c}\right)  ,\text{ \ }%
\mathbf{c}\in\mathbb{R}^{n},\text{ }\dim\mathbb{M}_{\mathbf{c}}=n
\label{leaves}%
\end{equation}
provide a foliation of $\mathcal{M}$ whose leaves are invariant manifolds for
the Hamiltonian flow (\ref{hamiltonian}). Moreover, if the leaves of the
foliation (\ref{leaves}) are compact and connected, then they are
diffeomorphic to $n$-dimensional tori, i.e.:%
\begin{equation}
\mathbb{M}_{\mathbf{c}}\approx\mathbb{T}^{n}=\underset{n\text{ }%
times}{\underbrace{\mathbb{S}^{1}\times...\times\mathbb{S}^{1}}}=\left\{
\phi\equiv\left(  \phi^{1},...,\phi^{n}\right)  \text{ }\operatorname{mod}%
2\pi\right\}  \label{tori}%
\end{equation}
and one can find a set of frequencies: $\nu\equiv\left(  \nu_{1},...,\nu
_{n}\right)  $, $\nu=\nu\left(  f\right)  $ such that the Hamiltonian flow on
the torus is given by\footnote{Such motions are called \textit{quasi-periodic}
or \textit{conditionally periodic}.}:%
\begin{equation}
\frac{d\phi_{i}}{dt}=\nu_{i}\Rightarrow\phi_{i}\left(  t\right)  =\phi
_{i}\left(  0\right)  +\nu_{i}t \label{motion}%
\end{equation}
and Hamilton's equations of motion are integrable by quadratures.

\bigskip

Let's summarize briefly how this leads to the well-known construction of
\textit{action-angle} variables.

Calling $X_{i}$ the Hamiltonian vector field associated with $f_{i}$,
$i=1,...,n$, Eqn.(\ref{poisson}) leads at once to:%
\begin{equation}
\left\{  f_{i},f_{j}\right\}  \equiv\mathcal{L}_{X_{j}}f_{i}\equiv
\omega\left(  X_{j},X_{i}\right)  =0 \label{poisson2}%
\end{equation}
Moreover, as:
\begin{equation}
i_{\left[  X,Y\right]  }=\mathcal{L}_{X}\cdot i_{Y}-i_{Y}\cdot\mathcal{L}_{X}%
\end{equation}
we obtain\footnote{As $X_{j}$ is Hamiltonian, $\mathcal{L}_{X_{j}}\omega=0$.}:%
\begin{equation}
i_{\left[  X_{i},X_{j}\right]  }\omega=\mathcal{L}_{X}\cdot(i_{X_{j}}%
\omega)-i_{X_{j}}\cdot(\mathcal{L}_{X_{i}}\omega)\equiv d\left(
\mathcal{L}_{X_{i}}f_{j}\right)  =0
\end{equation}
the final result following from Eqn.(\ref{poisson2}). Therefore, the $X_{i}$'s
commute pairwise. Moreover, it follows again from Eqn.(\ref{poisson2}) that
the invariant leaves (\ref{leaves}) of the foliation are Lagrangian
submanifolds. Defining the immersion: $i_{\mathbf{c}}:$ $\mathbb{M}%
_{\mathbf{c}}\hookrightarrow\mathcal{M}$, we have therefore:%
\begin{equation}
i_{\mathbf{c}}^{\ast}\omega=0
\end{equation}
Therefore, if we denote by $\theta$ the Cartan one-form ($\omega=-d\theta$),
its pull-back \ $i_{\mathbf{c}}^{\ast}\theta$ \ will be \textit{closed}:%
\begin{equation}
di_{\mathbf{c}}^{\ast}\theta=i_{\mathbf{c}}^{\ast}d\theta=0 \label{closure}%
\end{equation}
It need not be exact, though, as the invariant tori are not contractible.
Cycles on the torus need not be boundaries, and therefore the integral of
\ \ $i_{\mathbf{c}}^{\ast}\theta$ along a one-dimensional cycle need not
vanish. We can select a \textit{basis \ }$\left(  \gamma_{1},...,\gamma
_{n}\right)  $ of loops, i.e. $n$ one-dimensional cycles each one of which
winds around the torus exactly once and none of which is homologous \cite{Ahl}
to any other one (nor to the trivial loop), and define the \textit{action
variables} $I_{i}$ as:%
\begin{equation}
I_{i}=\frac{1}{2\pi}%
{\displaystyle\oint\limits_{\gamma_{i}}}
i_{\mathbf{c}}^{\ast}\theta,\text{ \ }i=1,...,n
\end{equation}
Of course: $I_{i}=I_{i}\left(  f\right)  $ depends only on the homology class
\cite{Ahl} of $\gamma_{i}$ and, provided the jacobian of the transformation
does not vanish or, equivalently:%
\begin{equation}
dI_{1}\wedge dI_{2}\wedge...\wedge dI_{n}\neq0
\end{equation}
invariant tori can be uniquely labelled by the set $I=\left(  I_{1}%
,...,I_{n}\right)  $ of the values of the action variables. Defining then:%
\begin{equation}
S=S\left(  I,q\right)  =%
{\displaystyle\int\limits_{q_{0}}^{q}}
i_{\mathbf{c}}^{\ast}\theta
\end{equation}
the integral being along a path $\gamma$ on the invariant torus labelled by
$I$ joining a fiducial point $q_{0}$ to the point $q$, $S$ will depend only on
the homology class of paths from $q_{0}$ to $q$ to which $\gamma$
belongs\footnote{This approach goes back to a paper \cite{Ein} by A.Einstein
of $1917$.}. Switching to a different homology class multiplying $\gamma$ by,
say, a loop $\gamma_{i}$ in the basis will change $S$ by a fixed amount:%
\begin{equation}
S\rightarrow S+\Delta S_{i};\text{ \ }\Delta S_{i}=2\pi I_{i}
\label{multivalued}%
\end{equation}

We can then use $S$ as the generator of a time-independent canonical
transformation:%
\begin{equation}\label{can}
\left(  q,p\right)  \rightarrow\left(  \phi,I\right)
\end{equation}
with the $I$'s playing the r\^{o}le of the new momenta, via\footnote{The
ambiguity expressed by Eqn.(\ref{multivalued}) tells us that the $\phi$'s are
actually defined "modulo" $2\pi$, i.e. that they are indeed angles.}:%
\begin{equation}
p_{i}=\frac{\partial S}{\partial q^{i}},\text{ \ }\phi^{i}=\frac{\partial
S}{\partial I_{i}}%
\end{equation}
and with the new Hamiltonian: $\mathcal{K}=\mathcal{H}$. Now, as $n$ is the
maximum allowed number of independent \ constants of the motion pairwise in
involution\footnote{If $\omega$ is non-degenerate, $n$ is the maximum allowed
dimension for an isotropic subspace.}, either the Hamiltonian is one of the
$f_{i}$'s or is a function thereof: $\mathcal{H}=\mathcal{H}\left(  f\right)
$ and therefore it is ultimately a function of the action variables alone.
Hamilton's equations become then:%
\begin{equation}
\frac{d}{dt}I_{i}=0,\text{ \ }\frac{d}{dt}\phi^{i}=\nu^{i};\text{ \ }\nu
^{i}=:\frac{\partial\mathcal{H}}{\partial I_{i}}=\nu^{i}\left(  I\right)
\label{tori2}%
\end{equation}
and we recover Eqn.(\ref{motion}). In the new coordinates the dynamical vector
field will be given by:%
\begin{equation}
\Gamma=%
{\displaystyle\sum\limits_{i=1}^{n}}
\nu^{i}\frac{\partial}{\partial\phi^{i}} \label{tori3}%
\end{equation}
and the symplectic structure will be:%
\begin{equation}
\omega=%
{\displaystyle\sum\limits_{i=1}^{n}}
d\phi^{i}\wedge dI_{i}%
\end{equation}
We notice that in these coordinates the dynamics is nilpotent of  index
two,i.e.:
\begin{equation}
\frac{d\phi^{i}}{dt}=\nu^{i};\text{ }\frac{d\nu^{i}}{dt}=0
\end{equation}

Moreover, in these coordinates the system is linear and associated with a
nilpotent matrix. It should be remarked that the transformation (\ref{can}) is not linear. Therefore, even if the system is linear in the $(q,p)$ coordinates, the transformation need not be isospectral, i.e. it may take us from a semisimple matrix to a nilpotent one.

\subsection{From Invariant Structures to Integrability}

\bigskip

In the case of Eqn.(\ref{tori2}), if we are in the so-called
\textit{non-resonant} case, i.e. if:%
\begin{equation}
d\nu_{1}\wedge d\nu_{2}\wedge...\wedge d\nu_{n}\neq0
\end{equation}
we can choose the $\nu_{i}$'s as new momenta (the transformation will be in
general \textit{not} canonical, however!). In the new coordinates the
dynamical system will be completely \textit{separated } into $n$ independent
systems, while the Hamiltonian and symplectic structure will become
respectively:%
\begin{equation}
H=\frac{1}{2}%
{\displaystyle\sum\limits_{i=1}^{n}}
\nu_{i}^{2}%
\end{equation}
and:%
\begin{equation}
\omega=%
{\displaystyle\sum\limits_{i=1}^{n}}
d\phi^{i}\wedge d\nu_{i}%
\end{equation}
\textit{Separability} of a dynamical system into a family of non-interacting subsystems appears therefore to be
intimately connected with integrability\footnote{See also
Refs.\cite{Fe1, Fe2} for a similar discussion in the Lagrangian
context.}. It is also well-known that a way to achieve (if possible)
integrability via separability occurs in the Hamilton-Jacobi theory
\cite{BCR1,BCR2,BCR3,IMM,Muk0}, a subject that we will not discuss here, though. Notice also that, in general, the two notions of separability do not in general coincide.

In this Subsection we will discuss a way to achieve separability (and
eventually integrability) with the aid of additional invariant structures
\cite{DSVM1,DSVM2}. We will not make reference, for the time being, to
symplectic structures and the like. What we are going to say generalizes to
vector fields, and hence also to non-linear situations, the familiar
block-diagonal form of matrices.

Let then $\mathcal{M}$ be \ a smooth manifold \ and let $\Gamma\in
\mathfrak{X}\left(  \mathcal{M}\right)  $ be a vector field. $\Gamma$ will be
said to be \textit{separable into dynamics of lower dimension} on an open set
$\mathcal{U}\subseteq\mathcal{M}$ if a holonomic frame $\left\{  e_{\left(
i,k\right)  }\right\}  $ can be found for the tangent bundle $T\mathcal{U}$,
with dual forms $\left\{  \theta^{\left(  i,k\right)  }\right\}  $, such that:%
\begin{equation}
\mathcal{L}_{e_{\left(  i,k\right)  }}\left\langle \theta^{\left(  j,h\right)
}|\Gamma\right\rangle \neq0\Leftrightarrow i=j \label{separable1}%
\end{equation}
This implies of course, in local coordinates, that we can choose coordinates
$x^{\left(  i,k\right)  }$ ($e_{\left(  i,k\right)  }=\partial/\partial
x^{\left(  i,k\right)  }$)\ in such a way that:%
\begin{equation}
\Gamma=\Gamma^{\left(  i,k\right)  }\frac{\partial}{\partial x^{\left(
i,k\right)  }} \label{separable2}%
\end{equation}
and:%
\begin{equation}
\Gamma^{\left(  i,k\right)  }=\Gamma^{\left(  i,k\right)  }\left(
x^{i}\right)  ;\text{ \ }x^{i}=:\left(  x^{\left(  i,1\right)  },x^{\left(
i,2\right)  },...,x^{\left(  i,k\right)  },...\right)  \label{separable3}%
\end{equation}
Finally, the vector field $\Gamma$ will be said to be \textit{separable }\ if
we can choose $\mathcal{U}=\mathcal{M}$ or, at least, $\mathcal{U}$ to be an
open dense set in $\mathcal{M}$.

Let us review briefly how one can achieve separation of the dynamics in the
presence of an \textit{invariant diagonalizable }$\left(  1,1\right)  $
\textit{tensor field }$T\in\mathcal{F}_{1}^{1}\left(  \mathcal{M}\right)  $
\textit{with at least two distinct eigenvalues and vanishing Nijenhuis
torsion}.

Recall\footnote{More properties of Nijenhuis torsions and tensors are briefly
reviewed in App.$A$.} that, given a $\left(  1,1\right)  $ tensor $T$, the
\textit{Nijenhuis torsion} \cite{FroNi,Mar0,Ni} associated with $T$ is the
$\left(  0,2\right)  $ tensor $\mathcal{N}_{T}$ defined by:%
\begin{equation}
\mathcal{N}_{T}\left(  \alpha,X,Y\right)  =:\left\langle \alpha|\mathcal{H}%
_{T}\left(  X,Y\right)  \right\rangle ;\text{ }\alpha\in\mathcal{X}^{\ast
}\left(  \mathcal{M}\right)  ,\text{ }X,Y\in\mathcal{X}\left(  \mathcal{M}%
\right)
\end{equation}
where:%
\begin{equation}
\mathcal{X}\left(  \mathcal{M}\right)  \ni\mathcal{H}_{T}\left(  X,Y\right)
=:\left[  TX,TY\right]  +T^{2}\left[  X,Y\right]  -T\left[  TX,Y\right]
-T\left[  X,TY\right]  \label{Nijenhuis}%
\end{equation}

Let's remark that, if $T$ is diagonalizable:%
\begin{equation}
Te_{i}=\lambda_{i}e_{i}%
\end{equation}
the\ eigenvectors $e_{i}$ are (locally at least) a basis of vector
fields\footnote{In fact, they are not only a \ vector space, but have in
addition the structure of an $\mathcal{F}\left(  \mathcal{M}\right)
$-module.}, and we will denote as $S_{\lambda_{i}}$ the eigenspace of the
eigenvalue $\lambda_{i}$. The $e_{i}$' being a basis implies:
\begin{equation}
\left[  e_{i},e_{j}\right]  =%
{\displaystyle\sum\limits_{k}}
c_{ij}^{k}e_{k};\text{ }c_{ij}^{k}=-c_{ji}^{k}%
\end{equation}
for some set of "structure constants" (actually in principle functions)
$c_{ij}^{k}$. The dual cobasis $\left\{  \theta^{i}\right\}  $, defined as
usual via:%
\begin{equation}
\left\langle \theta^{i}|e_{j}\right\rangle =\delta_{j}^{i} \label{dualform}%
\end{equation}
will be also a basis of eigenforms:
\begin{equation}
\widetilde{T}\theta^{i}=\lambda_{i}\theta^{i}%
\end{equation}
where $\widetilde{T}$ \ denotes the transpose action of $T$ on forms
($\left\langle \theta|TX\right\rangle =:\left\langle \widetilde{T}%
\theta|X\right\rangle $). Using then the identity \cite{CM}:%
\begin{equation}
d\theta\left(  X,Y\right)  =\mathcal{L}_{X}\left(  \theta\left(  Y\right)
\right)  -\mathcal{L}_{Y}\left(  \theta\left(  X\right)  \right)
-\left\langle \theta|\left[  X,Y\right]  \right\rangle
\end{equation}
it is easy to prove that:%
\begin{equation}
d\theta^{k}\left(  e_{i},e_{j}\right)  =-c_{ij}^{k} \label{structure1}%
\end{equation}
i.e. that:%
\begin{equation}
d\theta^{k}=-\frac{1}{2}%
{\displaystyle\sum\limits_{ij}}
c_{ij}^{k}\theta^{i}\wedge\theta^{j} \label{structure2}%
\end{equation}

Contracting the Nijenhuis torsion with the eigenvectors one finds, with some
algebra:%
\begin{equation}
\mathcal{H}_{T}\left(  e_{i},e_{j}\right)  =\left(  T-\lambda_{i}\right)
\left(  T-\lambda_{j}\right)  \left[  e_{i},e_{j}\right]  +\left(  \lambda
_{i}-\lambda_{j}\right)  \left\{  \left(  \mathcal{L}_{e_{i}}\lambda
_{j}\right)  e_{j}+\left(  \mathcal{L}_{e_{j}}\lambda_{i}\right)
e_{i}\right\}
\end{equation}
Let us remark first that:%
\begin{equation}
\left(  T-\lambda_{i}\right)  \left(  T-\lambda_{j}\right)  \left[
e_{i},e_{j}\right]  =%
{\displaystyle\sum\limits_{k}}
\left(  \lambda_{k}-\lambda_{i}\right)  \left(  \lambda_{k}-\lambda
_{j}\right)  c_{ij}^{k}e_{k}%
\end{equation}
has \textit{no} components in $S_{\lambda_{i}}\oplus S_{\lambda_{j}}$. \ If
the Nijenhuis torsion vanishes\footnote{i.e. $T$ is (see App.$A$) a Nijenhuis
tensor}, then the condition $\mathcal{H}_{T}\left(  e_{i},e_{j}\right)  =0$
separates into:
\begin{equation}
\left(  T-\lambda_{i}\right)  \left(  T-\lambda_{j}\right)  \left[
e_{i},e_{j}\right]  =0
\end{equation}
and:%
\begin{equation}
\left(  \lambda_{i}-\lambda_{j}\right)  \mathcal{L}_{e_{i}}\lambda_{j}%
\equiv\left(  \lambda_{i}-\lambda_{j}\right)  d\lambda_{i}\left(
e_{j}\right)  =0 \label{distribution0}%
\end{equation}
Contracting the first of the above equations with $\theta^{k}$ we obtain:%
\begin{equation}
\left(  \lambda_{k}-\lambda_{i}\right)  \left(  \lambda_{k}-\lambda
_{j}\right)  \left\langle \theta^{k}|\left[  e_{i},e_{j}\right]  \right\rangle
=0 \label{distribution1}%
\end{equation}
which implies: $\left\langle \theta^{k}|\left[  e_{i},e_{j}\right]
\right\rangle =0$ for $\lambda_{k}\neq\lambda_{i},\lambda_{j}$, i.e.:
\begin{equation}
\left[  e_{i},e_{j}\right]  \in S_{\lambda_{i}}\oplus S_{\lambda_{j}}
\label{distribution2}%
\end{equation}
and hence:%
\begin{equation}
c_{ij}^{k}=0 \mbox{ when } \lambda_{k}\neq \lambda_{i},\lambda_{j}
\label{structure3}%
\end{equation}

At this point we can somehow sharpen the analysis and make it a bit more
precise. If the eigenspaces are not one-dimensional (i.e. the eigenvalues of
$T$ have degeneracy), denoting by $\left\{  e_{\left(  i,r\right)  }\right\}
,r=1,2,...,d_{i}$, $d_{i}$ being the dimension of the $i$-th eigenspace, a
basis of eigenvectors in $S_{\lambda_{i}}$, it is not difficult to prove that
Eqn.(\ref{distribution1}) generalizes to:%
\begin{equation}
\left(  \lambda_{k}-\lambda_{i}\right)  \left(  \lambda_{k}-\lambda
_{j}\right)  \left\langle \theta^{\left(  k,r\right)  }|\left[  e_{\left(
i,p\right)  },e_{\left(  j,q\right)  }\right]  \right\rangle =0\text{ }\forall
r,p,q
\end{equation}
which holds in particular for $i=j$, thus leading to the conclusion that:%
\begin{equation}
\left[  e_{\left(  i,p\right)  },e_{\left(  i,q\right)  }\right]  \in
S_{\lambda_{i}}%
\end{equation}
i.e. that \textit{if }$T$ \textit{is diagonalizable and has vanishing
Nijenhuis torsion, the eigenvectors belonging to every eigenspace are an
involutive distribution}. As such, the distribution will be integrable by
Frobenius' theorem \cite{Mar10}, and we can speak (locally at least) of
\textit{eigenmanifolds}.

We can also reach the same conclusion in dual form as follows.
Eqn.(\ref{structure3}) implies that in Eqn.(\ref{structure2}) at least one of
the one-forms on the r.h.s. must be in the (dual) eigenspace of the eigenvalue
$\lambda_{k}$. To be more specific, if we denote by $\theta^{\left(
k,r\right)  },r=1,2,...$ the eigenforms belonging to the eigenvalue
$\lambda_{k}$ and by $c_{\left(  i,s\right)  \left(  j,p\right)  }^{\left(
k,r\right)  }$ the "structure constants", Eqns.(\ref{structure2}) and
(\ref{structure3}) imply:%
\begin{equation}
d\theta^{\left(  k,r\right)  }=-%
{\displaystyle\sum\limits_{\left(  i,p\right)  ,s}}
c_{(i,p)(k,s)}^{(k,s)}\theta^{\left(  i,p\right)  }\wedge\theta^{\left(
k,s\right)  }%
\end{equation}
\ But this is equivalent to the statement that:%
\begin{equation}
d\theta^{\left(  k,r\right)  }%
{\displaystyle\bigwedge\limits_{s}}
\theta^{\left(  k,s\right)  }=0
\end{equation}
which is again \cite{DSVM2} a statement of Frobenius' theorem.

The main conclusion is then that, under the stated assumptions, one can always
find a \textit{holonomic} frame (and coframe) that diagonalizes $T$ in the
form:%
\begin{equation}
T=%
{\displaystyle\sum\limits_{i}}
\lambda_{i}e_{i}\otimes\theta^{i}%
\end{equation}

Let us turn now to the consequences of the invariance of $T$ under the
dynamics. First of all, an invariant $\left(  1,1\right)  $ tensor $T$ will
generate an algebra $\mathcal{A}$ of vector fields all commuting with $\Gamma$
given by:%
\begin{equation}
\mathcal{A}=\left\{  \Gamma,T\Gamma,T^{2}\Gamma,..,T^{k}\Gamma,...\right\}
\end{equation}
If $\mathcal{L}_{\Gamma}T=0$, it can be proved \cite{DSVM2} that:%
\begin{equation}
\left[  T^{k}\Gamma,T^{k+h}\Gamma\right]  =%
{\displaystyle\sum\limits_{\underset{\alpha,\beta\geq0,\gamma\geq k}%
{\alpha+\beta+\gamma=2k+h-2}}}
T^{\alpha}\mathcal{H}_{T}\left(  T^{\beta}\Gamma,T^{\gamma}\Gamma\right)
\end{equation}
hence, if $\mathcal{H}_{T}=0$, $\mathcal{A}$ will be an abelian algebra of
vectors fields all commuting with $\Gamma$, i.e. an abelian algebra of
symmetries \cite{Mar10}.

Consider next the eigenvalue equation for $T$. Let $e$ and $\theta$ be an
eigenvector and an eigenform belonging to the same eigenvalue $\lambda$:%
\begin{equation}
Te=\lambda e,\text{ \ }\widetilde{T}\theta=\lambda\theta
\end{equation}
We can assume, without loss of generality: $\left\langle \theta|e\right\rangle
=1$.

If $T$ is invariant under the dynamics, $\mathcal{L}_{\Gamma}T=0$, then:%
\begin{equation}
T\left(  \mathcal{L}_{\Gamma}e\right)  =\mathcal{L}_{\Gamma}\left(  Te\right)
=\mathcal{L}_{\Gamma}\left(  \lambda e\right)  =\left(  \mathcal{L}_{\Gamma
}\lambda\right)  e+\lambda\left(  \mathcal{L}_{\Gamma}e\right)
\end{equation}
On the other hand:%
\begin{equation}
\left\langle \mathcal{L}_{\Gamma}e|\widetilde{T}\theta\right\rangle
=\left\langle T\left(  \mathcal{L}_{\Gamma}e\right)  |\theta\right\rangle
=\mathcal{L}_{\Gamma}\lambda+\lambda\left\langle \mathcal{L}_{\Gamma}%
e|\theta\right\rangle \equiv\mathcal{L}_{\Gamma}\lambda+\left\langle
\mathcal{L}_{\Gamma}e|\widetilde{T}\theta\right\rangle
\end{equation}
and hence:%
\begin{equation}
\mathcal{L}_{\Gamma}\lambda\equiv i_{\Gamma}d\lambda=0 \label{eigenvalues}%
\end{equation}
i.e., \textit{if }$T$ \textit{is invariant under the dynamics, so are the
eigenvalues of }$T$ .

Notice that, by Cartan's identity \cite{Mar10}:
\begin{equation}
\mathcal{L}_{e_{i}}\theta^{j}=\left\langle e_{i}|d\theta^{j}\right\rangle
+d\left\langle e_{i}|\theta^{j}\right\rangle
\end{equation}
and hence, if the (co)basis is holonomic, $d\theta^{j}=0$ (together with
$\left\langle e_{i}|\theta^{j}\right\rangle =\delta_{i}^{j}$) leads to:%
\begin{equation}
\mathcal{L}_{e_{i}}\theta^{j}=0\text{ }\forall i,j
\end{equation}
Then, for $i\neq  j$ we obtain:%
\begin{eqnarray}
\left(  \lambda_{i}-\lambda_{j}\right)  \mathcal{L}_{e_{i}}\left\langle
\Gamma|\theta^{j}\right\rangle 
&=&\lambda_{i}\left\langle \mathcal{L}_{e_{i}}\Gamma|\theta^{j}\right\rangle
-\left\langle \mathcal{L}_{e_{i}}\Gamma|\widetilde{T}\theta^{j}\right\rangle
=\\
&=&\lambda_{i}\,\left\langle \mathcal{L}_{e_{i}}\Gamma|\theta^{j}\right\rangle
-\left\langle T\left(  \mathcal{L}_{e_{i}}\Gamma\right)  |\theta
^{j}\right\rangle \nonumber =\\
&=&\lambda_{i}\left\langle \mathcal{L}_{e_{i}}\Gamma|\theta^{j}\right\rangle
-\left(  \mathcal{L}_{_{i}\Gamma}\lambda_{i}\right)  \left\langle e_{i}%
|\theta^{j}\right\rangle -\lambda_{i}\left\langle \mathcal{L}_{e_{i}}%
\Gamma|\theta^{j}\right\rangle =0 \nonumber
\end{eqnarray}
Hence:%
\begin{equation}
\mathcal{L}_{e_{i}}\left\langle \Gamma|\theta^{j}\right\rangle =0,\text{
\ }i\neq j
\end{equation}
and (cfr. Eqn.(\ref{separable1})) this proves \textit{separability} of
$\Gamma$. To be more explicit, we can write $T$ as:%
\begin{equation}
T=%
{\displaystyle\sum\limits_{i=1}^{n}}
\lambda_{i}%
{\displaystyle\sum\limits_{k=1}^{d_{i}}}
\frac{\partial}{\partial x^{\left(  i,k\right)  }}\otimes dx^{\left(
i,k\right)  }%
\end{equation}
where $n$ is the number of distinct eigenvalues and $d_{i}$ is the degeneracy
of the $i$-th eigenvalue. Finally, $\Gamma$ will be of the form already given
in Eqns.(\ref{separable2}) and (\ref{separable3}). On the eigenspaces of $T$
that are one-dimensional integrability of $\Gamma$ will be then essentially
trivial, and this case will not be considered further.

Proceeding further we obtain from Eqn.(\ref{distribution0}):%
\begin{equation}
0=\left(  \lambda_{i}-\lambda_{j}\right)  \left\langle e_{i}|d\lambda
_{j}\right\rangle =
\left\langle Te_{i}|d\lambda_{j}\right\rangle -\left\langle e_{i}|\lambda
_{j}d\lambda_{j}\right\rangle =
\left\langle e_{i}|\widetilde{T}d\lambda_{j}\right\rangle -\left\langle
e_{i}|\lambda_{j}d\lambda_{j}\right\rangle
\end{equation}
and hence:%
\begin{equation}
\widetilde{T}d\lambda_{j}=\lambda_{j}d\lambda_{j}%
\end{equation}
i.e. $d\lambda_{j}$ is an eigenform belonging to the eigenvalue $\lambda_{j}$.
Let us now assume the eigenvalues of $\Gamma$ to be \textit{doubly} degenerate
and functionally independent. This implies: $\dim\left(  \mathcal{M}\right)
=2n$ and:%

\begin{equation}
d\lambda_{1}\wedge d\lambda_{2}\wedge...\wedge d\lambda_{n}\neq0
\end{equation}
Then the $d\lambda_{i}$'s can be taken as half of the cobasis, and we can
write $T$ as:%
\begin{equation}
T=%
{\displaystyle\sum\limits_{i=1}^{n}}
\lambda_{i}\left(  e_{i}\otimes\theta^{i}+e_{n+i}\otimes d\lambda_{i}\right)
\end{equation}
With this choice, Eqn.(\ref{eigenvalues}) tells us that $\Gamma$ \ has no
components "along" the $d\lambda_{i}$'s, and that it is therefore of the form:%
\begin{equation}
\Gamma=%
{\displaystyle\sum\limits_{i=1}^{n}}
\Gamma^{i}e_{i} \label{gamma}%
\end{equation}
Proceeding further, closure of the $\theta^{i}$'s allow us to write:
$\theta^{i}=d\phi^{i}$, and hence: $e_{i}=\partial/\partial\phi^{i}$ for
$i=1,...,n$. The $\phi^{i}$'s are in general only locally defined (while the
$\lambda_{i}$'s are globally defined), and can be allowed to be angles. Hence
we can rewrite $T$ as:%
\begin{equation}
T=%
{\displaystyle\sum\limits_{i=1}^{n}}
\lambda_{i}\left(  \frac{\partial}{\partial\phi^{i}}\otimes d\phi^{i}%
+\frac{\partial}{\partial\lambda^{i}}\otimes d\lambda^{i}\right)
\end{equation}
and, in view of Eqn.(\ref{gamma}), $\Gamma$ will be of the form:%
\begin{equation}%
{\displaystyle\sum\limits_{i=1}^{n}}
\Gamma^{i}\left(  \lambda^{i},\phi^{i}\right)  \frac{\partial}{\partial
\phi^{i}}%
\end{equation}
The associated equations of motion will be:%
\begin{equation}
\begin{array}{l}
\frac{d}{dt}\phi^{i}=\Gamma^{i}\\
\\
\frac{d}{dt}\lambda^{i}=0
\end{array}
\; ;\text{ \ }i=1,..,n \label{dynamics}%
\end{equation}

Now, it is easy to show that the dynamical system (\ref{dynamics}) can be made
Hamiltonian with respect to a large family of symplectic structures. Indeed,
let's assume that no one of the $\Gamma^{i}$'s vanishes
identically\footnote{If they have isolated zeros, the closed set of the zeros,
which is an invariant subset, can be excluded from the manifold. The case in
which some component of $\Gamma$ vanishes has been discussed in
Ref.\cite{DSVM2}.}. Then, with any set of (smooth) functions $g_{i}%
=g_{i}\left(  \lambda^{i}\right)  $ we can associate the symplectic form:%
\begin{equation}
\omega=%
{\displaystyle\sum\limits_{i=1}^{n}}
f_{i}\left(  \lambda^{i},\phi^{i}\right)  d\phi^{i}\wedge d\lambda^{i}%
\end{equation}
where:%
\begin{equation}
f_{i}\left(  \lambda^{i},\phi^{i}\right)  =:\frac{g_{i}\left(  \lambda
^{i}\right)  }{\Gamma^{i}\left(  \lambda^{i},\phi^{i}\right)  }%
\end{equation}
and $\Gamma$ will be Hamiltonian:
\begin{equation}
i_{\Gamma}\omega=d\mathcal{H}%
\end{equation}
with:%
\begin{equation}
d\mathcal{H}=%
{\displaystyle\sum\limits_{i=1}^{n}}
g_{i}\left(  \lambda^{i}\right)  d\lambda^{i}%
\end{equation}
Therefore, under the assumption that there exists a $\left(  1,1\right)  $
diagonalizable tensor field $T$ \ invariant under the dynamics, with vanishing
Nijenhuis torsion and at most doubly degenerate and functionally independent
eigenvalues, what has been proved up to now is that \textit{the dynamical
vector field }$\Gamma$ \textit{is separable, integrable and, on the
eigenspaces of doubly degenerate eigenvalues, Hamiltonian. }$\blacksquare$

\bigskip

The equation $\mathcal{L}_{\Gamma}T=0$ expresses the invariance of the tensor
$T$ in intrinsic terms. It may be useful to write down the same condition in
the language of coordinates. If $\left(  x^{i},...,x^{m}\right)  $,
$m=(\dim\left(  \mathcal{M}\right)  )$ are local coordinates, and $T$ and
$\Gamma$ are given by:%
\begin{equation}
T=T^{i}\text{ }_{j}dx^{j}\otimes\frac{\partial}{\partial x^{i}};\text{
\ \ }\Gamma=\Gamma^{i}\frac{\partial}{\partial x^{i}}%
\end{equation}
then:%
\begin{equation}
\mathcal{L}_{\Gamma}T=\left\{  \mathcal{L}_{\Gamma}T^{i}\text{ }_{j}%
-\frac{\partial\Gamma^{i}}{\partial x^{k}}T^{k}\text{ }_{j}+T^{i}\text{ }%
_{k}\frac{\partial\Gamma^{k}}{\partial x^{j}}\right\}  dx^{j}\otimes
\frac{\partial}{\partial x^{i}}%
\end{equation}
and hence invariance under $\Gamma$ implies the \textit{matrix} equation:%
\begin{equation}
\mathcal{L}_{\Gamma}T=:\frac{d}{dt}T=\left[  C,T\right]  \label{Laxeq}%
\end{equation}
where, with abuse of notation, we have denoted by $T$ the $m\times m$ matrix:
$T=\left\Vert T^{i}\text{ }_{j}\right\Vert $and:%
\begin{equation}
C=\left\Vert C^{i}\text{ }_{j}\right\Vert ;\text{ \ }C^{i}\text{ }_{j}%
=:\frac{\partial\Gamma^{i}}{\partial x^{j}}%
\end{equation}
while $[.,.]$ denotes the usual commutator among matrices. Whenever two
matrices $C$ and $T$ satisfy Eqn.(\ref{Laxeq}) they are said to form a
\textit{Lax pair} \cite{Lax,Lax2,Lax3,Mar0,Vi}\footnote{We should notice that Eq.(\ref{Laxeq})
depends on the coordinate system we are using, and therefore has no intrinsic
meaning.}.

Whenever we may define a map $\mu$ from $\mathcal{M}$ to a space of matrices
such that the dynamics is $\mu$-related to a dynamics on the matrix space of
the form of Eq.(\ref{Laxeq}), we say that the original dynamics can be given a
"Lax form". This is what might be called also a "Heisenberg" form, and has
many general properties. For instance, the evolution of $T$ ruled by the
"Hamiltonian" $C$ is clearly isospectral.

Whenever it is possible to find a map from our carrier space to a
space of linear operators such that the dynamics on the carrier
space may be casted into the Heisenberg form we will say that our
dynamics may be put into the Lax form. 
As a matter of fact, by using the momentum map associated with the symplectic action of the unitary group on the Hilbert space or on the complex projective space (see below, Sect.\ref{se:gqm}), we may relate the Schr\"{o}dinger picture with the Heisenberg picture on the space of observables.

\subsection{From Liouville Integrability to Invariant Structures}
\bigskip

Reversing somehow our path, let's start by considering a dynamical system
$\Gamma$ \ that is Hamiltonian and completely integrable "a' la" Liouville.
Hence: $\dim\left(  \mathcal{M}\right)  =n$. Introducing action-angle
variables $\left(  I_{1},..,I_{n};\phi^{1},..,\phi^{n}\right)  $ in the
neighborhood of an Arnold- Liouville torus $\mathbb{T}^{n}$, we will have:%
\begin{equation}
dI_{1}\wedge dI_{2}\wedge...\wedge dI_{n}\neq0
\end{equation}
and the condition that the Hamiltonian $\mathcal{H}$ be a function of the
action variables alone can be written as:%
\begin{equation}
d\mathcal{H}\wedge dI_{1}\wedge...\wedge dI_{n}=0
\end{equation}
The symplectic form can be written as:%
\begin{equation}
\Omega=%
{\displaystyle\sum\limits_{k}}
d\phi^{k}\wedge dI_{k}
\end{equation}
and the vector field $\Gamma$ in action-angle variables will be given by:%
\begin{equation}
\Gamma=%
{\displaystyle\sum\limits_{k}}
\omega^{k}\frac{\partial}{\partial\phi^{k}};\text{ \ }\omega^{k}%
=:\frac{\partial\mathcal{H}}{\partial I_{k}}%
\end{equation}

Assume first that the Hamiltonian is separable:%
\begin{equation}
\mathcal{H}=%
{\displaystyle\sum\limits_{k}}
\mathcal{H}_{k}\left(  I_{k}\right)
\end{equation}
Then the class of $\left(  1,1\right)  $ tensor fields defined by:%
\begin{equation}
T=%
{\displaystyle\sum\limits_{k}}
\lambda_{k}\left(  I_{k}\right)  \left\{  dI_{k}\otimes\frac{\partial
}{\partial I_{k}}+d\phi^{k}\otimes\frac{\partial}{\partial\phi^{k}}\right\}
\end{equation}
with the $\lambda_{k}$'s \ arbitrary functions with nowhere vanishing
differential has all the required properties. Indeed:

\begin{itemize}
\item It is invariant under the dynamics;

\item It has doubly degenerate eigenvalues and:

\item It has vanishing Nijenhuis torsion.
\end{itemize}

This last property can be checked directly by testing Eqn.(\ref{Nijenhuis})
on: $\left(  X,Y\right)  =\left(  \partial/\partial I_{h},\partial/\partial
I_{k}\right)  ,\left(  \partial/\partial I_{h},\partial/\partial\phi
^{k}\right)  $ and $\left(  \partial/\partial\phi^{h},\partial/\partial
\phi^{k}\right)  \blacksquare$.

A second case in which an invariant $\left(  1,1\right)  $ tensor can be
constructed is the "non-resonant" case, i.e. when the Hamiltonian has a
non-vanishing Hessian:%
\begin{equation}
\det\left\Vert \frac{\partial^{2}\mathcal{H}}{\partial I_{h}\partial I_{k}%
}\right\Vert \neq0
\end{equation}
This means, of course:%
\begin{equation}
d\omega^{1}\wedge d\omega^{2}\wedge...\wedge d\omega^{n}\neq0
\end{equation}
Solving then for the $I$'s as functions of the $\omega$'s, we \ can use the
$\omega$'s as new coordinates and introduce\footnote{This change of variables
need not be a canonical transformation.} a new symplectic structure:%
\begin{equation}
\widetilde{\Omega}=%
{\displaystyle\sum\limits_{k}}
d\omega^{k}\wedge d\phi^{k}=%
{\displaystyle\sum\limits_{hk}}
\frac{\partial^{2}\mathcal{H}}{\partial I_{h}\partial I_{k}}dI_{h}\wedge
d\phi^{k}%
\end{equation}
and $\Gamma$ will be Hamiltonian with the separable Hamiltonian:%
\begin{equation}
\mathcal{H}=\frac{1}{2}%
{\displaystyle\sum\limits_{k}}
\left(  \omega^{k}\right)  ^{2}%
\end{equation}
The class of $\left(  1,1\right)  $ tensor fields will be given now by:%
\begin{equation}
T=%
{\displaystyle\sum\limits_{k}}
\lambda_{k}\left(  \omega^{k}\right)  \left\{  d\omega^{k}\otimes
\frac{\partial}{\partial\omega^{k}}+d\phi^{k}\otimes\frac{\partial}%
{\partial\phi^{k}}\right\}
\end{equation}

Complete integrability is also known to be related to the existence
of {\it recursion operators} \cite{DSVM1,LMV2,Mar17,ZK}. A brief account of
the latter is given in Appendix $B$. 

\newpage

%% file: Chapt3rev.tex
\section{Alternative Structures for Classical Systems} \label{ch:3}

\subsection{Preliminaries. A cursory look at the Inverse Problem in a classical
context} \label{se:look}
\bigskip

After having examined briefly in the previous Chapter the problem of the
integrability of a classical dynamical system, and before turning to the main
topic of this review, i.e. \textit{quantum} systems\footnote{What we
mean\ exactly by a "quantum" system will be specified in the next Chapter.},
we restate here in a very cursory way what is known in the literature as the
"Inverse Problem of Classical Dynamics".

Let then $\Gamma$ be a vector field on a (smooth) manifold $\mathcal{M}$. In a
nutshell, the Inverse Problem $\left(  IP\right)  $ can be formulated in (at
least\footnote{We will not consider here the Hamilton-Jacobi form of Classical
Dynamics, but see \cite{Muk0}}) three different, and often related,
contexts, namely:

\begin{itemize}
\item $IP1$: Lagrangian context \cite{Hav,Hel,MFVMR}. Let then $\mathcal{M}$ be the tangent bundle
of a smooth manifold $Q$, i.e. $\mathcal{M}=TQ$ equipped with tangent bundle
coordinates $\left(  q^{i},v^{i}\right)  $ such that $\Gamma\in\mathcal{X}%
\left(  TQ\right)  $ is a second-order vector field \cite{Mor}, i.e.:%
\begin{equation}
\Gamma=v^{i}\frac{\partial}{\partial q^{i}}+F^{i}\left(  q,v\right)
\frac{\partial}{\partial v^{i}}%
\end{equation}
The Lagrangian $IP$ amounts then to the following: find all the smooth
functions $\mathcal{L}=\mathcal{L}\left(  q,v\right)  \in\mathcal{F}\left(
TQ\right)  $ such that:%
\begin{equation}
\frac{\partial^{2}\mathcal{L}}{\partial v^{i}\partial v^{j}}F^{j}%
=\frac{\partial\mathcal{L}}{\partial q^{i}}-\frac{\partial^{2}\mathcal{L}%
}{\partial v^{i}\partial q^{j}}v^{j},\text{ \ }i=1,...,n=\dim Q \label{IP1}%
\end{equation}
It follows that if the \textit{Lagrangian} $\mathcal{L}$ is regular, i.e.:%
\begin{equation}
\det\left\Vert \frac{\partial^{2}\mathcal{L}}{\partial v^{i}\partial v^{j}%
}\right\Vert \neq0
\end{equation}
then the Euler-Lagrange equations can be put in normal form and,
via a Legendre transformation \cite{Ar,Mar10} one can go over to a
Hamiltonian description of the dynamical system on the cotangent
bundle $T^{\ast}Q$. We will not discuss this setting of the $IP$
any further, and refer for a full account of it to the literature
\ \cite{Mor}.

\item $IP2$: Hamiltonian context. Let instead $\mathcal{M}=T^{\ast}Q$ for some
smooth manifold $Q$ and $\Gamma\in\mathcal{X}\left(  T^{\ast}Q\right)  $. The
Hamiltonian $IP$ amounts then to finding all pairs $\left(  \omega
,\mathcal{H}\right)  $ with $\omega$ a symplectic form (a closed and
non-degenerate two-form) and $\mathcal{H}\in\mathcal{F}\left(  T^{\ast
}Q\right)  $ such that:%
\begin{equation}
i_{\Gamma}\omega=d\mathcal{H} \label{IP2}%
\end{equation}
At a local level, the problem reduces to finding all the closed and
non-degenerate two-forms $\omega$ such that:%
\begin{equation}
\mathcal{L}_{\Gamma}\omega=0 \label{IP3}%
\end{equation}
with $\mathcal{L}_{\Gamma}$ denoting the Lie derivative w.r.t. $\Gamma$, which
is a system of coupled $PDE^{\prime}s$ in \ $\left(
\begin{array}
[c]{c}%
n\\
2
\end{array}
\right)  =2n^{2}-n$ unknowns\footnote{Notice that, in  this as
well as in the previous case, $\mathcal{M}$ has obviously to be an \textit{even}%
-dimensional manifold.}. As a simple example, in a neighborhood
$U\subseteq M$ in which $\Gamma\neq0$ and defines a flow-box (the
"straightening-up-of-the-flux"
theorem \cite{Ar} holds) we can find coordinates $\left(  x_{0,}%
x_{1},...,x_{2n-1}\right)  $ such that $\Gamma=\partial/\partial x_{0}$ and
hence the problem has infinite solutions:%
\begin{equation}
\omega=dx_{0}\wedge df+a_{ij}df^{i}\wedge df^{j}%
\end{equation}
with: $a_{ij}=-a_{ji}\in\mathbb{R},$ $\det\left\Vert a_{ij}\right\Vert \neq0$
and: $\partial f/\partial x_{0}=\partial f^{i}/\partial x_{0}=0,$
$dx_{0}\wedge df\wedge df^{1}\wedge...\wedge df^{2n-2}\neq0$, and \ any such
$f$ will be an acceptable Hamiltonian ($i_{\Gamma}\omega=df$).

\item $IP3$: Poisson context \ \cite{CIMS,DGMS}. $\mathcal{M}$ is assumed here
to be a Poisson manifold \ \cite{Mar10}. In local coordinates $x^{i},$
$i=1,...,\dim\mathcal{M}$, and the $IP$ in this context amounts to finding all
pairs $\left(  \{.,\},\mathcal{H}\right)  $ with $\{.,\}$ a (possibly
degenerate\footnote{Which will be certainly the case if $\mathcal{M}$ is
\textit{odd}-dimensional.}) Poisson bracket \ and $\mathcal{H\in}$
$\mathcal{F}\{\mathcal{M}\}$ such that:%
\begin{equation}
\left\{  x^{i},\mathcal{H}\right\}  =\frac{dx^{i}}{dt};\text{ \ }\left\{
\left\{  x^{i},\mathcal{H}\right\}  ,\mathcal{H}\right\}  =F^{i}\left(
x,\left\{  x^{i},\mathcal{H}\right\}  \right)
\end{equation}

\end{itemize}

\subsection{\ The Hamiltonian Inverse Problem for linear vector fields} \label{se:invlin}
\bigskip

In view of the fact that what we are interested in this paper is a
theory that is usually casted in a linear setting, i.e. Quantum
Mechanics on Hilbert spaces, we will review here\cite{GMR} the
Inverse Problem in the Hamiltonian
context for linear vector fields, and we will assume: $\mathcal{M}%
=\mathbb{R}^{2n}$ for some $n$. In the appropriate coordinates, a
\textit{linear} vector field is then a vector field of the form:%
\begin{equation}
\Gamma=G^{i}\text{ }_{j}x^{j}\frac{\partial}{\partial x^{i}},\text{ \ }%
G^{i}\text{ }_{j}\in\mathbb{R} \label{linvec}%
\end{equation}
and the matrix $\left\Vert G^{i}\text{ }_{j}\right\Vert $ (which represents a
$\left(  1,1\right)  $-type tensor field) will be non-degenerate iff the
origin is an isolated fixed point of $\Gamma$.

\bigskip

\textbf{A Digression on: "Extracting the linear part" of a vector
field.}
In general, let $\mathcal{M}$ be a smooth manifold and $\Gamma\in
\mathfrak{X}\left(  \mathcal{M}\right)  $ be a vector field with
an isolated fixed point at $m_{0}\in\mathcal{M}$: $\Gamma\left(
m_{0}\right)  =0$. Considering then, for an arbitrary vector field
$Y\in\mathfrak{X}\left( \mathcal{M}\right)  $ and function
$f\in\mathcal{F}\left(  \mathcal{M}\right) $ the quantity
$\mathcal{L}_{Y}\left(  \mathcal{L}_{\Gamma}f\right)  \left(
m_{0}\right)  $, it is not hard to see that it is linear in $Y$
\textit{and}, by virtue of $\Gamma\left(  m_{0}\right)  =0$, in
$df$. it defines then a $\left(  1,1\right)  $
tensor\footnote{\textit{Not} a tensor field, in
general.} $\ T_{\Gamma}$ at $m_{0}$:%
\begin{equation}
\mathcal{L}_{Y}\left(  \mathcal{L}_{\Gamma}f\right)  \left(
m_{0}\right)
=T_{\Gamma}\left(  df,Y\right)  \left(  m_{0}\right)  \label{tensor}%
\end{equation}
Then, the \textit{linear part} of $\Gamma$ at $m_{0}$,
$\Gamma_{0}$, will be
defined as:%
\begin{equation}
\Gamma_{0}=T_{\Gamma}\left(  \Delta\right)
\end{equation}
with $\Delta$ the Liouville field.

Indeed, in the domain of a chart $\left(  x^{1},...,x^{n}\right)  $
$\left( n=\dim\mathcal{M}\right)  $ with the origin at $m_{0}$ and:
$\Gamma=\Gamma
^{i}\partial/\partial x^{i},Y=Y^{i}\partial/\partial x^{i}$: $\mathcal{L}%
_{Y}\left(  \mathcal{L}_{\Gamma}f\right)  =Y^{i}\partial\left(
\Gamma ^{j}\partial f/\partial x^{j}\right)  /\partial
x^{i}=Y^{i}\left(
\partial\Gamma^{j}/\partial x^{i}\right)  \left(  \partial f/\partial
x^{j}\right)  +Y^{i}\Gamma^{j}\left(  \partial^{2}f/\partial
x^{i}\partial
x^{j}\right)  T.$ But the second term vanishes at $m_{0}=0$, and hence:%
\begin{equation}
T_{\Gamma}=T^{j}\text{ }_{i}dx^{i}\otimes\frac{\partial}{\partial x^{j}%
};\text{ }T^{j}\text{ }_{i}=\frac{\partial\Gamma^{j}}{\partial x^{i}}|_{m_{0}%
}\text{\ }%
\end{equation}
and:%
\begin{equation}
\Gamma_{0}=T_{\Gamma}\left(  \Delta\right)  =\left(  \frac{\partial\Gamma^{j}%
}{\partial x^{i}}|_{m_{0}}\right)  x^{i}\frac{\partial}{\partial x^{j}}%
\end{equation}
This is of course what one would have guessed on much more
elementary grounds. The advantage of the definition (\ref{tensor})
is that it provides a tensorial characterization of the linear
part of a vector field at a critical point.

\bigskip

In a shorthand notation we can write $\Gamma$ as:%
\begin{equation}
\Gamma=\left(  \widetilde{\mathbb{G}x},\partial/\partial x\right)
\end{equation}
where: $\left(  \mathbb{G}x\right)  ^{i}=G^{i}$ $_{j}x^{j}$, "$\sim$" stands
for the transpose:
\begin{equation}
\frac{\partial}{\partial x}=\left\vert
\begin{array}
[c]{c}%
\partial/\partial x^{1}\\
.\\
.\\
.\\
\partial/\partial x^{2n}%
\end{array}
\right\vert
\end{equation}
and: $\left(  a,b\right)  =:a^{i}b_{i}$.

A symplectic form can be written as:%
\begin{equation}
\omega=\frac{1}{2}\Omega_{ij}dx^{i}\wedge dx^{j}%
\end{equation}
and the matrix: \ $\Omega\mathbb{=}\left\Vert \Omega_{ij}\right\Vert $ will be
(pointwise) skew-symmetric and non-degenerate. $\omega$ will be said to be a
\textit{constant} symplectic form iff the $\Omega_{ij}$'s are constant. If:%
\begin{equation}
\Omega=\left\vert
\begin{array}
[c]{cc}%
\mathbf{0}_{n\times n} & \mathbb{I}_{n\times n}\\
-\mathbb{I}_{n\times n} & \mathbf{0}_{n\times n}%
\end{array}
\right\vert
\end{equation}
$\omega$ will be said to be in the \textit{canonical (or Darboux) form.} If
$\Gamma$ is linear and Hamiltonian w.r.t. a constant symplectic form, then the
Hamiltonian is forced to be a quadratic function, i.e.:%
\begin{equation}
\mathcal{H}=\frac{1}{2}H_{ij}x^{i}x^{j},\text{ \ }H_{ij}\in\mathbb{R}%
\end{equation}
\begin{remark}
 The above is clearly a coordinate-dependent definition of a
quadratic function. A coordinate (and dimension)-free
characterization of quadratic functions, and one that is more suitable in the case of (infinite-dimensional) Hilbert spaces, can be given as follows. A
\ mapping: $\mathcal{H}:\mathbb{V}\rightarrow \mathbb{V}^{\prime}$
with $\mathbb{V},\mathbb{V}^{\prime}$  vector  spaces (over a
field $\mathbb{K}$, with $\mathbb{K}=$ $\mathbb{R}$ or
$\mathbb{C}$) is \textit{quadratic}(a \textit{quadratic function}
if $\mathbb{V}^{\prime }=\mathbb{R}$ or $\mathbb{C}$) if:

\begin{itemize}
\item
\begin{equation}
\mathcal{H}\left(  \lambda x\right)  =\lambda^{2}\mathcal{H}\left(
x\right)
,\text{ \ }\forall x\in\mathbb{V},\text{ }\lambda\in\mathbb{K}%
\end{equation}
and:
\item
\begin{equation}
b\left(  x,y\right)  =:\mathcal{H}\left(  x+y\right)
-\mathcal{H}\left( x\right)  -\mathcal{H}\left(  y\right)
\end{equation}
is a bilinear mapping for all $x,y\in\mathbb{V}$.
\end{itemize}
\end{remark}

\begin{remark}
Notice that, while $\mathbb{G}$ is a $\left(
1,1\right) $-type tensor (it "maps vectors to vectors") $\Omega$
and $\mathbb{H=}\left\Vert H_{ij}\right\Vert $ are $\left(
0,2\right) $-type tensors (they "map vectors to covectors" (and
viceversa in both cases)). This difference manifests itself in the
transformation under a general change of coordinates. If:
$x^{i}=T^{i}$
$_{j}y^{\text{,}j}$, then:%
\begin{equation}
\mathbb{G\rightarrow G}^{\prime}=\mathbb{T}^{-1}\mathbb{GT}%
\end{equation}
while ($\widetilde{\mathbb{T}}$ standing for the transpose of $\mathbb{T}$):%
\begin{equation}
\Omega\rightarrow\Omega^{\prime}=\widetilde{\mathbb{T}}\Omega\mathbb{T},\text{
\ }\mathbb{H}\rightarrow\text{\ }\mathbb{H}^{\prime}=\widetilde{\mathbb{T}%
}\mathbb{HT}%
\end{equation}
(the difference is not apparent when
$\mathbb{T}^{-1}=\widetilde{\mathbb{T}}$, i.e. $\mathbb{T}$ $\ $is
an orthogonal transformation, $\mathbb{T\in O}\left( 2n\right) $).

\end{remark}

Restricting from now on to linear vector fields and constant symplectic
structures, and omitting the superscripts and suffixes $"0"$, if $\Lambda$ is
the Poisson tensor ($\left\{  x^{i},x^{j}\right\}  =\Lambda^{ij}$,
$\Lambda^{ij}\Omega_{jk}=\delta^{i}$ $_{k}$), then if \ $\Gamma=G^{i}$
$_{j}x^{j}\partial/\partial x^{i}$ is Hamiltonian w.r.t. $\omega=\left(
1/2\right)  \Omega_{ij}dx^{i}\wedge dx^{j}$, this implies:%
\begin{equation}
\Omega G=-H\text{ \ }\label{product1}%
\end{equation}
and, equivalently:%
\begin{equation}
G=-\Lambda H\label{product2}%
\end{equation}

Hence: \textit{Looking for a Hamiltonian description w.r.t. a
constant symplectic structure for a linear vector field $\Gamma$ \
is therefore equivalent to looking for the decomposition of the
representative matrix $G$ into the product of an
\textbf{invertible} skew-symmetric matrix $\Lambda$ and a
symmetric matrix $H$. The former will provide a (non-degenerate)
Poisson structure, the latter a Hamiltonian adapted to the given
Poisson structure}\footnote{$\Lambda$ will be a $\left( 2,0\right)
$-type tensor, and under a general linear change of coordinates
(see above) will transform as: $\Lambda\rightarrow\Lambda^{\prime
}=T^{-1}\Lambda\widetilde{\left(  T^{-1}\right)  \text{.}}$}.

\bigskip

 At this point we can make contact with the discussion of
Ch.\ref{introduc}, where we dealt with linear Hamiltonian vector
fields on a finite-dimensional Hilbert space. There it was shown
that the Hermitian structure gives rise to both a metric tensor
and a symplectic form, and that the two are compatible in the
sense that they are connected to one another by a third structure,
the complex structure $J$. Here too we can reconstruct a
(compatible\footnote{See Ref.\cite{Mar7} and the following
Ch.\ref{QM}.}) complex structure starting from the tensors
$\Lambda$ and $H$, at least in the case when $H$ is
positive-definite.  If such is the case, we can find, as already
discussed elsewhere, a system of coordinates  in which the vector
field $\Gamma$ is given explicitly as a sum of independent
harmonic oscillators with proper
frequencies $\nu_{1},..,\nu_{n}$ (possibly not all distinct):%
\begin{equation}
\Gamma=%
{\displaystyle\sum\limits_{i=1}^{n}}
\nu_{i}\Gamma_{i}^{\left(  0\right)  };\text{ }\Gamma_{i}^{\left(
0\right) }=x_{i+n}\frac{\partial}{\partial
x_{i}}-x_{i}\frac{\partial}{\partial x_{i+n}},\text{ }i=1,...,n
\end{equation}
i.e.:%
\begin{equation}
G=\left\vert
\begin{array}
[c]{cc}%
\mathbf{0}_{n\times n} & \nu\\
-\nu & \mathbf{0}_{n\times n}%
\end{array}
\right\vert
\end{equation}
where: $\nu=diag\left(  \nu_{1},..,\nu_{n}\right)  $, with the
standard
Poisson tensor:%
\begin{equation}
\Lambda=\frac{1}{2}\Lambda_{ij}\frac{\partial}{\partial
x_{i}}\wedge
\frac{\partial}{\partial x_{j}}%
\end{equation}
whose representative matrix will be:%
\begin{equation}
\Lambda=\left\vert \Lambda_{ij}\right\vert =\left\vert
\begin{array}
[c]{cc}%
\mathbf{0}_{n\times n} & -\mathbb{I}_{n\times n}\\
\mathbb{I}_{n\times n} & \mathbf{0}_{n\times n}%
\end{array}
\right\vert \label{repq}%
\end{equation}
and Hamiltonian: $H=\left(  1/2\right)  \sum_{i}\nu_{i}\left(  x_{i}%
^{2}+x_{i+n}^{2}\right)  $. It is now clear that the vector field:%
\begin{equation}
\Gamma^{\left(  0\right)  }=%
{\displaystyle\sum\limits_{i=1}^{n}}
\Gamma_{i}^{\left(  0\right)  }%
\end{equation}
will be Hamiltonian with a new Hamiltonian: $H^{\prime}=$ $\left(
1/2\right) \sum_{i}\left(  x_{i}^{2}+x_{i+n}^{2}\right)  $ and
that, in terms of the
representative matrices:%
\begin{equation}
\left(  \Lambda H^{\prime}\right)  ^{2}=-\mathbb{I}%
\end{equation}
i.e. that the $\left(  1,1\right)  $ tensor $\Lambda H^{\prime}$
(whose representative matrix will coincide with the matrix
(\ref{repq})) will provide the required complex structure.

\bigskip

Some (necessary) consequences of $\Gamma$ being Hamiltonian have
been drawn in Ref.\cite{GMR}, namely:

\begin{enumerate}
\item As $\widetilde{G}=H\Lambda=\Lambda\left(  \Lambda^{-1}H\Lambda\right)
$, $\widetilde{G}$ is a representative of a vector field which is Hamiltonian
w.r.t. the same Poisson structure with Hamiltonian: $-\Lambda^{-1}H\Lambda$.
Indeed, in the basis in which $\Lambda$ has the standard form, i.e.:%
\begin{equation}
\Lambda=\left\vert
\begin{array}
[c]{cc}%
0_{n\times n} & -\mathbb{I}_{n\times n}\\
\mathbb{I}_{n\times n} & 0_{n\times n}%
\end{array}
\right\vert
\end{equation}
$\Lambda^{-1}=-\Lambda$. Hence: $-\Lambda^{-1}H\Lambda=\Lambda
H\Lambda$, which is symmetric. Notice, however, that in general
$\widetilde{G}$ and $G$ will not commute, nor will then the
associated vector fields.

\item $G^{3}=-\Lambda H\Lambda H\Lambda H=\Lambda\left(  H\Lambda\right)
H\left(  -\Lambda H\right)  =\Lambda\left(  \widetilde{G}HG\right)  $. More
generally, $G^{2k+1}$ can be written as:%
\begin{equation}
G^{2k+1}=-\underset{2k+1}{\underbrace{\Lambda H...\Lambda H}}\text{ }%
=-\Lambda\underset{k}{\underbrace{H\Lambda...H\Lambda}}H\underset
{k}{\underbrace{\Lambda H...\Lambda H}}%
\end{equation}
i.e.:%
\begin{equation}
G^{2k+1}=-\left(  -\right)  ^{k}\Lambda\left(  \widetilde{G}^{k}HG^{k}\right)
\end{equation}
Hence: $G^{2k+1}$ \textit{will represent a Hamiltonian vector field }%
$\Gamma_{k}$ \textit{with the Hamiltonian:}%
\begin{equation}
\mathcal{H}_{k}=\frac{1}{2}\left(  -\right)  ^{k}\left(  \widetilde{G}%
^{k}HG^{k}\right)  _{ij}x^{i}x^{j};\text{ }\mathcal{H}_{0}=\mathcal{H}%
\end{equation}
\textit{ } \textit{w.r.t. the }\textbf{same}\textit{ Poisson structure. As the
correspondence between matrices and linear vector fields is a Lie algebra
homomorphism, all these Hamiltonian vector fields will commute pairwise. As
the correspondence between linear vector fields and \ Hamiltonian functions is
a Lie algebra antihomomorphism\footnote{The Lie algebra on functions being
defined by the Poisson bracket. Recall that: $\left\{  f,g\right\}  =i_{X_{g}%
}i_{X_{f}}\omega=L_{X_{g}}f=-L_{X_{f}}g$, with $X_{f},X_{g}$ the associated
Hamiltonian vector fields, and that, for any two vector fields $X$ and $Y$:
$i_{[X,Y]}=L_{X}i_{Y}-i_{X}L_{Y}$. Therefore: $i_{[X_{f},X_{g}]}%
\omega=-d\left\{  f,g\right\}  $.}, in the linear case }$\mathcal{H}_{k}$
\textit{will be a constant of the motion for }$\Gamma_{k^{\prime}}$ $\forall
k,k^{\prime}$\textit{, and they will be pairwise in involution\footnote{Notice
that, in general (see the previous footnote): $i_{\left[  X_{f},X_{g}\right]
}\omega=-d\left\{  f,g\right\}  $ and that, therefore: $\left[  X_{f},X_{g}\right]  =0$ only implies in general: $\left\{  f,g\right\}  =const.$For
linear vector fields, however, both $f$ and $g$ will be quadratic functions.
The Poisson bracket $\left\{  f,g\right\}  $ will be quadratic as well, and it
will be constant iff it vanishes. }.}
\end{enumerate}

\begin{remark}
If $G$ is generic (and Hamiltonian), we will generate in this
way also a maximal set of (i.e. $n$) constants of the motion pairwise in
involution, and $\Gamma$ will be completely integrable a' la Liouville.
\end{remark}

$iii)$ As: $\widetilde{G}=H\Lambda=\Lambda^{-1}(\Lambda H)\Lambda=\Lambda
^{-1}(-G)\Lambda\Rightarrow TrG=0$ it follows that:%
\begin{equation}
TrG^{2k+1}=0\text{ \ }\forall k
\end{equation}

\textbf{Notes}. 

$a)$ That this is a necessary condition for the representative
matrix of a Hamiltonian vector field is pretty obvious. Indeed, for any vector
field $\Gamma$ on a symplectic $2n$-dimensional manifold, the divergence of
$\Gamma$ is defined by:%
\begin{equation}
\mathcal{L}_{\Gamma}\omega^{n}=:\left(  div\Gamma\right)  \omega^{n}%
\end{equation}
where $\omega$ is the symplectic form and $\omega^{n}$ the
symplectic volume. If the flow associated with $\Gamma$ is
Hamiltonian, it must be volume-preserving (Liouville's theorem
\cite{Ar}), and this implies: $div\Gamma=0$. But it is easy to
prove that, for a linear vector field and for a constant
symplectic structure: $div\Gamma=TrG$.

$b)$ The vanishing of the trace of odd powers of $G$ implies that the
characteristic polynomial $P\left(  \lambda\right)  $ will contain only
\textit{even }powers of $\lambda$ (i.e. $P(\lambda)$ will be actually a
polynomial in $\lambda^{2}$ of degree $n$). Real roots will appear then in
pairs $\left(  \lambda,-\lambda\right)  $ and (the coefficients of $P\left(
\lambda\right)  $ being real) complex roots will appear in quadruples $\left(
\lambda,\overline{\lambda},-\lambda,-\overline{\lambda}\right)  $.

$c)$ If $T$ is an invertible matrix:
\begin{equation}
T^{-1}GT=-T^{-1}\left(  \Lambda H\right)  T=-\left(  T^{-1}\Lambda
\widetilde{\left(  T^{-1}\right)  }\right)  \left(  \widetilde{T}HT\right)
\end{equation}
Then, if $T$ is in the \textit{commutant} of $G$ \ ($\left[  T,G\right]  =0$)
we find a new Hamiltonian description ($H^{\prime}=\widetilde{T}HT$) with a
new Poisson structure $(\Lambda^{\prime}=T^{-1}\Lambda\widetilde{\left(
T^{-1}\right)  })$  provided: $T^{-1}\Lambda\widetilde{\left(  T^{-1}\right)
}\neq\Lambda$. This implies that $T$ \ \ \textbf{be not} a canonical
transformation. \textit{Any "non-canonical" matrix }$T$ \textit{in the
commutant of }$G$ \textit{will provide a new Hamiltonian description for the
same vector field.}\\

Powers of $G$ are of course in the commutant of $G$. From:
$\Omega G=-H$ $\ $we obtain ($H$ being symmetric and $\Omega$ skew-symmetric):
$\widetilde{G}\Omega=H$ and hence:%
\begin{equation}
\widetilde{G}\Omega=-\Omega G
\end{equation}
It is then easy to prove that, in general:
\begin{equation}
\widetilde{G}^{h}\Omega=(-)^{h}\Omega G^{h}%
\end{equation}
Indeed, this holds for $h=1$. By induction: $\widetilde{G}^{h+1}\Omega
=(-)^{h}\widetilde{G}\Omega G^{h}=\left(  -\right)  ^{h}\left(  -\Omega
G\right)  G^{h}=\left(  -\right)  ^{h+1}\Omega G^{h+1}.$

As ($\Omega$ being skew-symmetric):
\begin{equation}
\Omega G^{h}=-\widetilde{\left(  \widetilde{G}^{h}\Omega\right)  }%
\end{equation}
this result implies:%
\begin{equation}
\widetilde{G}^{h}\Omega=(-)^{h+1}\widetilde{\left(  \widetilde{G}^{h}%
\Omega\right)  }%
\end{equation}
and hence $\widetilde{G}^{h}\Omega$ will be \textit{symmetric} for $h$ odd
(and, indeed, for $h=2k+1$, $\widetilde{G}^{2k+1}\Omega=-\mathcal{H}_{k}$) and
\ \textit{skew}-symmetric for even $h=2k$. Moreover:
\begin{equation}
\widetilde{G}(\widetilde{G}^{2k}\Omega)=-(\widetilde{G}^{2k}\Omega)G
\label{skew}%
\end{equation}
i.e. $\widetilde{G}^{2k}\Omega$ will be an admissible skew-symmetric factor in
the decomposition of $G$.

A slightly different way \cite{Man} to exploit even powers of $G$ to generate
alternative Hamiltonian descriptions is as follows (basically, we are
reverting from a finite to an infinitesimal description). Let, e.g.,
$\Gamma_{\left(  2\right)  }$ be the linear vector field associated with
$G^{2}$, i.e.:%
\begin{equation}
\Gamma_{\left(  2\right)  }=\left(  G^{2}\right)  ^{i}\text{ }_{j}x^{j}%
\frac{\partial}{\partial x^{i}}%
\end{equation}

If the Poisson structure is given by:%
\begin{equation}
\Lambda=\frac{1}{2}\Lambda^{hk}\frac{\partial}{\partial x^{h}}\wedge
\frac{\partial}{\partial x^{k}}%
\end{equation}
then:%
\begin{equation}
\mathcal{L}_{\Gamma_{\left(  2\right)  }}\Lambda=-\left(  G^{2}\Lambda\right)
^{ij}\frac{\partial}{\partial x^{i}}\wedge\frac{\partial}{\partial x^{j}%
}=-\left(  \Lambda H\Lambda H\Lambda\right)  ^{ij}\frac{\partial}{\partial
x^{i}}\wedge\frac{\partial}{\partial x^{j}}%
\end{equation}

Notice that $G^{2}\Lambda=\Lambda H\Lambda H\Lambda$ is manifestly
skew-symmetric. Therefore $\mathcal{L}_{\Gamma_{\left(  2\right)  }}\Lambda$,
if it does not vanish, defines a new Poisson structure:%
\begin{equation}
\Lambda_{\left(  2\right)  }=\frac{1}{2}\left(  \Lambda H\Lambda
H\Lambda\right)  ^{ij}\frac{\partial}{\partial x^{i}}\wedge\frac{\partial
}{\partial x^{j}}%
\end{equation}
and Poisson brackets:%
\begin{equation}
\left\{  f,g\right\}  _{\left(  2\right)  }=\left(  \Lambda H\Lambda
H\Lambda\right)  ^{ij}\frac{\partial f}{\partial x^{i}}\frac{\partial
g}{\partial x^{j}}%
\end{equation}

The new Poisson structure will be non-degenerate iff both $\Lambda$ and $H$
are invertible, i.e., as $G=-\Lambda H$, iff $G$ is invertible. Requiring then
that there exists a new Hamiltonian $\mathcal{H}_{\left(  2\right)  }$ s.t.
$\Gamma$ is \ again Hamiltonian w.r.t. the new Poisson structure, i.e.:%
\begin{equation}
\left\{  x^{i},\mathcal{H}_{\left(  2\right)  }\right\}  =\left(
\Lambda_{\left(  2\right)  }\right)  ^{ij}\frac{\partial\mathcal{H}_{\left(
2\right)  }}{\partial x^{j}}=G^{i}\text{ }_{j}x^{j}%
\end{equation}
together with $G=-\Lambda H$ leads to:%
\begin{equation}
\mathcal{H}_{\left(  2\right)  }=\frac{1}{2}H_{\left(  2\right)  ij}x^{i}%
x^{j};\text{ }H_{\left(  2\right)  }=\left(  \Lambda H\Lambda\right)  ^{-1}%
\end{equation}

If $G$ is not invertible, then one can proceed by exponentiation
\cite{GMR,Man}.

\bigskip

\begin{example}
We have seen in Ch.$1$ how the dynamics of a
quantum system separates into that of a set of
non-interacting\textbf{ }\ harmonic oscillators. All finite-level
quantum systems can be written as a family of harmonic oscillators
with frequencies related to the eigenvalues of the Hamiltonian. It
is therefore appropriate to consider here again the harmonic
oscillator. For this system the above procedure (i.e. taking Lie
derivatives of the Poisson structure) provides alternative
Hamiltonian descriptions.
Proceeding instead as in the previous discussion with $T=G^{2}$ and:%
\begin{equation}
G=\left\vert
\begin{array}
[c]{cccc}%
0 & 0 & 1/m & 0\\
0 & 0 & 0 & 1/m\\
-m\Omega_{1}^{2} & 0 & 0 & 0\\
0 & -m\Omega_{2}^{2} & 0 & 0
\end{array}
\right\vert
\end{equation}
and:%
\begin{equation}
\Lambda=\left\vert
\begin{array}
[c]{cccc}%
0 & 0 & 1 & 0\\
0 & 0 & 0 & 1\\
-1 & 0 & 0 & 0\\
0 & -1 & 0 & 0
\end{array}
\right\vert
\end{equation}
one finds:%
\begin{equation}
G^{2}=\left\vert
\begin{array}
[c]{cccc}%
-\Omega_{1}^{2} & 0 & 0 & 0\\
0 & -\Omega_{2}^{2} & 0 & 0\\
0 & 0 & -\Omega_{1}^{2} & 0\\
0 & 0 & 0 & -\Omega_{2}^{2}%
\end{array}
\right\vert
\end{equation}
and:%
\begin{equation}
G^{-2}\Lambda\widetilde{G}^{-2}=\left\vert
\begin{array}
[c]{cccc}%
0 & 0 & 1/\Omega_{1}^{4} & 0\\
0 & 0 & 0 & 1/\Omega_{2}^{4}\\
-1/\Omega_{1}^{4} & 0 & 0 & 0\\
0 & -1/\Omega_{2}^{4} & 0 & 0
\end{array}
\right\vert
\end{equation}
So, but for the isotropic case $\Omega_{1}=\Omega_{2}$ in which $G^{-2}%
\Lambda\widetilde{G}^{-2}$ and \ $\Lambda H\Lambda H\Lambda$ become
proportional, the two approaches appear to be genuinely different.

\end{example}

\begin{example}

As a last (almost trivial but explanatory) example let us take the most general linear vector field in ${\bf R}^2= \{(x,y)\}$:
\begin{equation}
\Gamma = (ax+by) \frac{\partial}{\partial x} +  (cx+dy) \frac{\partial}{\partial y}
\end{equation}
corresponding to the matrix
\begin{equation}
G = \left| \begin{array}{cc} a & b \\ c & d \end{array} \right| \; , \;  a,b,c,d \in \bf{R} . \label{mg}
\end{equation}
with $TrG^{2k+1}= 0$ if and only if $a=-d$. Given then the constant symplectic structure $\Omega = \alpha dx\wedge dy$ ($\alpha\in \bf{R}$):
\begin{equation}
\Omega = \alpha \left| \begin{array}{cc} 0 & 1 \\ -1 & 0 \end{array} \right|  .
\end{equation}

$\Gamma$ will be Hamiltonian with Hamiltonian: $H=\alpha
axy+\alpha\left(  by^{2}-cx^{2}\right)  /2$, corresponding to:
$H=-\Omega G$.

\bigskip

Three situations are possible:
\begin{enumerate}
\item The eigenvalues of $G$ are $\pm \lambda; \lambda \equiv \sqrt{ a^2 +bc} \in {\bf R}$. Then there exist coordinates $(x,y)$ such that the matrix (\ref{mg}) is of the form
\begin{equation}
G = \left| \begin{array}{cc} 0 & \lambda \\ \lambda & 0 \end{array} \right| .
\end{equation}
If we set:
\begin{equation}
x = A \cosh \Phi \; , \; y=A \sinh\Phi
\end{equation}
then:
\begin{eqnarray}
&\Omega = dH \wedge d\Phi \, \nonumber \\
& H = \frac{\lambda \alpha}{2} A^2. \label{wh}
\end{eqnarray}
\item The eigenvalues of $G$ are $\pm i \lambda; \lambda \equiv \sqrt{ |a^2 +bc|} \in {\bf R}$. Then $G$ may be put in the form:
\begin{equation}
G = \left| \begin{array}{cc} 0 & \lambda \\ -\lambda & 0 \end{array} \right| .
\end{equation}
We can now define
\begin{equation}
x = A \cos \Phi \; , \; y=A \sin\Phi
\end{equation}
that allow to write the symplectic form and the hamiltonian as in (\ref{wh}).
\item Finally we consider the case $a^2 +bc =0$, when there exist coordinates $(x,y)$ such that $G$ assumes the form
\begin{equation}
G = \left| \begin{array}{cc} 0 & 1 \\ 0 & 0 \end{array} \right| .
\end{equation}
Now $H= \alpha y^2 /2$ .

\end{enumerate}
\end{example}

Returning now to the general case, we have seen that a necessary condition for
a linear vector field $\Gamma$ with representative matrix $G$ to be
Hamiltonian is that the traces of odd powers of $G$ vanish. Whether or not
this is also sufficient requires a rather long analysis of the decomposition
of $G$ into Jordan blocks \cite{BM}, for whose details we refer to the
literature, and whose main result is contained in the following 
\cite{GMR}:

\begin{proposition}
A linear vector field $\Gamma$ is Hamiltonian iff the representative
matrix $G$ satisfies $TrG^{2k+1}=0$ and:

i) no further condition if the eigenvalues are
non-degenerate or purely imaginary,

ii) for degenerate real or genuinely complex (i.e not purely
imaginary) eigenvalues the Jordan block belonging to a given eigenvalue
$\lambda$ has the same structure as the block belonging to
$-\lambda$, this meaning that the Jordan block associated with the
eigenvalue $\lambda$ can be brought to the form:
\begin{equation}
G_{\left\{  \lambda\right\}  }=\left\Vert
\begin{array}
[c]{cc}%
\mathbb{J} & \mathbf{0}\\
\mathbf{0} & -\widetilde{\mathbb{J}}%
\end{array}
\right\Vert
\end{equation}

iii) zero eigenvalues have even multiplicity.
\end{proposition}

\bigskip

This solves the problem of under which conditions a linear vector field is
Hamiltonian, but does not tell us how many genuinely different Hamiltonians
(and symplectic structures) are permissible for a given vector field. A more
stringent result has also been proved in Ref.\cite{GMR} and precisely that:
\begin{proposition}
\textit{If $\Gamma$ has non-complex (i.e. either real or purely
imaginary) non-degenerate eigenvalues, then it has a minimal
family} ( a "pencil" \cite{GZ,IMM}) \textit{of equivalent
admissible symplectic forms parametrized by a number of parameters
equal to the number of couples $\left(  \lambda ,-\lambda\right) $
of eigenvalues minus one (i.e. a $\left(  n-1\right) $-parameter
family)}.
\end{proposition}

\bigskip

The case in which $\Gamma$ has (only) purely imaginary eigenvalues is of
particular interest for the analysis of (finite-dimensional, for the time
being) quantum systems. Indeed, we can remark that:

\begin{itemize}
\item If the eigenvalues are purely imaginary, then all the
motions of the system will be stable \cite{Ar,Ar1}. Considering
the decomposition: $G=-\Lambda H$ of Eq.(\ref{product2}), if $H$
is positive, it will define an Euclidean metric\footnote{Or \ a
pseudo-Euclidean one if it is non-degenerate but not necessarily
positive. } and, after possibly a rescaling that will be discussed
in the next Chapter, $\Lambda$ will define the Poisson tensor and
$G$ will become the complex structure. The system will become what
we will call a \textit{quantum system}, and that because the
evolution is unitary with respect to the Hermitian structure
associated with $G$ and $\Lambda$. In this sense, as we will see
shortly, the analysis of this Chapter provides also a way to
classify the possible, and alternative, Hamiltonian descriptions
for quantum systems.

\item With reference in particular to Ch.$1$, if the (quantum)
Hamiltonian $H$ has a real spectrum, then (cfr.
Eq.(\ref{vectorfield1})) (the realified of) $-iH/\hbar$ will turn
out to have purely imaginary eigenvalues. Even if $H$ is not
Hermitian w.r.t. the given Hermitian structure, one can always
find \cite{Bec,Mar141,Ve} a modified scalar product (see again Ch.$1$)
w.r.t. which $H$ turns out to be Hermitian.
\end{itemize}
All this material will be expanded and put into use in the next Chapter.

\subsection{Inequivalent Descriptions} \label{se:ineq}
\bigskip

In this section we discuss some methods to obtain inequivalent descriptions for a given classical system defined by a dynamical vector field $\Gamma$, not necessarily a linear one.

\subsubsection{Alternative Hamiltonian descriptions} \label{se:altham}
\bigskip

As explained in Appendix A, given any 1-1 tensor $T$, we can
define an antiderivation $d_T$ which acts on functions as
\begin{equation}
d_T f \equiv T(df) \; .
\end{equation}
In the sequel we will use extensively this construction with
$T=J$, the complex structure.
 Suppose now that the function $F$ be
a constant of motion and that the tensor $T$ be invariant under
the action of $\Gamma$ so that
\begin{equation}
L_{\Gamma} F =0 \; , \; L_{\Gamma} T =0 \; . \label{inv2}
\end{equation}
Then we can define a closed two-form
\begin{equation}
\omega_F \equiv d(d_T F)
\end{equation}
which is invariant under action of $\Gamma$ since
$L_{\Gamma}\omega_F = d(L_{\Gamma}d_T F)$ and $L_{\Gamma}d_T F =
0$ because of (\ref{inv2}). Assuming that
$\omega_F$  be non-degenerate, it will define a
new invariant symplectic structure. To obtain the alternative
Hamiltonian function H associated to $\omega_F$ it is sufficient
to notice that:
\begin{equation}
0 = L_{\Gamma}d_T F = i_{\Gamma}d(d_T F) + di_\Gamma (d_T F) =
i_\Gamma \omega_F + dF(T(\Gamma))=i_\Gamma \omega_F +
d(L_{T(\Gamma)} F)\; .
\end{equation}
Hence:

\begin{equation}
H=-L_{T\left(  \Gamma\right)  }F=-\left(  d_{T}F\right)  \left(
\Gamma\right)
\end{equation}
\bigskip

\begin{remark}
The above construction may  turn out to be empty if the function $F$ is
in the kernel of $dd_{T}:dd_{T}F=0$. For example, in $\mathbb{R}^{2}%
\approx\mathbb{C}$ with (real) coordinates $\left(  q,p\right)  $, take $T$
to be the complex structure :%
\begin{equation}
J=dp\otimes\frac{\partial}{\partial q}-dq\otimes\frac{\partial}{\partial p}%
\end{equation}
which is invariant under the dynamics of the $1D$ harmonic oscillator. Then,
it is immediate to check that:%
\begin{equation}
dd_{J}F=\left(  \frac{\partial^{2}F}{\partial q^{2}}+\frac{\partial^{2}%
F}{\partial p^{2}}\right)  dq\wedge dp
\end{equation}
and hence all the harmonic functions in the plane will be in the kernel of
\ $dd_{J}$.
\end{remark}

\begin{remark}
Suppose now that
$\Gamma=G^{i}\,_{j}x^{j}\frac{\partial
}{\partial x^{i}}$ be a linear vector field and $T=T^{i}\,_{j}dx^{j}%
\otimes\frac{\partial}{\partial x^{i}}$ a constant invariant 1-1
tensor. Then it is not difficult to check that $\omega_{F}$ is
constant if and only if $F$ is a quadratic function:
\begin{equation}
F=\frac{1}{2}F_{ij}x^{i}x^{j}\;,\;F_{ij}=F_{ji}\;.
\end{equation}
In this case, using the matrix notation of sect. \ref{se:invlin},
we have:
\begin{align}
H  & =\frac{1}{2}H_{ij}x^{i}x^{j}\;,\;H_{ij}=H_{ji}=-(FTG)_{ij}-(FTG)_{ji}%
\;,\\
\omega_{F}  & =\frac{1}{2}\Omega_{ij}dx^{i}\wedge
dx^{j}\;,\;\Omega _{ij}=-\Omega_{ji}=(FT)_{ij}-(FT)_{ji}\;.
\end{align}
Using the fact that Eqs. (\ref{inv2}) are equivalent to the
conditions: $(FG)_{ij}=-(FG)_{ji}$ and
$(GT)^{i}\,_{j}=(TG)^{i}\,_{j}$, one can show that, as it should
be, the relation $\Omega G=-H$ is trivially satisfied.
\end{remark}

As an example, let us consider the two-dimensional isotropic
harmonic oscillator whose dynamics is described by the vector
field
\begin{equation}
\Gamma=p^{a}\frac{\partial}{\partial
q^{a}}-q^{a}\frac{\partial}{\partial p^{a}}\;,
\end{equation}
where the summed-over index $a$ assumes the values: $a=1,2$. We
will take for $T$ \ the complex structure of the phase space
$\mathbb{R}^{4}$, i.e:
\begin{equation}
T=J=dp^{a}\otimes\frac{\partial}{\partial
q^{a}}-dq^{a}\otimes\frac{\partial }{\partial p^{a}}\;.
\end{equation}
Thus the representative matrices will be:
\begin{equation}
G=\left\vert
\begin{array}
[c]{cccc}%
0 & 0 & -1 & 0\\
0 & 0 & 0 & -1\\
1 & 0 & 0 & 0\\
0 & 1 & 0 & 0
\end{array}
\right\vert \;,\;T=\left\vert
\begin{array}
[c]{cccc}%
0 & 0 & 1 & 0\\
0 & 0 & 0 & 1\\
-1 & 0 & 0 & 0\\
0 & -1 & 0 & 0
\end{array}
\right\vert \;.
\end{equation}

With $T=J$ we have:%
\begin{equation}
-J(\Gamma)=\Delta\equiv q^{a}\frac{\partial}{\partial q^{a}}+p^{a}%
\frac{\partial}{\partial p^{a}}\label{jg}%
\end{equation}
with $\Delta$ the dilation (Liouville) field associated with the
standard linear structure on $\mathbb{R}^{4}$ and:
\begin{equation}
d_{J}F=\frac{\partial F}{\partial q^{a}}dp^{a}-\frac{\partial
F}{\partial
p^{a}}dq^{a}\Rightarrow\omega_{F}=dd_{J}F=\left(  \frac{\partial^{2}%
F}{\partial q^{a}\partial q^{b}}+\frac{\partial^{2}F}{\partial
p^{a}\partial
p^{b}}\right)  dq^{a}\wedge dp^{b}%
\end{equation}
as well as:%
\begin{equation}
H=-L_{J\left(  \Gamma\right)  }F=L_{\Delta}F
\end{equation}
for any function $F=F\left(  \mathbf{q},\mathbf{p}\right)  $.

\bigskip

It is well known that a basis of constants of motion is given, for
example, by the four independent functions
\begin{equation}%
\begin{array}
[c]{ll}%
F_{0}=\frac{1}{4}\left[  (p^{1})^{2}+(q^{1})^{2}+(p^{2})^{2}+(q^{2}%
)^{2}\right]  \;, & F_{1}=\frac{1}{4}\left[  (p^{1})^{2}+(q^{1})^{2}%
-(p^{2})^{2}-(q^{2})^{2}\right]  \;,\\
F_{2}=\frac{1}{2}\;\left[  p^{1}p^{2}+q^{1}q^{2}\right]  , & F_{3}=\frac{1}%
{2}\left[  q^{1}p^{2}-q^{2}p^{1}\right]  \;.
\end{array}
\end{equation}
All four functions \ being
quadratic\footnote{$L_{\Delta}F_{i}=2F_{i}$ for $i=0,1,2,3$.}, the
above construction yields then the following four alternative
hamiltonian descriptions:
\begin{equation}%
\begin{array}
[c]{ll}%
H_{0}=\frac{1}{2}\left[  (p^{1})^{2}+(q^{1})^{2}+(p^{2})^{2}+(q^{2}%
)^{2}\right]  \;, & \omega_{0}=dq^{1}\wedge dp^{1}+dq^{2}\wedge dp^{2}\;;\\
H_{1}=\frac{1}{2}\left[  (p^{1})^{2}+(q^{1})^{2}-(p^{2})^{2}-(q^{2}%
)^{2}\right]  \;, & \omega_{1}=dq^{1}\wedge dp^{1}-dq^{2}\wedge dp^{2}\;;\\
H_{2}=p^{1}p^{2}+q^{1}q^{2}\;, & \omega_{2}=dq^{1}\wedge
dp^{2}+dq^{2}\wedge
dp^{1}\;;\\
H_{3}=q^{1}p^{2}-q^{2}p^{1}\;, & \omega_{3}=-dq^{1}\wedge
dq^{2}+dp^{1}\wedge dp^{2}\;.
\end{array}
\label{hw}%
\end{equation}

Both the Liouville field $\Delta$ and the complex structure $J$
are associated with the standard linear structure on
$\mathbb{R}^{4}$. As we will now see, this observation may be
exploited to obtain alternative Hamiltonian descriptions by
defining inequivalent linear structures on phase space.

\subsubsection{Inequivalent Descriptions from Alternative Linear Structures}  \label{se:linstr}
\bigskip

We recall here \cite{EIMM} some known facts about the possibility of defining alternative (i.e. not linearly related)
linear structures on a vector space and/or of using the linear structure of a vector space to endow with a linear structure manifolds that are related to the given vector space.

Let  $E$ be a (real or complex) linear vector space with addition
$+$ and multiplication by scalars $\cdot$, and a nonlinear
diffeomorphism:
\begin{equation}
\phi:E\leftrightarrow E.
\end{equation}
We can define a new linear structure if we define:
\begin{itemize}
\item Addition of $u,v\in M$ as:%
\begin{equation}
u +_{\left(  \phi\right)  }v=:\phi(\phi^{-1}\left(  u\right)
+\phi^{-1}\left(  v\right)  ). \label{property1}%
\end{equation}
\item Multiplication by a scalar $\lambda\in\mathbb{R}$ or $\mathbb{C}$ of
$u\in M$ as:
\begin{equation}
\lambda\cdot_{\left(  \phi\right)  }u=:\phi\left(  \lambda\phi
^{-1}\left(  u\right)  \right) . \label{property2}%
\end{equation}
\end{itemize}
Obviously, the two linear spaces $(E,+,\cdot )$ and $(E,+_{(\phi )},\cdot
_{(\phi )})$ are finite dimensional vector spaces of the same dimension and
hence are isomorphic. However, the change of coordinates defined by $\phi $
that we are using to ``deform" the linear structure is a nonlinear
diffeomorphism. In other words, we are using two different (diffeomorphic
but not linearly related) global charts to describe the same manifold space $%
E$

Within  the framework of the new linear structure, it makes sense to consider the mapping:
\begin{equation}
\Psi: M\times\mathbb{R}\rightarrow M \; \; , \; \;
\Psi\left(  u,t\right)  =: e^{t}  \cdot_ {\left(  \phi\right)  } u=:u\left(  t\right),
\label{dilation}
\end{equation}
that defines a  one-parameter group as it can be easily checked. Its infinitesimal generator, the dilation (Liouville) field, is given by
\begin{equation}
\Delta\left(  u\right)  =\left[  \frac{d}{dt}u(t)\right]  _{t=0}=\left[
\frac{d}{dt}\phi\left(  e^{t}\phi^{-1}(u)\right)  \right]  _{t=0}.%
\end{equation}

As an example\footnote{More examples may be found in Ref.\cite{EIMM}.} consider $T^{\ast}\mathbb{R}$ with coordinates
$\left(  q,p\right)  $ and  linear structure defined by the dilation field:
\begin{equation}
\Delta=q\frac{\partial}{\partial q}+p\frac{\partial}{\partial p} ,
\end{equation}
which is such that $i_\Delta \omega=qdp-pdq$ with respect to the standard  symplectic form $\omega=dq\wedge dp$.\\
As it is well known the dynamics of the $1D$ harmonic oscillator is described, in appropriate
units,  by the vector field:
\begin{equation}\label{HO}
\Gamma=p\frac{\partial}{\partial q}-q\frac{\partial}{\partial p},
\end{equation}
which is $\omega$-Hamiltonian: $i_{\Gamma}\omega=dH$ with
Hamiltonian: $H=\left(  q^{2}+p^{2}\right)  /2$.
We can also define the complex structure:%
\begin{equation}
J=dp\otimes\frac{\partial}{\partial q}-dq\otimes\frac{\partial}{\partial p}%
\end{equation}
which is such that:%
\begin{equation}
J^{2}=-\mathbb{I},\text{ }J\left(  \Delta\right)  =\Gamma,\text{
}J\left(
\Gamma\right)  =-\Delta\text{ }%
\end{equation}
The composition of the symplectic and the complex structures gives
rise to  a compatible \cite{Mar7} metric tensor $g$:%
\begin{equation}
\omega\circ J=:-g,\text{ \ }g=dq\otimes dq+dp\otimes dp
\end{equation}

Notice also that the complex structure and the Hamiltonian are connected by:%
\begin{equation}
\omega=\frac{1}{2}dd_{J}H
\end{equation}

\bigskip

 Let us consider now the nonlinear change of coordinates on
$T^{\ast}\mathbb{R}$ \cite{Mar14}: $\left(  q,p\right)
\rightarrow\left(  Q,P\right)  $ with:
\begin{eqnarray}
&Q=q\left(  1+f\left(  H\right)  \right)  \\
 &P=p\left(  1+f\left(
H\right)  \right)  .\label{change}
\end{eqnarray}
Under very mild assumptions on the function $f\left(  H\right)  $
the mapping (\ref{change}) will be smooth and invertible with a
smooth inverse. One might assume, e.g., that $f\left(  \cdot
\right) $ be nonnegative and monotonically increasing for positive
argument.%
With the dynamics given by Eq.(\ref{HO}), it is immediate to check that:%
\begin{equation}
L_{\Gamma}Q=P,\text{ \ }L_{\Gamma}P=-Q
\end{equation}
Hence, although the two coordinates ystem are \textit{not} linearly related,
the vector field $\Gamma$ will be given, in the new coordinate system, by:%
\begin{equation}
\Gamma=P\frac{\partial}{\partial Q}-Q\frac{\partial}{\partial P}%
\end{equation}

which will be again Hamiltonian with respect to the symplectic
form $\omega^{\prime}=dQ\wedge dP$ with $H^{\prime}= \left(
Q^{2}+P^{2}\right) /2  =H\left(  1+f\left( H\right)  \right) ^{2}$
as Hamiltonian. Now the new Liouville field $\Delta'$, defined via
$i_{\Delta'} \omega'=QdP-PdQ$, is given by:
\begin{equation}
\Delta' = Q \frac{\partial}{\partial Q} + P \frac{\partial}{\partial P}\; ,
\end{equation}
Notice also that we can define a new 1-1 tensor (the new complex
structure):
\begin{equation}
J^{\prime}=dP\otimes\frac{\partial}{\partial
Q}-dQ\otimes\frac{\partial }{\partial P}\;,
\end{equation}
which is again such that $J^{\prime}(\Gamma)=-\Delta^{\prime}$.
$J^{\prime}$ and $\omega^{\prime}$ will generate then the new
metric tensor: $g^{\prime }=dQ\otimes dQ+dP\otimes dP$. Thus,
following the construction outlined in the previous section, we
might have obtained this alternative description of the dynamics
of the one-dimensional harmonic oscillator also by setting:
\begin{align}
& T=J^{\prime}\;,\\
& \omega^{\prime}=\frac{1}{2}dd_{J^{\prime}}H^{\prime}\;.
\end{align}
One obtains in this way a new linear structure, which is in some
sense "adapted" to the chosen Hamiltonian description. \newline

Finally, we observe that the above construction to obtain alternative descriptions may be easily generalized to the n-dimensional harmonic oscillator by defining
\begin{equation}
\omega_F \equiv \alpha_a d\left( \frac{\partial F}{\partial p^a}\right)  \wedge d\left( \frac{\partial F}{\partial q^a}\right)
\end{equation}
and
\begin{equation}
H_F \equiv \frac{1}{2}  \alpha_a \left[ \left( \frac{\partial F}{\partial p^a}\right)^2 + \left( \frac{\partial F}{\partial q^a}\right)^2 \right] \; ,
\end{equation}
where $F$ is a constant of the motion such that $\omega_F$ is
non-degenerate.

\subsubsection{Alternative Lagrangian Descriptions Coming from "Adapted" Linear Structures} \label{se:adapt}
\bigskip

Switching now to the Lagrangian framework, we recall \cite{MFVMR} that a regular Lagrangian $\mathcal{L}$ will define the symplectic structure on $TQ$:
\begin{equation}
\omega_{\mathcal{L}}=d\theta_{\mathcal{L}}=d\left(  \frac{\partial\mathcal{L}
}{\partial u^{i}}\right)  \wedge dq^{i};\; \theta_{\mathcal{L}}=\left(
\frac{\partial\mathcal{L}}{\partial u^{i}}\right)  dq^{i}.
\end{equation}
We look now \cite{Mar3}  for Hamiltonian vector
fields $X_{j},Y^{j}$ such that:
\begin{equation}
i_{X_{j}}\omega_{\mathcal{L}}=-d\left(
\frac{\partial\mathcal{L}}{\partial u^{j}}\right)  ,\;
i_{Y^{j}}\omega_{\mathcal{L}}=dq^{j}\label{equ3}
\end{equation}
Explicitly this implies:
\begin{eqnarray}
&L_{X_{j}}q^{i}=\delta_{j}^{i},\; L_{X_{j}}\frac{\partial\mathcal{L}
}{\partial u^{i}}=0 \label{condition1}, \\
&L_{Y^{j}}q^{i}=0,\;L_{Y^{j}}\frac{\partial\mathcal{L}}{\partial u^{i}%
}=\delta_{i}^{j} \label{condition2}.
\end{eqnarray}
Using then the identity $i_{\left[  Z,W\right]  }=L_{Z}\circ
i_{W}-i_{W}\circ L_{Z}$, and the fact that the Lie derivative of
the Hamiltonian of every field of the set (\ref{equ3}) 
with respect to any other of the fields is
either zero or a constant (actually unity), one can show that:
\begin{equation}
i_{\left[  Z,W\right]  }\omega_{\mathcal{L}}=0\; \mbox{\rm{whenever}} \; \left[  Z,W\right]  =\left[  X_{i},X_{j}\right]  ,\left[  X_{i},Y^{j}\right]
,\left[  Y^{i},Y^{j}\right],
\end{equation}
which proves that:
\begin{equation}
\left[  X_{i},X_{j}\right]  =\left[  X_{i},Y^{j}\right]  =\left[  Y^{i}
,Y^{j}\right]  =0 .
\end{equation}
This defines an infinitesimal action of an Abelian Lie group on $TQ$. If this
integrates to an action of the group $\mathbb{R}^{2n}$ ($\dim Q=n$) that is
free and transitive, this will define a new vector space structure on $TQ$
 that is "adapted" to the Lagrangian two-form $\omega_{\mathcal{L}}$.
More explicitly, defining  dual forms $\left(  \alpha^{i},\beta_{i}\right)  $ via:
$\alpha^{i}\left(  X_{j}\right)  =\delta_{j}^{i},\;\alpha^{i}\left(
Y^{j}\right)  =0 ; \;
\beta_{i}\left(  Y^{j}\right)  =\delta_{i}^{j},\;\beta_{i}\left(
X_{j}\right)  =0$, it is immediate to see that:
\begin{eqnarray}
\alpha^{i}&=&dq^{i} \\
\beta_{i}&=&d\left(  \frac{\partial\mathcal{L}}{\partial u^{i}}\right)
\end{eqnarray}
and that the symplectic form can be written as:
\begin{equation}
\omega_{\mathcal{L}}=\beta_{i}\wedge\alpha^{i}.
\end{equation}
Basically, what this means is that, to the extent that the definition of
vector fields and dual forms is global, we have found in this way a global
Darboux chart.

As an example of this construction, we may consider a particle in
a (time-independent) magnetic field $\textbf
{B}=\nabla\times\textbf{A}$. \ The corresponding second-order
vector field is given by ($e=m=c=1$):
\begin{equation}
\Gamma=u^{i}\frac{\partial}{\partial q^{i}}+\delta^{is}\epsilon_{ijk}
u^{j}B^{k}\frac{\partial}{\partial u^{s}} \label{vectorfield}.
\end{equation}
The Lagrangian is given in turn by :
\begin{equation}
\mathcal{L}=\frac{1}{2}\delta_{ij}u^{i}u^{j}+u^{i}A_{i}.
\end{equation}
while  the symplectic form is:
\begin{equation}
\omega_{\mathcal{L}}=\delta_{ij}dq^{i}\wedge
du^{j}-\frac{1}{2}\varepsilon_{ijk}B^{i}dq^{j}\wedge dq^{k}.
\end{equation}
The field \ $\Gamma$ is hamiltonian, the Hamiltonian being given
by:
\begin{equation}
H=\frac{1}{2}\delta_{ij}u^{i}u^{j}.
\end{equation}
Now it is easy to see that:
\begin{eqnarray}
&&X_{j}=\frac{\partial}{\partial q^{j}}-\delta^{ik}\frac{\partial A_{k}
}{\partial q^{j}}\frac{\partial}{\partial u^{i}},\\
&&Y^{j}=\delta^{jk}\frac{\partial}{\partial u^{k}}.
\end{eqnarray}
The dual forms\ $\alpha^{i},\beta_{i},i=1,...,n=\dim Q$ are given by:
\begin{eqnarray}
&&\alpha^{i}=dq^{i},\\
&&\beta_{i}=\delta_{ij}d(u^{j}+\delta^{jk}A_{k}).
\end{eqnarray}
Therefore the mapping
\begin{eqnarray}
Q^{i}&=&q^{i}\\
U^{i}&=&u^{i}+\delta^{ik}A_{k},
\label{mapping2}
\end{eqnarray}
provides us with a symplectomorphism that reduces $\omega_{\mathcal{L}}$ to the canonical form
\begin{equation}
\omega_{\mathcal{L}}=dq^{i}\wedge d\pi_{i} \label{symp-2},
\end{equation}
where  $\pi_{i}=\delta_{ij}U^{j}$. We may say that  the chart $\left(  Q,U\right)  $ is a Darboux chart "adapted" to the vector potential $\overrightarrow{A}$.

The  Liouville field will be{\footnote{We notice that $\Delta$ depends
on the gauge choice. The symplectic form will be however gauge-independent} then:
\begin{equation}
\Delta=Q^{i}\frac{\partial}{\partial Q^{i}}+\left[
U^{i}+\delta^{ik}\left(  Q^{j}\frac{\partial A_{k}}{\partial Q^{j}}
-A_{k}\right)  \right]  \frac{\partial}{\partial U^{i}} \label{newlinear}.
\end{equation}
Denoting collectively the old and new coordinates as $\left(
q,u\right)  $ and $\left(  Q,U\right)  $ respectively, Eq. (\ref{mapping2})
defines a mapping:%
\begin{equation}
\left(  q,u\right)  \stackrel{\phi}{\rightarrow}\left(  Q,U\right).
\end{equation}
It is then a straightforward application of the definitions (\ref{property1})
and (\ref{property2}) to show that the rules of addition and multiplication by
a constant become, in this specific case:%
\begin{equation}
\left(  Q,U\right)  +_{\left(  \phi\right)  }\left(
Q^{\prime},U^{\prime
}\right)  
=\left(  Q+Q^{\prime},U+U^{\prime}+\left[  A\left(
Q+Q^{\prime}\right) -\left(  A(Q)+A(Q^{\prime}\right)  )\right]
\right)\label{sum2} \\
\end{equation}
and:%
\begin{equation}
\lambda\cdot_{\left(  \phi\right)  }\left(  Q,U\right)  =\left(
\lambda Q,\lambda U+\left[  A\left(  \lambda Q\right)  -\lambda
A\left(  Q\right)
\right]  \right)  .\label{product3}%
\end{equation}

\bigskip
\subsection{Symmetries and Constants of the Motion for Systems Admitting of
Alternative Descriptions\label{symm_const}}

\bigskip

\subsubsection{Introduction\label{introd1}}
\bigskip

In our setting, according to which the primitive (or the more
physically relevant \cite{Mar10} ) object is the vector field
$\Gamma$ describing the dynamics on some carrier space
$\mathcal{M}$, a \textit{symmetry} will be defined as a
one-parameter group of diffeomorphisms of the carrier space that
maps solutions (i.e. integral curves of $\Gamma$) into solutions.
At the infinitesimal level, if $X$ $\in\mathfrak{X}\left(
\mathcal{M}\right)  $ is the associated infinitesimal generator of
the one-parameter group, this means
\cite{Mar10} that it must commute with $\Gamma$, i.e.:%
\begin{equation}
\left[  X,\Gamma\right]  =0 \label{symm1}%
\end{equation}

It is a straightforward consequence of the Jacobi identity on the
commutator bracket that\footnote{This is very much reminiscent of
Poisson's theorem of
Hamiltonian Mechanics.}:%
\begin{equation}
\left[  X_{1},\Gamma\right]  =0,\left[  X_{2},\Gamma\right]
=0\Rightarrow \left[  \left[  X_{1},X_{2}\right]  ,\Gamma\right]
=0
\end{equation}
(but not viceversa, of course). Hence: \textit{All the vector
fields satisfying the condition (\ref{symm1}) for a given
dynamical vector field }$\Gamma$ \textit{close on a Lie algebra,
the Lie algebra of (infinitesimal) symmetries of }$\Gamma$.

On the other hand, \textit{constants of the motion} are, as is
well known, functions $f\in\mathcal{F}\left(  \mathcal{M}\right)
$ that are invariant
under the flow of $\Gamma$, i.e.:%
\begin{equation}
L_{\Gamma}f=0
\end{equation}
where $L_{\Gamma}$ is the Lie derivative. A considerable effort is
usually devoted in textbooks (both in point-particle Mechanics
and/or in Field Theory, both elementary and more advanced) to try
and define a clear-cut procedure allowing to associate constants
of the motion (i.e. conserved quantities) with symmetries (and the
other way around). This goes usually through the use of
N\"{o}ther's Theorem\footnote{See however, e.g., Ref.\cite{MFVMR}
for the discussion of different approaches.}, that, for
completeness, we will revisit briefly here both in the Lagrangian
and Hamiltonian formulations of point-particle Mechanics.

\subsubsection{The N\"{o}ther Theorem\label{Nother}}
\bigskip

\begin{enumerate}
\item \textit{\underline{\textit{Lagrangian Formalism.}} }\ In
this case $\mathcal{M}=TQ$, with $Q$ a base manifold with (local)
coordinates $q^{1},...,q^{n}$, $n=\dim\left(  Q\right)  $. Before
proceeding, we recall how vector fields on the base manifold can
be lifted to vector fields on $TQ$.
Given:%
\begin{equation}
X=X^{i}\frac{\partial}{\partial q^{i}}\in\mathfrak{X}\left(
Q\right)  ,\text{
\ }X^{i}\in\mathcal{F}\left(  Q\right)  \label{ics}%
\end{equation}
the \textit{tangent lift }(sometimes called also the
\textit{complete lift}) $X^{c}$ of $X$ is defined
as\footnote{Here, with abuse of notation, we write $X^{i}$ for
what should be instead $\pi^{\ast}X^{i}$, with: $\pi:TQ\rightarrow
Q$ the canonical projection.}:%
\begin{equation}
X^{c}=X^{i}\frac{\partial}{\partial
q^{i}}+(L_{\Gamma_{0}}X^{i})\frac
{\partial}{\partial u^{i}}\in\mathfrak{X}\left(  TQ\right)  \label{ixc}%
\end{equation}
where the $u^{i}$'s are coordinates along the fibers and
$\Gamma_{0}$ is \textit{any} second-order vector field.

If $\mathcal{L}$ is a Lagrangian appropriate for the description,
via the Euler-Lagrange equations, of the dynamics associated with
a given second-order vector field $\Gamma$, a \textit{N\"{o}ther
symmetry} \cite{MFVMR} is, by definition, a tangent lift $X^{c}$
that is a symmetry for $\Gamma$, i.e. such
that:%
\begin{equation}
\left[  \Gamma,X^{c}\right]  =0
\end{equation}
and such that:%
\begin{equation}
L_{X^{c}}\mathcal{L}=L_{\Gamma}h
\end{equation}
where\footnote{Of course this is nothing but the familiar
statement that, under the action of $X^{c}$, the Lagrangian
changes by the total time derivative of a function of the $q$'s
alone.}: $h=\pi^{\ast}g,$ $g\in \mathcal{F}\left(  Q\right)  $
and: $\pi:TQ\rightarrow Q$ is the canonical projection. The
Lagrangian will be said to be \textit{strictly invariant} if $h=0$
(i.e. $g=0$)\footnote{Barring the trivial case $h$ (i.e.
$g$)$=const.$, a second-order vector field does not admit of
constants of the motion that are functions of the $q$'s alone.},
\textit{quasi-invariant} \cite{MMSS} if
$g\neq0$\textit{ } N\"{o}ther's theorem states then that:%
\begin{equation}
F_{X^{c}}=:i_{X^{c}}\theta_{\mathcal{L}}-h
\end{equation}
is a constant of the motion. Here:%
\begin{equation}
\theta_{\mathcal{L}}=\frac{\partial\mathcal{L}}{\partial u^{i}}dq^{i}%
\end{equation}
is the Lagrangian one-form associated with $\mathcal{L}$. In local
coordinates:%
\begin{equation}
F_{X^{c}}=X^{i}\frac{\partial\mathcal{L}}{\partial u^{i}}-h
\end{equation}

\item \underline{\textit{Hamiltonian Formalism.}} In this case $\mathcal{M}%
=T^{\ast}Q$, the cotangent bundle of the base manifold, with local
coordinates
$\left(  q^{i},p_{i}\right)  ,i=1,...,n$, equipped with the Cartan form:%
\begin{equation}
\theta_{0}=p_{i}dq^{i}%
\end{equation}
and the symplectic structure:%
\begin{equation}
\omega_{0}=-d\theta_{0}=dq^{i}\wedge dp_{i}%
\end{equation}
Here too there is a standard procedure for lifting vector fields
from $\mathfrak{X}\left(  Q\right)  $ to $\mathfrak{X}\left(
T^{\ast}Q\right)  $. namely, given a vector field
$X\in\mathfrak{X}\left(  Q\right)  $ of the form (\ref{ics}), the
\textit{cotangent lift } (sometimes called the
\textit{\ natural lift}) $X^{\ast}$ of $X$ is given by:%
\begin{equation}
X^{\ast}=X^{i}\frac{\partial}{\partial q^{i}}-\left(  \frac{\partial X^{j}%
}{\partial q^{i}}\right)  p_{j}\frac{\partial}{\partial p_{i}}\in
\mathfrak{X}\left(  T^{\ast}Q\right)
\end{equation}
and it is easy to show that it is the \textit{unique} vector field
that projects down to $X$ on the base manifold and that leaves the
Cartan form
invariant, i.e. such that:%
\begin{equation}
L_{X^{\ast}}\theta_{0}=0
\end{equation}

\end{enumerate}
\bigskip
\begin{remark}
In a more intrinsic way, both lifts can be defined \cite{Mar10} as
the infinitesimal generators of the tangent or, respectively,
cotangent lift of the one-parameter group of diffeomorphisms of
$Q$ that has $X$ as its infinitesimal generator.
\end{remark}

\begin{remark}
Symmetries for the dynamics that are (tangent or cotangent) lifts
of vector fields on the base manifold are also called
\underline{point symmetries}.
\end{remark}

\bigskip

A vector field $\Gamma\in\mathfrak{X}\left(  T^{\ast}Q\right)  $
is \textit{Hamiltonian} if there exists a (Hamiltonian) function
$H\in
\mathcal{F}\left(  T^{\ast}Q\right)  $ such that:%
\begin{equation}
i_{\Gamma}\omega_{0}=dH
\end{equation}
Given then a function $F\in\mathcal{F}\left(  TQ\right)  $, let
$X_{F}$ be the associated Hamiltonian vector field (not
necessarily a cotangent lift), i.e.:
$i_{X_{F}}\omega_{0}=dF$. Then:%
\begin{equation}
L_{X_{F}}H=i_{X_{F}}dH=i_{X_{F}}i_{\Gamma}\omega_{0}=-i_{\Gamma}i_{X_{F}%
}\omega_{0}=-i_{\Gamma}dF=-L_{\Gamma}F
\end{equation}
Hence:%
\begin{equation}
L_{\Gamma}F=0\Leftrightarrow L_{X_{F}}H=0
\end{equation}
Therefore, if $X_{F}$ is a symmetry for the Hamiltonian (i.e.: $L_{X_{F}}%
H=0$), then $F$ will be a constant of the motion and viceversa.
Moreover,
using the identity \cite{MFVMR}:%
\begin{equation}
i_{\left[  X,Y\right]  }=i_{X}\circ L_{Y}-L_{Y}\circ i_{X} \label{id}%
\end{equation}
valid for any pair of vector fields, it follows that, if $X$ is at
least
locally Hamiltonian (i.e.: $L_{X}\omega_{0}=0$), then:%
\begin{equation}
dL_{X}H=L_{X}i_{\Gamma}\omega_{0}=-i_{\left[  X,\Gamma\right]  }\omega_{0}%
\end{equation}
Hence, if $X$ is a symmetry for the Hamiltonian, and as
$\omega_{0}$ is
non-degenerate:%
\begin{equation}
L_{X}\omega_{0}=0\text{ \ }and\text{ \ }L_{X}H=0\Rightarrow\left[
X,\Gamma\right]  =0
\end{equation}
i.e. $X$ is also a symmetry for the dynamics. The converse however
is not true \cite{Mar10}: from: $\left[  X,\Gamma\right]  =0$ one
can only infer that: $L_{X}H=const.$, i.e. $X$ need not be a
symmetry for the Hamiltonian.

So far for the standard derivation of the N\"{o}ther Theorem. As a
simple example, considering, e.g., the $3D$ harmonic oscillator
with the standard Lagrangian: $\mathcal{L}=\left(  1/2\right)
\sum_{i=1}^{3}\left[  \left( u^{i}\right)  ^{2}-\left(
q^{i}\right)  ^{2}\right]  $ or the corresponding Hamiltonian
leads to the well-known association of (strict) rotational
invariance (of the Lagrangian and/or of the Hamiltonian)\ with the
conservation of angular momentum.

\bigskip

The motivation for having gone here to some length through
essentially standard material has been to emphasize the crucial
r\^{o}le that "intermediate" structures such as the Lagrangian or
the Hamiltonian, as well as the symplectic structure, play along
the way that leads to the association of symmetries with constants
of the motion. When these "intermediate" structures are not
unique, as it happens when more non-equivalent (Lagrangian (on
$TQ$) or Hamiltonian (on $T^{\ast}Q$)) descriptions are available
\cite{CMR,CS,LMR,Mar2,Mar3,MarMor,Mar6,MarSal,MFVMR,Ra}, the
connection becomes more ambiguous, and different (non-equivalent)
descriptions of the same dynamical system may lead to the
association of different constants of the motion with the same
group of symmetries, or of the same constants of the motion with
different groups of symmetry or to no association at all, as we
shall discuss now.

\subsubsection{Alternative Descriptions and Symmetries in the Lagrangian
Formalism\label{Lagr1}}
\bigskip

We will consider here some simple examples:

\begin{enumerate}
\item Let $Q=\mathbb{R}^{3}$, and let $\Gamma$ be the dynamics of
an isotropic harmonic oscillator (with unit mass and frequency for
simplicity). Then it is
immediate to show all the Lagrangians of the form:%
\begin{equation}
\mathcal{L}_{B}=\frac{1}{2}B_{ij}\left(
u^{i}u^{j}-q^{i}q^{j}\right)
\label{lagr1}%
\end{equation}
where: $B=\left\Vert B_{ij}\right\Vert $ is a real and
(necessarily) symmetric matrix are admissible Lagrangians for the
isotropic harmonic oscillator, and, moreover, regular ones iff the
matrix $B$ is non-singular. By "admissible" we mean obviously that
the Euler-Lagrange equations associated with any one of the
Lagrangians (\ref{lagr1}) reproduce the dynamics of the isotropic
harmonic oscillator. As we can always diagonalize $B$ with the aid
of an orthogonal transformation, we can limit ourselves to
considering only either the standard
Lagrangian:%
\begin{equation}
\mathcal{L}=\mathcal{L}_{1}+\mathcal{L}_{2}+\mathcal{L}_{3};\text{
}\mathcal{L}_{i}=\frac{1}{2}\left[  \left(  u^{i}\right)
^{2}-\left(
q^{i}\right)  ^{2}\right]  ,\text{ }i=1,2,3 \label{lagr2}%
\end{equation}
or (up to an overall sign and an overall factor):%
\begin{equation}
\mathcal{L}^{\prime}=\mathcal{L}_{1}+\mathcal{L}_{2}-\mathcal{L}_{3}
\label{lagr3}%
\end{equation}

Now, it is obvious that the Lagrangian (\ref{lagr2}) is (strictly)
invariant under the (lifted) action of $O\left(  3\right)  $,
while the Lagrangian (\ref{lagr3}) is (again, strictly) invariant
under the (lifted) action of $O\left(  2,1\right)  $, the Lorentz
group in $\left(  2+1\right)  $ dimensions. As it can be proved
\cite{MFVMR} that, in any number $n$ of dimensions, the most
general group of point symmetries for the dynamics of the
isotropic harmonic oscillator is $GL\left(  n,\mathbb{R}\right)
$, the above two groups are groups of N\"{o}ther symmetries. While
invariance under $O\left(  3\right)  $ associates, via
N\"{o}ther's theorem, the three components of the angular momentum
with the three generators of the group if the Lagrangian
(\ref{lagr2}) is chosen as the Lagrangian of the system, in the
case in which one chooses $\mathcal{L}^{\prime}$ as the Lagrangian
the situation is different. The three generators of $O\left(
2,1\right)  $ are
given by the tangent lifts of the vector fields:%
\begin{equation}
X_{1}=q^{3}\frac{\partial}{\partial
q^{1}}+q^{1}\frac{\partial}{\partial q^{3}},\text{
}X_{2}=q^{3}\frac{\partial}{\partial q^{2}}+q^{2}\frac{\partial
}{\partial q^{3}},\text{ }J=q^{1}\frac{\partial}{\partial q^{2}}-q^{2}%
\frac{\partial}{\partial q^{1}} \label{symmlift}%
\end{equation}
While $X_{1}$ and $X_{2}$ correspond to "boosts" in the $q^{1}$
and $q^{2}$
directions, $J$ \ represents ordinary rotations in the $\left(  q^{1}%
-q^{2}\right)  $ plane. They close on the Lie algebra
$\mathfrak{o}\left(
2,1\right)  $, namely:%
\begin{equation}
\left[  X_{1},X_{2}\right]  =J,\text{ }\left[  X_{1},J\right]
=X_{2},\text{
}\left[  J,X_{2}\right]  =X_{1}%
\end{equation}
and the same will hold true for the tangent lifts
$X_{1}^{c},X_{2}^{c}$ and $J^{c}$.

Applying now N\"{o}ther's theorem we find the following constants
of the
motion:%
\begin{equation}
F_{1}=:i_{X_{1}^{c}}\theta_{\mathcal{L}^{\prime}}=q^{3}u^{1}-q^{1}u^{3};\text{
}F_{2}=:i_{X_{2}^{c}}\theta_{\mathcal{L}^{\prime}}=q^{3}u^{2}-q^{2}u^{3}%
\end{equation}
while, as before: $i_{J^{c}}\theta_{\mathcal{L}^{\prime}}=q^{1}u^{2}%
-q^{2}u^{1}$. Therefore, we find that \textit{the angular momentum
is the (vector) constant of the motion associated not with the
rotation group but instead with the Lorentz group }$O\left(
2,1\right)  $.

\item Suppose however that we want to look for infinitesimal
(strict) symmetries of the Lagrangian (\ref{lagr1}) without
performing changes of coordinates (i.e. without diagonalizing the
matrix $B$). We will consider here only linear vector fields that
are generators of point symmetries, i.e. vector
fields of the form\footnote{This is the case of the symmetries (\ref{symmlift}).}:%
\begin{equation}
X=A^{i}\text{ }_{j}\left(  q^{j}\frac{\partial}{\partial q^{i}}+u^{j}%
\frac{\partial}{\partial u^{i}}\right)
\end{equation}
for some matrix $A=\left\Vert A^{i}\text{ }_{j}\right\Vert \in
End\left(
Q\right)  $. Then:%
\begin{equation}
L_{X}\mathcal{L}_{B}=\left(  BA\right)  _{jk}\left(  u^{j}u^{k}-q^{j}%
q^{k}\right)
\end{equation}
and strict invariance requires: $\left(  BA\right)  _{jk}+\left(
BA\right) _{kj}=0$, i.e.(as $B$ is symmetric):
\begin{equation}
A^{t}B+BA=0 \label{anticom}%
\end{equation}
which means that the matrix $AB$ has to be antisymmetric. For
example, with the Lagrangian (\ref{lagr3}): $B=diag\left(
1,1,-1\right)  $ and, e.g. for
the first symmetry $X_{1}$ of Eq.(\ref{symm}):%
\begin{equation}
A=\left\vert
\begin{array}
[c]{ccc}%
0 & 0 & 1\\
0 & 0 & 0\\
1 & 0 & 0
\end{array}
\right\vert
\end{equation}
($A=A^{t}$) and it is easy to check that the condition
(\ref{anticom}) is indeed satisfied. \

By assumption, the matrix $B$ in Eq.(\ref{lagr1}) can be
diagonalized with the aid of an orthogonal transformation:
$B=OB^{\prime}O^{t}$ with $B^{\prime}$ diagonal and:
$OO^{t}=O^{t}O=Id$. Then it is easy to see that
Eq.(\ref{anticom}) becomes:%
\begin{equation}
A^{^{\prime}t}B^{\prime}+B^{\prime}A^{\prime}=0 \label{anticom2}%
\end{equation}
with:%
\begin{equation}
A^{\prime}=O^{t}AO
\end{equation}
defining the transformed infinitesimal symmetry in the new
coordinate system.

\item Consider, as a further example, the (isotropic) harmonic
oscillator in
$2D$. Apart from the standard Lagrangian ($\mathcal{L}=\mathcal{L}%
_{1}+\mathcal{L}_{2}$ in the notation of Eq.(\ref{lagr2})) we may
consider the
(regular) Lagrangian:%
\begin{equation}
\mathcal{L}^{\prime}=u^{1}u^{2}-q^{1}q^{2} \label{lagr4}%
\end{equation}
This Lagrangian is (strictly) invariant under the "squeeze"
transformation,
i.e. the tangent lift of the one-parameter group:%
\begin{equation}
\left(  q^{1},q^{2}\right)  \mapsto\left(
q^{1}e^{t},q^{2}e^{-t}\right)
;\text{ }t\in\mathbb{R} \label{squeeze1}%
\end{equation}
whose infinitesimal generator is:%
\begin{equation}
S=q^{1}\frac{\partial}{\partial
q^{1}}-q^{2}\frac{\partial}{\partial q^{2}}
\label{squeeze2}%
\end{equation}
that lifts to:%
\begin{equation}
S^{c}=q^{1}\frac{\partial}{\partial
q^{1}}-q^{2}\frac{\partial}{\partial
q^{2}}+u^{1}\frac{\partial}{\partial
u^{1}}-u^{2}\frac{\partial}{\partial
u^{2}}%
\end{equation}
N\"{o}ther's theorem yields then the constant of the motion:%
\begin{equation}
F=:i_{s^{c}}\theta_{\mathcal{L}^{\prime}}=q^{1}u^{2}-q^{2}u^{1}%
\end{equation}
Hence: \textit{with the Lagrangian$\mathcal{\ L}^{\prime}$ angular
momentum is associated with invariance under squeeze.}

In the notation of the previous example, here: $B=\left\vert
\begin{array}
[c]{cc}%
0 & 1\\
1 & 0
\end{array}
\right\vert =\sigma_{1}$ and: $A=\left\vert
\begin{array}
[c]{cc}%
1 & 0\\
0 & -1
\end{array}
\right\vert =\sigma_{3}$, and, again, they satisfy the condition
(\ref{anticom}).

\item The Lagrangian (\ref{lagr4}) \ can be diagonalized via a
rotation of
$\pi/4$ to new coordinates:%
\begin{equation}
Q^{1}=\frac{q^{1}+q^{2}}{\sqrt{2}},\text{ }Q^{2}=\frac{q^{1}-q^{2}}{\sqrt{2}}%
\end{equation}
(and similarly for the velocities), whereby the Lagrangian becomes
(cfr.Eqs.(\ref{lagr2}) and (\ref{lagr3})):
\begin{equation}
\mathcal{L}\rightarrow\mathcal{L}_{1}-\mathcal{L}_{2}%
\end{equation}
Now, the "squeeze" transformation (\ref{squeeze1}) becomes:%
\begin{equation}
\left\vert
\begin{array}
[c]{c}%
Q^{1}\\
Q^{2}%
\end{array}
\right\vert \longrightarrow\left\vert
\begin{array}
[c]{cc}%
\cosh t & \sinh t\\
\sinh t & \cosh t
\end{array}
\right\vert \cdot\left\vert
\begin{array}
[c]{c}%
Q^{1}\\
Q^{2}%
\end{array}
\right\vert
\end{equation}
whose infinitesimal generator is:
\begin{equation}
X=Q^{2}\frac{\partial}{\partial Q^{1}}+Q^{1}\frac{\partial}{\partial Q^{2}}%
\end{equation}
(corresponding to the matrix: $A=\sigma_{1}$) i.e., as expected, a
(the unique) Lorentz boost with the parameter $t$ playing the
r\^{o}le of the rapidity of the boost.
\end{enumerate}

\subsection{The Transition to the Hamiltonian Formalism\label{Ham1}}
\bigskip

\subsubsection{Preliminaries and Recollections\label{s:prelim}}
\bigskip

Restricting ourselves for simplicity to dynamical systems
described by regular Lagrangians, we recall
\cite{Mar10},\cite{MFVMR} that the Euler-Lagrange equations for
the second-order vector field $\Gamma$ associated with a regular
Lagrangian $\mathcal{L}$ can be written, in intrinsic terms, as:
\begin{equation}
L_{\Gamma}\theta_{\mathcal{L}}-d\mathcal{L}=0
\end{equation}
where:%
\begin{equation}
\theta_{\mathcal{L}}=:\frac{\partial\mathcal{L}}{\partial u^{i}}dq^{i}%
\end{equation}
is the Lagrangian one-form or in the equivalent, "Hamiltonian" form:%
\begin{equation}
i_{\Gamma}\Omega_{\mathcal{L}}=dE_{\mathcal{L}}%
\end{equation}
where:%
\begin{equation}
\Omega_{\mathcal{L}}=:-d\theta_{\mathcal{L}}%
\end{equation}
is the "Lagrangian two-form", which is symplectic if $\mathcal{L}$
is regular,
and:%
\begin{equation}
E_{\mathcal{L}}=:i_{\Gamma}\theta_{\mathcal{L}}-\mathcal{L}%
\end{equation}
is known \cite{Mar10},\cite{MFVMR} as the "energy function"
associated with the Lagrangian $\mathcal{L}$.

The transition to the Hamiltonian formulation on $T^{\ast}Q$ is
accomplished, as is well known\cite{Mar10}, with the aid of the
"fiber derivative" (or
"Legendre map"): $F\mathcal{L}:TQ\rightarrow T^{\ast}Q$ that is defined by:%
\begin{equation}
F\mathcal{L}:\left(  q^{i},u^{i}\right)  \mapsto\left(  q^{i},p_{i}%
=\partial\mathcal{L}/\partial u^{i}\right)
\end{equation}

If, as assumed here, the Lagrangian is regular, the fiber
derivative is invertible and has the following properties (see
Ref.\cite{MFVMR} for details):

\begin{itemize}
\item $\left(  F\mathcal{L}\right)
_{\ast}\theta_{\mathcal{L}}=\theta_{0}$ \ and: $\left(
F\mathcal{L}\right)  _{\ast}\Omega_{\mathcal{L}}=\omega_{0}$
\end{itemize}

where $\left(  F\mathcal{L}\right)  _{\ast}$ denotes the
"push-forward" associated with the fiber derivative, i.e.: $\left(
F\mathcal{L}\right) _{\ast}=\left(  \left(  F\mathcal{L}\right)
^{-1}\right)  ^{\ast}$;

\begin{itemize}
\item Via push-forward, the vector field $\Gamma$ is mapped onto a
vector field $\widetilde{\Gamma}\in\mathfrak{X}\left(
T^{\ast}Q\right)  $ that is Hamiltonian with respect to the
canonical symplectic form $\omega_{0}$ with an Hamiltonian $H$
given by:
\begin{equation}
H=:\left(  F\mathcal{L}\right)  _{\ast}E_{\mathcal{L}}=E_{\mathcal{L}}%
\circ\left(  F\mathcal{L}\right)  ^{-1}%
\end{equation}

\item Explicitly (and locally):%
\begin{equation}
\widetilde{\Gamma}=\frac{\partial H}{\partial
p_{i}}\frac{\partial}{\partial
q^{i}}-\frac{\partial H}{\partial q^{i}}\frac{\partial}{\partial p_{i}}%
\end{equation}

\end{itemize}

All this can be summarized in the following scheme:%
\begin{equation}
\left(  \Gamma,\Omega_{\mathcal{L}},dE_{\mathcal{L}}\right)
\overset {F{\mathcal{L}}}{\longrightarrow}\left(
\widetilde{\Gamma},\omega
_{0},dH\right)  \label{scheme1}%
\end{equation}

\bigskip
\subsubsection{Consequences of the existence of alternative
descriptions\label{ham2}}
\bigskip

It is clear that the transition to $T^{\ast}Q$ summarized in the
scheme (\ref{scheme1}) will be non-ambiguous and unique if and
only if, apart from trivial equivalencies, the Lagrangian is
unique.

When more than one Lagrangian description is available, the
situation can become more involved. To be more specific, let, say,
$\mathcal{L}^{\left( 1\right)  }$ and $\mathcal{L}^{\left(
2\right)  }$ be two alternative Lagrangians for the same dynamical
system, $\Gamma$, on $TQ$. Each one defining its own fiber
derivative, we can obtain different Hamiltonian descriptions on
$T^{\ast}Q$ with different vector fields and Hamiltonians but
\textit{the same} symplectic structure (i.e. $\omega_{0}$) using
alternatively the two fiber derivatives according to the scheme:

\begin{equation}
\underset{TQ}{\underbrace{%
\begin{array}
[c]{ccc} &  & \Omega_{\mathcal{L}^{\left(  1\right)
}},dE_{\mathcal{L}^{\left(
1\right)  }}\\
& \nearrow & \\
\Gamma &  & \\
& \searrow & \\
&  & \Omega_{\mathcal{L}^{\left(  2\right)
}},dE_{\mathcal{L}^{\left(
2\right)  }}%
\end{array}
}}%
\begin{array}
[c]{c}%
\overset{F\mathcal{L}^{\left(  1\right)  }}{\longrightarrow}\\
\\
\\
\\
\overset{F\mathcal{L}^{\left(  2\right)  }}{\longrightarrow}%
\end{array}
\underset{T^{\ast}Q}{\underbrace{%
\begin{array}
[c]{ccccc}%
\widetilde{\Gamma}^{\left(  1\right)  } &  &  &  & dH^{\left(  2\right)  }\\
& \searrow &  & \nearrow & \\
&  & \omega_{0} &  & \\
& \nearrow &  & \searrow & \\
\widetilde{\Gamma}^{\left(  2\right)  } &  &  &  & dH^{\left(  1\right)  }%
\end{array}
}} \label{scheme2}%
\end{equation}
Although the vector fields $\widetilde{\Gamma}^{\left(  1\right)
}$ and $\widetilde{\Gamma}^{\left(  2\right)  }$ may look
different, it is worth stressing that nonetheless they offer
different descriptions of the same dynamical system. Indeed, in
both cases their trajectories in $T^{\ast}Q$ project down to the
\textit{same} set of trajectories in the physical space $Q$.
Stated otherwise, the two sets of first-order differential
equations on $T^{\ast}Q$ associated with
$\widetilde{\Gamma}^{\left(  1\right)  }$ and
$\widetilde{\Gamma}^{\left(  2\right)  }$ give rise to the
\textit{same} set of second-order differential equations on $Q$.

\begin{example}
\label{Ex1}Let $\Gamma$ represent, as in Sect.\ref{Lagr1}, the
dynamics of the
two-dimensional isotropic harmonic oscillator:%
\begin{equation}
\Gamma=u^{1}\frac{\partial}{\partial
q^{1}}+u^{2}\frac{\partial}{\partial
q^{2}}-q^{1}\frac{\partial}{\partial
u^{1}}-q^{2}\frac{\partial}{\partial
u^{2}}%
\end{equation}
and let, again in the notation of Sect.\ref{Lagr1}, the two
Lagrangians be: $\mathcal{L}^{\left(  1\right)
}=\mathcal{L=L}_{1}+\mathcal{L}_{2}$ (the
standard Lagrangian) and: $\mathcal{L}^{\left(  2\right)  }=\mathcal{L}%
^{\prime}$ (cfr. Eq.(\ref{lagr4})). Then, omitting unnecessary
details,
$H^{\left(  1\right)  }$ has the standard form:%
\begin{equation}
H^{\left(  1\right)  }=\frac{1}{2}\left[  \left(  p^{1}\right)
^{2}+\left( p^{2}\right)  ^{2}+\left(  q^{1}\right)  ^{2}+\left(
q^{2}\right)
^{2}\right]  \label{harmosc0}%
\end{equation}
and:%
\begin{equation}
\widetilde{\Gamma}^{\left(  1\right)  }=p^{1}\frac{\partial}{\partial q^{1}%
}+p^{2}\frac{\partial}{\partial q^{2}}-q^{1}\frac{\partial}{\partial p^{1}%
}-q^{2}\frac{\partial}{\partial p^{2}} \label{harmosc1}%
\end{equation}
As to $\mathcal{L}^{\left(  2\right)  }$, we find instead:%
\begin{equation}
F\mathcal{L}^{\left(  2\right)  }:\left(
q^{1},q^{2},u^{1},u^{2}\right) \mapsto\left(
q^{1},q^{2},p^{2},p^{1}\right)
\end{equation}
and:
\end{example}

\begin{equation}
\widetilde{\Gamma}^{\left(  2\right)  }=p_{2}\frac{\partial}{\partial q^{1}%
}+p_{1}\frac{\partial}{\partial q^{2}}-q^{2}\frac{\partial}{\partial p_{1}%
}-q^{1}\frac{\partial}{\partial p_{2}} \label{harmosc2}%
\end{equation}
\textit{with the Hamiltonian: }%

\begin{equation}
H^{\left(  2\right)  }=p_{1}p_{2}+q^{1}q^{2} \label{hamsqueeze}%
\end{equation}
\textit{Concerning symmetries, while the Hamiltonian
(\ref{harmosc0}) is rotationally-invariant and we obtain, via
N\"{o}ther's theorem, the usual association of the angular
momentum with rotations, The Hamiltonian \ (\ref{hamsqueeze}) is
squeeze-invariant, the squeeze transformation being
generated by the cotangent lift of the vector field (\ref{squeeze2}), i.e.}:%
\begin{equation}
S^{\ast}=S-p_{1}\frac{\partial}{\partial
p_{1}}+p_{2}\frac{\partial}{\partial
p_{2}} \label{squeeze3}%
\end{equation}
\textit{Now:}%

\begin{equation}
i_{S^{\ast}}\omega_{0}=dF
\end{equation}
\textit{where now the (Hamiltonian) constant of the motion is:}%
\begin{equation}
F=q^{1}p_{1}-q^{2}p_{2}%
\end{equation}
\textit{which, although it doesn't look such at first sight, is
again the
(only component of the) angular momentum,as: }%
\begin{equation}
\left(  F\mathcal{L}^{\left(  2\right)  }\right)  ^{\ast}F=q^{1}u^{2}%
-q^{2}u^{1}%
\end{equation}
\textit{ }

$\mathit{\bigskip}$

The scheme (\ref{scheme2}) outlined above is not the only possible
one, though. We might decide instead to perform the Legendre map
by using only one
of the two fiber derivatives in both cases. If we select, e.g., $F\mathcal{L}%
^{\left(  1\right)  }$, we obtain the following scheme for the
transition from
$TQ$ to $T^{\ast}Q$:%

\begin{equation}
\underset{TQ}{\underbrace{%
\begin{array}
[c]{ccc} &  & \Omega_{\mathcal{L}^{\left(  1\right)
}},dE_{\mathcal{L}^{\left(
1\right)  }}\\
& \nearrow & \\
\Gamma &  & \\
& \searrow & \\
&  & \Omega_{\mathcal{L}^{\left(  2\right)
}},dE_{\mathcal{L}^{\left(
2\right)  }}%
\end{array}
}}%
\begin{array}
[c]{cc}%
\overset{F\mathcal{L}^{\left(  1\right)  }}{\longrightarrow} & \\
& \searrow\\
& \\
& \nearrow\\
\overset{F\mathcal{L}^{\left(  1\right)  }}{\longrightarrow} &
\end{array}
\underset{T^{\ast}Q}{\underbrace{%
\begin{array}
[c]{ccc}
&  & \omega^{\left(  1,2\right)  },dH^{\left(  1,2\right)  }\\
& \nearrow & \\
\widetilde{\Gamma}^{\left(  1\right)  } &  & \\
& \searrow & \\
&  & \omega_{0},dH^{\left(  1\right)  }%
\end{array}
}} \label{scheme3}%
\end{equation}
where now:%
\begin{equation}
\omega^{\left(  1,2\right)  }=\left(  F\mathcal{L}^{\left(
1\right) }\right)  _{\ast}\Omega_{\mathcal{L}^{\left(  2\right)
}}=\left( F\mathcal{L}^{\left(  1\right)  }\right)  _{\ast}\left(
F\mathcal{L}^{\left( 2\right)  }\right)  ^{\ast}\omega_{0}=\left(
F\mathcal{L}^{\left(  2\right) }\circ\left(  F\mathcal{L}^{\left(
1\right)  }\right)  ^{-1}\right)  ^{\ast
}\omega_{0}%
\end{equation}
and similarly for $H^{\left(  1,2\right)  }$.

\begin{remark}
If we forget about the "$TQ$ part" of the scheme (\ref{scheme3})
and retain only the "$T^{\ast}Q$ part", we see that this procedure
exhibits an example of a given dynamical system
($\widetilde{\Gamma}^{\left(  1\right)  }$) on $T^{\ast}Q$ that is
bihamiltonian.
\end{remark}

\begin{example}
\label{Ex2} For the same system as in Example \ref{Ex1} above,
$\widetilde {\Gamma}^{\left(  1\right)  }$ is again given by
Eq.(\ref{harmosc1}), but we
find instead:%
\begin{equation}
\omega^{\left(  1,2\right)  }=dq^{1}\wedge dp_{2}+dq^{2}\wedge dp_{1}%
\end{equation}
while:%
\begin{equation}
H^{\left(  1,2\right)  }=p_{1}p_{2}+q^{1}q^{2}%
\end{equation}
as in the previous example. However, now:%
\begin{equation}
i_{S^{\ast}}\omega^{\left(  1,2\right)  }=q^{1}dp_{2}-p_{2}dq^{1}+p_{1}%
dq^{2}-q^{2}dp_{1}%
\end{equation}
and:%
\begin{equation}
d\left(  i_{S^{\ast}}\omega^{\left(  1,2\right)  }\right)  =L_{S^{\ast}}%
\omega^{\left(  1,2\right)  }=2\left(  dq^{1}\wedge
dp_{2}-dq^{2}\wedge dp_{1}\right)  \neq0
\end{equation}
Therefore, although: $L_{S^{\ast}}H^{\left(  1,2\right)  }=0$ and
hence $S^{\ast}$ is a symmetry for the Hamiltonian $H^{\left(
1,2\right)  }$, \ it is not Hamiltonian with respect to the
symplectic form $\omega^{\left( 1,2\right)  }$, and ceases
therefore to be the generator of a N\"{o}ther symmetry.
\end{example}

\bigskip

To conclude this Section, we would like to "re-visit", in the
Hamiltonian formalism, the consequences of the use, for the
isotropic harmonic oscillator, of one of the Lagrangians
(\ref{lagr1}), parametrized by the family of symmetric and
nonsingular matrices: $B=\left\Vert B_{ij}\right\Vert $.

Let us specialize here too to $n=3$. The canonical momenta are
defined by:
\begin{equation}
p_{i}=B_{ij}u^{j}\Leftrightarrow u^{i}=A^{ij}p_{j},\text{ }i=1,2,3
\end{equation}
where: $A=\left\Vert A^{ij}\right\Vert $ is the matrix inverse of
$B:A^{ij}B_{jk}=\delta_{k}^{i}$. The Hamiltonian is therefore:%
\begin{equation}
H=\frac{1}{2}\left(  A^{ij}p_{i}p_{j}+B_{ij}q^{i}q^{j}\right)
\end{equation}
while the three components of the angular momentum:
$J_{i}=\varepsilon
_{ijk}q^{j}u^{k}$ are given, in the canonical formalism on $T^{\ast}%
\mathbb{R}^{3}$, by:
\begin{equation}
J_{i}=\varepsilon_{ijk}A^{kl}q^{j}p_{l}%
\end{equation}

The $J_{i}$'s are of course constants of the motion, i.e.:%
\begin{equation}
\left\{  J_{i},H\right\}  =0,\text{ }i=1,2,3 \label{PB1}%
\end{equation}
where $\left\{  .,.\right\}  $ is the canonical Poisson bracket on
$T^{\ast }\mathbb{R}^{3}$. \ Now, some long but straightforward
algebra \cite{MarSal}
shows that the Poisson brackets among the $J_{i}$'s are given by:%
\begin{equation}
\left\{  J_{h},J_{k}\right\}  =\varepsilon_{hkr}A^{rs}J_{s} \label{PB2}%
\end{equation}

Eq.(\ref{PB2}) defines a Lie algebra whose derived algebra is
spanned by the vectors of the form:
$J_{hk}=:\varepsilon_{hkr}A^{rs}J_{s}$. As the Ricci tensor is
antisymmetric, there are only three independent such vectors and,
as the matrix $A$ is symmetric, they are independent. Therefore,
the derived algebra is three-dimensional, and the Lie algebra can
be only \cite{Kir} \ (apart from a sign) that of $O\left(
3\right)  $ or that of $O\left( 2,1\right)  $\footnote{These are
called $su\left(  2\right)  $ and $su\left( 1,1\right)  $ in
Ref.\cite{Kir}, but the Lie algebras are isomorphic.}. Denoting by
$X_{i}$ and $X_{hk}$ the associated Hamiltonian vector fields,
defined by:%
\begin{equation}
i_{X_{i}}\omega_{0}=dJ_{i};\text{ }i_{X_{hk}}\omega_{0}=dJ_{hk}%
\end{equation}
($X_{hk}=\varepsilon_{hkr}A^{rs}X_{s}$) which implies, in
particular: $\mathcal{L}_{X_{i}}\omega_{0}=0$, Eq.(\ref{PB1}) is
equivalent to the statement that: $\mathcal{L}_{X_{i}}H=0$. Hence
(see Sect.\ref{Nother}), the $X_{i}$'s are also symmetries for the
dynamics. Moreover, using the identity (\ref{id}), one sees at
once that:
\begin{equation}
i_{\left[  X_{h},X_{k}\right]  }\omega_{0}=-\mathcal{L}_{X_{h}}i_{X_{k}}%
\omega_{0}=-\mathcal{L}_{X_{h}}dJ_{k}=-d\mathcal{L}_{X_{h}}J_{k}%
\end{equation}
i.e. that:%
\begin{equation}
i_{\left[  X_{h},X_{k}\right]  }\omega_{0}=d\left\{
J_{h},J_{k}\right\}
=dJ_{hk}=i_{X_{hk}}\omega_{0}%
\end{equation}
which implies in turn, as $\omega_{0}$ is nondegenerate:%
\begin{equation}
\left[  X_{h},X_{k}\right]  =\varepsilon_{hkr}A^{rs}X_{s}%
\end{equation}
Hence, the $X_{i}$'s generate the same algebra of symmetries (that
of $O\left(  3\right)  $ or that of $O\left(  2,1\right)  $), and
this is in agreement with the results of Sect.\ref{Lagr1}.

\newpage

%% file: Chapt4rev.tex
\section{Geometry of Quantum Mechanics and Alternative Structures}

\label{QM}

\subsection{Introduction\label{introd:geometry}}
\bigskip

Alternative descriptions for both classical and quantum systems have been
discussed already all along the previous Chapters. In particular, in Sect.
\ref{SE} we have discussed how one can obtain alternative descriptions both in
the Schr\"{o}dinger and Heisenberg pictures either by modifying the Hermitian
structure using constants of the motion (Sect. \ref{SE1}) or, in the
infinite-dimensional case (Sect. \ref{SE2}) by changing the symplectic
structure (as well as the Hamiltonian) using powers of the original Hamiltonian.

The discussion was carried on systematically within the framework of the
description of \ states as vectors on some (finite- or infinite-dimensional)
complex Hilbert space $\mathcal{H}$ (with the associated Hermitian structure
$\left\langle .|.\right\rangle $) and of observables as self-adjoint linear
operators on $\mathcal{H}$. \ 

Hilbert spaces were introduced and used in a systematic way first by Dirac
\cite{Dir3} as a consequence of the fact that one needs a superposition rule
(and hence a linear structure) in order to accommodate a consistent
description of the interference phenomena that are fundamental for Quantum
Mechanics. \ Parenthetically, we should note that a \textit{complex} Hilbert
space carries with it in a natural way a "complex structure" (multiplication
of vectors by the imaginary unit). The r\^{o}le of the latter was discussed in
the early Forties by Reichenbach \cite{Re}. Later on St\"{u}ckelberg
\cite{Stu} emphasized the r\^{o}le of the complex structure in deducing in a
consistent way the uncertainty relations of Quantum Mechanics (see also the
discussion in Refs.\cite{EM} and \cite{Mar12}).

However, it is well known that a "complete" measurement in Quantum Mechanics
(a simultaneous measurement of a complete set of commuting
observables\footnote{We will not worry at this stage about the technical
complications that can arise, in the infinite-dimensional case, when the
spectrum of some observable has a continuum part.} \cite{Dir3,EM,Mes}) does
not provide us with an uniquely defined vector in some Hilbert space, but
rather with a "ray", i.e. an equivalence class of vectors differing by
multiplication through a nonzero complex number. Even fixing the
normalization, an overall phase\footnote{Not a \textit{relative} phase in a
superposition of vectors, of course.} will remain unobservable. \ Quotienting
w.r.t. both multiplications leads, for a finite-dimensional Hilbert
space\ $\mathcal{H}$ ($\dim_{\mathbb{C}}\mathcal{H}=n$), to the following
double fibration:%

\begin{equation}%
\begin{array}
[c]{ccc}%
\mathbb{R}_{+} & \longrightarrow & \mathcal{H}_{0}=\mathcal{H-}\left\{
\mathbf{0}\right\} \\
&  & \downarrow\\
U\left(  1\right)  & \longrightarrow & \mathbb{S}^{2n-1}\\
&  & \downarrow\\
&  & P\left(  \mathcal{H}\right)
\end{array}
\label{quotient}%
\end{equation}
whose final result is the \textit{projective Hilbert space }$P\mathcal{H}$,
and it is clear that:%
\begin{equation}%
\begin{array}
[c]{c}%
P(\mathcal{H})\simeq\mathbb{C}P^{n-1}=\left\{  [|\psi\rangle]\;:|\psi
\rangle,|\psi^{\prime}\rangle\in\left[  |\psi\rangle\right]  \Leftrightarrow
\;|\psi\rangle=\lambda|\psi^{\prime}\rangle\right\} \\
|\psi\rangle,|\psi^{\prime}\rangle\in\mathcal{H-}\left\{  \mathbf{0}\right\}
,\;\lambda\in\mathbb{C}_{0}=\mathbb{C}\mathbf{-}\left\{  \mathbf{0}\right\}
\}
\end{array}
\label{PH}%
\end{equation}
where $\left[  |\psi\rangle\right]  $ denotes the equivalence class to which
$|\psi\rangle\in\mathcal{H}$ belongs under multiplication by a non-zero
complex number.

\bigskip

\begin{remark} 
Notice that in this way the Hilbert space $\mathcal{H}$
acquires the structure of a principal fiber bundle \cite{Hus,Mar10,Ste}, with
base $P\mathcal{H}$ and typical fiber $\mathbb{C}_{0}$.
\end{remark}

The self-duality of $\mathcal{H}$\textbf{ }determined by the Hermitian
structure allows for the (unique) association of every equivalence class
$[|\psi\rangle]$ with the rank-one projector:%
\begin{equation}
\rho_{\psi}=\frac{|\psi\rangle\langle\psi|}{\left\langle \psi|\psi
\right\rangle }%
\end{equation}
with the known properties:%
\begin{equation}%
\begin{array}{l}%
\rho_{\psi}^{\dag}=\rho_{\psi}\\
Tr\rho_{\psi}=1\\
\rho_{\psi}^{2}=\rho_{\psi}%
\end{array}
\end{equation}
It is clear by construction that the association depends on the Hermitian
structure we consider.

The space of rank-one projectors is usually denoted \cite{GKM} as
$\mathcal{D}_{1}^{1}\left(  \mathcal{H}\right)  $. It is then clear that in
this way we can identify it with the projective Hilbert space $P\mathcal{H}$.
Hence, what the best of measurements will yield will be always (no more and
not less than) a \textit{rank-one projector }(also called a \textit{pure state
}\cite{Ha}).

\ \ \ \ \ Also, transition probabilities that, together with the expectation
values of self-adjoint linear operators that represent dynamical variables,
are among the only observable quantities one can think of, will be insensitive
to overall phases, i.e. they will depend only on the (rank-one) projectors
associated with the states. If $A=A^{\dag}$ is any such observable, then the
expectation value $\left\langle A\right\rangle _{\psi}$in the state
$|\psi\rangle$ will be given by:%
\begin{equation}
\left\langle A\right\rangle _{\psi}=\frac{\left\langle \psi|A|\psi
\right\rangle }{\left\langle \psi|\psi\right\rangle }\equiv Tr\left\{
\rho_{\psi}A\right\}  \label{expvalue}%
\end{equation}
Transition probabilities are in turn expressed via a binary product that can
be defined on pure states. Again, if $|\psi\rangle$ and $|\phi\rangle$ are any
two states, then the (normalized) transition probability from $|\psi\rangle$
to $|\phi\rangle$ will be given by:%
\begin{equation}
\frac{|\left\langle \phi|\psi\right\rangle |^{2}}{\left\langle \psi
|\psi\right\rangle \left\langle \phi|\phi\right\rangle }=Tr\left\{  \rho
_{\psi}\rho_{\phi}\right\}  \label{1:trace1}%
\end{equation}
and the trace on the r.h.s. of Eq.(\ref{trace1}) will define the binary
product among pure states (but more on this shortly below).

It appears therefore that the most natural setting for Quantum Mechanics is
not primarily the Hilbert space itself but rather the projective Hilbert
space, or, equivalently, the space of rank-one projectors $\mathcal{D}_{1}%
^{1}\left(  \mathcal{H}\right)  $, whose convex hull will provide us with the
set of all density \ states. \cite{Neu1,Neu2,Fano}.

On the other hand, the superposition rule, which leads to interference
phenomena, remains one of the fundamental building blocks of Quantum
Mechanics, one that, among other things, lies at the very heart of the modern
formulation of Quantum Mechanics in terms of path integrals
\cite{Br2,Fe,FH,GJ}, an approach that goes actually back to earlier
suggestions by Dirac \cite{Dir3,Dir1}.

To begin with, if we consider, for simplicity, two orthonormal states:%
\begin{equation}
|\psi_{1}\rangle,|\psi_{2}\rangle\in\mathcal{H},\left\langle \psi_{i}|\psi
_{j}\right\rangle =\delta_{ij},i,j=1,2
\end{equation}
with the associated projection operators:%
\begin{equation}
\rho_{1}=|\psi_{1}\rangle\langle\psi_{1}|,\text{ }\rho_{2}=|\psi_{2}%
\rangle\langle\psi_{2}|
\end{equation}
a linear superposition with (complex) coefficients $c_{1}$ and $c_{2}$ with:
$\left\vert c_{1}\right\vert ^{2}+\left\vert c_{2}\right\vert ^{2}=1$ will
yield the normalized vector:%
\begin{equation}
|\psi\rangle=c_{1}|\psi_{1}\rangle+c_{2}|\psi_{2}\rangle\label{superp}%
\end{equation}
and the associated projector:%
\begin{equation}
\rho_{\psi}=|\psi\rangle\langle\psi|=\left\vert c_{1}\right\vert ^{2}\rho
_{1}+\left\vert c_{2}\right\vert ^{2}\rho_{2}+\left(  c_{1}c_{2}^{\ast}%
\rho_{12}+h.c.\right)  \label{purif0}%
\end{equation}
where: $\rho_{12}=:|\psi_{1}\rangle\langle\psi_{2}|$, which cannot however be
expressed directly in terms of the initial projectors.

A procedure to overcome this difficulty by retaining at the same time the
information concerning the relative phase of the coefficients can be
summarized as follows \cite{CGM,MMSZ4,MMSZ6,MMSZ5,MMSZ2,Mar12}.

Considering a third, fiducial vector $|\psi_{0}\rangle$ with the only
requirement that it be not orthogonal\footnote{In terms of the associated
rank-one projections, we require: $Tr\left(  \rho_{i}\rho_{0}\right)
\neq0,i=1,2,$ with: $\rho_{0}=|\psi_{0}\rangle\langle\psi_{0}|$.} neither to
$|\psi_{1}\rangle$ nor to $|\psi_{2}\rangle$, it is possible to associate
normalized vectors $|\phi_{i}\rangle$ with the projectors $\rho_{i}$ $\left(
i=1,2\right)  $ by setting:%
\begin{equation}
|\phi_{i}\rangle=\frac{\rho_{i}|\psi_{0}\rangle}{\sqrt{Tr\left(  \rho_{i}%
\rho_{0}\right)  }},\text{ }i=1,2
\end{equation}

\begin{remark}
Note that, as all the $\rho$'s involved are rank-one
projectors\footnote{The proof of Eqs.(\ref{trace1}) and (\ref{trace2})is
elementary and will not be given here.}:

\begin{itemize}
\item
\begin{equation}
Tr\left(  \rho_{i}\rho_{0}\right)  Tr\left(  \rho_{j}\rho_{0}\right)
=Tr\left(  \rho_{i}\rho_{0}\rho_{j}\rho_{0}\right)  \text{ }\forall i,j
\label{trace1}%
\end{equation}
and that:

\item
\begin{equation}
|\phi_{i}\rangle\langle\phi_{i}|=\frac{\rho_{i}\rho_{0}\rho_{i}}%
{\sqrt{Tr\left(  \rho_{i}\rho_{0}\rho_{i}\rho_{0}\right)  }}\equiv\rho
_{i},\text{ }i=1,2 \label{trace2}%
\end{equation}

\end{itemize}
\end{remark}

Forming now the linear superposition: $|\phi\rangle=c_{1}|\phi_{1}%
\rangle+c_{2}|\phi_{2}\rangle$ and the associated projector: $\rho
=|\phi\rangle\langle\phi|$, one finds easily, using also Eqs.(\ref{trace1})
and (\ref{trace2}), that:
\begin{equation}
\rho=\left\vert c_{1}\right\vert ^{2}\rho_{1}+\left\vert c_{2}\right\vert
^{2}\rho_{2}+\frac{c_{1}c_{2}^{\ast}\rho_{1}\rho_{0}\rho_{2}+h.c.}%
{\sqrt{Tr\left(  \rho_{1}\rho_{0}\rho_{2}\rho_{0}\right)  }} \label{purif1}%
\end{equation}
which can be written in a compact form as:%
\begin{equation}
\rho=%
{\displaystyle\sum\limits_{i,j=1}^{2}}
c_{i}c_{j}^{\ast}\frac{\rho_{i}\rho_{0}\rho_{j}}{\sqrt{Tr\left(  \rho_{i}%
\rho_{0}\rho_{j}\rho_{0}\right)  }} \label{purif2}%
\end{equation}

The results (\ref{purif1}) and (\ref{purif2}) are now written entirely in
terms of rank-one projectors. Thus, a superposition of rank-one projectors
which yields another rank-one projector is possible, but requires the
arbitrary choice of the fiducial projector $\rho_{0}$. This procedure is
equivalent to the introduction of a connection on the bundle, usually called
the Pancharatnam connection \cite{Mor2,Pan}.

\bigskip

\begin{remark}
If the (normalized) probabilities $\left\vert c_{1}\right\vert ^{2}$ and
$\left\vert c_{2}\right\vert ^{2}$ are given, Eq.(\ref{superp}) describes a
one-parameter family of linear superposition of states, and the same will be
true in the case of Eq.(\ref{purif1}). Both families will be parametrized by
the relative phase of the coefficients.

\end{remark}

\begin{remark}
Comparison of Eqs.(\ref{purif0}) and (\ref{purif1}) shows that, while
the first two terms on the r.h.s. of both are identical, the last terms of the
two differ by an extra (fixed) phase, namely that:%
\begin{equation}
\frac{\rho_{1}\rho_{0}\rho_{2}}{\sqrt{Tr\left(  \rho_{1}\rho_{0}\rho_{2}%
\rho_{0}\right)  }}=\rho_{12}\exp\left\{  i\left[  \arg\left(  \left\langle
\psi_{1}|\psi_{0}\right\rangle -\arg\left(  \left\langle \psi_{2}|\psi
_{0}\right\rangle \right)  \right)  \right]  \right\}
\end{equation}
\end{remark}

\begin{remark}
The result of Eq.(\ref{purif2}) can be generalized in an obvious way to
the case of an arbitrary number, say $n$, of orthonormal states none of which
is orthogonal to the fiducial state. The corresponding family of rank-one
projectors will be parametrized in this case by the $\left(  n-1\right)  $
relative phases.
\end{remark}

\bigskip

If, now, we are given two\footnote{Or more, with an obvious generalization.}
(rank-one) projectors and only the relative probabilities are given, we are
led to conclude that the system is described by the convex combination (a
rank-two density matrix): $\rho=\left\vert c_{1}\right\vert ^{2}\rho
_{1}+\left\vert c_{2}\right\vert ^{2}\rho_{2}$, \ which is again Hermitian and
of trace one, but now: $\rho-\rho^{2}>0$ (strictly). The procedure leading
from this "impure" state to one of the pure states given by, say,
Eq.(\ref{purif2}), i.e. the procedure that associates a pure state with a pair
of pure states, is a composition law for pure states that has been termed in
the literature \cite{MMSZ6} as a "purification" of "impure" states.

In the Hilbert space formulation of Quantum Mechanics one needs also to find
the spectral family associated with any observable, represented by a
self-adjoint operator on the Hilbert space of states. Limiting ourselves for
simplicity to observables with a pure point-spectrum, these notions can be
made easily to "descend" to the projective Hilbert space $P\mathcal{H}$ by
noticing that, if $A=A^{\dag}$ is an observable, and considering from now on
only normalized vectors, the expectation value (\ref{expvalue}) associates
with the observable $A$ a (real) functional on $P\mathcal{H}$. The standard
variational principle of Quantum Mechanics \cite{EM,Mes} can be rephrased
\cite{CJM,CGM} by saying that the critical points of this functional are the
eigenprojectors of $A$ and that the critical values yield the corresponding eigenvalues.

Unitary (and, as a matter of fact, also anti-unitary\footnote{Think of the
operation \cite{EM,Mes} of time-reversal.}) operators play also a relevant
r\^{o}le in Quantum Mechanics \cite{EM,Mes}. In particular, self-adjoint
operators can act as infinitesimal generators of one-parameter groups of
unitaries. Both unitary and anti-unitary operators share the property of
leaving all transition probabilities invariant. At the level of the projective
Hilbert space they represent then \textit{isometries }of the binary product
(\ref{1:trace1}). The converse is \ also true. Indeed, it was proved long ago
by Wigner \cite{Wig,Wig2} that \ bijective maps on $P\mathcal{H}$ that
preserve transition probabilities (i.e., isometries of the projective Hilbert
space) are associated with unitary or anti-unitary transformations on the
original Hilbert space\footnote{The association being up to a phase, this may
lead to the appearance of "ray" (or "projective") representations
\cite{Barg,EM,Ha,Mac1,Mac2,Mes,Sam} of unitary groups on the Hilbert space
instead of ordinary ones, a problem that we will not discuss here, though.
\par
\bigskip}. For a recent version of this theorem, see Ref.\cite{GKM2}.

To summarize the content of this Section, we have argued that all the relevant
building blocks of Quantum Mechanics can be re-formulated in terms of parent
objects that "live" in the projective Hilbert space $P\mathcal{H}$. The
latter, however, is no more a linear vector space. As will be discussed in the
following Sections, it carries instead a rich manifold structure. In this
context, the very notion of linear transformations looses meaning, and we are
led in a natural way to consider a non-linear manifold and (non-linear)
diffeomorphisms thereof. This given, only objects that have a tensorial
character will be allowed. We will have then, as a preliminary step, to
proceed to, so-to-speak, "tensorialize" all the notions that have been
established in the context of the linear Hilbert space. We will do that in the
second part of this Chapter, where we will discuss the geometry of Quantum
Mechanics. In the last part of the Chapter, having achieved this goal, we will
re-discuss the problem of alternative structures in the context of Quantum Mechanics.

\subsection{The Geometry of Quantum Mechanics} \label{se:gqm}
\bigskip
\subsubsection{Some Preliminaries\label{s:geoprelim}}
\bigskip

We recall here some basic notions, in order mainly to fix the language and
notations to be employed in what follows.

\begin{enumerate}
\item Given an $n$-dimensional vector space $\mathcal{H}$ over the field
$\mathbb{C}$ of the complex numbers, the \textit{realified }\cite{Ar1}
$\mathcal{H}_{\mathbb{R}}$ of $\mathcal{H}$ is a real vector space that
coincides with $\mathcal{H}$ $\ $as a group (abelian group under
addition) but in which only multiplication by real scalars is allowed. The
realified of $\mathcal{H}$ can be constructed as follows. Let $\left(
e_{1},...,e_{n}\right)  $ be a basis for $\mathcal{H}$. Then, a basis for
$\mathcal{H}_{\mathbb{R}}$ will be provided by $\left(  e_{1},...,e_{n}%
,ie_{1},...,ie_{n}\right)  $ and $\mathcal{H}_{\mathbb{R}}\approx
\mathbb{R}^{2n}$. Once a basis has been chosen, $\mathcal{H\approx}%
\mathbb{C}^{n}$. If: $x=x^{k}e_{k},x^{k}=u^{k}+iv^{k};u^{k},v^{k}\in
\mathbb{R}$ (in short: $x=u+iv;u,v\in\mathbb{R}^{n}$), then the corresponding
vector in $\mathcal{H}_{\mathbb{R}}$ is represented by $\left(  u^{1}%
,...,u^{n},v^{1},...,v^{n}\right)  $, or $\left(  u,v\right)  $, again for
short, and it is immediate to check that the group property is satisfied. Let
now: $A:\mathcal{H}\rightarrow\mathcal{H}$ be a linear operator on
$\mathcal{H}$. The \textit{realified} of $A$ will be the linear operator:
$A_{\mathbb{R}}:\mathcal{H}_{\mathbb{R}}\rightarrow\mathcal{H}_{\mathbb{R}}$
that coincides with $A$ pointwise, i.e., if: $Ax=x^{\prime}$, $x=u+iv,
x^{\prime}=u^{\prime}+iv^{\prime}$, then: $A_{\mathbb{R}}(u,v)=(u^{\prime
},v^{\prime})$. In any given basis for $\mathcal{H}$, $A$ will be represented
by a matrix of the form: $A=\alpha+i\beta$, with $\alpha,\beta$ real $n\times
n$ matrices. Then it is also immediate to check that $A_{\mathbb{R}}$ will be
represented by the $2n\times2n$ real matrix:%
\begin{equation}
A_{\mathbb{R}}=\left\vert
\begin{array}
[c]{cc}%
\alpha & -\beta\\
\beta & \alpha
\end{array}
\right\vert \label{realif1}%
\end{equation}
It is also immediate to check that: $\left(  A+B\right)  _{\mathbb{R}%
}=A_{\mathbb{R}}+B_{\mathbb{R}}$, as well as that: $\left(  AB\right)
_{\mathbb{R}}=A_{\mathbb{R}}B_{\mathbb{R}}$, and hence the set of the linear
operators that are realifications of complex operators on $\mathcal{H}$ is
both a subspace of the vector space of all linear operators on $\mathcal{H}%
_{\mathbb{R}}$ as well as a subalgebra of the associative algebra
$\mathfrak{gl}\left(  2n,\mathbb{R}\right)  $. In particular, multiplication
in $\mathcal{H}$ by the imaginary unit will be represented by the linear
operator:%
\begin{equation}
J=\left\vert
\begin{array}
[c]{cc}%
\mathbf{0}_{n\times n} & -\mathbb{I}_{n\times n}\\
\mathbb{I}_{n\times n} & \mathbf{0}_{n\times n}%
\end{array}
\right\vert \label{complex0}%
\end{equation}
(or: $\left(  u,v\right)  \rightarrow\left(  -v,u\right)  $) with the
property:%
\begin{equation}
J^{2}=-\mathbb{I}_{2n\times2n}%
\end{equation}

\item A \textit{complex} \textit{manifold} \cite{Ch,SCH} is a manifold $Z$
that can be locally modeled on $\mathbb{C}^{n}$ for some $n$, and for which
the chart-compatibility conditions are required to be $\mathbb{C}^{\omega}$
diffeomorphisms. Then, on the tangent bundle $TZ$ one can define the
\textit{complex structure} $J_{0}$ via:%
\begin{equation}
J_{0}:TZ\rightarrow TZ;\text{ \ }J_{0}\left(  v\right)  =:iv,\text{ \ }v\in
TZ. \label{complex1}%
\end{equation}
Clearly: $J_{0}^{2}=-\mathbb{I}$. Also:

\item An \textit{almost complex }manifold \cite{MFVMR} is an even-dimensional
real manifold $M$ endowed with a $(1,1)$-type tensor field $J$ , called an
\textit{almost complex structure,} satisfying:%
\begin{equation}
J^{2}=-\mathbb{I} \label{complex3}%
\end{equation}
It was proved in Ref.\cite{NN} that an almost complex manifold becomes a
complex one iff the almost complex structure $J$ satisfies the
\textit{Nijenhuis condition} $N_{J}=0$, where $N_{J}$ is the Nijenhuis torsion
associated with $J$.

\item Finally, let $\mathcal{K}$ be a real, even-dimensional, manifold with a
complex structure and a closed two-form satisfying the compatibility
condition:%
\begin{equation}
\omega\left(  x,Jy\right)  +\omega\left(  Jx,y\right)  =0;\text{ }x,y\in
T\mathcal{K} \label{compat1}%
\end{equation}
Notice that this implies that:%
\begin{equation}
g\left(  .,.\right)  =:\omega\left(  .,J\left(  .\right)  \right)  ;\text{
}\left(  x,y\right)  \mapsto g\left(  x,y\right)  =:\omega\left(  x,Jy\right)
\label{compat2}%
\end{equation}
is \textit{symmetric} $\left(  g\left(  x,y\right)  =g\left(  y,x\right)
\forall x,y\right)  $ and nondegenerate iff $\omega$ is, hence a metric. When
$g$ is positive, then $\mathcal{K}$ is a \textit{K\"{a}hler manifold}
\cite{Ch,SCH,Weil}\footnote{If not, then $\mathcal{K}$ is also called
\cite{MFVMR} a \textit{pseudo-K\"{a}hler }manifold.}. Also, $J^{2}%
=-\mathbb{I}$ implies:%
\begin{equation}
\omega\left(  Jx,Jy\right)  =\omega\left(  x,y\right)  ;\text{ }g\left(
Jx,Jy\right)  =g\left(  x,y\right)  \text{ }\forall x,y \label{compat3}%
\end{equation}
Notice that Eq.(\ref{compat2}) implies the analog of Eq.(\ref{compat1}) for
$g$, namely:%
\begin{equation}
g\left(  x,Jy\right)  +g\left(  Jx,y\right)  =0 \label{compat4}%
\end{equation}

\end{enumerate}

A tensorial triple $\left(  g,J,\omega\right)  $, with $g$ a metric, $J$ a
complex structure and $\omega$ a symplectic structure satisfying the
conditions (\ref{compat1}),(\ref{compat2}) and (\ref{compat3}) will be called
an \textit{admissible triple.} Eq.(\ref{compat2}) and the parent equation,
obtained by substituting: $y\rightarrow Jy$ in it tell us also that:%
\begin{equation}
\omega\left(  .,.\right)  =-g\left(  .,J\left(  .\right)  \right)
\label{compat5}%
\end{equation}

Coming back now to the complex vector space $\mathcal{H}$, let it be endowed
also with an Hermitian structure $h(.,.)=\left\langle .|.\right\rangle $, i.e.
a positive-definite sesquilinear form,nondegenerate, linear in the second
factor and antilinear in the first one. Then $\mathcal{H}$ will become a
(finite-dimensional: $\dim_{\mathbb{C}}\mathcal{H}=n$) Hilbert space.\ We will
keep denoting vectors in $\mathcal{H}$ with Latin letters (i.e.: $x,y$ etc.)
and we will use Dirac's notation ($|x\rangle,|y\rangle$ etc.) only when
convenient. Separating real and imaginary parts, we can write:%
\begin{equation}%
\begin{array}
[c]{c}%
h\left(  x,y\right)  =g\left(  x,y\right)  +i\omega\left(  x,y\right) \\
g\left(  x,y\right)  =\operatorname{Re}h\left(  x,y\right) \\
\omega\left(  x,y\right)  =\operatorname{Im}h\left(  x,y\right)
\end{array}
\end{equation}
$g$ is clearly symmetric, positive and nondegenerate, while $\omega$ is
antisymmetric and nondegenerate too.

Now we can consider $\mathcal{H}_{\mathbb{R}}$ together with its tangent
bundle $T\mathcal{H}_{\mathbb{R}}\approx\mathcal{H}_{\mathbb{R}}%
\times\mathcal{H}_{\mathbb{R}}$. Points in $\mathcal{H}_{\mathbb{R}}$, i.e. in
the first factor, will be again denoted by the same Latin
letters\footnote{With reference\ to a basis, $x=$ $u+iv$ will stand (see item
$1$ above) for the (real) pair $\left(  u,v\right)  $}, and we will use Greek
letters for the second factor. Then, e.g., $\left(  x,\psi\right)  $ will
denote a point in $\mathcal{H}_{\mathbb{R}}$ and a tangent vector at $x$:
$\psi\in T_{x}\mathcal{H}_{\mathbb{R}}\approx\mathcal{H}_{\mathbb{R}}$. We can
associate with every point $x\in\mathcal{H}_{\mathbb{R}}$ the constant vector
field:%
\begin{equation}
X_{\psi}=:\left(  x,\psi\right)  \label{dilat1}%
\end{equation}
Then, we can "promote" $g$ and $\omega$ to $\left(  0,2\right)  $ tensor
fields by defining:%
\begin{equation}
g\left(  x\right)  \left(  X_{\psi},X_{\phi}\right)  =:g\left(  \psi
,\phi\right)
\end{equation}
and similarly for $\omega$. In this way, $g$ becomes a Riemannian metric and
$\omega$ a symplectic structure. Proceeding in a similar way, we define:%
\begin{equation}
J\left(  x\right)  \left(  X_{\psi}\right)  =\left(  x,J\psi\right)
\end{equation}
where: $J\psi=i\psi$ (i.e.: $J\left(  u,v\right)  =\left(  -v,u\right)  $) and
in this way $J$ too is "promoted" to a $\left(  1,1\right)  $ tensor field. As
all these tensors fields are translationally invariant, and hence the
Nijenhuis condition for $J$ is trivially satisfied, and as all the
compatibility conditions are also satisfied, $\mathcal{H}_{\mathbb{R}}$
becomes in this way a \textit{linear} K\"{a}hler manifold, with $J$ playing
the r\^{o}le of the complex structure. Explicitly, if $\left(  e_{1}%
,...,e_{n}\right)  $ is an orthonormal basis for $\mathcal{H}$, and:
$x=\left(  u,v\right)  ,y=\left(  u^{\prime},v^{\prime}\right)  $, then:%
\begin{equation}%
\begin{array}
[c]{c}%
g\left(  x,y\right)  =u\cdot u^{\prime}+v\cdot v^{\prime}\\
\omega\left(  x,y\right)  =u\cdot v^{\prime}-v\cdot u^{\prime}%
\end{array}
\end{equation}

It may be convenient to give explicit expressions by introducing real
coordinates $x^{1},...,x^{2n}$ on $\mathcal{H}_{\mathbb{R}}\approx
\mathbb{R}^{2n}$. Then, e.g., $g$ and $J$ will be explicitly represented as:%

\begin{equation}
g=g_{ij}dx^{i}\otimes dx^{j} \label{metric}%
\end{equation}
and\footnote{Here: $Jx=\left\{  \left(  Jx\right)  ^{i}\right\}  _{1}%
^{2n};\left(  Jx\right)  ^{i}=J^{i}$ $_{j}x^{j}$.
\par
\label{J}}:%
\begin{equation}
J=J_{j}^{i}dx^{j}\otimes\frac{\partial}{\partial x^{i}}%
\end{equation}
Hence:

\bigskip%

\begin{equation}
J^{2}=-\mathbb{I\Longleftrightarrow}J^{i}\text{ }_{k}J^{k}\text{ }_{j}%
=-\delta^{i}\text{ }_{j} \label{jj2}%
\end{equation}

\begin{remark}
 With the given metric, \textit{orthogonal}
matrices will be those leaving the scalar product invariant, and they will
provide a representation of $O\left(  2n\right)  $ which need not be the
standard one. Eq.(\ref{compat3}) tells us that $J$ is what we might call a
"$g$-orthogonal" matrix. In this context, it is worth recalling that the
\textit{adjoint} $\mathbb{A}^{\dag}$ w.r.t. $g$ of any linear operator
$\mathbb{A}$ (a $\left(  1,1\right)  $ tensor)is defined by:%
\begin{equation}
g\left(  x,\mathbb{A}y\right)  =g\left(  \mathbb{A}^{\dag}x,y\right)
\end{equation}
In terms of matrices:%
\begin{equation}
\mathbb{A}^{\dag}=g^{-1}\widetilde{\mathbb{A}}g
\end{equation}
where $\widetilde{\mathbb{A}}$ stands for the transpose matrix and hence, for
a generic metric tensor, (real) symmetric matrices need not be self-adjoint.
Eq.(\ref{compat4}) tells us then that $J$ is \textit{skew-adjoint} w.r.t. $g$,
i.e. that: $J^{\dag}=-J$, which \ implies, according to Eq.(\ref{complex2}):%
\begin{equation}
J^{\dag}J=\mathbb{I}%
\end{equation}
\end{remark}

\begin{remark}
$ii)$ If we consider a one-parameter group $\left\{  \exp\left(
t\mathbb{A}\right)  \right\}  _{t\in\mathbb{R}}$ of $g$-orthogonal matrices,
then: $g\left(  e^{t\mathbb{A}}x,e^{t\mathbb{A}}y\right)  =g\left(
x,y\right)  $ implies, at the infinitesimal level:%
\begin{equation}
g\left(  \mathbb{A}x,y\right)  +g\left(  x,\mathbb{A}y\right)  =0
\end{equation}
Hence, $J$ \ acts at \ the same time as a generator of finite and
infinitesimal orthogonal transformations (rotations).\newline$iii)$ in terms
of the representative matrices, the condition $g\left(  Jx,y\right)  +g\left(
x,Jy\right)  =0$ can be written as:%
\begin{equation}
\widetilde{J}\circ g+g\circ J=0
\end{equation}
i.e., as $g$ is symmetric: $\widetilde{\left(  g\circ J\right)  }=-g\circ J$,
i.e. $g\circ J$ must be a skew-symmetric matrix.
\end{remark}

Using $g$ and $J$ we can construct, as discussed before, the skew-symmetric
tensor $\omega$ (cfr Eq.(\ref{compat5})). $\omega$ will be nondegenerate iff
$g$ is, hence a symplectic form. In terms of matrices:

\begin{equation}
\omega=-g\circ J
\end{equation}
($\omega_{ij}=-g_{ik}J^{k}$ $_{j}$), Moreover. Eqs.(\ref{compat3}) and
(\ref{compat1}), i.e.:
\begin{equation}
\omega\left(  Jx,Jy\right)  =\omega\left(  x,y\right)  \forall x,y
\end{equation}
and:%
\begin{equation}
\omega\left(  Jx,y\right)  +\omega\left(  x,Jy\right)  =0\forall x,y
\end{equation}
tell us that $J$ will generate (both finite and infinitesimal) symplectic
transformations as well. Notice that, for $y=Jx$:%
\begin{equation}
\omega\left(  x,Jx\right)  =g\left(  x,x\right)
\end{equation}
and hence: $\omega\left(  x,Jx\right)  >0$ if $g$ is positive-definite.

One could start instead from the datum of a symplectic form and of a complex
structure, requiring the admissibility condition $\omega\left(  Jx,y\right)
+\omega\left(  x,Jy\right)  =0$ (which implies $\omega\left(  Jx,Jy\right)
=\omega\left(  x,y\right)  $ and viceversa), and define then:
\begin{equation}
g\left(  x,y\right)  =:\omega\left(  x,Jy\right)
\end{equation}
($g=\omega\circ J$ in terms of representative matrices), the only difference
being that, although $g$ will be still nondegenerate iff $\omega$ is, it need
not be positive unless \ $\omega\left(  x,Jx\right)  >0\forall x$.

Finally, one could start \ from $g$ and $\omega$ and require the admissibility
condition that: $J=:g^{-1}\circ\omega$ be a complex structure,i.e.:
$J^{2}=-\mathbb{I}$. \ In conclusion, a third tensor is determined whenever
any other admissible two are given.

\bigskip

\begin{remark}

We have already encountered examples of admissible triples $(g,\omega,J)$ in
Sect. \ref{se:ineq}. E.g., for the isotropic two-dimensional harmonic
oscillator we may consider $(H_{0},\omega_{0},J)$ or $(H_{3},\omega_{3},J)$ as
given in Eqns. (\ref{jg}) and (\ref{hw}), while for the one-dimensional
harmonic oscillator we may choose (see again Sect.\ref{se:ineq})
$(H,\omega,J)$ or $H^{\prime},\omega^{\prime},J^{\prime})$, as long as the
Hamiltonian is positive definite.
\end{remark}

\subsubsection{Geometric Quantum Mechanics\label{GQM}}
\bigskip

Here and in the following we will exploit the already-discussed connection
between the space $P(\mathcal{H})$ of rays and the space $\mathcal{D}_{1}%
^{1}(\mathcal{H})$ of density states of rank one to see how it is possible to
use symplectic methods to study quantum systems. This geometric approach is
based on some observations that will be developed in the following.

We have just proved that the realification $\mathcal{H}_{{\mathbb{R}}}$ of the
Hilbert space $\mathcal{H}$ \ (the space of states) is a linear K\"{a}lher
manifold, equipped with an admissible triple $(J,g,\omega)$. Now, taking into
account that $P(\mathcal{H})$ is not a linear space, we will have to use a
tensorial description of these structures. Via a momentum map on $P\left(
\mathcal{H}\right)  $ that we shall define shortly below, the space of
Hermitian operators \ (the observables) will be identified with the dual
$u^{\ast}(\mathcal{H})$ of the Lie algebra of the unitary group $U(\mathcal{H}%
)$, which can be thought of as the intersection of the Lie algebras of the
symplectic and orthogonal groups. By exploiting the fact that the action of
the latter is Hamiltonian, we will use the momentum map to define
contravariant metric and Poisson tensors on $u^{\ast}(\mathcal{H})$. Finally
we will study how these structures behave under the $U(\mathcal{H})$-action on
$u^{\ast}(\mathcal{H})$ and see how $\mathcal{D}_{1}^{1}(\mathcal{H})$ itself
becomes a K\"{a}lher manifold.

\subsubsection{Tensors on Hilbert spaces}
\label{s:ths}
\bigskip

We have seen how we can construct the tensor fields $g,J$ and $\omega$ on
$T\mathcal{H}_{\mathbb{R}}$. The $\left(  0,2\right)  $-tensors $g$ and
$\omega$ define maps from $T\mathcal{H}_{\mathbb{R}}$ to $T^{\ast}%
\mathcal{H}_{\mathbb{R}}$. The two being both non-degenerate, we can also
consider their inverses, i.e. the $\left(  2,0\right)  $ contravariant tensors
$G$ (a metric tensor) and $\Lambda$ (a Poisson tensor) mapping $T^{\ast
}\mathcal{H}_{\mathbb{R}}$ to $T\mathcal{H}_{\mathbb{R}}$ and such that:%
\begin{equation}
G\circ g=\Lambda\circ\omega=\mathbb{I}_{T\mathcal{H}_{\mathbb{R}}}%
\end{equation}
i.e., in short: $G=g^{-1},\Lambda=\omega^{-1}$. $G$ and $\Lambda$ can be used
together to define an Hermitian product between any two $\alpha,\beta$ in the
dual $\mathcal{H}_{{\mathbb{R}}}^{\ast}$ equipped with the dual complex
structure $J^{\ast}$\footnote{Which will act (see Footnote \ref{J}) via the
transpose matrix of $J$.}:
\begin{equation}
\langle\alpha,\beta\rangle_{\mathcal{H}_{{\mathbb{R}}}^{\ast}}=G(\alpha
,\beta)+i\Lambda(\alpha,\beta).
\end{equation}
This induces two (non-associative) real brackets on smooth, real-valued
functions on $\mathcal{H}_{{\mathbb{R}}}$: \newline$\bullet$ the (symmetric)
Jordan bracket $\{f,h\}_{g}=:G(df,dh)$, and:\newline$\bullet$ the
(antisymmetric) Poisson bracket $\{f,h\}_{\omega}=:\Lambda(df,dh)$.

By extending both these brackets to complex functions via complex linearity we
obtain eventually a complex bracket $\left\{  .,.\right\}  _{\mathcal{H}}$
\ defined as:
\begin{equation}
\{f,h\}_{\mathcal{H}}=\langle df,dh\rangle_{\mathcal{H}_{{\mathbb{R}}}^{\ast}%
}=:\{f,h\}_{g}+i\{f,h\}_{\omega}. \label{bra}%
\end{equation}

To make these structures more explicit, we may introduce an orthonormal basis
$\{e_{k}\}_{k=1,\cdots,n}$ in $\mathcal{H}$ and global coordinates
$(q^{k},p^{k})$ for $k=1,\cdots,n$ on $\mathcal{H}_{{\mathbb{R}}}$ defined as
\begin{equation}
\langle e_{k},x\rangle=(q^{k}+ip^{k})(x),\;\forall x\in\mathcal{H}.
\end{equation}
Then\footnote{Summation over repeated indices being understood here and in the
rest of the Section.}:%
\begin{equation}
J=dp^{k}\otimes\frac{\partial}{\partial q^{k}}-dq^{k}\otimes\frac{\partial
}{\partial p^{k}}%
\end{equation}%
\begin{equation}
g=:dq^{k}\otimes dq^{k}+dp^{k}\otimes dp^{k}%
\end{equation}%
\begin{equation}
\omega=:dq^{k}\otimes dp^{k}-dp^{k}\otimes dq^{k}%
\end{equation}
as well as:%
\begin{equation}
G=\frac{\partial}{\partial q^{k}}\otimes\frac{\partial}{\partial q^{k}}%
+\frac{\partial}{\partial p^{k}}\otimes\frac{\partial}{\partial p^{k}}
\label{tens1}%
\end{equation}%
\begin{equation}
\Lambda=\frac{\partial}{\partial p^{k}}\otimes\frac{\partial}{\partial q^{k}%
}-\frac{\partial}{\partial q^{k}}\otimes\frac{\partial}{\partial p^{k}}
\label{Poisson0}%
\end{equation}
and hence:%
\begin{equation}
\{f,h\}_{g}=\frac{\partial f}{\partial q^{k}}\frac{\partial h}{\partial q^{k}%
}+\frac{\partial f}{\partial p^{k}}\frac{\partial h}{\partial p^{k}}%
\end{equation}%
\begin{equation}
\{f,h\}_{\omega}=\frac{\partial f}{\partial p^{k}}\frac{\partial h}{\partial
q^{k}}-\frac{\partial f}{\partial q^{k}}\frac{\partial h}{\partial p^{k}}%
\end{equation}
Introducing complex coordinates: $z^{k}=:q^{k}+ip^{k}$, $\bar{z}^{k}%
=:q^{k}-ip^{k}$, we can also write
\begin{equation}
G+i\cdot\Lambda=4\frac{\partial}{\partial z^{k}}\otimes\frac{\partial
}{\partial\bar{z}^{k}}, \label{tens2}%
\end{equation}
where
\begin{equation}
\frac{\partial}{\partial z^{k}}=:\frac{1}{2}\left(  \frac{\partial}{\partial
q^{k}}-i\frac{\partial}{\partial p^{k}}\right)  ,\;\frac{\partial}%
{\partial\bar{z}^{k}}=:\frac{1}{2}\left(  \frac{\partial}{\partial q^{k}%
}+i\frac{\partial}{\partial p^{k}}\right)  . \label{vecfield}%
\end{equation}

\bigskip

Complex coordinates are employed here and also elsewhere in
this paper only as a convenient shorthand or as a stenographic notation. Their
use does not mean at all that vector fields like those in Eq.(\ref{vecfield})
should operate on functions that are holomorphic (or anti-holomorphic) in the
$z^{k}$'s. They must rather be seen as complex-valued vector fields that
operate on (smooth) complex-valued functions defined on a real differentiable manifold.

\bigskip

With this in mind, we have :
\begin{equation}
\{f,h\}_{\mathcal{H}}=4\frac{\partial f}{\partial z^{k}}\frac{\partial
h}{\partial\bar{z}^{k}},
\end{equation}
or, in more detail:%
\begin{equation}
\left\{  f,h\right\}  _{g}=2\left(  \frac{\partial f}{\partial z^{k}}%
\frac{\partial h}{\partial\bar{z}^{k}}+\frac{\partial h}{\partial z^{k}}%
\frac{\partial f}{\partial\bar{z}^{k}}\right)  ;\text{ \ }\left\{
f,h\right\}  _{\omega}=\frac{2}{i}\left(  \frac{\partial f}{\partial z^{k}%
}\frac{\partial h}{\partial\bar{z}^{k}}-\frac{\partial h}{\partial z^{k}}%
\frac{\partial f}{\partial\bar{z}^{k}}\right)  \label{det}%
\end{equation}
Notice also that:
\begin{equation}
J=-i\left(  dz^{k}\otimes\frac{\partial}{\partial z^{k}}-d\bar{z}^{k}%
\otimes\frac{\partial}{\partial\bar{z}^{k}}\right)  \label{complex}%
\end{equation}

In particular, for any $A\in gl(\mathcal{H})$ we can define the quadratic
function:
\begin{equation}
f_{A}(x)=\frac{1}{2}\langle x,Ax\rangle=\frac{1}{2}z^{\dagger}Az \label{qfa}%
\end{equation}
where $z$ is the column vector $\left(  z_{1},...,z_{n}\right)  $. It follows
immediately from Eq.(\ref{det}) that, for any $A,B\in gl(\mathcal{H})$:%
\begin{equation}
\{f_{A},f_{B}\}_{g}=f_{AB+BA} \label{bs}%
\end{equation}%
\begin{equation}
\{f_{A},f_{B}\}_{\omega}=f_{\frac{AB-BA}{i}} \label{ba}%
\end{equation}
So, the Jordan bracket of any two quadratic functions $f_{A}$ and $f_{B}$ is
related to the (commutative) Jordan bracket of $A$ and $B$, $\left[
A,B\right]  _{+}$, defined\footnote{This is actually \textit{twice} the Jordan
Bracket as it is usually defined in the literature \cite{Em}, but we find here
more convenient to employ this slightly different definition.} as:%
\begin{equation}
\left[  A,B\right]  _{+}=:AB+BA \label{Jor}%
\end{equation}
while their Poisson bracket is related to the commutator product ( the Lie
bracket) $\left[  A,B\right]  _{-}$ defined as:
\begin{equation}
\left[  A,B\right]  _{-}=:\frac{1}{i}\left(  AB-BA\right)  \label{Liecomm}%
\end{equation}
In particular, if $A$ and $B$ are Hermitian, their Jordan product
\ (\ref{Jor}) and their Lie bracket will be Hermitian as well. \ Hence, the
set of Hermitian operators on $\mathcal{H}_{\mathbb{R}}$, equipped with the
binary operations (\ref{Jor}) and (\ref{Liecomm}), becomes a \ \textit{Lie-Jordan
algebra} \cite{Em,Jo,JNW}, and the binary product \cite{Em}:
\begin{equation}
\left(  A,B\right)  =\frac{1}{2}\left(  \left[  A,B\right]  _{+}+i\left[
A,B\right]  _{-}\right)  \label{assoc1}%
\end{equation}
is an associative product (Indeed: $\left(  A,B\right)  \equiv AB$). We remark
parenthetically that all this extends without modifications \cite{Em} to the
infinite-dimensional case, if we assume: $A,B\in\mathcal{B}_{sa}\left(
\mathcal{H}\right)  $, the set of bounded self-adjoint operators on the
Hilbert space $\mathcal{H}$.

Coming back to quadratic functions, it is not hard to check that:
\begin{equation}
\{f_{A},f_{B}\}_{\mathcal{H}}=2f_{AB},
\end{equation}
which proves the associativity of the bracket (\ref{bra}) on quadratic
functions, i.e.:%
\begin{equation}
\left\{  \{f_{A},f_{B}\}_{\mathcal{H}},f_{C}\right\}  _{\mathcal{H}}=\left\{
f_{A},\left\{  f_{B},f_{C}\right\}  _{\mathcal{H}}\right\}  _{\mathcal{H}%
}=4f_{ABC},\text{\ \ }\forall A,B,C\in gl(\mathcal{H}). \label{assoc2}%
\end{equation}

We look now at real, smooth functions on $\mathcal{H}_{{\mathbb{R}}}$.\ 

First of all, it is clear that \ $f_{A}$ will be a real function iff $A$ is
Hermitian.\ The Jordan and Poisson brackets will define then a Lie-Jordan
algebra structure on the set of real, quadratic functions, and, according to
Eq.(\ref{assoc2}), the bracket $\{\mathbf{\cdot},\mathbf{\cdot}\}_{\mathcal{H}%
}$ will be an associative bracket.

For any such $f\in\mathcal{F}\left(  \mathcal{H}_{{\mathbb{R}}}\right)  $ we
may define two vector fields, the \textit{gradient} $\nabla f$ of $f$ and the
\textit{Hamiltonian vector field} $X_{f}$ associated with $f$, defined by:
\begin{equation}%
\begin{array}
[c]{ll}%
g(\cdot,\nabla f)=df & \\
\omega(\cdot,X_{f})=df &
\end{array}
\;\text{\ \ }or\text{ \ }\;\;%
\begin{array}
[c]{ll}%
G(\cdot,df)=\nabla f, & \\
\Lambda(\cdot,df)=X_{f} &
\end{array}
.
\end{equation}
which allow us also to obtain the Jordan and the Poisson brackets as:
\begin{align}
\{f,h\}_{g}  &  =g(\nabla f,\nabla h),\\
\{f,h\}_{\omega}  &  =\omega(X_{f},X_{h}).
\end{align}
Explicitly, in coordinates:%
\begin{equation}
\nabla f=\frac{\partial f}{\partial q^{k}}\frac{\partial}{\partial q^{k}%
}+\frac{\partial f}{\partial p^{k}}\frac{\partial}{\partial p^{k}}=2\left(
\frac{\partial f}{\partial z^{k}}\frac{\partial}{\partial\bar{z}^{k}}%
+\frac{\partial f}{\partial\bar{z}^{k}}\frac{\partial}{\partial z^{k}}\right)
\end{equation}%
\begin{equation}
X_{f}=\frac{\partial f}{\partial p^{k}}\frac{\partial}{\partial q^{k}}%
-\frac{\partial f}{\partial q^{k}}\frac{\partial}{\partial p^{k}}=2i\left(
\frac{\partial f}{\partial z^{k}}\frac{\partial}{\partial\bar{z}^{k}}%
-\frac{\partial f}{\partial\bar{z}^{k}}\frac{\partial}{\partial z^{k}}\right)
\end{equation}
which are such that $J(\nabla f)=X_{f}$.

Turning to linear operators, to any $A:\mathcal{H}\rightarrow\mathcal{H}$ we
can associate:

\begin{enumerate}
\item A quadratic function as in Eq. (\ref{qfa}), and (cfr. also below,
Sect.\ref{Hermitian}),

\item A vector field: $X_{A}:\mathcal{H}\rightarrow T\mathcal{H}$ \ via:
$\ x\longmapsto\left(  x,Ax\right)  ,$ and:

\item A $(1,1)$ tensor field: $T_{A}:T_{x}\mathcal{H}\ni\left(  x,y\right)
\longmapsto\left(  x,Ay\right)  \in T_{x}\mathcal{H}$. Clearly, as already
remarked, $f_{A}$ is real if and only if $A$ is Hermitian. In this case:%
\begin{equation}
\nabla f_{A}=X_{A} \label{grad}%
\end{equation}
and:%
\begin{equation}
X_{f_{A}}=J(X_{A}) \label{comp}%
\end{equation}
Indeed, denoting with $(\cdot,\cdot)$ the pairing between vectors and
covectors, \ Eq.(\ref{grad}) holds because:%
\begin{eqnarray}
g\left(  y,X_{A}\left(  x\right)  \right)  &=&g\left(  y,Ax\right)  =\frac{1}%
{2}\left(  \left\langle y,Ax\right\rangle _{\mathcal{H}}+\left\langle
Ax,y\right\rangle _{\mathcal{H}}\right)  =\nonumber \\
&=&\left(  df_{A}\left(  x\right)  ,y\right)
\end{eqnarray}
while Eq.(\ref{comp}) follows from the second expression in Eq.(\ref{compat2}%
), i.e. from : $g\left(  y,Ax\right)  =\omega\left(  y,(JX_{A})(x)\right)
=\omega\left(  y,iAx\right)  $. $\blacksquare$
\end{enumerate}

Thus, we will write:%
\begin{equation}
\nabla f_{A}=A\text{ \ and\textit{: \ }}X_{f_{A}}=iA
\end{equation}

In particular, if we consider the identity operator $\mathbb{I}$, \ we obtain
the dilation (or \textit{Liouville}) field (cfr. also Eq.(\ref{dilat1})):
\begin{equation}
\Delta:x\longmapsto\left(  x,x\right)  \label{dilat2}%
\end{equation}
or, in real coordinates:%
\begin{equation}
\Delta=q^{k}\frac{\partial}{\partial q^{k}}+p^{k}\frac{\partial}{\partial
p^{k}}%
\end{equation}
which is such that:
\begin{equation}
X_{A}=T_{A}(\Delta).
\end{equation}
Finally we can also define the \textit{phase vector field}:
\begin{equation}
\Gamma=J(\Delta)=p^{k}\frac{\partial}{\partial q^{k}}-q^{k}\frac{\partial
}{\partial p^{k}} \label{phase}%
\end{equation}
that will be considered in the next Section.

\subsubsection{The complex projective space}
\label{se:cps} 
\bigskip

We would like now to discuss in some detail the structure of
the complex projective Hilbert space $P\mathcal{H}$, which, as we have already
mentioned, represents the right context to describe a geometric formulation of
Quantum Mechanics. Indeed, given any vector $|x\rangle\in\mathcal{H-}\left\{
\mathbf{0}\right\}  $, the corresponding element in $P\mathcal{H}$ may be
represented by the rank-one projector: \ $\widehat{\rho}_{x}=:$ $|x\rangle
\langle x|/\left\langle x|x\right\rangle $ in $D_{1}^{1}\left(  \mathcal{H}%
\right)  $ (or simply: $\widehat{\rho}_{x}=:$ $|x\rangle\langle x|$ if the
vector is already normalized), and this will encode all the relevant physical
information contained in $|x\rangle$.

In more geometric terms, we can consider the distribution generated by the
dilation field $\Delta$ and the phase field $\Gamma=J\left(  \Delta\right)  $,
which is involutive as $\left[  \Delta,J\left(  \Delta\right)  \right]  =0$.
Going to the quotient with respect to the foliation associated with this
distribution (cfr.Eq.(\ref{quotient})) will be a way of generating the ray
space $P\mathcal{H}$ which is independent on any Hermitian structure.
Contravariant tensorial objects on $\mathcal{H}$ will "pass to the quotient"
(i.e. will be projectable) if and only if they are left invariant by both
$\Delta$ and $\Gamma$, i.e. if they are homogeneous of degree zero and
invariant under multiplication of vectors by a phase. Typical quadratic
functions that "pass to the quotient" will be normalized expectation values of
the form:%
\begin{equation}
\rho_{x}\left(  A\right)  =:Tr\left\{  \widehat{\rho}_{x}A\right\}
=\frac{\left\langle x|A|x\right\rangle }{\left\langle x|x\right\rangle }%
\end{equation}
with $A$ any linear operator and for any Hermitian structure on $\mathcal{H}$.
We note parenthetically that the subalgebra of functions on $\mathcal{H}$ that
are invariant under $\Gamma$ and $\Delta$ will define, via the construction of
the Gel'fand-Kolmogoroff theorem \cite{MarMor2}, a manifold which can again be
identified with $P \mathcal{H}$.

Concerning projectability of tensors, the complex structure $J$, being (cfr.,
e.g., Eq.(\ref{complex})) homogeneous of degree zero and phase-invariant, will
be a projectable tensor, while it is clear that the Jordan and Poisson tensors
$G$ and $\Lambda$ defined respectively in Eq.(\ref{tens1}) or, for that
matter, the complex-valued tensor of Eq.(\ref{tens2}) will \textit{not} be
projectable (as they are phase-invariant but homogeneous of degree $-2$). To
turn them into projectable objects we will have to multiply them \cite{GKM} by
the "conformal factor": $\theta\left(  z\right)  =:$ $z^{\dag}z$, thus
defining new tensors:%
\begin{equation}
\widetilde{\Lambda}\left(  z\right)  =:\theta\left(  z\right)  \Lambda\left(
z\right)  \label{newtensor}%
\end{equation}
and similarly for $G$.

Let us examine these structures directly on $P\mathcal{H}$ more
closely\footnote{In the following of this Section, we will use the
$(0,2)$-tensors $g,\omega$ instead of their (inverse) $(2,0)$-tensors
$G,\Lambda$ since calculations result to be more easily performed.}. Recall
that, in the finite dimensional case, $P\mathcal{H}$ is homeomorphic to
${\mathbb{C}}{\mathbb{P}}^{n}$ and it is therefore made up of the equivalence
classes of vectors $\mathbf{Z}=(Z^{0},Z^{1},\cdots,Z^{n})\in{\mathbb{C}}%
^{n+1}$ w.r.t. the equivalence relation $Z\approx\lambda Z$; $\lambda
\in{\mathbb{C}}-\{0\}$. The space ${\mathbb{C}}{\mathbb{P}}^{n}$ is a
K\"{a}hler manifold when endowed with the 
Fubini-Study metric \cite{Ben,Huy}, whose pull-back to ${\mathbb{C}}^{n+1}$ is given by:
\begin{equation}
g_{FS}=\frac{1}{(\mathbf{Z}\cdot\bar{\mathbf{Z}})^{2}}\left[  (\mathbf{Z}%
\cdot\bar{\mathbf{Z}})d\mathbf{Z}\otimes_{S}d\bar{\mathbf{Z}}-(d\mathbf{Z}%
\cdot\bar{\mathbf{Z}})\otimes_{S}(\mathbf{Z}\cdot d\bar{\mathbf{Z}})\right]
\end{equation}
where $\mathbf{Z}\cdot\bar{\mathbf{Z}}=Z^{a}\bar{Z}^{a}$, $d\mathbf{Z}%
\cdot\bar{\mathbf{Z}}=dZ^{a}\bar{Z}^{a}$, $d\mathbf{Z}\otimes_{S}%
d\bar{\mathbf{Z}}=dZ^{a}d\bar{Z}^{a}+d\bar{Z}^{a}dZ^{a}$, and so on (the sum
over repeated indices has to be understood), together with the compatible
symplectic form:
\begin{equation}
\omega_{FS}=\frac{i}{(\mathbf{Z}\cdot\bar{\mathbf{Z}})^{2}}\left[
(\mathbf{Z}\cdot\bar{\mathbf{Z}})d\mathbf{Z}\wedge d\bar{\mathbf{Z}%
}-(d\mathbf{Z}\cdot\bar{\mathbf{Z}})\wedge(\mathbf{Z}\cdot d\bar{\mathbf{Z}%
})\right]  =d\theta_{FS}%
\end{equation}
where:
\begin{equation}
\theta_{FS}=\frac{1}{2i}\frac{\overline{\mathbf{Z}}d\mathbf{Z}-\mathbf{Z}%
d\overline{\mathbf{Z}}}{\mathbf{Z\cdot}\overline{\mathbf{Z}}}%
\end{equation}

The isometries are just the usual unitary transformations which, in
infinitesimal form, are written as:
\begin{equation}
\dot{Z}^{a}=iA^{ab}Z^{b}%
\end{equation}
where $A=[A^{ab}]$ is a Hermitian matrix. These are the equations for the flow
of a generic Killing vector field, which therefore has the
form\footnote{Notice that these are exactly the Killing vector fields of
$S^{2n+1}$. In particular, for $A={\mathbb{I}}$ we obtain $X_{k}=\Gamma$ which
is a vertical vector field w.r.t. the Hopf projection $\pi_{H}:S^{2n+1}%
\rightarrow{\mathbb{C}}{\mathbb{P}}^{n}$.}:
\begin{equation}
X_{A}=\dot{Z}^{a}\partial_{Z^{a}}-\dot{\bar{Z}}^{a}\partial_{\bar{Z}^{a}%
}=iA^{ab}(Z^{b}\partial_{Z^{a}}-\bar{Z}^{a}\partial_{\bar{Z}^{b}})
\end{equation}
A straightforward calculation shows that:%
\begin{eqnarray}%
\omega_{FS}(\cdot,X_{A})&=&\frac{1}{\mathbf{Z}\cdot\bar{\mathbf{Z}}}[d\bar
{Z}^{a}A^{ab}Z^{b}+\bar{Z}^{a}A^{ab}dZ^{b}]-\frac{\bar{Z}^{a}A^{ab}Z^{b}%
}{(\mathbf{Z}\cdot\bar{\mathbf{Z}})^{2}}[dZ^{c}\bar{Z}^{c}+Z^{c}d\bar{Z}%
^{c}]=\nonumber \\
&=&d\left(  i_{X_{A}}\theta_{FS}\right)
\end{eqnarray}
i.e. that $X_{A}$ is  the Hamiltonian vector field $X_{f_{A}}$, $\omega
_{FS}(\cdot,X_{f_{A}})=df_{A}$ associated with the (real) quadratic function:
\begin{equation}
f_{A}=\frac{\bar{\mathbf{Z}}\cdot A\mathbf{Z}^{b}}{\mathbf{Z}\cdot
\bar{\mathbf{Z}}}=\frac{\bar{Z}^{a}A^{ab}Z^{b}}{Z^{c}\bar{Z}^{c}}=i_{X_{A}%
}\theta_{FS}\label{qf}%
\end{equation}
for the Hermitian matrix $A$. Also, some algebra shows that, given any two
real quadratic functions $f_{A},f_{B}$ ($A,B$ being Hermitian matrices), their
corresponding Hamiltonian vector fields satisfy:
\begin{equation}
\omega_{FS}(X_{f_{A}},X_{f_{B}})=X_{f_{A}}(df_{B})=f_{\frac{AB-BA}{i}}%
\end{equation}
Therefore, the Poisson brackets associated with the symplectic form:
\begin{equation}
\{f,g\}_{\omega_{FS}}:=-\omega(X_{f},X_{g})
\end{equation}
are such that:
\begin{equation}
\{f_{A},f_{B}\}_{\omega_{FS}}=f_{\frac{AB-BA}{i}}%
\end{equation}
In a similar way, one can prove that the gradient vector field $\nabla_{f_{A}%
}$, $g_{FS}(\cdot,\nabla_{f_{A}})=df_{A}$, of $f_{A}$ has the form:
\begin{equation}
\nabla_{A}=A^{ab}(Z^{b}\partial_{Z^{a}}+\bar{Z}^{a}\partial_{\bar{Z}^{b}})
\end{equation}
so that
\begin{equation}
g_{FS}(\nabla_{f_{A}},\nabla_{f_{B}})=\nabla_{f_{A}}(df_{B})=f_{AB+BA}%
-f_{A}\cdot f_{B}%
\end{equation}
Given any two real quadratic functions $f_{A},f_{B}$, we can therefore
define a Jordan bracket by setting:
\begin{equation}
\{f_{A},f_{B}\}_{g}:=g_{FS}(\nabla_{f_{A}},\nabla_{f_{B}})+f_{A}\cdot
f_{B}=f_{AB+BA}%
\end{equation}

One says \cite{chru} that a real function on $P\mathcal{H}$ is K\"{a}hlerian
iff its Hamiltonian vector field is also Killing. Such  functions represent
quantum observables. The above calculations show that the space $\mathcal{F}%
(P\mathcal{H})$ of real quadratic functions on $P\mathcal{H}$ consists exactly
of all K\"{a}hlerian functions. To extend this concept to the complex case,
one says that a complex valued function on $P\mathcal{H}$ is K\"{a}hlerian iff
are so its real and imaginary parts. Clearly, any such $f$ is a quadratic
function of the form (\ref{qf}) with now $A\in\mathcal{B}(\mathcal{H})$. Also,
on the space, $\mathcal{F}^{\mathbb{C}}(P\mathcal{H})$, of K\"{a}hlerian
complex functions one can define both an Hermitian two-form:
\begin{equation}
h(\cdot,\cdot)=g_{FS}(\cdot,\cdot)+i\omega_{FS}(\cdot,\cdot)
\end{equation}
and and associative bilinear product (star-product) via:
\begin{equation}
f\star g:=f\cdot g+\frac{1}{2}h(df,dg)=\frac{1}{2}\left[  \{f,g\}_{g}%
+i\{f,g\}_{\omega}\right]+f\cdot g  \label{spc}%
\end{equation}
under which the space $\mathcal{F}^{\mathbb{C}}(P\mathcal{H})$ is closed since
$f_{A}\star f_{B}=f_{AB}$, thus obtaining a particular realization of the
${\mathbb{C}}^{\ast}$-algebra of bounded operators $\mathcal{B}(\mathcal{H})$.

Let us suppose now that $(\mathcal{M},\tilde{h})$ be a generic K\"{a}hler
manifold. Also in this generic case, given any two functions $f,g$ in the
space of K\"{a}hlerian (w.r.t. the metric $\tilde{g}=Re(\tilde{h})$) complex
functions $\mathcal{F}^{\mathbb{C}}(\mathcal{M})$ one can define a $\star
$-product:
\begin{equation}
f\star g:=f\cdot g+\frac{1}{2}\tilde{h}(df,dg)
\end{equation}
but now this product, although inner, will be not in general associative
unless the functions are K\"{a}hlerian.
The condition that $\mathcal{F}^{\mathbb{C}}(\mathcal{M})$  be closed puts
very restrictive conditions on the K\"{a}hler structure of $\mathcal{M}$ which
imply \cite{Cir} that $\mathcal{M}$ be a projective Hilbert space
$P\mathcal{H}$. At the end of Sect. (\ref{se:GNS}), after the discussion of
the so called GNS construction, we will see how realizations of a
${\mathbb{C}}^{\ast}$-algebra as bounded operators on a suitable Hilbert space
are in one-to-one correspondence with the action of the unitary group on the
K\"{a}hler manifold.

\subsubsection{The momentum map\label{mom_map}}
\bigskip

We shall consider now the action of the unitary group $\mathcal{U}%
(\mathcal{H})$ on $\mathcal{H}$, which is the group of linear transformations
that preserve the triple $(g, \omega,J)$. In the following, we will denote
with $u(\mathcal{H})$ the Lie algebra of $\mathcal{U}(\mathcal{H})$ of
anti-Hermitian operators and identify the space of all Hermitian operators
with the dual $u^{\ast}(\mathcal{H})$ of $u(\mathcal{H})$ via the pairing:
\begin{equation}
\langle A,T\rangle=:\frac{i}{2}Tr(AT),A\in u^{\ast}(\mathcal{H}),T\in
u(\mathcal{H}) \label{pair}%
\end{equation}
On $u^{\ast}(\mathcal{H})$ we can define a Lie bracket (cfr.also
Sect.\ref{s:ths}):
\begin{equation}
\lbrack A,B]_{-}=:\frac{1}{i}(AB-BA),
\end{equation}
with respect to which it becomes a Lie algebra, and also a \ Jordan bracket:
\begin{equation}
\lbrack A,B]_{+}=:AB+BA.
\end{equation}
with the two together \ giving $u^{\ast}(\mathcal{H})$ the structure of \ a
Lie-Jordan algebra \cite{Em}.

In addition, $u^{\ast}(\mathcal{H})$ is equipped with the scalar product
\begin{equation}
\langle A,B\rangle_{u^{\ast}}=\frac{1}{2}Tr(AB) \label{scalar}%
\end{equation}
which satisfies:%
\begin{equation}
\langle\lbrack A,\xi]_{-},B\rangle_{u^{\ast}}=\frac{1}{2}Tr([A,\xi
]_{-}B)=\frac{1}{2}Tr(A,[\xi,B]_{-})=\langle A,[\xi,B]_{-}\rangle_{u^{\ast}}
\label{is}%
\end{equation}%
\begin{equation}
\langle\lbrack A,\xi]_{+},B\rangle_{u^{\ast}}=\frac{1}{2}Tr([A,\xi
]_{+}B)=\frac{1}{2}Tr(A,[\xi,B]_{+})=\langle A,[\xi,B]_{+}\rangle_{u^{\ast}}
\label{ia}%
\end{equation}

With any $A\in u^{\ast}(\mathcal{H})$, we can associate the fundamental vector
field $X_{A}$ on the Hilbert space corresponding to the element $\frac{1}%
{i}A\in u(\mathcal{H})$ defined by the formula:
\begin{equation}
\frac{d}{dt}e^{-\frac{t}{i}A}(x)|_{t=0}=iA(x),\;\forall x\in\mathcal{H}%
\end{equation}
In other words, $X_{A}=iA$. We already know from Sect. \ref{s:ths} that $iA$ has
$f_{A}$ as its Hamiltonian function: $\omega(\cdot,X_{A})=df_{A}$. Thus, for
any $x\in\mathcal{H}_{\mathbb{R}}$ we obtain a $\mu(x)\in u^{\ast}%
(\mathcal{H})$ such that:
\begin{equation}
\langle\mu(x),\frac{1}{i}A\rangle=f_{A}(x)=\frac{1}{2}\langle x,Ax\rangle
_{\mathcal{H}} \label{mom}%
\end{equation}
In such a way we obtain a mapping:
\begin{equation}
\mu:\mathcal{H}_{{\mathbb{R}}}\rightarrow u^{\ast}(\mathcal{H})
\end{equation}
which is called the momentum map \cite{Mar10}.

More explicitly, it follows from Eq.(\ref{pair}) that:
\begin{equation}
\langle\mu(x),\frac{1}{i}A\rangle=\frac{1}{2}Tr(\mu(x)A)
\end{equation}
which, \ when compared with Eq.(\ref{mom}), yields:
\begin{equation}
\mu(x)=|x\rangle\langle x|\, \label{mu}%
\end{equation}
We may therefore conclude that the unit sphere in $\mathcal{H}$ can be
projected onto $u^{\ast}(\mathcal{H}) $ in an equivariant way with respect to
the coadjoint action of $\mathcal{U} (\mathcal{H})$. Also, in finite
dimensions, the unit sphere is odd dimensional and the orbit in $u^{\ast
}\left(  \mathcal{H}\right)  $ is symplectic.

With every $A\in u^{\ast}\left(  \mathcal{H}\right)  $ we can associate, with
the by now familiar identification (as with every other linear vector space)
of the tangent space at every point of $u^{\ast}\left(  \mathcal{H}\right)  $
with $u^{\ast}\left(  \mathcal{H}\right)  $ itself, the linear function (hence
a one-form) $\hat{A}:$ $u^{\ast}\left(  \mathcal{H}\right)  \rightarrow
\mathbb{R}$ defined as: $\hat{A}=:\left\langle A,\mathbf{\cdot}\right\rangle
_{u^{\ast}}$. Then, we can define two contravariant tensors, a symmetric
(Jordan) \ tensor:%
\begin{equation}
R(\hat{A},\hat{B})\left(  \xi\right)  =:\langle\xi,[A,B]_{+}\rangle_{u^{\ast}}
\label{Jord}%
\end{equation}
and a Poisson (Konstant-Kirillov-Souriau \cite{Kir,Kir2,Ko,So}) tensor:%
\begin{equation}
I(\hat{A},\hat{B})\left(  \xi\right)  =\langle\xi,[A,B]_{-}\rangle_{u^{\ast}}
\label{KKS}%
\end{equation}
($A,B,\xi\in u^{\ast}\left(  \mathcal{H}\right)  $). We notice that the
quadratic function $f_{A}$ is the pull-back of $\hat{A}$ \ via the momentum
map since, for all $x\in\mathcal{H}$:
\begin{equation}
\mu^{\ast}(\hat{A})(x)=\hat{A}\circ\mu(x)=\langle A,\mu(x)\rangle_{u^{\ast}%
}=\frac{1}{2}\langle x,Ax\rangle_{\mathcal{H}}=f_{A}(x)
\end{equation}
This means also that, if: $\xi=\mu(x)$:
\begin{equation}
(\mu_{\ast}G)(\hat{A},\hat{B})\left(  \xi\right)  =G(df_{A},df_{B})\left(
x\right)  =\{f_{A},f_{B}\}_{g}(x)=f_{[A,B]_{+}}(x)=R(\hat{A},\hat{B})\left(
\xi\right)
\end{equation}
where the last equality follows from Eq.(\ref{bs}), i.e.:%
\begin{equation}
\mu_{\ast}G=R
\end{equation}
Similarly, by using now Eq.(\ref{ba}), we find:
\begin{equation}
(\mu_{\ast}\Lambda)(\hat{A},\hat{B})\left(  \xi\right)  =\Lambda(df_{A}%
,df_{B})\left(  x\right)  =\{f_{A},f_{B}\}_{\omega}(x)=f_{\left[  A,B\right]
_{-}}(x)=I(\hat{A},\hat{B})\left(  \xi\right)
\end{equation}
i.e.:%
\begin{equation}
\mu_{\ast}\Lambda=I
\end{equation}
Thus, the momentum map relates the contravariant metric tensor $G$ and the
Poisson tensor $\Lambda$ with the corresponding contravariant tensors $R$ and
$I$ . Together they form the complex tensor:
\begin{equation}
(R+iI)(\hat{A},\hat{B})\left(  \xi\right)  =2\langle\xi,AB\rangle_{u^{\ast}}%
\end{equation}
which is related to the Hermitian product on $u^{\ast}(\mathcal{H})$.

\bigskip

\begin{example}
Let $\mathcal{H}=\mathbb{C}^{2}$ (the Hilbert space
appropriate for a two-level system). We can write any $A\in u^{\ast
}(\mathbb{C}^{2})$ as:%
\begin{equation}
A=y^{0}\mathbb{I}+\mathbf{y}\cdot\mathbf{\sigma} \label{ustar}%
\end{equation}
where $\mathbb{I}$ is the $2\times2$ identity, $\mathbf{y}\cdot\mathbf{\sigma
}=y^{1}\sigma_{1}+y^{2}\sigma_{2}+y^{3}\sigma_{3}$ and: $\mathbf{\sigma
=}(\sigma_{1},\sigma_{2},\sigma_{3})$ are the Pauli matrices:%
\begin{equation}
\sigma_{1}=\left\vert
\begin{array}
[c]{cc}%
0 & 1\\
1 & 0
\end{array}
\right\vert ,\sigma_{2}=\left\vert
\begin{array}
[c]{cc}%
0 & -i\\
i & 0
\end{array}
\right\vert ,\sigma_{3}=\left\vert
\begin{array}
[c]{cc}%
1 & 0\\
0 & -1
\end{array}
\right\vert
\end{equation}
with the well-known identities \cite{Mes}:%
\begin{equation}
\sigma_{h}\sigma_{k}=\delta_{hk}\mathbb{I}+i\varepsilon_{hkl}\sigma_{l}
\label{sigmaid1}%
\end{equation}
$\left(  h,k,l=1,2,3\right)  $ and:%
\begin{equation}
\sigma_{j}\sigma_{k}\sigma_{l}=i\varepsilon_{jkl}\mathbb{I}+\sigma_{j}%
\delta_{kl}-\sigma_{k}\delta_{jl}+\sigma_{l}\delta_{jk} \label{sigmaid2}%
\end{equation}
Every $A\in u^{\ast}(\mathbb{C}^{2})$ is then represented by the (real)
"four-vector" $\left(  y_{A}^{0},\mathbf{y}_{A}\right)  $, and:%
\begin{equation}
y_{A}^{0}=\frac{1}{2}Tr\left(  A\right)  ;\text{ }y_{A}^{k}=\frac{1}%
{2}Tr\left(  \sigma_{k}A\right)  ;\text{ }k=1,2,3
\end{equation}
or, in short:%
\begin{equation}
y_{\mu}(A)=\left\langle A|\sigma_{\mu}\right\rangle ,\text{ }\mu=0,1,2,3,
\sigma_{0}=\mathbb{I}%
\end{equation}

\end{example}
\bigskip

\textbf{Digression.}

Rank-one projectors $\left(  A=\rho,\rho^{\dag}=\rho,\text{ }Tr\rho=1,\text{
}\rho^{2}=\rho\right)  $ can be parametrized as \cite{MNE}:%
\begin{equation}
\rho=\rho\left(  \theta,\phi\right)  =\left\vert
\begin{array}
[c]{cc}%
\sin^{2}\frac{\theta}{2} & \frac{1}{2}e^{i\phi}\sin\theta\\
\frac{1}{2}e^{-i\phi}\sin\theta & \cos^{2}\frac{\theta}{2}%
\end{array}
\right\vert ;\text{ }0\leq\theta<\pi,0\leq\phi<2\pi
\end{equation}
Then, they correspond to:%
\begin{equation}
y^{0}=\frac{1}{2},\text{ }y^{1}=\frac{1}{2}\sin\theta\cos\phi,\text{ }%
y^{2}=-\frac{1}{2}\sin\theta\sin\phi,\text{ }y^{3}=-\frac{1}{2}\cos
\theta\label{dmc2}%
\end{equation}
(hence: $\mathbf{y}^{2}=1/4$ for all rank-one projectors). \bigskip As already
discussed elsewhere, we can associate with every $A\equiv\left(  y_{A}%
^{0},\mathbf{y}_{A}\right)  $ the vector field: $y^{0}(A)\partial_{0}%
+y^{1}(A)\partial_{1}+y^{2}(A)\partial_{2}+y^{3}(A)\partial_{3}$ $\left(
\partial_{0}=\partial/\partial y^{0}\text{ and so on}\right)  $. Also (see the
discussion immediately above Eq.(\ref{Jord})), $\hat{A}=\left\langle
A,\mathbf{\cdot}\right\rangle _{u^{\ast}}$ will be represented by the
one-form:%
\begin{equation}
\hat{A}=y^{0}(A)dy^{0}+y^{1}(A)dy^{1}+y^{2}(A)dy^{2}+y^{3}(A)dy^{3}%
\end{equation}

Using then Eq.(\ref{ustar}) one proves easily that:%
\begin{equation}
AB=\left(  y_{A}^{0}y_{B}^{0}+\mathbf{y}_{A}\cdot\mathbf{y}_{B}\right)
\mathbb{I}+\left(  y_{A}^{0}\mathbf{y}_{B}+y_{B}^{0}\mathbf{y}_{A}%
+i\mathbf{y}_{A}\mathbf{\times y}_{B}\right)  \cdot\mathbf{\sigma}%
\end{equation}
(with $"\times"$ denoting the standard cross-product of three-vectors) and
hence\footnote{In particular: $\left\langle \rho\left(  \theta,\phi\right)
\rho\left(  \theta^{\prime},\phi^{\prime}\right)  \right\rangle _{u^{\ast}}=$
\par
$\left\{  1+\sin\theta\sin\theta^{\prime}\cos\left(  \phi-\phi^{\prime
}\right)  +\cos\theta\cos\theta^{\prime}\right\}  /4$ \ for rank-one
projectors.}:%
\begin{equation}
\left\langle AB\right\rangle _{u^{\ast}}=\frac{1}{2}Tr\left(  AB\right)
=y_{A}^{0}y_{B}^{0}+\mathbf{y}_{A}\cdot\mathbf{y}_{B}%
\end{equation}
Moreover:%

\begin{equation}
\left[  A,B\right]  _{+}=2\left\{  \left(  y_{A}^{0}y_{B}^{0}+\mathbf{y}%
_{A}\cdot\mathbf{y}_{B}\right)  \mathbb{I+(}y_{A}^{0}\mathbf{y}_{B}+y_{B}%
^{0}\mathbf{y}_{A})\cdot\mathbf{\sigma}\right\}
\end{equation}
while:%
\begin{equation}
\left[  A,B\right]  _{-}=2\mathbf{y}_{A}\mathbf{\times y}_{B}\cdot
\mathbf{\sigma}%
\end{equation}

Then:%
\begin{equation}%
\begin{array}
[c]{c}%
R(\hat{A},\hat{B})\left(  \xi\right)  =\langle\xi,[A,B]_{+}\rangle_{u^{\ast}%
}=\left\langle \left[  \xi,A\right]  _{+},B\right\rangle =\\
2\xi^{0}\left(  y_{A}^{0}y_{B}^{0}+\mathbf{y}_{A}\cdot\mathbf{y}_{B}\right)
+2\left(  y_{A}^{0}\mathbf{y}_{B}+y_{B}^{0}\mathbf{y}_{A}\right)
\cdot\mathbf{\xi=}\\
2\left(  y_{A}^{0}\xi^{0}+\mathbf{y}_{A}\cdot\mathbf{\xi}\right)  y_{B}%
^{0}+2\left(  y_{A}^{0}\mathbf{\xi}+\xi^{0}\mathbf{y}_{A}\right)
\cdot\mathbf{y}_{B}%
\end{array}
\label{Jord2}%
\end{equation}
and hence, explicitly \cite{GKM}:%
\begin{equation}%
\begin{array}
[c]{c}%
R\left(  \xi\right)  =2\partial_{0}\otimes\left(  \xi^{1}\partial_{1}+\xi
^{2}\partial_{2}+\xi^{3}\partial_{3}\right)  +2\left(  \xi^{1}\partial_{1}%
+\xi^{2}\partial_{2}+\xi^{3}\partial_{3}\right)  \otimes\partial_{0}+\\
2\xi^{0}\left(  \partial_{0}\otimes\partial_{0}+\partial_{1}\otimes
\partial_{1}+\partial_{2}\otimes\partial_{2}+\partial_{3}\otimes\partial
_{3}\right)
\end{array}
\label{Jord3}%
\end{equation}
Quite similarly, one finds:%
\begin{equation}
I(\hat{A},\hat{B})\left(  \xi\right)  =2(\xi\times\mathbf{y}_{A}%
)\cdot\mathbf{y}_{B}=2(\mathbf{y}_{A}\times\mathbf{y}_{B})\cdot\xi\label{iab}%
\end{equation}
and:
\begin{equation}
I\left(  \xi\right)  =2\left(  \xi^{1}\partial_{2}\wedge\partial_{3}+\xi
^{2}\partial_{3}\wedge\partial_{1}+\xi^{3}\partial_{1}\wedge\partial
_{2}\right)  \label{KKS2}%
\end{equation}
We thus find the following tensor:
\begin{align}
R+iI = 2\left[  \right.  \partial_{0} \otimes y^{k} \partial_{k} + y^{k}
\partial_{k} \otimes\partial_{0}  &  +\nonumber\\
y^{0} ( \partial_{0} \otimes\partial_{0} + \partial_{k} \otimes\partial_{k} )
&  + \left.  i \epsilon_{hkl} y^{h} \partial_{k} \otimes\partial_{l} \right]
\end{align}
\bigskip

To conclude this Section, we define also two $\left(  1,1\right)  $ tensors,
$\widetilde{\mathcal{R}}$ and $\widetilde{\mathcal{J}}:Tu^{\ast}\left(
\mathcal{H}\right)  \rightarrow Tu^{\ast}\left(  \mathcal{H}\right)  $ that
will be employed below in Sect.\ref{density} via:%
\begin{equation}
\widetilde{\mathcal{R}}_{\xi}\left(  A\right)  =:\left[  \xi,A\right]
_{+}=R\left(  \widehat{A},.\right)  \left(  \xi\right)
\end{equation}
and:%
\begin{equation}
\widetilde{\mathcal{J}}_{\xi}\left(  A\right)  =:\left[  \xi,A\right]
_{-}=I\left(  \widehat{A},.\right)  \left(  \xi\right)
\end{equation}
for any $A\in T_{\xi}u^{\ast}\left(  \mathcal{H}\right)  \approx u^{\ast
}\left(  \mathcal{H}\right)  $, the last passage in both equations following
from Eqns.(\ref{is}) and (\ref{ia}).

In the previous example ($\mathcal{H}\approx\mathbb{C}^{2}$) we find
explicitly, in coordinates:%
\begin{equation}
\widetilde{\mathcal{R}}_{\xi}\left(  A\right)  =2\left(  y_{A}^{0}\xi
^{0}+\mathbf{y}_{A}\cdot\mathbf{\xi}\right)  \partial_{0}+2\left(  y_{A}%
^{0}\xi^{i}+\xi^{0}y_{A}^{i}\right)  \partial_{i}%
\end{equation}
or:%
\begin{equation}
\widetilde{\mathcal{R}}_{\xi}=2\left(  \xi^{0}dy^{0}+\mathbf{\xi\cdot
dy}\right)  \otimes\partial_{0}+2\left(  \xi^{i}dy^{0}+\xi^{0}dy^{i}\right)
\otimes\partial_{i}%
\end{equation}
and:%
\begin{equation}
\widetilde{\mathcal{J}}_{\xi}\left(  A\right)  =2\varepsilon_{ijk}\xi^{i}%
y_{A}^{j}\partial_{k}%
\end{equation}
or:%
\begin{equation}
\widetilde{\mathcal{J}}_{\xi}=2\varepsilon_{ijk}\xi^{i}dy^{j}\otimes
\partial_{k}%
\end{equation}

\subsubsection{The space of density states\label{density}}
\bigskip

We have seen in Sect. \ref{se:cps} that it is possible to obtain
$\mathcal{P}({\mathcal{H}})$ as a quotient of $\mathcal{H}-\left\{
\mathbf{0}\right\}  $ with respect to the involutive distribution associated
with $\Delta$ and $J(\Delta)$. Eq. (\ref{mu}) shows that the image of
$\mathcal{H}-\left\{  \mathbf{0}\right\}  $ under the momentum map consists of
the set of all non-negative Hermitian operators of rank one, that will be
denoted as $\mathcal{P}^{1}(\mathcal{H})$, i.e.\footnote{Note that here the
vectors are not necessarily normalized.}:%
\begin{equation}
\mathcal{P}^{1}(\mathcal{H})=\left\{  |x\rangle\langle x|;\text{ }%
x\in\mathcal{H},\text{ }x\neq0\right\}
\end{equation}

On the other hand, the coadjoint action of $\mathcal{U}(\mathcal{H})$:
$(U,\rho)\mapsto U\rho U^{\dagger}$ ($\rho\in\mathcal{P}^{1}(\mathcal{H}%
),U\in\mathcal{U}(\mathcal{H})$) foliates $\mathcal{P}^{1}(\mathcal{H})$ into
the spaces $\mathcal{D}_{r}^{1}(\mathcal{H})=\{|x\rangle\langle x|\,:\,\langle
x,x\rangle_{\mathcal{H}}=r\}$. In particular we have already denoted with
$\mathcal{D}_{1}^{1}(\mathcal{H})$ the space of one-dimensional projection
operators, which is the image via the momentum map of the sphere
$S_{\mathcal{H}}=\{x\in\mathcal{H}\,;\,\langle x,x\rangle_{\mathcal{H}}=1\}$
and can be identified with the complex projective space $P(\mathcal{H})$ via
the identification:
\begin{equation}
\lbrack x]\in P(\mathcal{H})\leftrightarrow\frac{|x\rangle\langle
x|}{\left\langle x,x\right\rangle }\in\mathcal{D}_{1}^{1}(\mathcal{H})
\end{equation}
We have also argued that $P(\mathcal{H})$ is a K\"{a}hler
manifold. In the following we will examine this fact in more detail, by
showing explicitly that $\mathcal{D}_{1}^{1}(\mathcal{H})$ is a K\"{a}hler manifold.

Let $\xi\in u^{\ast}(\mathcal{H}) $ be the image through the momentum map of a
unit vector $x\in S_{\mathcal{H}}$, i.e. $\xi=|x\rangle\langle x|$ with
$\langle x|x\rangle=1$, so that $\xi^{2}=\xi$. The tangent space of the
coadjoint $\mathcal{U}(\mathcal{H})$-orbit at $\xi$ is generated by vectors of
the form $[A,\xi]_{-}$, for any Hermitian $A$. From Eq.(\ref{is}), it follows
that the Poisson tensor $I$ defined in (\ref{KKS}) satisfies:
\begin{equation}
I(\hat{A},\hat{B})\left(  \xi\right)  =\langle\xi,[A,B]_{-}\rangle_{u^{\ast}%
}=\langle\lbrack\xi,A]_{-},B\rangle_{u^{\ast}}%
\end{equation}
This defines an invertible map $\tilde{I}$ that associates to any one-form
$\hat{A}$ the tangent vector at $\xi$: $\tilde{I}(\hat{A}) =:I(\hat{A}, \cdot)
= [\xi, A]_{-} $. We will denote with $\tilde{\eta}_{\xi}$ its inverse:
$\tilde{\eta}_{\xi}([\xi,A]_{-})=\hat{A} $. This allows us to define, on
$u^{\ast}(\mathcal{H}) $, a canonical two-form which is given by:
\begin{equation}
\eta_{\xi}([A,\xi]_{-},[B,\xi]_{-}) =:(\tilde{\eta}_{\xi}([\xi,A]_{-}),
[B,\xi]_{-}) = (\hat{A}, [B,\xi]_{-}) \label{form}%
\end{equation}
for all $[A,\xi]_{-},[B,\xi]_{-} \in T_{\xi}u^{\ast}(\mathcal{H})$.

It is also easy to check that $\eta$ satisfies the equalities: $\eta_{\xi
}([A,\xi]_{-},[B,\xi]_{-})=-\left(  \hat{A},[B,\xi]_{-}\right)  =-\langle
A,[B,\xi]_{-}\rangle_{u^{\ast}}=-\langle\xi,[A,B]_{-}\rangle_{u^{\ast}%
}=\langle\lbrack A,\xi]_{-},B\rangle_{u^{\ast}}$, for any $A,B\in u^{\ast
}(\mathcal{H})$.

We can summarize these results in the following:

\begin{theorem}
\textit{The restriction of the two-form (\ref{form}) to the $\mathcal{U}%
(\mathcal{H})$-orbit $\mathcal{D}_{1}^{1}(\mathcal{H})$ defines a canonical
symplectic form $\eta$ characterized by the property }
\begin{equation}
\eta_{\xi}([A,\xi]_{-},[B,\xi]_{-})=\left\langle \left[  A,\xi\right]
_{-},B\right\rangle _{u^{\ast}} = - \langle\xi,[A,B]_{-}\rangle_{u^{\ast}}
\label{pa}%
\end{equation}

\end{theorem}

In a very similar way, starting from the symmetric Jordan tensor $R$ given in
(\ref{Jord}) , one can construct a $(1,1)$ tensor $\tilde{R} (\hat{A}) =:
R(\hat{A},\cdot) = [\xi,A]_{+}$ and its inverse: $\tilde{\sigma} ([\xi,A]_{+})
= \hat{A}$. Thus we obtain a covariant tensor $\sigma$ such that:
\begin{equation}
\sigma_{\xi}([A,\xi]_{+},[B,\xi]_{+})=\langle\lbrack A,\xi]_{+},B\rangle
_{u^{\ast}}=\langle\xi,[A,B]_{+}\rangle_{u^{\ast}}. \label{p1}%
\end{equation}
Notice that, at this stage, $\sigma_{\xi}$ is only a partial tensor, being
defined on vectors of the form $[A,\xi]_{+}$, which belong to the image of the
map $\tilde{R}$. However, on $\mathcal{D}_{1}^{1}(\mathcal{H})$, we have
$[A,\xi]_{-}=[A,\xi^{2}]_{-}=[[A,\xi],\xi]_{+}$, so that, after some algebra,
one can also prove that:\newline$\sigma_{\xi}([A,\xi]_{-},[B,\xi]_{-}%
)=\sigma_{\xi}([[A,\xi]_{-},\xi]_{+},[[B,\xi]_{-},\xi]_{+})=\langle\xi
,[[A,\xi]_{-},[B,\xi]_{-}]_{+}\rangle_{u^{\ast}}=$ \newline$=\frac{1}{2}%
Tr(\xi\lbrack\lbrack A,\xi]_{-},[B,\xi]_{-}]_{+})=\frac{1}{2}Tr(\xi[
A,\xi]_{-}[B,\xi]_{-})=\langle\lbrack A,\xi]_{-},[B,\xi]_{-}\rangle_{u^{\ast}%
}$.

Therefore we have also the following:

\begin{theorem}
\textit{On the $\mathcal{U}(\mathcal{H})$-orbit $\mathcal{D}_{1}%
^{1}(\mathcal{H})$ we can define a symmetric covariant tensor }$\sigma
$\textit{ such that:}
\begin{equation}
\sigma_{\xi}([A,\xi]_{-},[B,\xi]_{-})=\langle\lbrack A,\xi]_{-},[B,\xi
]_{-}\rangle_{u^{\ast}}. \label{ps}%
\end{equation}
holds.
\end{theorem}

Moreover, going back to the the $(1,1)$ tensor $\tilde{I}$ given above, one
has the following result \cite{GKM}:

\begin{theorem}
\textit{When restricted to $\mathcal{D}_{1}^{1}(\mathcal{H})$, the $(1,1)$
tensor $\tilde{I}$ , which satisfies:
\begin{equation}
\tilde{I}^{3}=-\tilde{I}\label{jjj}%
\end{equation}
will become invertible. Hence: }$\tilde{I}^{2}=-\mathbb{I}$\textit{ and
therefore it will define a complex structure $\jmath$ such that:%
\begin{equation}
\eta_{\xi}([A,\xi]_{-},\jmath_{\xi}([B,\xi]_{-}))=\sigma_{\xi}([A,\xi
]_{-},[B,\xi]_{-})\label{pc}%
\end{equation}%
\begin{equation}
\eta_{\xi}(\jmath_{\xi}([A,\xi]_{-}),\jmath_{\xi}([B,\xi]_{-}))=\eta_{\xi
}([A,\xi]_{-},[B,\xi]_{-})
\end{equation}
}
\end{theorem}

Eq. (\ref{jjj}) follows from a direct calculation by taking into account that
$\xi^{2}=\xi$. The last two expressions follow by combining Eqs.(\ref{pa}) and
(\ref{ps}). To prove that $\jmath$ is a complex structure one has first to
show that it defines an almost complex structure (which follows easily from
the fact that $[[[A,\xi]_{-},\xi]_{-},\xi]_{-}=-[A,\xi]_{-}$) and then that
its Nijenhuis torsion vanishes. Detailed calculations of this can be found in
Ref.\cite{GKM}.

Putting everything together, we can now conclude that, as expected:

\begin{theorem}
\ \textit{$(\mathcal{D}_{1}^{1}(\mathcal{H}),\jmath,\sigma,\eta)$ is a
K\"{a}hler manifold.}\newline
\end{theorem}

At last, we notice that there is an identification of the orthogonal
complement of any unit vector $x\in\mathcal{H}$ with the tangent space of the
$\mathcal{U}(\mathcal{H})$-orbit in $u^{\ast}(\mathcal{H})$ at $\xi
=|x\rangle\langle x|$. Indeed, for any $y$ perpendicular to $x$ ($\Vert
x\Vert^{2}=1$) the operators:
\begin{equation}
P_{y}^{x}=:(\mu_{\ast})_{x}(y)=|y\rangle\langle x|+|x\rangle\langle y|
\end{equation}
can be written as $P_{y}^{x}=[A_{y},\xi]$, where $A_{y}$ is a Hermitian
operator such that $A_{y}x=iy$, $A_{y}y=-i\Vert y\Vert^{2}x$ and $A_{y}z=0$
for any $z$ perpendicular to both $x$ and $y$, as it can be directly checked
by applying both expressions to a generic vector in $\mathcal{H}$ which can be
written as $ax+by+cz$ with $a,b,c\in\mathbb{C}$. Then, from Eqs.(\ref{pa}) and
(\ref{ps}), it follows immediately that, for any $y,y^{\prime}$ orthogonal to
$x$:%
\begin{equation}
\eta_{\xi}(P_{y}^{x},P_{y^{\prime}}^{x})=-\frac{1}{2}Tr(\xi\lbrack
A_{y},A_{y^{\prime}}]_{-})=-\frac{1}{2i}(\langle y,y^{\prime}\rangle-\langle
y^{\prime},y\rangle)=-\omega(y,y^{\prime})
\end{equation}

\begin{equation}
\sigma_{\xi}(P_{y}^{x},P_{y^{\prime}}^{x})=\frac{1}{2}Tr(\xi\lbrack
A_{y},A_{y^{\prime}}]_{-})=-\frac{1}{2}(\langle y,y^{\prime}\rangle+\langle
y^{\prime},y\rangle)=g(y,y^{\prime})
\end{equation}
In conclusion, we have the following:\newline


\begin{theorem}
\textit{For any $y,y^{\prime}\in\mathcal{H}$, the vectors $(\mu_{\ast}%
)_{x}(y),(\mu_{\ast})_{x}(y)$ are tangent to the $\mathcal{U}(\mathcal{H}%
)$-orbit in $u^{\ast}(\mathcal{H})$ at $\xi=\mu(x)$ and:}%
\begin{equation}
\sigma_{\xi}((\mu_{\ast})_{x}(y),(\mu_{\ast})_{x}(y))=g(y,y^{\prime})
\end{equation}%
\begin{equation}
\eta_{\xi}((\mu_{\ast})_{x}(y),(\mu_{\ast})_{x}(y))=-\omega(y,y^{\prime})
\end{equation}%
\begin{equation}
\jmath_{\xi}(\mu_{\ast})_{x}(y))=(\mu_{\ast})_{x}(Jy)
\end{equation}
where the last formula follows from Eq.(\ref{pc}).
\end{theorem}

More generally, with minor changes, we can reconstruct similar structures for
any $\mathcal{D}_{r}^{1}(\mathcal{H})$, obtaining K\"{a}hler manifolds
$(\mathcal{D}_{r}^{1}(\mathcal{H}),\jmath^{r},\sigma^{r},\eta^{r})$. The
analog of above theorem shows then that the latter can be obtained from a sort
of \textquotedblleft K\"{a}hler reduction" starting from the original linear
K\"{a}hler manifold $(\mathcal{H}_{{\mathbb{R}}},J,g,\omega)$.

\begin{example}
Let us go back to the previous example of rank-one
projectors on $\mathcal{H}=\mathbb{C}^{2}$. According to (\ref{dmc2}), the
latter are described by three dimensional vectors $\mathbf{\xi}=(y^{1}%
,y^{2},y^{3})$ such that $\mathbf{\xi}^{2}=1/4$ ($y_{0}=1/2$ always), which
form a 2-dimensional sphere of radius $1/2$. A generic tangent vector $X_{A}$
and a generic one form $\hat{A}$ at $\xi$ are of the form $X_{A}=y_{A}%
^{0}\partial_{0}+y_{A}^{1}\partial_{1}+y_{A}^{2}\partial_{2}+y_{A}^{3}%
\partial_{3}$ and $\hat{A}=y_{A}^{0}dy^{0}+y_{A}^{1}dy^{1}+y_{A}^{2}%
dy^{2}+y_{A}^{3}dy^{3}$ with $y_{A}^{0}=0$ and $\mathbf{y}_{A}\cdot\xi=0$.

It is clear from (\ref{iab}) that the map $\tilde{I}$ that associates to any
one-form $\hat{A}$ the tangent vector at $\xi$: $\tilde{I}(\hat{A})=:I(\hat
{A},\cdot)=[A,\xi]_{-}$ is manifestly invariant and given by: $\tilde{I}%
(\hat{A})=2(\xi\times\mathbf{y}_{A})\cdot\vec{\partial}$, where we have set
$\vec{\partial}=(\partial_{1},\partial_{2},\partial_{3})$. It follows that the
two-form $\eta_{\xi}$ is such that:
\begin{equation}
\eta_{\xi}([A,\xi]_{-},[B,\xi]_{-})=2\xi\cdot(\mathbf{y}_{A}\times
\mathbf{y}_{B})
\end{equation}
so that
\begin{equation}
\eta_{\xi}=2\epsilon^{ijk}y^{i}dy^{j}\wedge dy^{k}%
\end{equation}
which is proportional \ by a factor $\left(  y_{1}^{2}+y_{2}^{2}+y_{3}%
^{2}\right)  ^{-\frac{3}{2}}$ to the symplectic two-form on a 2-dimensional
sphere\footnote{This is also the volume element of a 2-dimensional sphere of
radius $r=1/2$, as it should be.}, when pulled back to the sphere.

In a similar way, from (\ref{Jord2}), one can prove that $\tilde{R}(\hat
{A})=:R(\hat{A},\cdot)=[\xi,A]_{+}=2(y_{A}^{0}y^{0}+\mathbf{y}_{A}\cdot
\xi)\partial_{0}+2(y_{A}^{0}\xi+y_{0}\mathbf{y}_{A})\cdot\vec{\partial}$.
Thus, because of (\ref{ps}), we have:
\begin{equation}
\sigma_{\xi}([A,\xi]_{-},[B,\xi]_{-})=4(\xi\times\mathbf{y}_{A})\cdot
(\xi\times\mathbf{y}_{B})=\mathbf{y}_{A}\cdot\mathbf{y}_{B}%
\end{equation}
where the last equality follows from the fact that $\xi^{2}=1/4$ and $\xi$ is
orthogonal to both $\mathbf{y}_{A}$ and $\mathbf{y}_{B}$.

Finally, starting for example from Eq. (\ref{pc}), it is not difficult to check
that
\begin{equation}
\jmath_{\xi}([B,\xi]_{-})=y_{B}^{\prime}\cdot\vec{\partial}\;\text{\ }%
with:\;\mathbf{y}_{B}^{\prime}=\xi\times\mathbf{y}_{B}%
\end{equation}
A direct calculation shows that $\jmath_{\xi}^{3}=-\jmath_{\xi}$.
\end{example}

\subsection{The geometry of quantum mechanics and the $GNS$ construction}

\label{se:GNS}
\bigskip

In the previous Sections of this Chapter, we have worked out the geometrical
structures that naturally arise in the standard approach to quantum mechanics,
which starts from the Hilbert space and identifies the space of physical
states with the associated complex projective space. In this framework,
algebraic notions, such that of the ${\mathbb{C}}^{*}$-algebra that contains
observables as real elements, arises only as a derived concept.

In this Section, we would like to see how geometrical structures emerge also
in a more algebraic setting, where one starts from the very beginning with an
abstract ${\mathbb{C}}^{\ast}$-algebra containing the algebra of quantum
observables as real elements to obtain the Hilbert space of states is a
derived concept via the so called Gelfand-Naimark-Segal ($GNS$) construction
\cite{bratteli}. A detailed discussion can be found in Ref. \cite{chru}.

\subsubsection{The $GNS$ construction}
\bigskip

The algebraic approach known as the $GNS$ construction started with the work
of Haag and Kastler \cite{kastler}, and is also at the basis of the
mathematical approach to quantum field theory \cite{Ha}.

The starting point of this construction is an abstract ${\mathbb{C}}^{\ast}%
$-algebra $\mathcal{A}$ \cite{bratteli,Em} with unity, the latter being
denoted as ${\mathbb{I}}$. The elements $a\in\mathcal{A}$ such that:
$a=a^{\ast\text{ }}$constitute the set $\mathcal{A}_{re}$ (a vector space over
the reals) of the \textit{real elements}\footnote{Also called the
\textit{observables}.} of the algebra. In particular: ${\mathbb{I\in}%
}\mathcal{A}_{re}$. The obvious decomposition: $a=a_{1}+ia_{2}$, with:%
\begin{equation}
a_{1}=\frac{a+a^{\ast}}{2};\text{ \ }a_{2}=\frac{a-a^{\ast}}{2i}%
\end{equation}
means that, as a vector space, $\mathcal{A}$ is the direct sum of
$\mathcal{A}_{re}$ and of the set $\mathcal{A}_{im}$ (also a vector space over
the reals) of the \textit{imaginary elements}, i.e. of the elements of the
form $ia,$ $a\in\mathcal{A}_{re}$. $\mathcal{A}_{re}$ can be given \cite{chru}
the structure of a Lie-Jordan algebra \cite{Em}, where, using here the
conventions of Sect.\ref{s:ths}, the Lie product is defined as:%
\begin{equation}
\left[  a,b\right]  =:\frac{1}{2i}\left(  ab-ba\right)
\end{equation}
while the Jordan product is given by:%
\begin{equation}
a\circ b=\frac{1}{2}\left(  ab+ba\right)
\end{equation}
for all $a,b\in\mathcal{A}_{re}$. The product in the algebra is then recovered
as:%
\begin{equation}
ab=a\circ b+i\left[  a,b\right]
\end{equation}

\begin{remark}
A typical example of a ${\mathbb{C}}^{\ast}$-algebra is the algebra $\mathcal{B}\left(  \mathcal{H}\right)  $
of the bounded operators on a Hilbert space $\mathcal{H}$. In
this case \cite{Em}: $\mathcal{A}_{re}\equiv\mathcal{B}_{sa}\left(
\mathcal{H}\right)  $, the set of the bounded self-adjoint operators
on $\mathcal{H}$.
\end{remark}

The space $\mathcal{D}(\mathcal{A})$ of the \textit{states} \ over the
${\mathbb{C}}^{\ast}$-algebra $\mathcal{A}$ is the space of the linear
functionals $\omega:$ $\mathcal{A}\rightarrow\mathbb{C}$ that are \cite{Ha}:

\begin{itemize}
\item \textit{real}: $\omega\left(  a^{\ast}\right)  =\overline{\omega\left(
a\right)  }$ $\forall a\in\mathcal{A}$,

\item \textit{positive: }$\omega\left(  a^{\ast}a\right)  \geq0$ $\forall
a\in\mathcal{A}$ and

\item \textit{normalized}: $\omega\left(  \mathbb{I}\right)  =1$
\end{itemize}

\bigskip

Each functional $\omega$ defines a non-negative pairing $\left\langle
\cdot|\cdot\right\rangle _{\omega}$ between any two elements $a,b\in
\mathcal{A}$ via:
\begin{equation}
\left\langle a|b\right\rangle _{\omega}:=\omega(a^{\ast}b) \label{4:pairing}%
\end{equation}

Reality and positivity of the state guarantee that the pairing
(\ref{4:pairing}) satisfies the Schwartz inequality, i.e.:%
\begin{equation}
\left\vert \left\langle a|b\right\rangle _{\omega}\right\vert \leq
\sqrt{\left\langle a|a\right\rangle _{\omega}}\sqrt{\left\langle
b|b\right\rangle _{\omega}}\label{Schwartz}%
\end{equation}
but the pairing might be degenerate. We are thus led to consider the "Gelfand
ideal" \cite{Em,Ha} $\mathcal{I}_{\omega}$ consisting of all elements
$j\in\mathcal{A}$ such that $\omega(j^{\ast}j)=0$ and to define the set
$\mathcal{A}/\mathcal{I}_{\omega}$ of equivalence classes:
\begin{equation}
\Psi_{a}=:[a+\mathcal{I}_{\omega}]\label{equiv}%
\end{equation}
It is immediate to see that $\mathcal{A}/\mathcal{I}_{\omega}$ is a
pre-Hilbert space with respect to the scalar product\footnote{The Schwartz
inequality (\ref{Schwartz}) implies: $\left\langle i|a\right\rangle _{\omega
}=\left\langle a|i\right\rangle _{\omega}=0$ $\forall a\in\mathcal{A}%
,i\in\mathcal{I}_{\omega}$, and hence that the scalar product (\ref{scalprod})
does indeed depend only on the equivalence classes of $a$ and $b$ and not on
the specific representatives chosen.}:
\begin{equation}
\langle\Psi_{a},\Psi_{b}\rangle=\omega(a^{\ast}b)\label{scalprod}%
\end{equation}

Completing this space with respect to the topology defined by the scalar
product, one obtains a Hilbert space $\mathcal{H}_{\omega}$ on which the
original ${\mathbb{C}}^{\ast}$-algebra $\mathcal{A}$ acts via the following
representation\footnote{Notice that if such a representation is faithful, i.e.
the map $\pi_{\omega}:a\mapsto\pi_{\omega}(a)$ is an isomorphism, the operator
norm of $\pi_{\omega}(a)$ equals the ${\mathbb{C}}^{*}$-norm of $a$
\cite{bratteli}.}:
\begin{equation}
\pi_{\omega}(a)\Psi_{b}=\Psi_{ab}%
\end{equation}

Clearly the equivalence class of the unit element in $\mathcal{A}$, i.e.
$\Omega=\Psi_{{\mathbb{I}}}$, satisfies: $\left\Vert \Psi_{{\mathbb{I}}%
}\right\Vert :=\sqrt{\left\langle \Psi_{{\mathbb{I}}}|\Psi_{{\mathbb{I}}%
}\right\rangle }=1$ \ and provides a cyclic vector\footnote{We recall
\cite{Ha} that a vector $\Omega\in\mathcal{H}_{\omega}$ is called
\textit{cyclic} if $\pi_{\omega}\left(  \mathcal{A}\right)  $ is dense in
$\mathcal{H}_{\omega}$.} for the representation $\pi_{\omega}$. Moreover:
\begin{equation}
\langle\Omega|\pi_{\omega}(a)|\Omega\rangle=\omega(a)
\end{equation}

This tells us that, if we consider that $\mathcal{A}$ acts \ by duality \ on
$\mathcal{D}(\mathcal{A})$, the Hilbert space corresponding to a given state
$\omega$ is the orbit of $\mathcal{A}$ through $\omega$ itself. Notice that
any other element $b\in\mathcal{A}$ such that the vector $\Psi=\pi_{\omega
}(b)\Omega$ is of unit norm, defines a new state $\omega_{\Psi}$ by:
\begin{equation}
\omega_{\Psi}(a)=\langle\Psi|\pi_{\omega}(a)|\Psi\rangle=\omega(b^{\ast}ab)
\end{equation}
These states are called vector states of the representation $\pi_{\omega}$,
and are particular examples of more general states of the form:
\begin{equation}
\omega_{\rho}(a)=Tr[\rho\pi_{\omega}(a)] \label{folium}%
\end{equation}
where $\rho\in\mathcal{B}(\mathcal{H}_{\omega})$ is a density operator
\cite{Em,Ha}. States of the form (\ref{folium}) are called a "folium" of the
representation $\pi_{\omega}$. Also, one says that a state $\omega$ is pure
iff it cannot be written as a convex combination of other states in
$\mathcal{D}(\mathcal{A})$, so that the set of pure states $\mathcal{D}%
^{1}(\mathcal{A})$ defines a set of extremal points in $\mathcal{D}%
(\mathcal{A})$.

The universality and uniqueness of the $GNS$ construction is guaranteed
\cite{bratteli} by the following:
\begin{theorem}~\\
\begin{enumerate}
\item If $\pi_{\omega}$ is a cyclic
representation of $A$ on $H$, any vector representation
$\omega_{\Psi}$ for a normalized $\Psi$, see
Eq.(\ref{folium}), ie equivalent to $\pi_{\omega}$.
\item A $GNS$ representation $\pi_{\omega}$ of $\mathcal{A}$
is irreducible iff $\omega$  is a pure state.
\end{enumerate}
\end{theorem}

\bigskip

\begin{example} 
The $GNS$ construction can be very simple
for finite- dimensional $C^{\ast}$-algebras. Consider, e.g., the
algebra $A=B(C^{n})$ of linear operators on $C^{n}$, i.e. of
the $n\times n$ matrices with complex entries. Any non-negative
operator $\omega\in B(C^{n})$ defines a state by: 
\begin{equation}
\omega(A)=Tr[\omega A]\;,\;\forall A\in\mathcal{A}%
\end{equation}
while we can define the scalar product in $H_{\omega}$ as: 
\begin{equation}
\langle A|B\rangle=\omega(A^{\ast}B)=Tr[B\omega A^{\ast}] \label{4:scalProd}%
\end{equation}
If $\omega$  is a rank-1 projector and $\{e_{k}\}$ is
an orthonormal basis for which $\omega=|e_{1}\rangle\langle e_{1}|$,
writing $A_{km}$ for the matrix elements of $A$  in such a
basis, the scalar product assumes the form: 
\begin{equation}
\langle A|B\rangle=\sum_{k=1}^{n}\bar{A}_{k1}B_{k1}%
\end{equation}
while the Gelfand ideal $I_{\omega}$ is given by: 
\begin{equation}
\mathcal{I}_{\omega}=\{X\in\mathcal{A}\;:\;X_{k1}=0\,,\,k=1,\cdots,n\}
\end{equation}
Thus $H_{\omega}=A/I_{\omega}$ is nothing but $C^{n}$
 itself and $\pi_{\omega}$ is the defining
representation.\newline If $\omega$ is a rank-$m$ density
operator: $\omega=p_{1}|e_{1}\rangle\langle e_{1}|+\cdots+p_{m}|e_{m}%
\rangle\langle e_{m}|$\textit{ with }$p_{1},\cdots p_{m}>0$ and
$p_{1}+\cdots p_{m}=1$, the scalar product is given by: 
\begin{equation}
\langle A|B\rangle=\sum_{k=1}^{n}\sum_{j=1}^{m}p_{m}\bar{A}_{kj}B_{kj}%
\end{equation}
and the Gelfand ideal is given by: 
\begin{equation}
\mathcal{I}_{\omega}=\{X\in\mathcal{A}\;:\;X_{kj}=0\,,\,k=1,\cdots,n;\text{
}j=1,\cdots,m\}
\end{equation}
showing that $H_{\omega}$ is the direct sum of $m$
copies of $C^{n}$. Now the representation $\pi_{\omega}$  is
no longer irreducible, decomposing into the direct sum of $m$ copies
of the defining representation:
\begin{equation}
\pi_{\omega}(A)=\mathbb{I}_{m}\otimes A
\end{equation}
where $\mathbb{I}_{m}$ is the $m\times m$ identity
matrix.
\end{example}

\subsubsection{Geometric structures over a ${\mathbb{C}}^{\ast}$-algebra}
\bigskip

Let $V$ be a vector space and $V^{\ast}$ its dual. To any element $v\in V$,
there is a corresponding element in the bi-dual $\hat{v}\in(V^{\ast})^{\ast}$
given by:
\begin{equation}
\hat{v}(\alpha)=\alpha(v)\;,\;\forall\alpha\in V^{\ast}%
\end{equation}
Thus any multilinear function on $V^{\ast}$, $f:V^{\ast}\times\cdots V^{\ast
}\rightarrow{\mathbb{R}}$ defines, by restricting it to the diagonal, a
polynomial function $\tilde{f}\in\mathcal{F}(V^{\ast})$, $\tilde{f}%
(\alpha)=f(\alpha,...,\alpha)$ , which can be obtained from the "monomials of
degree one", $\hat{v}\in(V^{\ast})^{\ast}$, on which one has defined the
(commutative) product:
\begin{equation}
(\hat{v}_{1}\cdot\hat{v}_{2})(\alpha):=\hat{v}_{1}(\alpha)\,\hat{v}_{2}%
(\alpha) \label{commpro}%
\end{equation}
Suppose now that on $V$ there is defined an additional bilinear operation:
\begin{equation}
B:V\times V\rightarrow V
\end{equation}
which induces a (in general noncommutative) product $\times_{B}$ on
$V\subset\mathcal{F}(V^{\ast})$ by:
\begin{equation}
\hat{v}_{1}\times_{B}\hat{v}_{2}=\widehat{B(v_{1},v_{2})} \label{pro}%
\end{equation}
Then we can define a 2-tensor $\tau_{B}$ in $\mathcal{F}(V^{\ast})$, at the
point $\alpha$, by the relation:
\begin{equation}
\tau_{B}(d\hat{v}_{1},d\hat{v}_{2})(\alpha):=\alpha(B(v_{1},v_{2}))
\end{equation}
which satisfies the Leibniz rule:
\begin{equation}
\tau_{B}(d\hat{v},d(\hat{v}_{1}\cdot\hat{v}_{2}))=\tau_{B}(d\hat{v},\hat
{v}_{1}\cdot d\hat{v}_{2}+d\hat{v}_{1}\cdot\hat{v}_{2})=\hat{v}_{1}\cdot
\tau_{B}(d\hat{v},\hat{v}_{2})+\tau_{B}(d\hat{v},\hat{v}_{1})\cdot\hat{v}_{2}%
\end{equation}
Thus, $\tau_{B}(d\hat{v},\cdot)$ defines a derivation on $V\subset
\mathcal{F}(V^{\ast})$ with respect to the commutative product (\ref{commpro}).

In particular, suppose that $B$ is a skew-symmetric bilinear operation which
satisfies the Jacobi identity, so that $g=(V,B)$ is a Lie algebra. The
corresponding 2-tensor $\Lambda:=\tau_{B}$:
\begin{equation}
\Lambda(d\hat{v}_{1},d\hat{v}_{2})=\widehat{B(v_{1},v_{2})}%
\end{equation}
is a Poisson tensor in $\mathcal{F}(V^{\ast})$ and $\Lambda(d\hat{v},\cdot)$
is a derivation with respect to the commutative product (\ref{commpro}).
Moreover, $\Lambda(d\hat{v},\cdot)$ is a derivation also with respect to the
product (\ref{pro}). Indeed, by using the fact that $B$ is antisymmetric and
satisfies the Jacobi identity, one has:%
\begin{eqnarray}
\Lambda(d\hat{v},d(\hat{v_{1}}\mathbf{\cdot}\hat{v_{2}}))&=&\widehat
{B(v,B(v_{1},v_{2}))} = \\
&=&\widehat{B(v_{1},B(v,v_{2}))}+\widehat{B(B(v,v_{1}),v_{2})} = \nonumber\\
&=&\hat{v}_{1}\mathbf{\cdot}\Lambda(d\hat{v},d\hat{v}_{2})+\Lambda(d\hat
{v},d\hat{v}_{1})\mathbf{\cdot}\hat{v}_{2} \nonumber
\end{eqnarray}

Similarly, if on $V$ one has a Jordan product $B^{\prime}$, the corresponding
2-tensor $G :=\tau_{B^{\prime}}$ is a metric tensor and $G(d\hat{v},\cdot)$ is
a derivation with respect to the commutative product (\ref{commpro}), but not
with respect to the product (\ref{pro}).

If now $V=\mathcal{A}$ is a ${\mathbb{C}}^{\ast}$-algebra, where we have
defined both a Lie product and a Jordan product as:
\begin{equation}
B(a_{1},a_{2}):=[a_{1},a_{2}]=\frac{1}{2i}(a_{1}a_{2}-a_{2}a_{1})\;,\;\forall
a_{1},a_{2}\in\mathcal{A}\label{Lie}%
\end{equation}
and a Jordan product
\begin{equation}
B^{\prime}(a_{1},a_{2}):=a_{1}\circ a_{2}=\frac{1}{2}(a_{1}a_{2}+a_{2}%
a_{1})\;,\;\forall a_{1},a_{2}\in\mathcal{A}\label{Jordan}%
\end{equation}
in $\mathcal{F}(\mathcal{A}^{\ast})$ we have defined both a Poisson tensor
$\Lambda$ and a metric tensor $G$ such that $\Lambda(d\hat{a},\cdot)$ and
$G(d\hat{a},\cdot)$ are both derivations with respect to the pointwise
commutative product, with the former being also a derivation with respect to
the Lie product. It is also not difficult to check that the subalgebra
$\mathcal{B}\subset\mathcal{A}$ composed of all real elements, when embedded
in $\mathcal{F}(\mathcal{A}^{\ast})$, comes equipped with an antisymmetric and
a symmetric product, denoted by $[\cdot,\cdot]$ and $\circ$ respectively, such
that:
\begin{enumerate}
\item The Leibniz rule is satisfied: $[a,b\circ c]=[a,b]\circ c+b\circ\lbrack
a,c]$,

\item The Jacobi identity is satisfied: $[a,[b,c]]=[[a,b],c]+[b,[a,c]]$, and

\item The identity: $(a\circ b)\circ c-a\circ(b\circ c)=[[a,c],b]$ holds.
\end{enumerate}
meaning that $(\mathcal{B},[\cdot,\cdot],\circ)$ is a Lie-Jordan algebra
\cite{Em}
Finally, we notice that the Hamiltonian vector fields:
\begin{equation}
X_{\hat{a}}:=\Lambda(\cdot,d\hat{a})=-[\hat{a},\cdot]
\end{equation}
are derivations with respect to the Jordan product, since, by using the
properties above:
\begin{align}
X_{\hat{a}}(d(\hat{a}_{1}\circ\hat{a}_{2}))  &  =-[\hat{a},\hat{a}_{1}%
\circ\hat{a}_{2}]=-[\hat{a},\hat{a}_{1}]\circ\hat{a}_{2}+-\hat{a}_{1}%
\circ\lbrack\hat{a},\hat{a}_{2}]\nonumber\\
&  =X_{\hat{a}}(d\hat{a}_{1})\circ\hat{a}_{2}+\hat{a}_{1}\circ X_{\hat{a}%
}(d\hat{a}_{2})
\end{align}

Let us go back now to the $GNS$ construction and consider first a pure state
$\omega$ over $\mathcal{A}$, which gives rise to the irreducible
representation $\pi_{\omega}$ in the Hilbert space $\mathcal{H}_{\omega}$. We
have already seen (see Sect. 4.2.5) that self-adjoint operators, that
correspond to the real elements of $\mathcal{A}$, may be identified with the
dual $u^{\ast}(\mathcal{H}_{\omega})$ of the Lie algebra $u(\mathcal{H}%
_{\omega})$ of the unitary group $U(\mathcal{H}_{\omega})$ and how the
momentum map
\begin{equation}
\mu_{\omega}:\mathcal{H}_{\omega}\rightarrow u^{\ast}(\mathcal{H}_{\omega
})\;,\;\mu_{\omega}(\psi)=|\psi\rangle\langle\psi|
\end{equation}
relates the Poisson tensors on $u^{\ast}(\mathcal{H}_{\omega})$ with those on
$\mathcal{H}_{\omega}$, via the pull-back. We will say that a Poisson map
$\Phi:S\rightarrow M$, with $(S,\Omega)$ a Poisson manifold, is a symplectic
realization of a Poisson manifold $(M,\Lambda)$. When $S$ is a vector space we
call $\Phi$ a \textit{classical Jordan-Schwinger map} \cite{JS}; when $S$ is a
Hilbert space, as in the case we are considering, we say it is a Hermitian realization.

We have also seen that the unit sphere in $\mathcal{H}_{\omega}-\left\{
\mathbf{0}\right\}  $ can be projected onto $u^{\ast}(\mathcal{H}_{\omega})$
in an equivariant way, in such a way that the Poisson and the Riemann tensor
in $\mathcal{P}(\mathcal{H}_{\omega})$ are both related to the same tensors
defined on $u^{\ast}(\mathcal{H}_{\omega})$ by using the Lie and the Jordan
product that are defined on it. Thus the momentum map provides a symplectic
realization, which we call a K\"{a}hlerian realization where $S$ is the
complex projective space.

\subsection{Recovering a Hilbert Space out of ${\mathbb{R}}^{2n}$}

\label{Hermitian}
\bigskip

Given now $\mathbb{A\in}\mathfrak{gl}\left(  2n,\mathbb{R}\right)  \equiv
End(\mathbb{R}^{2n})$, $\mathbb{A=}\left\Vert A^{i}\text{ }_{j}\right\Vert $
we can make two distinct associations, namely:

$i)$ $\mathfrak{gl}\left(  2n,\mathbb{R}\right)  \rightarrow\left(
1,1\right)  $ tensor fields, via:%
\begin{equation}
\mathbb{A\rightarrow}T_{\mathbb{A}}=A^{i}\text{ }_{j}dx^{j}\otimes
\frac{\partial}{\partial x^{i}}%
\end{equation}
The correspondence is an isomorphism of associative algebras, i.e.:%
\begin{equation}
T_{\mathbb{A}}\circ T_{\mathbb{B}}=T_{\mathbb{AB}}%
\end{equation}
and $T_{\mathbb{A}}$ is homogeneous of degree zero, i.e.:%
\begin{equation}
\mathcal{L}_{\Delta}T_{\mathbb{A}}=0
\end{equation}
where $\Delta$ is the dilation (Liouville) vector field associated with the
linear structure of $\mathbb{R}^{2n}$:%
\begin{equation}
\Delta=x^{i}\frac{\partial}{\partial x^{i}}%
\end{equation}

$ii)$ $\mathfrak{gl}\left(  2n,\mathbb{R}\right)  \rightarrow$ $\{$linear
vector fields$\}$, via:%
\begin{equation}
\mathbb{A\rightarrow}X_{\mathbb{A}}=A^{i}\text{ }_{j}x^{j}\frac{\partial
}{\partial x^{i}}%
\end{equation}

The latter is only a Lie algebra (anti)isomorphism, i.e.:%
\begin{equation}
\left[  X_{\mathbb{A}},X_{\mathbb{B}}\right]  =-X_{\left[  \mathbb{A}%
,\mathbb{B}\right]  }%
\end{equation}
$X_{\mathbb{A}}$ is also homogeneous of degree zero:%
\begin{equation}
\left[  \Delta,X_{\mathbb{A}}\right]  =0\text{ }\forall\mathbb{A}
\label{hom_vector}%
\end{equation}

$i)$ and $ii)$ are connected by:%
\begin{equation}
T_{\mathbb{A}}\left(  \Delta\right)  =X_{\mathbb{A}}%
\end{equation}
Moreover, for any $\mathbb{A},\mathbb{B\in}\mathfrak{gl}\left(  2n,\mathbb{R}%
\right)  $:%
\begin{equation}
\mathcal{L}_{X_{\mathbb{A}}}T_{\mathbb{B}}=-T_{\left[  \mathbb{A}%
,\mathbb{B}\right]  }%
\end{equation}

\begin{remark} 
Going back to the compatibility condition between, say, $g$
and $J$, and defining the linear vector field: $X_{J}=J^{i}$ $_{j}x^{j}\left(
\partial/\partial x^{i}\right)  $, one checks easily that the compatibility
condition $\widetilde{J}\circ g+g\circ J=0$ is identical to requiring:%
\begin{equation}
\mathcal{L}_{X_{J}}g=0
\end{equation}
This clarifies also why $J$ can be associated with infinitesimal
$g$-orthogonal transformations.
\end{remark}
\bigskip 

Given now a triple, a
\textit{Hermitian structure} on $\mathbb{R}^{2n}$ will be a map:%
\begin{equation}
h:\mathbb{R}^{2n}\rightarrow\mathbb{R}^{2};\text{ \ }h\left(  x,y\right)
=\left(  g\left(  x,y\right)  ,\omega\left(  x,y\right)  \right)
\equiv\left(  g\left(  x,y\right)  ,g\left(  x,Jy\right)  \right)
\end{equation}

$\mathbb{R}^{2n}$ can be given a complex vector space structure by defining,
for $z=\alpha+i\beta\in\mathbb{C}$:%
\begin{equation}
\left(  \alpha+i\beta\right)  \cdot x=:\alpha x+\beta Jx
\end{equation}

\begin{remark}
Notice that, e.g., $g\left(  x,Jx\right)  =0$
$\forall x$, i.e. $x,Jx\in\mathbb{R}^{2n}$ are orthogonal and hence
$\mathbb{R}$-linearly independent\footnote{Indeed, if $x\neq0$ and $\alpha
x+\beta Jx=0$ with $\alpha,\beta\in\mathbb{R}$, then: $0=g\left(  \alpha
x+\beta Jx,\alpha x+\beta Jx\right)  =(\alpha^{2}+\beta^{2})g\left(
x,x\right)  $, implying $\alpha=\beta=0$.}, but they are \textit{not} linearly
independent when linear combinations with complex coefficients are allowed,
as: $Jx=:ix$. This means that the \textit{complex} dimension is reduced from
$2n$ to $n$, and $\mathbb{R}^{2n}\approx\mathbb{C}^{n}$ as a complex vector
space. One possible (non-canonical i.e. not unique) way of "mapping"
$\mathbb{R}^{2n}$onto $\mathbb{C}^{n}$ is to choose a basis in $\mathbb{R}%
^{2n}$, to pick up $n$ vectors $\left(  e^{1},...,e^{n}\right)  $ of the basis
and to construct $\mathbb{C}^{n}$ by taking complex linear combinations
thereof with the rule given above (i.e.: $ze^{i}=:\alpha e^{i}+$ $\beta
Je^{i}$).
\end{remark}

Then, we can write:%
\begin{equation}
h\left(  x,y\right)  =g\left(  x,y\right)  +i\omega\left(  x,y\right)  \equiv
g\left(  x,y\right)  +ig\left(  x,Jy\right)  \label{herm}%
\end{equation}
or:%
\begin{equation}
h\left(  x,y\right)  =\omega\left(  Jx,y\right)  +i\omega\left(  x,y\right)
\end{equation}
and in this way $h$ will be a Hermitian scalar product linear in the first
factor and antilinear in the \textit{second }factor\footnote{Had we been
using: $\omega\left(  x,y\right)  =g\left(  Jx,y\right)  $ instead of
$\omega\left(  x,y\right)  =g\left(  x,Jy\right)  $ we would have obtained the
opposite, which is the most common convention \cite{Dir3,Mes} among
physicists.}.

For the alternative descriptions obtained in the previous chapter, we get a
new Hermitian scalar product by replacing $\omega$ in (\ref{herm}) with
$\omega_{F}$.

Let now an admissible triple $\left(  g,J,\omega\right)  $ be given on
$\mathbb{R}^{2n}$. First of all we can construct the quadratic function:%
\begin{equation}
\mathbf{g}=:\frac{1}{2}g\left(  \Delta,\Delta\right)
\end{equation}
and the associated Hamiltonian vector field $\Gamma$ via:%
\begin{equation}
i_{\Gamma}\omega=-d\mathbf{g}%
\end{equation}
Explicit calculation shows that, with $\omega$ and $g$ (admissible and)
constant, $\Gamma$ s forced to be a linear vector field:%
\begin{equation}
\Gamma=\Gamma^{i}\text{ }_{j}x^{j}\frac{\partial}{\partial x^{i}}%
\end{equation}
and that:
\begin{equation}
\Gamma^{i}\text{ }_{j}=J^{i}\text{ }_{j}%
\end{equation}
i.e.\footnote{In terms of representative matrices.}: $\Gamma=J$, for short.
This can be written in coordinate-free language as:
\begin{equation}
\Gamma=J\left(  \Delta\right)  \text{ \ and: }\Delta=-J\left(  \Gamma\right)
\end{equation}
Notice that $\Gamma$ is symplectic:%
\begin{equation}
\mathcal{L}_{\Gamma}\omega=0
\end{equation}
with Hamiltonian function $\mathbf{g}$. Therefore:%
\begin{equation}
0=\mathcal{L}_{\Gamma}\mathbf{g=}\frac{1}{2}\left(  \mathcal{L}_{\Gamma
}g\right)  \left(  \Delta,\Delta\right)  +g\left(  \Delta,\left[
\Gamma,\Delta\right]  \right)
\end{equation}
But $\left[  \Gamma,\Delta\right]  =0$, so $\Gamma$ is also a Killing vector
field:
\begin{equation}
\mathcal{L}_{\Gamma}g=0
\end{equation}
Thus $\Gamma$ will preserve both the metric, the symplectic structure and (of
course) the complex structure, i.e. all the tensors of the admissible triple.
So, there will be two linear vector fields "canonically" associated with every
admissible triple, one of them defining the linear structure.\newline

Of course:%
\begin{equation}
\mathcal{L}_{\Gamma}h=0
\end{equation}
which is a complex condition equivalent to the two real ones: $\mathcal{L}%
_{\Gamma}g=0$ and: $\mathcal{L}_{\Gamma}J=0$. As the linear transformations
that leave the Hermitian scalar product unchanged are those of the unitary
group on $\mathbb{C}^{n}$, $\Gamma$ will be an infinitesimal transformation of
this group, and the representative matrix \ (i.e. $J$) will belong to its Lie
algebra. All the vector fields with this property will be called
\textit{quantum systems.} A quantum system will be therefore any linear vector
field:%
\begin{equation}
X_{\mathbb{A}}=A^{i}\text{ }_{j}x^{j}\frac{\partial}{\partial x^{i}}%
\end{equation}
such that:
\begin{equation}
\mathcal{L}_{X_{\mathbb{A}}}h=0 \label{killing1}%
\end{equation}
In terms of the defining matrices. The matrix $\mathbb{A}$ belongs then both
to the Lie algebra of the orthogonal ($g$-orthogonal) group and to the Lie
algebra of the symplectic group, i.e. Eq.(\ref{killing1}) splits into the two
real conditions:%
\begin{equation}
\mathcal{L}_{X_{\mathbb{A}}}g=0 \mbox{ and: } \mathcal{L}_{X_{\mathbb{A}%
}}\omega=0 \label{killing2}%
\end{equation}
The intersection of these algebras is the Lie algebra of \ the unitary group.
At the finite level (i.e. by exponentiation) the one-parameter group
$\exp\{t\mathbb{A\}}$ will belong to a \textit{real} realization of the
unitary group $U(n)$ in $\mathbb{R}^{2n}$. Notice also that\ the first of
Eqs.(\ref{killing2}) implies, together with Eq.(\ref{hom_vector}), that:%
\begin{equation}
\mathcal{L}_{X_{\mathbb{A}}}g\left(  \Delta,\Delta\right)  =0
\end{equation}

\begin{example}
Consider, e.g., $SU(2)$ in the defining representation, i.e.:%
\begin{equation}
SU(2)\ni U=\left\vert
\begin{array}
[c]{cc}%
\alpha & \beta\\
-\overline{\beta} & \overline{\alpha}%
\end{array}
\right\vert :\mathbb{C}^{2}\rightarrow\mathbb{C}^{2},\text{ \ }\left\vert
\alpha\right\vert ^{2}+\left\vert \beta\right\vert ^{2}=1
\end{equation}
(i.e. we are viewing $U$ \ as a $\left(  1,1\right)  $ tensor). \ Writing:
$U=a+ib$, with $a$ and $b$ real $2\times2$ matrices, the unitarity condition
$U^{\dag}U=\mathbb{I}$ becomes:
\begin{equation}
\widetilde{a}a+\widetilde{b}b=\mathbb{I}\text{ };\text{ \ }a\widetilde
{b}-b\widetilde{a}=0
\end{equation}
(i.e. $a\widetilde{b}$ must be a symmetric matrix). \ We can
realify\footnote{See, e.g., Ref.\cite{Ar1} Sect.$18$.} $\ \mathbb{C}^{2}$ onto
$\mathbb{R}^{4}$ as ($z=x+iy$ etc.):%
\begin{equation}
z=\left\vert
\begin{array}
[c]{c}%
z_{1}\\
z_{2}%
\end{array}
\right\vert \rightarrow x=\left\vert
\begin{array}
[c]{c}%
x_{1}\\
x_{2}\\
y_{1}\\
y_{2}%
\end{array}
\right\vert
\end{equation}
and $U$ \ as the $4\times4$ real matrix:%
\begin{equation}
G=\left\vert
\begin{array}
[c]{cc}%
a & -b\\
b & a
\end{array}
\right\vert
\end{equation}
Assume for simplicity the metric to be the standard Euclidean metric. Then it
can be checked at once that the unitarity condition leads both to:%
\begin{equation}
\widetilde{G}G=\mathbb{I}%
\end{equation}
and to:%
\begin{equation}
\widetilde{G}\mathbb{J}G=\mathbb{J}%
\end{equation}
where:%
\begin{equation}
\mathbb{J}=\left\vert
\begin{array}
[c]{cc}%
\mathbf{0} & -\mathbb{I}\\
\mathbb{I} & \mathbf{0}%
\end{array}
\right\vert
\end{equation}
with $\mathbb{I}$ the $2\times2$ identity matrix, i.e. $\mathbb{J}$ is the
realification of the multiplication by the imaginary unit $i$ in
$\mathbb{C}^{2}$. In this case, as matrices: $\omega=\mathbb{J}$ (we stress
however that $\omega$ is a $(0,2)$ tensor, while $J$ is a $(1,1)$ tensor), and
one checks easily that: $h\left(  x,x^{\prime}\right)  =g\left(  x,x^{\prime
}\right)  +i\omega(x,x^{\prime})\Leftrightarrow z\overline{z^{\prime}}$ which
is the Hermitian scalar product in $\mathbb{C}^{2}$ antilinear in the
\textit{second } factor. $G$ \ provides then also a realization of both
$SO(4)$ and of $Sp(4)$, and hence of: $SU(2)=SO(4)\cap Sp(4)$. \ Explicitly,
the vector field associated with $\mathbb{J}$ will be:%
\begin{equation}
\Gamma=x_{1}\frac{\partial}{\partial y_{1}}-y_{1}\frac{\partial}{\partial
x_{1}}+x_{2}\frac{\partial}{\partial y_{2}}-y_{2}\frac{\partial}{\partial
x_{2}}%
\end{equation}
This is the dynamical vector field for the $2D$ harmonic oscillator. In
$\mathbb{C}^{2}$ it corresponds of course to: $\overset{\cdot}{z}_{j}=iz_{j}$,
$j=1,2$.
\end{example}

\subsection{Compatible Hermitian structures and Bihamiltonian vector fields}

\label{se:bicomp}
\bigskip

Consider two different Hermitian structures, $h_{1}$ and $h_{2}$, on
$\mathbb{R}^{2n}$, with associated quadratic functions $\mathbf{g}_{a}\left(
\Delta,\Delta\right)  $ and Hamiltonian vector fields $\Gamma_{a}$
($\Gamma_{a}=X_{J_{a}}$), $a=1,2.$ The two structures will be called
\textit{compatible} iff:%
\begin{equation}
\mathcal{L}_{\Gamma_{1}}h_{2}=\mathcal{L}_{\Gamma_{2}}h_{1}=0
\end{equation}
which implies, of course, that the $\Gamma$'s will be \textit{bi}Hamiltonian.
In more detail, this implies: $\mathcal{L}_{\Gamma_{1}}\omega_{2}%
=\mathcal{L}_{\Gamma_{1}}g_{2}=0$ as well as: $\mathcal{L}_{\Gamma_{1}%
}\mathbf{g}_{2}=0$ (and similarly by interchanging indices).

As already recalled, given a symplectic form $\omega$ and/or a metric tensor
$g$ and a linear vector field $X_{\mathbb{A}}$ , the following statements are
equivalent:%
\begin{equation}
\mathcal{L}_{X_{\mathbb{A}}}\omega=0;\text{ }\omega\left(  \mathbb{A}%
x,y\right)  +\omega\left(  x,\mathbb{A}y\right)  =0;\text{ \ }\omega
\mathbb{A=}\widetilde{\left(  \omega\mathbb{A}\right)  }%
\end{equation}
as well as:%
\begin{equation}
\mathcal{L}_{X_{\mathbb{A}}}g=0;\text{ }g\left(  \mathbb{A}x,y\right)
+g\left(  x,\mathbb{A}y\right)  =0;\text{ \ }g\mathbb{A=-}\widetilde{\left(
g\mathbb{A}\right)  }%
\end{equation}
(remember that $\omega$ is skew-symmetric: $\widetilde{\omega}=-\omega$, while
$g$ is symmetric: $\widetilde{g}=g$). So, $X_{\mathbb{A}}$ will leave $\omega$
invariant iff $\omega\mathbb{A}$ is symmetric\footnote{Compare Ch.$3$.}, and
it will leave $g$ invariant iff $g\mathbb{A}$ is skew-symmetric.

Now, as $\mathcal{L}_{\Gamma_{1}}\omega_{2}=0=\mathcal{L}_{\Gamma_{1}}g_{2}$
and: $i_{\Gamma_{2}}\omega_{2}=-d\mathbf{g}_{2}$:%
\begin{equation}
0=\mathcal{L}_{\Gamma_{1}}\left(  i_{\Gamma_{2}}\omega_{2}\right)
=\mathcal{L}_{\Gamma_{1}}\omega_{2}\left(  \Gamma_{2},.\right)  =\omega
_{2}\left(  \left[  \Gamma_{1},\Gamma_{2}\right]  ,.\right)
\end{equation}
and, as the symplectic forms are non-degenerate:%
\begin{equation}
\left[  \Gamma_{1},\Gamma_{2}\right]  =0
\end{equation}
which, in view of the fact that: $\Gamma_{a}=X_{J_{a}},a=1,2$ implies (and is
implied by):%
\begin{equation}
\left[  J_{1},J_{2}\right]  =0
\end{equation}

Given a symplectic form $\omega$, the Poisson bracket of any two functions $f$
and $g$ is given by:%
\begin{equation}
\left\{  f,g\right\}  =\omega\left(  X_{g},X_{f}\right)
\end{equation}
where $X_{f}$ and $X_{g}$ are the Hamiltonian vector fields associated with
$f$ and $g$ respectively. Hence, denoting with $\left\{  .,.\right\}  _{a}$
the Poisson bracket associated with $\omega_{a}$ ($a=1,2$) we have, e.g.:
\begin{equation}
\left\{  \mathbf{g}_{1},\mathbf{g}_{2}\right\}  _{2}=\omega_{2}\left(
\Gamma_{2},\Gamma_{1}\right)  =-d\mathbf{g}_{2}\left(  \Gamma_{1}\right)
=-\mathcal{L}_{\Gamma_{1}}\mathbf{g}_{2}=0
\end{equation}
and similarly with the other Poisson bracket. All in all:%
\begin{equation}
\left\{  \mathbf{g}_{1},\mathbf{g}_{2}\right\}  _{1}=\left\{  \mathbf{g}%
_{1},\mathbf{g}_{2}\right\}  _{2}=0
\end{equation}

\bigskip

Out of the metric tensors and symplectic structures one can form the $\left(
1,1\right)  $ tensors:%
\begin{equation}
G=g_{1}^{-1}\circ g_{2}%
\end{equation}
(not to be confused with the $(2,0)$ tensor $G$ introduced in Sect.\ref{s:ths})
and:
\begin{equation}
T=\omega_{1}^{-1}\circ\omega_{2}%
\end{equation}
In intrinsic terms: $G\left(  X\right)  =g_{1}^{-1}\left(  g_{2}\left(
X\right)  \right)  $, i.e.:%

\begin{equation}
G=G^{i}\text{ }_{j}dx^{j}\otimes\frac{\partial}{\partial x^{i}};\text{
\ }G^{i}\text{ }_{j}=\left(  g_{1}\right)  ^{ik}\left(  g_{2}\right)  _{kj}%
\end{equation}
and similarly for $T$. The two are not independent, though. Indeed, using:
$J_{a}=\left(  g_{a}\right)  ^{-1}\circ\omega_{a}$ $\left(  a=1,2\right)  $
\ and: $J_{a}^{-1}=-J_{a}$:%
\begin{equation}
G=-J_{1}\circ T\circ J_{2}\Leftrightarrow T=-J_{1}\circ G\circ J_{2}%
\end{equation}

Having been built out of invariant tensors, it is clear that: $\mathcal{L}%
_{\Gamma_{a}}G=\mathcal{L}_{\Gamma_{a}}T=0$. In terms of the defining
matrices, this implies (see the previous Section):%
\begin{equation}
\left[  G,J_{a}\right]  =\left[  T,J_{a}\right]  =0,\text{ \ }a=1,2
\end{equation}
Hence: $GT=-J_{1}\circ T\circ J_{2}\circ T=-T^{2}\circ J_{1}\circ J_{2}=TG$,
i.e.:%
\begin{equation}
\left[  G,T\right]  =0
\end{equation}

By direct calculation, using the representative matrices and the symmetry of
the metric tensors, one proves immediately that: $g_{1}\left(  Gx,y\right)
=g_{2}\left(  x,y\right)  =g_{1}\left(  x,Gy\right)  $. Also, by direct
calculation: $g_{2}\left(  Gx,y\right)  =\left(  g_{1}\right)  ^{-1}\left(
g_{2}\left(  x,.\right)  ,g_{2}\left(  y,.\right)  \right)  =g_{2}\left(
x,Gy\right)  $. Hence, $G$ is \textit{self-adjoint} w.r.t. both metrics:%
\begin{equation}
g_{a}\left(  Gx,y\right)  =g_{a}\left(  x,Gy\right)  ,\text{ \ }a=1,2
\end{equation}

Furthermore, the compatibility condition implies: $\mathcal{L}_{\Gamma_{1}%
}\omega_{2}=0.$ In terms of the representative matrices, this implies (see
above): $\omega_{2}J_{1}=\widetilde{\left(  \omega_{2}J_{1}\right)  }$. As:
$\widetilde{\omega}=-\omega$ and $\widetilde{J_{1}}=-\omega_{1}\circ
g_{1}^{-1}$, we obtain: $\omega_{2}\circ g_{1}^{-1}\circ\omega_{1}=\omega
_{1}\circ g_{1}^{-1}\circ\omega_{2}$. This implies: $\left(  \omega_{1}%
^{-1}\circ\omega_{2}\right)  \circ g_{1}^{-1}\circ\omega_{1}=g_{1}^{-1}%
\circ\left(  \omega_{2}\circ\omega_{1}^{-1}\right)  \circ\omega_{1}$ or
(multiplying on the right by $\omega_{1}^{-1}$ and remembering that:
$T=\omega_{1}^{-1}\circ\omega_{2}$): $T\circ g_{1}^{-1}=g_{1}^{-1}%
\circ\widetilde{T}$ . Remembering the definition of the adjoint of a $\left(
1,1\right)  $ tensor we have then:%
\begin{equation}
T=g_{1}^{-1}\circ\widetilde{T}\circ g_{1}\equiv\left(  T^{\dag}\right)  _{1}%
\end{equation}
i.e., $T$ is self-adjoint w.r.t. the metric $g_{1}$. Interchanging indices,
one proves that: $\left(  T^{\dag}\right)  _{2}=T$ as well. Finally, each
$J_{a}$ ($a=1,2$) is \textit{skew}-adjoint w.r.t. the respective metric
tensor: $J_{a}=-\left(  J_{a}^{\dag}\right)  _{a}=-g_{a}^{-1}\circ
\widetilde{J_{a}}\circ g_{a}$. On top of that we have also, e.g.: $\left(
J_{1}^{\dag}\right)  _{2}=g_{2}^{-1}\circ\widetilde{J_{1}}\circ g_{2}%
=-g_{2}^{-1}\circ g_{1}\circ J_{1}\circ g_{1}^{-1}\circ g_{2}=-G^{-1}\circ
J_{1}\circ G=-J_{1}$, as $G$ and the $J$'s commute. Interchanging indices, one
proves a similar result for $J_{2}$. All in all:%
\begin{equation}
\left(  J_{a}^{\dag}\right)  _{b}=-J_{a},\text{ \ }a,b=1,2
\end{equation}
In summary, $\mathit{G,T,J}_{1}$ \textit{and }$\mathit{J}_{2}$ \textit{are a
set of mutually commuting operators. }$\mathit{G}$ \textit{and }$\mathit{T}$
\textit{are self-adjoint, while }$\mathit{J}_{1}\mathit{\ }$\textit{and
}$\mathit{J}_{2}$ \textit{are skew-adjoint w.r.t. both metric tensors.}

\bigskip

$G$ being self-adjoint, one can proceed to diagonalize it, and $\mathbb{V}%
=\mathbb{R}^{2n}$ will split into an orthogonal sum\footnote{The sum will be
orthogonal w.r.t. \textit{both} metrics.} of eigenspaces: $\mathbb{V}=%
{\displaystyle\bigoplus\limits_{k=1,...,r}}
\mathbb{V}_{k}$ where: $G|_{\mathbb{V}_{k}}=\lambda_{k}\mathbb{I}_{k}$ and the
$\lambda_{k}$'s ($k=1,...,r\leq2n$) are the distinct eigenvalues of $G$, and
$\lambda_{k}>0$. Notice that, as: $G=g_{1}^{-1}\circ g_{2}$, this implies:
\begin{equation}
g_{2}|_{\mathbb{V}_{k}}=\lambda_{k}g_{1}|_{\mathbb{V}_{k}}%
\end{equation}

$T$ commutes with $G$ and is self-adjoint as well. Then $\mathbb{V}_{k}$ will
decompose further into the (bi)orthogonal sum:
\begin{equation}
\mathbb{V}_{k}=%
{\displaystyle\bigoplus\limits_{\alpha}}
\mathbb{W}_{k,\alpha}%
\end{equation}
where, denoting as $\mu_{k,\alpha}$ the distinct eigenvalues of $T$ in
$\mathbb{V}_{k}$ (labeled by the index $\ \alpha$), $\mathbb{W}_{k,\alpha}$
will be the eigenspace of the eigenvalue $\mu_{k,\alpha}$. Once again:
$T|_{\mathbb{W}_{k,\alpha}}=\mu_{k,\alpha}\mathbb{I}_{k,\alpha}$, and hence:%
\begin{equation}
\omega_{2}|_{\mathbb{W}_{k,\alpha}}=\mu_{k,\alpha}\omega_{1}|_{\mathbb{W}%
_{k,\alpha}}%
\end{equation}
Notice that, neither symplectic form being degenerate by assumption, each
$\mathbb{W}_{k,\alpha}$ will be necessarily even-dimensional. The dimension of
each $\mathbb{W}_{k,\alpha}$ will be then at least two.

The complex structures $J_{1}$ and $J_{2}$ commute with both $G$ and $T$. So,
they will leave the subspaces $\mathbb{W}_{k,\alpha}$ invariant.
Reconstructing them from the $g$'s and $\omega$'s we find:%
\begin{equation}
J_{2}|_{\mathbb{W}_{k,\alpha}}=\frac{\mu_{k,\alpha}}{\lambda_{k}}%
J_{1}|_{\mathbb{W}_{k,\alpha}}%
\end{equation}
and, as: $J_{1}^{2}=J_{2}^{2}=-\mathbb{I}$: $\left(  \mu_{k,\alpha}%
/\lambda_{k}\right)  ^{2}=1$, i.e.: $\mu_{k,\alpha}=\pm\lambda_{k}$,
implying:
\begin{equation}
J_{2}|_{\mathbb{W}_{k,\alpha}}=\pm J_{1}|_{\mathbb{W}_{k,\alpha}}%
\end{equation}
Therefore, the index $\alpha$ can assume only \textit{at most } two values,
corresponding to $\pm\lambda_{k}$, i.e.: $\mathbb{V}_{k}=%
{\displaystyle\bigoplus\limits_{\alpha=\pm}}
\mathbb{W}_{k,\alpha}$ \textit{at most}, with \ $\mathbb{W}_{k,\pm}$
corresponding to the eigenvalues $\pm\lambda_{k}$ respectively. The dimension
of each eigenspace $\mathbb{V}_{k}$ will be then at least two if only one of
the possible eigenvalues $\pm\lambda_{k}$ of $T$ is present, at least four if
both are present. Hence, the maximum number of distinct eigenvalues of $G$
will be $r\leq n$.

\bigskip

In general, a $\left(  0,2\right)  $ and a $\left(  2,0\right)  $ tensors
(such as, say, $g_{2}$ and $g_{1}^{-1}$) can be composed to yield a $\left(
1,1\right)  $ tensor. They will be said to be \textit{"in a generic position"}
iff the resulting $\left(  1,1\right)  $ tensor has eigenvalues of minimum
degeneracy. In the present context, we will say that $h_{1}$ and $h_{2}$ are
in a generic position iff the eigenvalues of both $G$ and $T$ have minimum
degeneracy, which means \textit{double} degeneracy. Then: $r=n$ and we will
have the (bi)orthogonal decomposition:%
\begin{equation}
\mathbb{V}=%
{\displaystyle\bigoplus\limits_{k=1,...,n}}
\mathbb{E}_{k}%
\end{equation}
where: $\dim\mathbb{E}_{k}=2$ and either $\mathbb{E}_{k}=\mathbb{W}_{k,+}$ or
\ $\mathbb{E}_{k}=\mathbb{W}_{k,-}$ (only one can be present but \textit{not}
both, otherwise $\lambda_{k}$ would be fourfold degenerate). One can choose in
$E_{k}$ a $g_{1}$-orthogonal basis $\left(  e_{1},e_{2}\right)  $ in such a
way that:
\begin{equation}
g_{1}|_{E_{k}}=e_{1}^{\ast}\otimes e_{1}^{\ast}+e_{2}^{\ast}\otimes
e_{2}^{\ast}%
\end{equation}
the $e^{\ast}$'s being the dual basis: $e_{i}^{\ast}\left(  e_{j}\right)
=\delta_{ij}$. Then the condition: $g_{1}\left(  x,J_{1}y\right)
+g_{1}\left(  J_{1}x,y\right)  =0$ will imply:%
\begin{equation}
J_{1}|_{E_{k}}=e_{2}\otimes e_{1}^{\ast}-e_{1}\otimes e_{2}^{\ast}%
\end{equation}
or the opposite \ (i.e.: $J_{1}e_{1}=e_{2},J_{1}e_{2}=-e_{1}$), and hence
that:%
\begin{equation}
\omega_{1}|_{E_{k}}=e_{1}^{\ast}\wedge e_{2}^{\ast}%
\end{equation}

Correspondingly, we will have:%
\begin{equation}
g_{2}|_{E_{k}}=\lambda_{k}g_{1}|_{E_{k}};\text{ \ }J_{2}|_{E_{k}}=\pm
J_{1}|_{E_{k}};\text{ \ }\omega_{2}|_{E_{k}}=\pm\lambda_{k}\omega_{1}|_{E_{k}}%
\end{equation}

\bigskip

Coming now to the general problem of bihamiltonian fields, every linear vector
field $\Gamma$ preserving both $h_{1}$ and $h_{2}$ will have a representative
matrix commuting with those of $G$ and $T$. Therefore, it will be
block-diagonal in the common eigenspaces of both tensors. In the generic
(linear) case, the analysis can be restricted to the two-dimensional
eigenspaces $E_{k}$. On each one of these $\Gamma$ will preserve both a
symplectic structure and a positive-definite metric. Therefore it will be in
$sp\left(  2\right)  \cap so\left(  2\right)  =u\left(  1\right)  $ and it
will represent a harmonic oscillator, with a frequency possibly depending on
$E_{k}$.
\bigskip

Using, say, $\Gamma_{1}$ and $T$, one can construct the $n$ vectors:
$\Gamma_{k+1}=T^{k}\Gamma_{1},$ $k=0,1,...,n-1$. First of all one sees
immediately, by looking at the representative matrices, that, as that of
$\Gamma_{1}$ is $J_{1}$, which commutes with $T$, the $\Gamma_{k}$'s will
commute pairwise,
i.e.:%
\begin{equation}
\left[  \Gamma_{r},\Gamma_{s}\right]  =0\text{ }\forall r,s=1,2,...,n
\end{equation}

Moreover, we have shown that $T$ can be brought into the diagonal form:%
\begin{equation}
T=%
{\displaystyle\bigoplus\limits_{k=1,...,n}}
\rho_{k}\mathbb{I}_{k}%
\end{equation}
with $\rho_{k}=\pm\lambda_{k}$ and $\rho_{k}\neq\rho_{r}$ for $k\neq r$. If
the $\Gamma$'s were linearly dependent, there would exist a linear combination
such that:%
\begin{equation}%
{\displaystyle\sum\limits_{r=0}^{n-1}}
\alpha_{r}T^{r}=0
\end{equation}
But on each $E_{k}$ this would reduce to:%
\begin{equation}%
{\displaystyle\sum\limits_{r=0}^{n-1}}
\alpha_{r}\left(  \rho_{k}\right)  ^{r}=0,\text{ \ }k=1,...n
\end{equation}

The determinant of the coefficients of this system of linear equations being
the Vandermonde determinant of the $\rho$'s, it will be nonzero, and hence the
$\alpha$'s must all vanish, which proves that the $\Gamma$'s are linearly
independent, and hence a basis. As $T$ is a constant tensor, its Nijenhuis
torsion vanishes identically. Therefore, as discussed in Sect.\ref{recursion},
$T$ is a \textit{strong recursion operator}.$\blacksquare$\newline

What has been proved up to now is the following. Given two admissible triples:
$\left(  g_{1},\omega_{1},J_{1}\right)  $ and $\left(  g_{2},\omega_{2}%
,J_{2}\right)  $, on $V\approx\mathbb{R}^{2n}$, each triple defines a
$2n$-dimensional real representation $U_{r}\left(  2n,g_{a},\omega_{a}\right)
,$ $a=1,2$, of the group that leaves simultaneously invariant both $g_{a}$ and
$\omega_{a}$ (and hence $J_{a}$), i.e. of the unitary group. The intersection:%
\begin{equation}
W_{r}=U_{r}\left(  2n,g_{1},\omega_{1}\right)  \cap U_{r}\left(
2n,g_{2},\omega_{2}\right)
\end{equation}
will be the common invariance group of both triples. As shown in a $2D$
example in Ref.\cite{Mar7} and as emerges from the previous analysis, the
compatibility condition implies that $W_{r}$ does not reduce to the identity
alone. Any "quantum" bihamiltonian (linear) vector field $\Gamma$, i.e. a
field such that: $\mathcal{L}_{\Gamma}\omega_{a}=0$ \textit{and }%
$\mathcal{L}_{\Gamma}g_{a}=0$ will be in the Lie algebra of $W_{r}$. In the
generic case:%
\begin{equation}
W_{r}=\underset{n\text{ \ \ }times}{\underbrace{SO(2)\times SO\left(
2\right)  \times...\times SO(2)}}%
\end{equation}
otherwise:%
\begin{equation}
W_{r}=U_{r}\left(  2r_{1};g,\omega\right)  \times U_{r}\left(  2r_{2}%
;g,\omega\right)  \times...\times U_{r}\left(  2r_{k};g,\omega\right)
\end{equation}
where $\left(  g,\omega\right)  $ is any one of \ the pairs $\left(
g_{a},\omega_{a}\right)  $ and: $r_{1}+...+r_{k}=n$. Quite a similar analysis
can be done by complexifying $V$ in two different ways using the two complex
structures and reasoning in terms of the two Hermitian structures. In the
generic case, then:%
\begin{equation}
W_{r}=\underset{n\text{ \ \ }times}{\underbrace{U\left(  1\right)  \times
U\left(  1\right)  \times...\times U\left(  1\right)  }}%
\end{equation}
For further details, see Ref.\cite{Mar7}.\newline

To end this Section, we will like to rephrase the previous results in a way
more suitable to be generalized to the infinite dimensional case.

We first notice that, going back to the original complex n-dimensional Hilbert
space $\mathbb{H}$, there exist two positive constants $\alpha$ and $\beta$,
such that:
\begin{equation}
\alpha\Vert x\Vert_{1}\leq\Vert x\Vert_{2}\leq\beta\Vert x\Vert_{1}%
\;,\;\forall x\in\mathbb{H} \label{riesz}%
\end{equation}
This implies, by Riesz's theorem \cite{Kato,Na,RN}, that there exists a
bounded\footnote{With respect to both Hermitian structures.} positive and
self-adjoint operator $F$ such that:
\begin{equation}
h_{2}(x,y)=h_{1}(Fx,y)\;,\;\forall x,y\in\mathbb{H} \label{deff}%
\end{equation}
Formally ($h_{a}=g_{a}+i\omega_{a}$, $a=1,2$):
\begin{equation}
F=h_{1}^{-1}\circ h_{2}%
\end{equation}
and $F$ replaces the previous $G$ and $T$.

Then \cite{Mar112,Mar13} a necessary and sufficient condition for $h_{1}$ and
$h_{2}$ to be in generic position is that $F$ be a cyclic operator, i.e. that
there exists a vector $x_{0}$ such that the vectors $x_{0},Fx_{0}%
,\cdots,F^{n-1}x_{0}$ span the whole Hilbert space. Indeed, when $h_{1}$ and
$h_{2}$ are in generic position, $F$ has $n$ distinct eigenvalues,
$\lambda_{k}$. If we now denote with $\{f_{k}\}$ its eigenvector basis and
with $\{\mu^{(k)}\}$ a set of $n$ nonzero complex numbers, we can construct
the vectors
\begin{equation}
F^{m}x_{0}=\sum_{k}\mu^{(k)}\lambda_{k}^{m}f_{k}\;,\;m=0,1,\cdots,n-1.
\end{equation}
They are linear independent because the determinant of their components is
given by $(\prod_{k}\mu^{(k)})V(\lambda_{1},\cdots,\lambda_{n})$, where the
Vandermonde determinant $V$ is nonzero, the eigenvalues $\lambda_{k}$'s being
distinct. Clearly, the converse is also true.

Also, it has been argued in Ref.\cite{Mar7}, that "bi-unitary" operators, i.e.
operators that are unitary w.r.t. both Hermitian structures\footnote{Of
course, any linear vector field that leaves both $h$'s invariant will generate
a one-parameter group of bi-unitary transformations.}, must commute with $F$
(the proof is simple and we refer to the above reference for it), i.e.
bi-unitary operators are in the commutant $F^{\prime}$ of $F$\footnote{The
commutant $F^{\prime}$ of $F$ is the set of all operators that commute with
$F$. It is of course closed under commutation because of the Jacobi identity,
i.e. it is a Lie algebra. The \textit{bicommutant} $F^{\prime\prime}$ is the
set of all operators that commute with all those in the commutant. In
particular, they will commute with $F$ itself, and hence: $F^{\prime\prime
}\subset F^{\prime}$. Moreover, any two operators in $F^{\prime\prime}$ must
commute among themselves. $F^{\prime\prime}$ is therefore a (maximal) Abelian
subalgebra of $F^{\prime}$, i.e. $F^{\prime\prime}$ is the \textit{center} of
$F^{\prime}$}.

The results of this discussion can be summarized in the following:
\begin{proposition}
Two Hermitian forms are
in a generic position iff the bicommutant of $F$ coincides with the commutant:
$F^{\prime\prime}=F^{\prime}$.
\end{proposition}

It should be
clear from our presentation that many results will carry over to the
infinite-dimensional case, although new problems may arise because the
algebraic properties do not "control" properties such as continuity and
differentiability in infinite dimensions.

\subsection{The infinite-dimensional case}

\label{se:infex}
\bigskip

In the (genuinely) infinite-dimensional case of a Hilbert space $\mathbb{H}$
there arise two difficulties, namely:\newline i) Given two Hermitian
structures, $(\cdot,\cdot)_{1}$ and $(\cdot,\cdot)_{2}$ on $\mathbb{H}$
defining two complex scalar products (both linear in, say, the second factor
and antilinear in the first, but this is not a crucial point), they might
define two non-equivalent topologies on $\mathbb{H}$, and:\newline ii) The
spectra of self-adjoint operators may have both a point part and a continuum part.

Point i) is taken care of in an almost standard way, assuming that there exist
two positive constants $\alpha$ and $\beta$, such that formula (\ref{riesz})
holds. It follows that we can define the operator $F$ as in (\ref{deff}). But
now, due to point $ii)$, we have to better specify what we mean, for example,
by requiring $F$ to have nondegenerate eigenvalues. On the other side, the
definitions of the commutant and the bicommutant of $F$ are of purely
algebraic character and can therefore be generalized to the infinite
dimensional situation. Then, following Refs. \cite{Mar112} and \cite{Mar13},
we will adopt the following definition:\newline

\textbf{Definition.} \textit{Two Hermitian structures $h_{1}$ and $h_{2}$ are
said to be in generic position iff $F^{\prime\prime}=F^{\prime}$, $F$ being
their connecting operator.}

To proceed further in understanding the situation in which $F$ has also a
continuous spectrum, one needs suitable mathematical tools such as the
spectral theory and the theory of rings of operators in Hilbert spaces
\cite{Na}. We first observe that $F^{\prime}$ and $F^{\prime\prime}\subset
F^{\prime}$ are both (weakly closed) rings of bounded operators on
$\mathbb{H}$. Now, given any set $S\in\mathcal{B}(\mathbb{H})$, it can be
proved \cite{Na} that the minimal weakly closed ring $R(S)$ containing $S$
contains only those elements $A\in S^{\prime\prime}$ such that
\begin{equation}
E_{0}A=AE_{0}=A
\end{equation}
where $E_{0}$ is the so called principal identity of the set $S$, i.e. the
projection operator on $(kerS\cap kerS^{\dagger})^{\perp}$. If $S=\{F\}$, $F$
being self-adjoint and positive, we have that $\mathbb{I}\in R(F)$ and
$R(F)=F^{\prime\prime}$, which is therefore commutative.

If we decompose now $F$ in terms of its spectral family $\{P(\lambda)\}$:
\begin{equation}
F=\int_{\Delta}\lambda\,dP(\lambda)
\end{equation}
where $\Delta=[a,b]$ is a closed interval containing the spectrum of $F$, it
is possible to show that:\newline$a)$ The weakly closed commutative ring
$R(F)$ corresponds to a decomposition of the Hilbert space $\mathbb{H}$ into
the direct integral
\begin{equation}
\mathbb{H}=\int_{\Delta}H_{\lambda}\,d\sigma(\lambda)
\end{equation}
where the measure $\sigma(\lambda)$ is obtained from the spectral family
$\{P(\lambda)\}$ of $F$. \newline$b)$ Any operator $A\in F^{\prime}$ can be
represented as
\begin{equation}
A=\int_{\Delta}A(\lambda)\,d\sigma(\lambda)
\end{equation}
where $A(\lambda)$ is a bounded operator on $H_{\lambda}$, for almost all
$\lambda$. \newline$c)$ Every $B\in F^{\prime\prime}=R(F)$ is a multiplication
by a number $b(\lambda)$ on $H_{\lambda}$, for almost all $\lambda$.\newline
Moreover, since $R(F)$ is a maximal commutative ring by itself, the family
$F^{\prime}(\lambda)$ of all operators $A(\lambda)$ corresponding to
$F^{\prime}$, for a fixed $\lambda$, is irreducible so that we can rewrite
$a,b)$ above as:\newline$a^{\prime})$ The spectrum $\Delta$ of $F$ is the
union of a countable number of measurable sets $\Delta_{k}$ such that, for
$\lambda\in\Delta_{k}$, the spaces $H_{\lambda}$ have the same dimension
$n_{k}$ (finite or infinite) and:
\begin{equation}
\mathbb{H}=\bigoplus_{k}\int_{\Delta_{k}}H_{\lambda}\,d\sigma(\lambda)
\end{equation}
$b^{\prime})$ Any $A\in F^{\prime}$ can be written as
\begin{equation}
A=\bigoplus_{k}\int_{\Delta_{k}}A(\lambda)\,d\sigma(\lambda)
\end{equation}
Now, going back to the two Hermitian structures $h_{1}$ and $h_{2}$ on
$\mathbb{H}$, since the connecting operator $F$ acts on each $H_{\lambda}$ as
a multiplication by the number $\lambda$, we can easily derive the following
result generalizing the finite-dimensional situation.

\begin{proposition}
There exists a decomposition of $\mathbb{H}$ as
direct integral of Hilbert spaces $H_{\lambda}$, of dimension $n_{k}$ such
that in each $H_{\lambda}$: $h_{2}=\lambda h_{1}$.
\end{proposition}

It follows that the elements of the unitary group that leave simultaneously
invariant $h_{1}$ and $h_{2}$ have the form (see Eq.(4.5.3)):
\begin{equation}
U=\bigoplus_{k}\int_{\Delta_{k}}U_{k}(\lambda)\,d\sigma(\lambda) \label{Uk}%
\end{equation}
where $U_{k}(\lambda)$ is an element of the unitary group $U(n_{k})$, for each
$\lambda\in\Delta_{k}$.

Also, it is now immediate to prove that definition (1) is equivalent to:

\textbf{Definition.} \textit{ Two Hermitian structures $h_{1}$ and $h_{2}$ are
said to be in generic position iff the spaces $H_{\lambda}$ are
one-dimensional.}

Indeed, if $h_{1}$ and $h_{2}$ are in generic position, then $R(F)=F^{\prime
\prime}=F^{\prime}$, so that the latter is commutative and $A(\lambda)$, for
almost all $\lambda\in\Delta$, acts on a one-dimensional Hilbert space
$H_{\lambda}$. Conversely, if $R(F)=F^{\prime\prime}\neq F^{\prime}$,
$F^{\prime}$ is non-commutative and hence there is a subset $\Delta_{0}%
\subset\Delta$ such that $H_{\lambda}$ has dimension greater than one for
$\lambda\in\Delta_{0}$. $\blacksquare$

Notice also that, in the generic case, the operators $U_{k}(\lambda)$ in
(\ref{Uk}) are one-dimensional and reduces to a multiplication by a phase
factor $\exp[i\theta(\lambda)]$.

Finally, we may prove the following equivalence between the genericity
condition and the cyclicity of the operator $F$:

\textbf{Definition.}\textit{ $F$ is cyclic iff $F^{\prime\prime}=F^{\prime}$.}

This follows from the fact that, if $F^{\prime\prime}=F^{\prime}$, the latter
is commutative and each space $H(\lambda)$, where $F$ acts as a multiplication
by $\lambda$, is one-dimensional. So the vector $x_{0}=1/\lambda$ is a cyclic
vector. Viceversa, if we suppose now that $F$ is cyclic, each $H(\lambda)$ is
one-dimensional and any $A\in F^{\prime}$ acts as a multiplication by a
number. Hence $F^{\prime}=F^{\prime\prime}=R(F)$. $\blacksquare$\newline

\begin{example}
\textbf{ A particle in a box.} \textit{We consider the operator
}$F=1+X^{2}$\textit{ where }$X$\textit{ is the position operator which acts as
multiplication by }$x$\textit{ on the Hilbert space }$L^{2}([-\alpha
,\alpha],dx)$\textit{. From the spectrum }$\Delta_{X}=[-\alpha,\alpha
]$\textit{ and the spectral family }$\{P_{X}(\lambda)=\chi_{\lbrack
-\alpha,\lambda]}\}$\textit{ of }$X$\textit{ ( }$\chi_{\lbrack-\alpha
,\lambda]}$\textit{ being the characteristic function on }$[-\alpha,\lambda
]$\textit{), one easily sees that the spectrum of }$F$\textit{ is }$\Delta
_{F}=[1,1+\alpha^{2}]$\textit{ while its spectral family }$\{P_{F}(\lambda
)\}$\textit{ is given by }%
\begin{equation}
P_{F}(\lambda)=P(\sqrt{\lambda-1})-P(-\sqrt{\lambda-1})
\end{equation}
In fact, t is easy to check that:%
\begin{equation}
P_{F}^{2}=P_{F};\text{ }P_{F}\left(  1\right)  =0;\text{ }P_{F}\left(
1+\alpha^{2}\right)  =\mathbb{I}%
\end{equation}
\textit{We can write }$F$\textit{ as: }%
\begin{equation}
F=\int_{[-\alpha,\alpha]}(1+\lambda^{2})\,dP(\lambda)
\end{equation}
\textit{If we now divide the interval as }$[-\alpha,\alpha]=[-\alpha
,0]\cup\lbrack0,\alpha]$\textit{ and change variable by setting }%
$\lambda=-\sqrt{\mu-1}$\textit{ or }$\lambda=\sqrt{\mu-1}$\textit{ in the
negative or positive parts of the interval respectively, we get: }%
\begin{equation}
F=\int_{[1,1+\alpha^{2}]}\lambda\,dP_{F}(\lambda)
\end{equation}

\textit{Now }$F$\textit{ has no cyclic vector on the whole }$L^{2}%
([-\alpha,\alpha])$\textit{ since }$G^{\prime}$\textit{, which contains both
}$X$\textit{ and the parity operator is not commutative. On the contrary,
}$\chi_{\lbrack-\alpha,0]}$\textit{ is cyclic on }$L^{2}([-\alpha,0])$\textit{
and, similarly, }$\chi_{\lbrack0,\alpha]}$\textit{ is so on }$L^{2}%
([0,\alpha])$\textit{. Thus the Hilbert space splits in two }$F$%
\textit{-cyclic spaces: }$L^{2}([-\alpha,\alpha])=L^{2}([-\alpha,0])\oplus
L^{2}([0,\alpha])$\textit{ and we obtain the decomposition }%
\begin{equation}
\mathbb{H}=\int_{[1,1+\alpha^{2}]}H_{\lambda}\,d\sigma(\lambda)
\end{equation}
\textit{where the measure is obtained from: }%
\begin{equation}
\sigma(\lambda)=P_{F}(\lambda)\chi_{\lbrack-\alpha,0]}=P_{F}(\lambda
)\chi_{\lbrack0,\alpha]}=\sqrt{\lambda-1}%
\end{equation}
\textit{Notice that the spaces }$H_{\lambda}$\textit{ are one-dimensional if
we work in the interval }$[0,\alpha]$\textit{ or bidimensional if we consider
}$[-\alpha,\alpha]$\textit{. Also, the bi-unitary transformations read,
respectively, as: }%
\begin{align}
&  U=\int_{[1,1+\alpha^{2}]}e^{i\phi(\lambda)}\,d\sigma(\lambda)\\
&  U=\int_{[1,1+\alpha^{2}]}U_{2}(\lambda)\,d\sigma(\lambda)
\end{align}
\end{example}

%% file: Chapt5rev.tex
\newpage

\section{From Finite to Infinite Dimensions. Weyl Systems}\label{Weyl0}

\subsection{An Abstract Setting for Weyl Systems}\label{Weyl-sys}

\bigskip

A known theorem by A.Wintner \cite{Win} states that if, say, $\widehat{q}$ and
$\widehat{p}$ are quantum-mechanical operators on an infinite-dimensional
Hilbert space satisfying a commutation relation of the form: $\left[
\widehat{q},\widehat{p}\right]  =c\widehat{\mathbb{I}}$ (or, better: $\left[
\widehat{q},\widehat{p}\right]  \subseteq c\widehat{\mathbb{I}}$), with $c$ a
constant and $\widehat{\mathbb{I}}$ the identity operator, then at least one
of them must be unbounded.

Motivated then by the need of formulating Quantum Mechanics without having to
do with unbounded operators, it was apparently H.Weyl \cite{We} (see also
\cite{Sud}) who proposed first a different scheme of quantization that goes as follows:

Let $\mathcal{S}$ be a (real) linear vector space endowed with a
constant\footnote{I.e. translationally-invariant.} symplectic
structure\footnote{Hence, necessarily: $\dim\left(  \mathcal{S}\right)  $ will
be \textit{even}, and: $\mathcal{S}\approx\mathbb{R}^{2n}$ for some $n$.}
$\omega$. Weyl's approach consists in the following:

\begin{itemize}
\item It is a map $W$ \ from $\mathcal{S}$ \ to the set of \textit{unitary
operators} on a \textbf{(}so far unspecified\footnote{That's why the setting
we are describing here has been defined as "abstract".}\textbf{)} Hilbert
space $\mathcal{H}$:%
\begin{equation}
W:\mathcal{S}\rightarrow\mathcal{U}(\mathcal{H})
\end{equation}
via:%
\begin{equation}
\mathcal{S\ni}\text{ }z\rightarrow\widehat{W}\left(  z\right)  \in
\mathcal{U}(\mathcal{H}),\text{ \ }\widehat{W}\left(  z\right)  \widehat
{W}^{\dag}\left(  z\right)  =\widehat{W}^{\dag}\left(  z\right)  \widehat
{W}\left(  z\right)  =\widehat{\mathbb{I}}%
\end{equation}
with the following specifications:

\item $W$ is a strongly continuous map, and

\item For any $z,z^{\prime}\in\mathcal{S}$:%
\begin{equation}
\widehat{W}\left(  z+z^{\prime}\right)  =\widehat{W}\left(  z\right)
\widehat{W}\left(  z^{\prime}\right)  \exp\left\{  -i\omega\left(
z,z^{\prime}\right)  /2\hbar\right\}  \label{Weyl1}%
\end{equation}
with $\hbar$ the reduced Planck constant. It follows then that:%
\begin{equation}
\widehat{W}\left(  z\right)  \widehat{W}\left(  z^{\prime}\right)
=\widehat{W}\left(  z^{\prime}\right)  \widehat{W}\left(  z\right)
\exp\left\{  i\omega\left(  z,z^{\prime}\right)  /\hbar\right\}  ,\text{
\ }\forall z,z^{\prime} \label{Weyl2}%
\end{equation}

\end{itemize}

Moreover, setting $z^{\prime}=0$ in (\ref{Weyl1}) we obtain: $\widehat{W}%
^{-1}\left(  z\right)  \widehat{W}\left(  z\right)  =\widehat{W}\left(
0\right)  $, and hence: $\widehat{W}\left(  0\right)  =\widehat{\mathbb{I}}$,
while setting $z^{\prime}=-z$ we obtain: $\widehat{W}^{-1}\left(  z\right)
=\widehat{W}\left(  -z\right)  $, and hence:%

\begin{equation}
\widehat{W}^{\dag}\left(  z\right)  =\widehat{W}\left(  -z\right)
\end{equation}

Then, \textit{a Weyl system is a projective unitary representation of the
linear vector space }$\mathcal{S}$ \textit{(thought of as the group manifold
of the translation group) in the Hilbert space }$\mathcal{H}$.

As a running example we shall consider $\mathcal{S}=\mathbb{R}^{2}$ with
coordinates $\left(  q,p\right)  $ and the standard symplectic form:
$\omega=dq\wedge dp$, which is represented by the matrix:%
\begin{equation}
\omega=\left\vert
\begin{array}
[c]{cc}%
0 & 1\\
-1 & 0
\end{array}
\right\vert
\end{equation}
Hence:%
\begin{equation}
\omega\left(  \left(  q,p\right)  ,\left(  q^{\prime},p^{\prime}\right)
\right)  =\left\vert
\begin{array}
[c]{cc}%
q & p
\end{array}
\right\vert \left\vert
\begin{array}
[c]{cc}%
0 & 1\\
-1 & 0
\end{array}
\right\vert \left\vert
\begin{array}
[c]{c}%
q^{\prime}\\
p^{\prime}%
\end{array}
\right\vert =qp^{\prime}-q^{\prime}p
\end{equation}
and therefore:%
\begin{equation}
\widehat{W}\left(  \left(  q,p\right)  +\left(  q^{\prime},p^{\prime}\right)
\right)  =\widehat{W}\left(  q,p\right)  \widehat{W}\left(  q^{\prime
},p^{\prime}\right)  \exp\left\{  -\frac{i}{2\hbar}\left(  qp^{\prime
}-q^{\prime}p\right)  \right\}  \label{Weyl22}%
\end{equation}

In the general case, we can decompose $\mathcal{S}$ into the direct sum of two
Lagrangian subspaces: $\mathcal{S}=\mathcal{S}_{1}\oplus\mathcal{S}_{2}$, and
hence any vector $z$ as: $z=(z_{1},0)+(0,z_{2})$, $z_{1}\in\mathcal{S}_{1},$
$z_{2}\in\mathcal{S}_{2}$. \ We can consider then the restrictions of $W$ to
the Lagrangian subspaces, i.e.:%
\begin{equation}
U=W|_{\mathcal{S}_{1}}:\mathcal{S}_{1}\rightarrow\mathcal{H}%
\end{equation}
and:%
\begin{equation}
V=W|_{\mathcal{S}_{2}}:\mathcal{S}_{2}\rightarrow\mathcal{H}%
\end{equation}
As: $\omega|_{\mathcal{S}_{1}}=\omega|_{\mathcal{S}_{2}}=0$, $U$ and $V$ are
\textit{faithful} representations of the corresponding Lagrangian subspaces:%
\begin{equation}
\widehat{U}\left(  z_{1}+z_{1}^{\prime}\right)  =\widehat{U}\left(
z_{1}\right)  \widehat{U}\left(  z_{1}^{\prime}\right)  ;\text{ }z_{1}%
,z_{1}^{\prime}\in\mathcal{S}_{1}%
\end{equation}
and similarly for $V$. Moreover:
\begin{equation}
\widehat{U}\left(  z_{1}\right)  \widehat{V}\left(  z_{2}\right)  =\widehat
{V}\left(  z_{2}\right)  \widehat{U}\left(  z_{1}\right)  \exp\left\{
i\omega\left(  \left(  z_{1},0\right)  ,\left(  0,z_{2}\right)  \right)
/\hbar\right\}  \label{Weyl3}%
\end{equation}

Viceversa, we have the following:

\bigskip

\textbf{Proposition:} \textit{Given two faithful representations }$U$
\textit{and }$V$ \ \textit{of two transversal Lagrangian subspaces of a symplectic vector
space }$\mathcal{S}$\textit{ satisfying} (\ref{Weyl3}), \textit{the map:}%
\begin{equation}
z\longrightarrow\widehat{W}\left(  z\right)  =\widehat{U}\left(  z_{1}\right)
\widehat{V}\left(  z_{2}\right)  \exp\left\{  -i\omega\left(  \left(
z_{1},0\right)  ,\left(  0,z_{2}\right)  \right)  /2\hbar\right\}
\label{Weyl4}%
\end{equation}
\textit{is a Weyl system.}

The proof that (\ref{Weyl4}) does indeed satisfy the defining property
(\ref{Weyl1}) can be done by direct calculation, and will be omitted
here.$\blacksquare$

\bigskip

Consider now a one-dimensional subspace of $\mathcal{H}$ spanned by a fixed
vector $z$. From (\ref{Weyl1}) we have, with $\alpha,\beta$ real numbers:%
\begin{equation}
\widehat{W}\left(  \alpha z\right)  \widehat{W}\left(  \beta z\right)
=\widehat{W}\left(  \left(  \alpha+\beta\right)  z\right)
\end{equation}
Therefore, $\left\{  \widehat{W}\left(  \alpha z\right)  \right\}  _{\alpha
\in\mathbb{R}\text{ }}$ is a strongly continuous one-parameter group of
unitaries and, by Stone's theorem \cite{RS}:%
\begin{equation}
\widehat{W}\left(  \alpha z\right)  =\exp\left\{  i\alpha\widehat{G}\left(
z\right)  /\hbar\right\}
\end{equation}
with an infinitesimal generator $\widehat{G}\left(  z\right)  $ which is
(essentially) self-adjoint. Furthermore, $\left\{  \widehat{W}\left(
\alpha\beta z\right)  \right\}  _{\beta\in\mathbb{R}}$ is also a strongly
continuous one-parameter group, and therefore:%
\begin{equation}
\widehat{W}\left(  \alpha\beta z\right)  =\exp\left\{  i\beta\widehat
{G}\left(  \alpha z\right)  /\hbar\right\}
\end{equation}
and, setting $\beta=1$, we find:%
\begin{equation}
\widehat{G}\left(  \alpha z\right)  =\alpha\widehat{G}\left(  z\right)
\end{equation}
In terms of infinitesimal generators and setting: $z\rightarrow\alpha
z,z^{\prime}\rightarrow\beta z^{\prime}$, \ Eq. (\ref{Weyl2}) \ reads:%
\begin{equation}
e^{i\alpha\widehat{G}\left(  z\right)  /\hbar}e^{i\beta\widehat{G}\left(
z^{\prime}\right)  /\hbar}=e^{i\alpha\beta\omega\left(  z,z^{\prime}\right)
/\hbar}e^{i\alpha\widehat{G}\left(  z\right)  /\hbar}e^{i\beta\widehat
{G}\left(  z^{\prime}\right)  /\hbar}%
\end{equation}
and, for $\alpha$ and $\beta$ infinitesimal, this yields, to the lowest
nontrivial order:
\begin{equation}
\left[  \widehat{G}\left(  z\right)  ,\widehat{G}\left(  z^{\prime}\right)
\right]  =-i\hbar\omega\left(  z,z^{\prime}\right)  \label{commutator}%
\end{equation}

\subsection{Von Neumann's Representation Theorem}\label{sec:Neumann}
\bigskip

What is lacking in the "abstract" presentation of the previous Section is a
concrete realization of the Hilbert space $\mathcal{H}$ on which the mapping
$W$ should operate.

Before discussing von Neumann's theorem, let us resume our running example on
$\mathbb{R}^{2}\approx T^{\ast}\mathbb{R}$. Writing $\left(  q,p\right)  $ as:
$\left(  q,p\right)  =\left(  q,0\right)  +\left(  0,p\right)  $, whence:
$\omega\left(  \left(  q,0\right)  ,\left(  0,p\right)  \right)  =qp$, our
Weyl system becomes ( $z=\left(  q,p\right)  ,z_{1}=\left(  q,0\right)
,z_{2}=\left(  0,p\right)  $) (see Eq.(\ref{Weyl22})):%
\begin{equation}
\widehat{W}\left(  q,p\right)  =\widehat{W}\left(  \left(  q,0\right)
+\left(  0,p\right)  \right)  =\widehat{W}\left(  q,0\right)  \widehat
{W}\left(  0,p\right)  \exp\left\{  -iqp/2\hbar\right\}
\end{equation}
while:%
\begin{equation}
\widehat{W}\left(  q+q^{\prime},0\right)  =\widehat{W}\left(  q,0\right)
\widehat{W}\left(  q^{\prime},0\right)
\end{equation}
and similarly for $\widehat{W}\left(  0,p\right)  $. Define then:%
\begin{equation}
\widehat{W}\left(  q,0\right)  =\exp\left\{  iq\widehat{P}/\hbar\right\}
;\text{ }\widehat{W}\left(  0,p\right)  =\exp\left\{  ip\widehat{Q}%
/\hbar\right\}
\end{equation}
In other words, as: $\left(  q,0\right)  =q\left(  1,0\right)  ,\left(
0,p\right)  =p\left(  0,1\right)  $, we are defining:
\begin{equation}
\widehat{G}\left(  0,1\right)  =\widehat{Q},\text{ \ }\widehat{G}\left(
1,0\right)  =\widehat{P}%
\end{equation}
with (cfr. Eq. (\ref{commutator})):%
\begin{equation}
\left[  \widehat{Q},\widehat{P}\right]  =i\hbar\mathbb{I}\label{5:comm1}
\end{equation}
Moreover, using the truncated Baker-Campbell-Hausdorff \cite{RS}
formula\footnote{$e^{a+b}=e^{a}e^{b}e^{-\left[  a,b\right]  /2}$ whenever:
$\left[  a,\left[  a,b\right]  \right]  =\left[  b,\left[  a,b\right]
\right]  =0$.} one finds easily:%
\begin{equation}
\widehat{W}\left(  q,p\right)  =\exp\left\{  i\left(  q\widehat{P}%
+p\widehat{Q}\right)  /\hbar\right\}  \label{Weyl5}%
\end{equation}

Consider now $L_{2}\left(  \mathbb{R},dx\right)  $ with the Lebesgue measure,
and define the families of operators $\left\{  \widehat{U}\left(  q\right)
\right\}  _{q\in\mathbb{R}}$ and $\left\{  \widehat{V}\left(  p\right)
\right\}  _{p\in\mathbb{R}}$ via:%
\begin{equation}
\left(  \widehat{U}\left(  q\right)  \psi\right)  \left(  x\right)
=\psi\left(  x+q\right)  \label{action1}%
\end{equation}
and:%
\begin{equation}
\left(  \widehat{V}\left(  p\right)  \psi\right)  \left(  x\right)
=\exp\left\{  ipx/\hbar\right\}  \psi\left(  x\right)  \label{action2}%
\end{equation}
for $\psi\in L_{2}\left(  \mathbb{R},dx\right)  $. It is easy to show that
both families are actually one-parameter, strongly continuous groups of
unitaries, and that:%
\begin{equation}
\left(  \widehat{U}\left(  q\right)  \widehat{V}\left(  p\right)  \psi\right)
\left(  x\right)  =\exp\left\{  iqp/\hbar\right\}  \left(  \widehat{V}\left(
p\right)  \widehat{U}\left(  q\right)  \psi\right)  \left(  x\right)
\end{equation}
Then:%
\begin{equation}
\widehat{W}\left(  q,p\right)  =\widehat{U}\left(  q\right)  \widehat
{V}\left(  p\right)  \exp\left\{  -iqp/\hbar\right\}
\end{equation}
is a \textit{concrete} realization of a Weyl system. Defining again:
$\widehat{U}\left(  q\right)  =\exp\left\{  iq\widehat{P}/\hbar\right\}  $
and: $\widehat{V}\left(  p\right)  =\exp\left\{  ip\widehat{Q}/\hbar\right\}
$, we find both Eq.(\ref{Weyl5}) and, at the infinitesimal level\footnote{And
in the appropriate domains.}:%
\begin{equation}
\left(  \widehat{Q}\psi\right)  \left(  x\right)  =x\psi\left(  x\right)
,\text{ \ }\left(  \widehat{P}\psi\right)  \left(  x\right)  =-i\hbar
\frac{d\psi}{dx}%
\end{equation}
Moreover:
\begin{equation}
\left(  \widehat{W}\left(  q,p\right)  \psi\right)  \left(  x\right)
=\exp\left\{  ip\left[  x+q/2\right]  /\hbar\right\}  \psi\left(  x+q\right)
\end{equation}

A generic matrix element of $\widehat{W}\left(  q,p\right)  $ will be given
then by:%
\begin{equation}
\left\langle \phi,\widehat{W}\left(  q,p\right)  \psi\right\rangle
=\exp\left\{  iqp/2\hbar\right\}
{\displaystyle\int\limits_{-\infty}^{+\infty}}
dx\overline{\phi\left(  x\right)  }\exp\left\{  ipx/\hbar\right\}  \psi\left(
x+q\right)
\end{equation}

\begin{remark}

\textit{Viewed as a function on }$T^{\ast}Q$\textit{, }$\left\langle
\phi,\widehat{W}\left(  q,p\right)  \psi\right\rangle $\textit{ is
square-integrable for all }$\phi,\psi\in L^{2}\left(  \mathbb{R}\right)
$\textit{. Indeed, defining the Lebesgue measure on }$R^{2}$\textit{ as
}$dqdp/2\pi\hbar$\textit{, a direct calculation shows that:}%
\begin{equation}
\left\Vert \left\langle \phi,\widehat{W}\left(  q,p\right)  \psi\right\rangle
\right\Vert ^{2}=:%
{\displaystyle\iint}
\frac{dqdp}{2\pi\hbar}\left\vert \left\langle \phi,\widehat{W}\left(
q,p\right)  \psi\right\rangle \right\vert ^{2}=\left\Vert \phi\right\Vert
^{2}\left\Vert \psi\right\Vert ^{2} \label{norm}%
\end{equation}
\end{remark}§

Instead, for plane-wave states: 

\begin{equation}
{\phi\left(  x\right)  =(1/\sqrt{2\pi})\exp(ik^{\prime}x),\psi\left(
x\right)  =(1/\sqrt{2\pi})\exp(ikx)}
\end{equation}
and denoting as $\left\langle k^{\prime}|\widehat{W}\left(  q,p\right)
|k\right\rangle $ the matrix elements of $\widehat{W}\left(  q,p\right)  $
between these states, we obtain:%

\begin{equation}
\left\langle k^{\prime}|\widehat{W}\left(  q,p\right)  |k\right\rangle
=\delta\left(  k-k^{\prime}+p/\hbar\right)  \exp\left(  iq\left(  k+k^{\prime
}\right)  /2\right)  \label{matelement}%
\end{equation}
and, in particular:
\begin{equation}
\left\langle k|\widehat{W}\left(  q,p\right)  |k\right\rangle =\hbar
\delta\left(  p\right)  \exp\left\{  ikq\right\}  \label{expectation}%
\end{equation}

\bigskip Integrating Eq.(\ref{expectation}) over $k$, we obtain for the
\textit{trace} of $W$\footnote{Actually, we can define the trace only if we
admit distribution-valued traces. Strictly speaking \cite{DHS}, and as
Eq.(\ref{norm}) shows, $\widehat{W}$ is bounded but \textit{not}
trace-class.}:%
\begin{equation}
Tr\left\{  \widehat{W}\left(  q,p\right)  \right\}  =2\pi\hbar\delta\left(
q\right)  \delta\left(  p\right)  \label{wtrace}%
\end{equation}

\bigskip

Coming now to the general case, let's assume that we are given a symplectic
vector space $\left(  \mathcal{S},\omega\right)  $ and a decomposition of
$\mathcal{S}$ as the direct sum:
\begin{equation}
\mathcal{S}=\mathcal{S}_{1}\oplus\mathcal{S}_{2}%
\end{equation}
with $\mathcal{S}_{1}$ and $\mathcal{S}_{2}$ Lagrangian subspaces. Every
vector $z\in\mathcal{S}$ can then be decomposed in a unique way as: $z=\left(
z_{1},0\right)  +\left(  0,z_{2}\right)  ,z_{i}\in\mathcal{S}_{i},i=1,2$. Let
us remark first of all that the symplectic structure allows each one of the
two subspaces to be identified with the dual of the other. Indeed, we can
define a pairing:%
\begin{equation}
\left\langle .,.\right\rangle :\mathcal{S}_{2}\times\mathcal{S}_{1}%
\rightarrow\mathbb{R}%
\end{equation}
via:%
\begin{equation}
\left\langle z_{2},z_{1}\right\rangle :\omega\left(  \left(  z_{1},0\right)
,\left(  0,z_{2}\right)  \right)  \label{dual}%
\end{equation}
The details of the proof that in this way \ $\mathcal{S}_{2}\approx
\mathcal{S}_{1}^{\ast}$ (and viceversa, of course) can be found in Ref.
\cite{Man}.

Assume now $\mathcal{H}$ to be a separable Hilbert space and let:%
\begin{equation}%
\begin{array}
[c]{c}%
U:\mathcal{S}_{1}\rightarrow\mathcal{H}\\
V:\mathcal{S}_{2}\rightarrow\mathcal{H}%
\end{array}
\end{equation}
be \textit{unitary, irreducible and strongly continuous} representations of
\ $\mathcal{S}_{1}$ and $\mathcal{S}_{2}$ respectively on $\mathcal{H}$,
satisfying the additional condition that defines the \textit{"Weyl form"} of
the commutation relations:%
\begin{equation}
\widehat{U}\left(  z_{1}\right)  \widehat{V}\left(  z_{2}\right)  =\widehat
{V}\left(  z_{2}\right)  \widehat{U}\left(  z_{1}\right)  \exp\left\{
i\omega\left(  \left(  z_{1},0\right)  ,\left(  0,z_{2}\right)  \right)
/\hbar\right\}
\end{equation}
Then we can define:%
\begin{equation}
\widehat{W}\left(  z\right)  =\widehat{U}\left(  z_{1}\right)  \widehat
{V}\left(  z_{2}\right)  \exp\left\{  -i\omega\left(  \left(  z_{1},0\right)
,\left(  0,z_{2}\right)  \right)  /2\hbar\right\}  \label{Weyl6}%
\end{equation}
which is a Weyl system. Let us denote $z_{1}$ and $z_{2}$ as $\left(
q,0\right)  $ and $\left(  0,p\right)  $ respectively, with $q$ and $p$
$n$-dimensional vectors ($n=\dim S_{1}=\dim S_{2}$). \ Correspondingly, we
will denote \ $\widehat{U}\left(  z_{1}\right)  $ and $\widehat{V}\left(
z_{2}\right)  $ as $\widehat{U}\left(  q\right)  $ and $\widehat{V}\left(
p\right)  $ respectively.

Von Neumann's theorem \cite{Neu2} states then that there exists a unitary map:%
\begin{equation}
T:\mathcal{H}\rightarrow\mathcal{L}_{2}\left(  \mathbb{R}^{n},d\mu\right)
\end{equation}
such that:%
\begin{equation}
\left(  T\widehat{U}\left(  q\right)  T^{-1}\psi\right)  \left(  x\right)
=\psi\left(  x+q\right)
\end{equation}
and (cfr.Eqn.(\ref{dual})):%
\begin{equation}
\left(  T\widehat{V}\left(  p\right)  T^{-1}\psi\right)  \left(  x\right)
=e^{i\left\langle x,p\right\rangle }\psi\left(  x\right)
\end{equation}
This theorem proves that all the representations of the Weyl commutation
relations are unitarily equivalent to the Schr\"{o}dinger representation, and
hence are unitarily equivalent among themselves (but see below,
Sect.\ref{evade}).

\bigskip

\begin{example}

\textit{In the case of }$L_{2}\left(  \mathbb{R}\right)  $\textit{, setting
}$\hbar=1$\textit{ and using the Fourier transform:}%
\begin{equation}
\psi\left(  x\right)  =%
{\displaystyle\int\limits_{-\infty}^{\infty}}
\frac{dp}{\sqrt{2\pi}}\widetilde{\psi}\left(  p\right)  \exp\left\{
ipx\right\}
\end{equation}
\textit{one finds easily that:}%
\begin{equation}
\widetilde{\left(  \exp\left(  ix\widehat{P}\right)  \psi\right)  }\left(
p\right)  =e^{ixp}\widetilde{\psi}\left(  p\right)
\end{equation}

\textit{(i.e.: }$\left(  \widehat{P}\widetilde{\psi}\right)  \left(  p\right)
=p\widetilde{\psi}\left(  p\right)  $\textit{, and:}%
\begin{equation}
\widetilde{\left(  \exp\left(  i\pi\widehat{Q}\right)  \psi\right)  }\left(
p\right)  =\widetilde{\psi}\left(  p-\pi\right)
\end{equation}

\textit{(}$\left(  \widehat{Q}\widetilde{\psi}\right)  \left(  p\right)
=id\widetilde{\psi}\left(  p\right)  /dp$\textit{). Denoting by:}%
\begin{equation}
\mathcal{F}:\mathcal{L}_{2}\left(  \mathbb{R}\right)  \rightarrow
\mathcal{L}_{2}\left(  \mathbb{R}\right)
\end{equation}
\textit{the unitary operator defined by the Fourier transform, we can conclude
that:}%
\begin{equation}
\mathcal{F}^{\dag}\widehat{Q}\mathcal{F=-}\widehat{P}%
\end{equation}
\textit{and:}%
\begin{equation}
\mathcal{F}^{\dag}\widehat{P}\mathcal{F=}\widehat{Q}%
\end{equation}
\end{example}

\subsection{Weyl Systems and Linear Transformations}

\bigskip

Let's begin by considering linear transformations that preserve the
symplectic structure, i.e. linear maps: $T:S\rightarrow S$ such that:%
\begin{equation}
\omega\left(  Tz,Tz^{\prime}\right)  =\omega\left(  z,z^{\prime}\right)
\forall z,z^{\prime}\in S
\end{equation}
In terms of matrices this means:%
\begin{equation}
\widetilde{T}\omega T=\omega
\end{equation}
(where $\widetilde{T}$ stands for the transpose of the matrix $T$), and this
defines a realization of the symplectic group $Sp\left(  2n,\mathbb{R}\right)
$ associated with the symplectic structure $\omega$.

Then, we can define:%
\begin{equation}
\widehat{W}_{T}:\mathcal{S}\rightarrow\mathcal{H}%
\end{equation}
via:%
\begin{equation}
\widehat{W}_{T}\left(  z\right)  =:\widehat{W}\left(  Tz\right)
\end{equation}
and, as:%
\begin{eqnarray}
\widehat{W}\left(  T\left(  z+z^{\prime}\right)  \right)  &=&\widehat{W}\left(
Tz\right)  \widehat{W}\left(  Tz^{\prime}\right)  \exp\left\{  -i\omega\left(
Tz,Tz^{\prime}\right)  /2\hbar\right\}  = \nonumber \\
&=&\widehat{W}\left(  Tz\right)  \widehat{W}\left(  Tz^{\prime}\right)
\exp\left\{  -i\omega\left(  z,z^{\prime}\right)  /2\hbar\right\}
\end{eqnarray}
we find:$\mathcal{\ }$:%
\begin{equation}
\widehat{W}_{T}\left(  z+z^{\prime}\right)  =\widehat{W}_{T}\left(  z\right)
\widehat{W}_{T}\left(  z^{\prime}\right)  \exp\left\{  -i\omega\left(
z,z^{\prime}\right)  /2\hbar\right\}
\end{equation}
i.e. $\widehat{W}_{T}$ is also a Weyl system, and hence, by von Neumann's
theorem, it is unitarily equivalent to $\widehat{W}$.

As a simple example, consider, in $\mathbb{R}^{2}$, the map:
\begin{equation}
\left(  q,p\right)  \rightarrow\left(  -p,q\right)
\end{equation}
which is realized via the transformation\footnote{The \textit{matrix}
representing $T$ is simply minus that of the complex structure. However, the two
have different transformation properties (see Chapt.$1$).}:%
\begin{equation}
T=\left\vert
\begin{array}
[c]{cc}%
0 & -1\\
1 & 0
\end{array}
\right\vert
\end{equation}
Then it is clear that:%
\begin{equation}
\widehat{U}\left(  q\right)  =\widehat{W}\left(  \left(  q,0\right)  \right)
\rightarrow\widehat{W}\left(  \left(  0,-p\right)  \right)  =\widehat
{V}\left(  -p\right)
\end{equation}
and:%
\begin{equation}
\widehat{V}\left(  p\right)  =\widehat{W}\left(  \left(  0,p\right)  \right)
\rightarrow\widehat{W}\left(  \left(  q,0\right)  \right)  =\widehat{U}\left(
q\right)
\end{equation}
which is precisely (see the end of the previous Section) what the Fourier
transform does.

\bigskip

As \ $\widehat{W}_{T}$ is unitarily equivalent to $\widehat{W}$, to the map
$T$ there is associated an automorphism of the group $\mathcal{U}\left(
\mathcal{H}\right)  $ of the unitary operators. As every automorphism of
$\mathcal{U}\left(  \mathcal{H}\right)  $ is inner, there is a unitary
operator $\widehat{U}_{T}$ such that:%
\begin{equation}
\widehat{W}_{T}\left(  z\right)  =\widehat{U}_{T}^{\dag}\left(  \widehat
{W}\left(  z\right)  \right)  \widehat{U}_{T}%
\end{equation}

More generally, we can consider a one-parameter group $\left\{  T_{\lambda
}\right\}  _{\lambda\in\mathbb{R}}$ of linear symplectic transformations.
Calling $\Gamma$ the linear vector field that is the infinitesimal generator
of the group, the condition:%
\begin{equation}
\omega\left(  T_{\lambda}z,T_{\lambda}z^{\prime}\right)  =\omega\left(
z,z^{\prime}\right)  \forall z,z^{\prime}\in S,\forall\lambda\in\mathbb{R}%
\end{equation}
becomes:%
\begin{equation}
L_{\Gamma}\omega=0
\end{equation}
with \ $L_{\Gamma}$ the Lie derivative. There exists then (globally on a
vector space) a function $g$ such that:%
\begin{equation}
i_{\Gamma}\omega=dg \label{vectorfi}%
\end{equation}
and, for linear transformations, $g$ \ will be a quadratic function of the coordinates.

According to what has been said above, the family \ $\left\{  T_{\lambda
}\right\}  $ defines a (strongly continuous) one-parameter group $\left\{
U_{\lambda}\right\}  _{\lambda\in\mathbb{R}}$ of unitary operators such that:%
\begin{equation}
\widehat{W}\left(  z\left(  \lambda\right)  \right)  =\widehat{U}_{\lambda
}^{\dag}\widehat{W}\left(  z\right)  \widehat{U}_{\lambda}%
\end{equation}
where: $z\left(  \lambda\right)  =T_{\lambda}\left(  z\right)  $. By Stone's
theorem, then:%
\begin{equation}
\widehat{U}_{\lambda}=\exp\left\{  -i\lambda\widehat{G}/\hbar\right\}
\end{equation}
with $\widehat{G}$ self-adjoint. \textit{The self-adjoint operator }%
$\widehat{G}$ \textit{is the quantum counterpart of the quadratic function
}$g$. In this way we have achieved a way to quantize all the quadratic
functions: given $G$, we can define via Eq.(\ref{vectorfi}) the associated
Hamiltonian vector field. This in turns defines a one-parameter group of
(linear) symplectic transformations, and the corresponding Weyl system allows
us to find the (self-adjoint) quantum operator to be associated with $g$.

Let's consider now \ a general linear transformation $T\in GL\left(
2n,\mathbb{R}\right)  $, not necessarily a symplectic one. We will denote for
clarity as $\omega_{0}$ a reference (comparison) symplectic
structure,  written in a Darboux chart as:
\begin{equation}
\omega_{0}=\left\vert
\begin{array}
[c]{cc}%
0 & \mathbb{I}\\
-\mathbb{I} & 0
\end{array}
\right\vert
\end{equation}
We define then a new symplectic structure $\omega_{T}$ via:%
\begin{equation}
\omega_{T}\left(  z,z^{\prime}\right)  =:\omega_{0}\left(  Tz,Tz^{\prime
}\right)
\end{equation}
That $\omega_{T}$ is a symplectic structure is obvious. It is represented by
the matrix:%
\begin{equation}
\omega_{T}=\widetilde{T}\omega_{0}T
\end{equation}

Now, if we define again:%
\begin{equation}
\widehat{W}_{T}\left(  z\right)  =:\widehat{W}\left(  Tz\right)
\end{equation}
it is easy to prove that:%
\begin{equation}
\widehat{W}_{T}\left(  z+z^{\prime}\right)  =\widehat{W}_{T}\left(  z\right)
\widehat{W}_{T}\left(  z^{\prime}\right)  \exp\left\{  -i\omega_{T}\left(
z,z^{\prime}\right)  /2\hbar\right\}
\end{equation}
Therefore, $\widehat{W}_{T}$ defines a Weyl system, but for $\left(
\mathcal{S},\omega_{T}\right)  $ an not for $\left(  \mathcal{S},\omega
_{D}\right)  $. Mimicking the analysis that has been done previously, we
conclude that:%
\begin{equation}
\widehat{W}_{T}\left(  \lambda z\right)  =\exp\{i\lambda\widehat{G}\left(
z\right)  \}
\end{equation}
and that:%
\begin{equation}
\left[  \widehat{G}\left(  z\right)  ,\widehat{G}\left(  z^{\prime}\right)
\right]  =-i\hbar\omega_{T}\left(  z,z^{\prime}\right)
\end{equation}

Now we are in a position to consider Weyl systems for a vector space with an
arbitrary and translationally invariant symplectic structure $\omega$. By Darboux' theorem
\cite{AM,Ar}, there exists always an invertible linear transformation $T$ such
that:%
\begin{equation}
\omega=\widetilde{T}\omega_{0}T \label{T2}%
\end{equation}
Then, the sequence of transformations:%
\begin{equation}
\left(  \mathcal{S},\omega\right)  \overset{T}{\rightarrow}\left(
\mathcal{S},\omega_{0}\right)  \overset{W}{\rightarrow}\mathcal{U}\left(
\mathcal{H}\right)
\end{equation}
defines a Weyl system $W\circ T=W_{T}~\ $for \ $\left(  \mathcal{S}%
,\omega\right)  $ such that:%
\begin{equation}
\widehat{W}_{T}\left(  z\right)  =:\widehat{W}\left(  Tz\right)
\end{equation}

\begin{remark}

The matrix $T$ in Eq.(\ref{T2}) is clearly ambiguous by \textit{left }
multiplication by any matrix $T^{\prime}$ such that $\widetilde{T^{\prime}%
}\omega_{0}T=\omega_{0}$. However, as:%
\begin{equation}
\omega_{0}\left(  T^{\prime}Tz,T^{\prime}Tz^{\prime}\right)  =\omega
_{0}\left(  Tz,Tz^{\prime}\right)  =\omega\left(  z,z^{\prime}\right)
\end{equation}
the Weyl systems associated with $T$ and $T^{\prime}T$ are unitarily equivalent.
\end{remark}

\subsection{Some Examples}
\bigskip

As is well known \cite{Ar}, a conspicuous example of a one-parameter group of
symplectic transformations is provided by the time evolution of a Hamiltonian
system. So, let's study some simple examples.

\subsubsection{The free particle}
\bigskip

In this case, the one-parameter group is given by: $\left(  q,p\right)
\rightarrow\left(  q+tp/m,p\right)  $ and is represented by the matrix:%
\begin{equation}
\left\vert
\begin{array}
[c]{c}%
q\left(  t\right) \\
p\left(  t\right)
\end{array}
\right\vert =F\left(  t\right)  \left\vert
\begin{array}
[c]{c}%
q\\
p
\end{array}
\right\vert ;\text{ \ }F\left(  t\right)  =\left\vert
\begin{array}
[c]{cc}%
1 & t/m\\
0 & 1
\end{array}
\right\vert ;\text{ \ }F\left(  t\right)  F\left(  t^{\prime}\right)
=F\left(  t+t^{\prime}\right)
\end{equation}
Then:%
\begin{eqnarray}
\widehat{W}_{t}\left(  q,p\right) & =&\widehat{W}\left(  q\left(  t\right)
,p\left(  t\right)  \right)  =
\exp\left\{  (i/\hbar)\left[  q\left(  t\right)  \widehat{P}+p\left(
t\right)  \widehat{Q}\right]  \right\}  \nonumber \\
&=:&\exp\left\{  (i/\hbar)\left[
q\widehat{P}_{t}+p\widehat{Q}_{t}\right]  \right\}
\end{eqnarray}
where:%
\begin{equation}
\widehat{P}_{t}=\widehat{P},\text{ \ }\widehat{Q}_{t}=\widehat{Q}+t\widehat
{P}/m \label{operators1}%
\end{equation}
There exists therefore a one-parameter family $\left\{  \widehat{F}\left(
t\right)  \right\}  _{t\in\mathbb{R}}$ of unitary operators such that:%
\begin{equation}
\exp\left\{  ip\widehat{Q}_{t}/\hbar\right\}  =\widehat{F}^{\dag}\left(
t\right)  \exp\left\{  ip\widehat{Q}/\hbar\right\}  \widehat{F}\left(
t\right)
\end{equation}
and:%
\begin{equation}
\exp\left\{  iq\widehat{P}_{t}/\hbar\right\}  =\widehat{F}^{\dag}\left(
t\right)  \exp\left\{  iq\widehat{P}/\hbar\right\}  \widehat{F}\left(
t\right)
\end{equation}
Setting then:%
\begin{equation}
\widehat{F}\left(  t\right)  =\exp\left\{  -i\widehat{H}t/\hbar\right\}
\end{equation}
using Eq.(\ref{operators1}) and expanding for small $q,p$ and $t$, one finds
the commutation relations:%
\begin{equation}
\left[  \widehat{P},\widehat{H}\right]  =0,\text{ \ }\left[  \widehat
{Q},\widehat{H}\right]  =\frac{i\hbar}{m}\widehat{P} \label{FreePart1}%
\end{equation}
\begin{remark}
Note that the previous equation does not specify what are the basic commutation relations between $\widehat{Q}$ and $\widehat{P}$. Stated otherwise, we are not yet specifying what should be the symplectic structure that appears on the r.h.s. of Eq.(\ref{commutator}), and this is very much in the spirit \cite{Wig3} of Wigner's approach. In what follows, however, and as we are dealing with this and the following Examples only to exhibit simple instances of Weyl systems, we shall assume that $q$ and $p$ are Darboux coordinates, and hence that the basic commutation relations are of the standard form of Eq.(\ref{5:comm1}). The only unknown quantity in Eq.(\ref{FreePart1}) will be then the Hamiltonian $\widehat{H}$.
\end{remark}

As the generators of linear  and homogeneous canonical transformations are quadratic functions,
it is natural to look for a quantum operator $\widehat{H}$ \ that is also a
quadratic function:%
\begin{equation}
\widehat{H}=a\widehat{P}^{2}+b\widehat{Q}^{2}+c\left(  \widehat{P}\widehat
{Q}+\widehat{Q}\widehat{P}\right)
\end{equation}
Then the solution of the previous commutation relations is precisely:
\begin{equation}
\widehat{H}=\frac{\widehat{P}^{2}}{2m}+\lambda\widehat{\mathbb{I}}%
\label{free}
\end{equation}
where $\widehat{\mathbb{I}}$ is the identity operator and $\lambda$ and arbitrary (real) constant. Apart from this, the quantum
operator associated with the time evolution is the standard quantum Hamiltonian.

\bigskip

\subsubsection{The Harmonic Oscillator}
\bigskip

The classical Hamiltonian is:%
\begin{equation}
\mathcal{H}=\frac{p^{2}}{2m}+\frac{1}{2}m\omega^{2}q^{2}%
\end{equation}
and the solution of the equations of motion is:%
\begin{eqnarray}
&&q\left(  t\right)  =q\cos\omega t+p\frac{\sin\omega t}{m\omega}\\
&& p\left(  t\right)  =p\cos\omega t-qm\omega\sin\omega t \nonumber 
\end{eqnarray}
The matrix $F\left(  t\right)  $ \ is then:%
\begin{equation}
F\left(  t\right)  =\left\vert
\begin{array}
[c]{cc}%
\cos\omega t & \frac{\sin\omega t}{m\omega}\\
-m\omega\sin\omega t & \cos\omega t
\end{array}
\right\vert
\end{equation}

Proceeding just as in the previous case we find again:%
\begin{equation}
\widehat{W}_{t}\left(  q,p\right)  =\exp\left\{  (i/\hbar)\left[  q\widehat
{P}_{t}+p\widehat{Q}_{t}\right]  \right\}
\end{equation}
with, now:%
\begin{equation}
\widehat{Q}_{t}=\widehat{Q}\cos\omega t+\widehat{P}\frac{\sin\omega t}%
{m\omega}%
\end{equation}
and:%
\begin{equation}
\widehat{P}_{t}=\widehat{P}\cos\omega t-\widehat{Q}m\omega\sin\omega t
\end{equation}

Defining again: $\widehat{F}\left(  t\right)  =\exp\left\{  -i\widehat
{H}t/\hbar\right\}  $ and working out the commutation relations of
$\widehat{H}$ with $\widehat{Q}$ and $\widehat{P}$ , that read now:%
\begin{equation}
\left[  \widehat{Q},\widehat{H}\right]  =\frac{i\hbar}{m}\widehat{P}%
\end{equation}
just as before, and:%
\begin{equation}
\left[  \widehat{P},\widehat{H}\right]  =-i\hbar m\omega^{2}\widehat{Q}%
\end{equation}
one finds :%
\begin{equation}
\widehat{H}=\frac{\widehat{P}^{2}}{2m}+\frac{1}{2}m\omega^{2}\widehat{Q}^{2}+\lambda\widehat{\mathbb{I}}%
\end{equation}
i.e., again "modulo" an additive multiple of the identity, the standard quantum Hamiltonian.

\bigskip

\subsubsection{A Charged Particle in a Constant Magnetic Field}
\bigskip

The equations of motion for a particle of mass $m$ and charge $q$ in a
constant magnetic field $\mathbf{B}$ are \cite{Ba,Mor}\footnote{For an
analysis at the quantum level, see \cite{Br,DF,DZ,Ho,LSU}}(in units $c=1$):%
\begin{equation}
\frac{d\mathbf{x}}{dt}=\mathbf{v}%
\end{equation}%
\begin{equation}
m\frac{d\mathbf{v}}{dt}=q\mathbf{v\times B}%
\end{equation}
The vector field is therefore:%
\begin{equation}
\Gamma=\mathbf{v\cdot}\frac{\partial}{\partial\mathbf{x}}+\frac{q}%
{m}\mathbf{v\times B\cdot}\frac{\partial}{\partial\mathbf{v}} \label{dyn}%
\end{equation}
The equations of motion can be derived either from the Lagrangian:%
\begin{equation}
\mathcal{L}=\frac{1}{2}m\mathbf{v}^{2}+q\mathbf{v\cdot A}%
\end{equation}
where $\mathbf{A}$ is the vector potential: $\nabla\times\mathbf{A}%
=\mathbf{B}$, or from the Hamiltonian:%
\begin{equation}
\mathcal{H}=\frac{\mathbf{p}^{2}}{2m} \label{magham}%
\end{equation}
with the symplectic form:%

\begin{equation}
\omega_{B}=-\frac{1}{2}q\varepsilon_{ijk}B^{k}dx^{i}\wedge dx^{j}+dx^{i}\wedge
dp_{i}%
\end{equation}
where:%

\begin{equation}
\mathbf{p}=\mathbf{\pi}-q\mathbf{A}%
\end{equation}
$\mathbf{\pi}$ is the canonical momentum:%
\begin{equation}
\mathbf{\pi=}\frac{\partial\mathcal{L}}{\partial\mathbf{v}}%
\end{equation}
and $\mathbf{p}=m\mathbf{v}$ is the \textit{kinetic} momentum.

We will consider here a field: $B=\left(  0,0,B\right)  $. As the motion along
$x^{3}$ is trivial and decouples, we will ignore it and concentrate on the
dynamics in the $\left(  x^{1},x^{2}\right)  $ plane. Among the various gauges
that one can employ the most popular are the \textit{Landau gauges}:%
\begin{equation}
\mathbf{A}_{1}=B\left(  x^{2},0,0\right)  ;\text{ \ }\mathbf{A}_{2}=B\left(
0,-x^{1},0\right)
\end{equation}
or the \textit{symmetric gauge}:%
\begin{equation}
\mathbf{A}_{s}=\frac{B}{2}\left(  x^{2},-x^{1},0\right)  =\frac{1}%
{2}\mathbf{B}\times\mathbf{r}=\frac{\mathbf{A}_{1}+\mathbf{A}_{2}}{2}%
\end{equation}
Introducing collective coordinates: $z=\left(  z^{1},...,z^{4}\right)  $ with:
$\left(  z^{1},z^{2}\right)  =\left(  x^{1},x^{2}\right)  $, $\left(
z^{3},z^{4}\right)  =\left(  p_{1},p_{2}\right)  $ and setting $q=m=1$, the
symplectic form can be written as:%
\begin{equation}
\omega_{B}=\frac{1}{2}\Omega_{ij}dz^{i}\wedge dz^{j}%
\end{equation}
where $\Omega$ is the matrix:%
\begin{equation}
\Omega=\left\vert \allowbreak%
\begin{array}
[c]{cccc}%
0 & -B & 1 & 0\\
B & 0 & 0 & 1\\
-1 & 0 & 0 & 0\\
0 & -1 & 0 & 0
\end{array}
\right\vert
\end{equation}
Explicitly:%
\begin{equation}
\omega_{B}=-Bdx^{1}\wedge dx^{2}+dx^{1}\wedge dp_{1}+dx^{2}\wedge dp_{2}%
\end{equation}
The inverse of $\Omega:$%
\begin{equation}
\Lambda=-\Omega^{-1}=\left\vert
\begin{array}
[c]{cccc}%
0 & 0 & 1 & 0\\
0 & 0 & 0 & 1\\
-1 & 0 & 0 & B\\
0 & -1 & -B & 0
\end{array}
\right\vert \allowbreak
\end{equation}
defines the Poisson tensor:%
\begin{equation}
\Lambda=\frac{1}{2}\Lambda_{ij}\frac{\partial}{\partial z^{i}}\wedge
\frac{\partial}{\partial z^{j}}%
\end{equation}
or, explicitly:%
\begin{equation}
\Lambda=\frac{\partial}{\partial x^{1}}\wedge\frac{\partial}{\partial p_{1}%
}+\frac{\partial}{\partial x^{2}}\wedge\frac{\partial}{\partial p^{2}}%
+B\frac{\partial}{\partial p_{1}}\wedge\frac{\partial}{\partial p_{2}}%
\end{equation}

A transformation that reduces $\omega_{B}$ to the standard Darboux form,
defined by the matrix:%
\begin{equation}
\Omega_{0}=\left\vert
\begin{array}
[c]{cc}%
\mathbf{0}_{2\times2} & \mathbb{I}_{2\times2}\\
-\mathbb{I}_{2\times2} & \mathbf{0}_{2\times2}%
\end{array}
\right\vert
\end{equation}
is: $z\rightarrow\widetilde{z}=Tz$ \ with:%
\begin{equation}
T=\left\vert
\begin{array}
[c]{cccc}%
1 & 0 & 0 & 0\\
0 & 1 & 0 & 0\\
0 & -B & 1 & 0\\
0 & 0 & 0 & 1
\end{array}
\right\vert \label{matrix1}%
\end{equation}
i.e., explicitly:%
\begin{equation}
p_{1}\rightarrow\widetilde{p}_{1}=p_{1}-Bx^{2}%
\end{equation}
with all the other coordinates unchanged and:%
\begin{equation}
\omega_{B}\left(  z,z^{\prime}\right)  =\omega_{D}\left(  Tz,Tz^{\prime
}\right)  =\omega_{0}\left(  \widetilde{z},\widetilde{z}^{\prime}\right)
\end{equation}
which implies, as can also be checked by direct calculation on the
representative matrices:%

\begin{equation}
\widetilde{T}\omega_{0}T=\Omega\label{matrix2}%
\end{equation}

Notice that this amounts to the transformation:%
\begin{equation}
\text{ }\mathbf{r}\rightarrow\mathbf{r,}\text{ \ \ }\mathbf{p}\rightarrow
\mathbf{p}-\mathbf{A}_{1}%
\end{equation}
One could have used instead, e.g., the transformation:%
\begin{equation}
\mathbf{r}\rightarrow\mathbf{r,}\text{ \ \ }\mathbf{p}\rightarrow
\mathbf{p}-\mathbf{A}_{s}%
\end{equation}
that is defined by the matrix:%
\begin{equation}
T^{\prime}=\left\vert
\begin{array}
[c]{cccc}%
1 & 0 & 0 & 0\\
0 & 1 & 0 & 0\\
0 & B/2 & 1 & 0\\
-B/2 & 0 & 0 & 1
\end{array}
\right\vert \label{matrix3}%
\end{equation}
and here too:%
\begin{equation}
\widetilde{T}^{\prime}\omega_{0}T^{\prime}=\Omega\label{matrix4}%
\end{equation}
Notice that, defining: $V=:T^{\prime}T^{-1}$, Eqns. (\ref{matrix2}) and
(\ref{matrix4}) imply:%
\begin{equation}
\widetilde{V}\omega_{0}V=\omega_{0}%
\end{equation}
i.e.: $V\in Sp\left(  \mathbb{R}^{6}\right)  $, and this too can be checked by
direct calculation.

Concentrating now on the transformation defined by Eq.(\ref{matrix1}) and
following the procedure outlined in Sect.$4.3$, we define the Weyl system:%
\begin{eqnarray}
\widehat{W}_{T}\left(  z\right) & =&\widehat{W}\left(  Tz\right)  =\\
&=&\exp\left\{  i\left[  \widetilde{x}^{1}\widehat{P}_{1}+\text{ }\widetilde
{x}^{2}\widehat{P}_{2}+\widetilde{p}_{1}\widehat{Q}^{1}+\widetilde{p}%
_{2}\widehat{Q}^{2}\right]  \right\} \nonumber
\end{eqnarray}
or, explicitly:%
\begin{equation}
\widehat{W}_{T}\left(  z\right)  =\exp\left\{  i%
{\displaystyle\sum\limits_{i=1,2}}
\left[  x_{i}\widehat{P}_{i}^{\left(  T\right)  }+p_{i}\widehat{Q}%
_{i}^{\left(  T\right)  }\right]  \right\}
\end{equation}
where:%
\begin{equation}
\widehat{Q}_{i}^{\left(  T\right)  }=\widehat{Q}_{i},\text{~}i=1,2,\text{
\ }\widehat{P}_{1}^{\left(  T\right)  }=\widehat{P}_{1},\text{ }\widehat
{P}_{2}^{\left(  T\right)  }=\widehat{P}_{2}+B\widehat{Q}^{1} \label{hat}%
\end{equation}
Notice that:%
\begin{equation}
\left[  \widehat{Q}_{i}^{\left(  T\right)  },\widehat{Q}_{j}^{\left(
T\right)  }\right]  =0;\text{ \ }\left[  \widehat{Q}_{i}^{\left(  T\right)
},\widehat{P}_{j}^{\left(  T\right)  }\right]  =i\delta_{ij}%
\end{equation}
while:%
\begin{equation}
\left[  \widehat{P}_{1}^{\left(  T\right)  },\widehat{P}_{2}^{\left(
T\right)  }\right]  =-iB
\end{equation}

Time evolution is given by:%
\begin{equation}
\left\vert
\begin{array}
[c]{c}%
x^{1}\left(  t\right) \\
x^{2}\left(  t\right) \\
p_{1}\left(  t\right) \\
p_{2}\left(  t\right)
\end{array}
\right\vert =F\left(  t\right)  \left\vert
\begin{array}
[c]{c}%
x^{1}\\
x^{2}\\
p_{1}\\
p_{2}%
\end{array}
\right\vert
\end{equation}
where:%
\begin{equation}
F\left(  t\right)  =\left\vert
\begin{array}
[c]{cccc}%
1 & 0 & \frac{\sin\left(  Bt\right)  }{B} & \frac{1-\cos\left(  Bt\right)
}{B}\\
0 & 1 & -\frac{1-\cos\left(  Bt\right)  }{B} & \frac{\sin\left(  Bt\right)
}{B}\\
0 & 0 & \cos\left(  Bt\right)  & \sin\left(  Bt\right) \\
0 & 0 & -\sin\left(  Bt\right)  & \cos\left(  Bt\right)
\end{array}
\right\vert
\end{equation}
is a linear symplectic map. Explicitly:%
\begin{eqnarray}
&& x^{1}\left(  t\right)  =x^{1}+p_{1}\frac{\sin\left(  Bt\right)  }{B}%
+p_{2}\frac{1-\cos\left(  Bt\right)  }{B}   \label{dynamics1} \\
&&x^{2}\left(  t\right)
=x^{2}-p_{1}\frac{1-\cos\left(  Bt\right)  }{B}+p_{2}\frac{\sin\left(
Bt\right)  }{B} \nonumber
\end{eqnarray} 
and:%
\begin{eqnarray}
&&p_{1}\left(  t\right)  =p_{1}\cos\left(  Bt\right)  +p_{2}\sin\left(
Bt\right) \label{dynamics2}\\
&& p_{2}\left(  t\right)  =-p_{1}\sin\left(  Bt\right)
+p_{2}\cos\left(  Bt\right)  \nonumber
\end{eqnarray}

Following the procedure outlined in the previous examples, we define then the
Weyl system:
\begin{equation}
\widehat{W}_{T}^{\left(  t\right)  }\left(  z\right)  =\widehat{W}\left(
Tz\left(  t\right)  \right)  =\exp\left\{  i%
{\displaystyle\sum\limits_{i=1,2}}
\left[  x_{i}\left(  t\right)  \widehat{P}_{i}^{\left(  T\right)  }%
+p_{i}\left(  t\right)  \widehat{Q}_{i}^{\left(  T\right)  }\right]  \right\}
\end{equation}
or:%
\begin{equation}
\widehat{W}_{T}^{\left(  t\right)  }\left(  z\right)  =\exp\left\{  i%
{\displaystyle\sum\limits_{i=1,2}}
\left[  x_{i}\widehat{P}_{i}^{\left(  T\right)  }\left(  t\right)
+p_{i}\widehat{Q}_{i}^{\left(  T\right)  }\left(  t\right)  \right]  \right\}
\end{equation}
where the $\widehat{P}_{i}^{\left(  T\right)  }\left(  t\right)  $'s and
$\widehat{Q}_{i}^{\left(  T\right)  }\left(  t\right)  $'s are defined by:%
\begin{equation}%
{\displaystyle\sum\limits_{i=1,2}}
\left[  x_{i}\widehat{P}_{i}^{\left(  T\right)  }\left(  t\right)
+p_{i}\widehat{Q}_{i}^{\left(  T\right)  }\left(  t\right)  \right]  =:%
{\displaystyle\sum\limits_{i=1,2}}
\left[  x_{i}\left(  t\right)  \widehat{P}_{i}^{\left(  T\right)  }%
+p_{i}\left(  t\right)  \widehat{Q}_{i}^{\left(  T\right)  }\right]
\end{equation}
and Eqns.(\ref{dynamics1}) and (\ref{dynamics2}) have to be used on the r.h.s.
Here too we conclude that there exists a unitary operator $\widehat{F}\left(
t\right)  =\exp\left\{  -it\mathcal{H}/\hbar\right\}  $ such that:
\begin{eqnarray}
&&
\exp\left\{  ip\widehat{Q}_{i}^{\left(  T\right)  }\left(  t\right)
/\hbar\right\}  =\widehat{F}^{\dag}\left(  t\right)  \exp\left\{
ip\widehat{Q}_{i}^{\left(  T\right)  }/\hbar\right\}  \widehat{F}\left(
t\right) \\
&& \exp\left\{  iq\widehat{P}_{i}^{\left(  T\right)  }\left(  t\right)
/\hbar\right\}  =\widehat{F}^{\dag}\left(  t\right)  \exp\left\{
iq\widehat{P}_{i}^{\left(  T\right)  }/\hbar\right\}  \widehat{F}\left(
t\right)  ;\text{ }i=1,2 \nonumber
\end{eqnarray}

Expanding again for small $q,p,t$ and using Eq.(\ref{hat}) we find the
commutation relations:%
\begin{equation}
\left[  \widehat{P}_{1},\mathcal{H}\right]  =0,\text{ \ }\left[  \widehat
{P}_{2},\mathcal{H}\right]  =i\hbar B\left(  \widehat{P}_{1}-B\widehat{Q}%
_{2}\right)  \text{ \ }%
\end{equation}
and:%
\begin{equation}
\left[  \widehat{Q}_{1},\mathcal{H}\right]  =i\hbar\left(  \widehat{P}%
_{1}-B\widehat{Q}_{2}\right)  ,\text{ }\left[  \widehat{Q}_{2},\mathcal{H}%
\right]  =i\hbar\widehat{P}_{2}%
\end{equation}
and it is easy to conclude that the Hamiltonian operator is now:%
\begin{equation}
\mathcal{H}=\frac{1}{2}\left[  \left(  \widehat{P}_{1}-B\widehat{Q}%
_{2}\right)  ^{2}+\widehat{P}_{2}^{2}\right]
\end{equation}
which corresponds to the "minimal coupling" prescription:%
\begin{equation}
\widehat{\mathbf{P}}\rightarrow\widehat{\mathbf{P}}-\mathbf{A}%
\end{equation}
with: $\mathbf{A}=\mathbf{A}_{1}$.

\subsubsection{Magnetic Translation Groups and Weyl Systems}
\bigskip

We will exhibit in this final Subsection another example \cite{Mor} of a Weyl
system, which is provided by the implementation at the quantum level of the
group of translations in a two-dimensional electron gas in a constant
(perpendicular) magnetic field that has been studied in the previous Subsection.

\bigskip Reinstating the constants $\left(  c,m,q\right)  $ in the proper
places, the Hamiltonian is given by \ (cfr.Eq.(\ref{magham})):
\begin{equation}
H=\frac{1}{2m}{\LARGE [}\mathbf{\pi}-\frac{q}{c}\mathbf{A}{\LARGE ]}^{2}%
\end{equation}

Introducing complex coordinates: $\zeta=x+iy$, the equations of motion
become:
\begin{equation}
\frac{d}{dt}\{\overset{\cdot}{\zeta}+i\Omega\zeta\}=0;\text{ \ \ }\Omega
=\frac{qB}{mc}%
\end{equation}
and they have the solution:
\begin{equation}
\zeta(t)=X+A\exp\{-i\Omega t\}
\end{equation}
where:
\begin{equation}
X=\zeta-\frac{i}{\Omega}\overset{\cdot}{\zeta}=const.
\end{equation}
The associated total energy is:
\begin{equation}
E=\frac{\mathbf{p}^{2}}{2m}\equiv\frac{1}{2}m|\overset{\cdot}{\zeta}%
|^{2}\equiv\frac{1}{2}m\Omega^{2}|A|^{2}%
\end{equation}
and the orbits are circles of radius $|A|$ and center:
\begin{equation}
X=x_{0}+iy_{0};\text{ \ \ }x_{0}=x+\frac{p_{y}}{m\Omega},\text{ \ }%
y_{0}=y-\frac{p_{x}}{m\Omega}%
\end{equation}

The Poisson brackets for the components of the kinetic momentum are:
\begin{equation}
\{p_{i},p_{j}\}=\frac{q}{c}(\partial_{i}A_{j}-\partial_{j}A_{i})\equiv\frac
{q}{c}\varepsilon_{ijk}B_{k}%
\end{equation}
i.e.:
\begin{equation}
\{p_{x},p_{y}\}=m\Omega
\end{equation}
and hence:
\begin{equation}
\{x_{0},y_{0}|=-\frac{1}{m\Omega}=-\frac{c}{qB}%
\end{equation}

The Cartan form:
\begin{equation}
\theta_{\mathcal{L}}=\frac{\partial\mathcal{L}}{\partial\mathbf{v}}\cdot
d\mathbf{q}\equiv\{p_{i}+\frac{q}{c}A_{i}\}dx_{i}%
\end{equation}
leads to:
\begin{equation}
\omega_{\mathcal{L}}=:-d\theta_{\mathcal{L}}=dx_{i}\wedge dp_{i}+\frac{q}%
{2c}(\partial_{i}A_{j}-\partial_{j}A_{i})dx_{i}\wedge dx_{j}%
\end{equation}
i.e.:
\begin{equation}
\omega_{\mathcal{L}}=dx_{i}\wedge dp_{i}+\frac{q}{2c}\varepsilon_{ijk}%
B_{i}dx_{j}\wedge dx_{k}%
\end{equation}

The dynamical vector field is given by (cfr. Eq.(\ref{dyn})):
\begin{equation}
\Gamma=\frac{p_{i}}{m}\frac{\partial}{\partial q_{i}}+\frac{q}{mc}%
\varepsilon_{ijk}p_{j}B_{k}\frac{\partial}{\partial v_{i}}%
\end{equation}

\bigskip

N\"{o}ther's theorem \cite{MFVMR} states that, if $X^{c}$ is a tangent
lift\footnote{We recall \cite{MFVMR} that,if: $X=X^{i}\partial/\partial
q^{i},$ $X^{i}=X^{i}\left(  q\right)  $ is a vector field on some manifold
$M$, its \textit{tangent lift} $X^{c}$ is the vector field on $TM$ defined by:
$X^{c}=X^{i}\partial/\partial q^{i}+L_{\Gamma}(X^{i})\partial/\partial v^{i}$,
with $\Gamma$ \textit{any} second-order vector field.}, and:
\begin{equation}
L_{X^{c}}\mathcal{L=}L_{\Gamma}h,\text{ \ \ }h=\pi^{\ast}\widehat{G},\text{
}i.e.:h=h(\mathbf{r})
\end{equation}
then the associated constant of the motion is:
\begin{equation}
\chi=i_{X^{c}}\theta_{\mathcal{L}}-h
\end{equation}

For translations in the plane:
\begin{equation}
X_{i}\equiv X_{i}^{c}=\frac{\partial}{\partial x}_{i}%
\end{equation}
Hence:
\begin{equation}
L_{X_{i}}\mathcal{L}=\frac{q}{mc}(\partial_{i}A_{j})p_{j}\equiv\frac{q}%
{c}(\partial_{i}A_{j}-\partial_{j}A_{i})+\frac{q}{c}\partial_{j}A_{i}%
\end{equation}
i.e.:
\begin{equation}
L_{X_{i}}\mathcal{L=}\frac{q}{c}\frac{d}{dt}(\mathbf{A}+\mathbf{r}%
\times\mathbf{B})_{i}%
\end{equation}
($\mathbf{h}=\frac{q}{c}(\mathbf{A}+\mathbf{r}\times\mathbf{B})$), and the
associated N\"{o}ther's constants of the motion are:
\begin{equation}
\chi_{i}=i_{X_{i}}\theta_{\mathcal{L}}-h_{i}=(\mathbf{p}+\frac{q}{c}%
\mathbf{B}\times\mathbf{r})_{i}%
\end{equation}

Notice that the coordinates of the center of the Larmor orbit are:
\begin{eqnarray}
&& x_{0}=\frac{c}{qB}(p_{y}+\frac{qB}{c}x)\equiv\frac{c}{qB}(\mathbf{p}+\frac
{q}{c}\mathbf{B}\times\mathbf{r})_{y}\\
&& y_{0}=-\frac{c}{qB}(mv_{x}-\frac{qB}{c}y)\equiv-(\mathbf{p}+\frac{q}%
{c}\mathbf{B}\times\mathbf{r})_{x} \nonumber
\end{eqnarray}
Hence:
\begin{equation}
\chi_{x}=-\frac{qB}{c}y_{0},\text{ \ }\chi_{y}=\frac{qB}{c}x_{0}%
\end{equation}

It follows then that the $P.B.$'s among the N\"{o}ther's constants of the motion
are:
\begin{equation}
\{\chi_{i},\chi_{j}\}=-\frac{q}{c}\varepsilon_{ijk}B_{k}%
\end{equation}

\bigskip

Following then the standard rules for the implementation of \ symmetries at
the quantum level, we associate with a (finite) translation by a vector
$\mathbf{a}$ in the plane the \textit{magnetic translation operator
\ }$\widehat{T}\left(  \mathbf{a}\right)  $ \ \cite{Fl,Mor,Zak1,Zak2} defined
by:%
\begin{equation}
\widehat{T}\left(  \mathbf{a}\right)  =\exp\left\{  \frac{i}{\hbar}%
\widehat{\mathbf{\chi}}_{op}\cdot\mathbf{a}\right\}
\end{equation}
where ($\mathbf{p}=\pi-q\mathbf{A}/c$):%
\begin{equation}
\widehat{\mathbf{\chi}}_{op}=\widehat{\pi}-\frac{q}{c}\left(  \mathbf{B}%
\times\mathbf{r}-\mathbf{A}\right)  ;\text{ \ }\widehat{\pi}=\frac{\hbar}%
{i}\triangledown
\end{equation}
with the commutation relations:%
\begin{equation}
\left[  \widehat{\chi}_{i},\widehat{\chi}_{j}\right]  =i\hbar\frac{q}%
{c}\varepsilon_{ijk}B_{k}%
\end{equation}
Of course, $\widehat{\mathbf{\chi}}_{op}$ commutes with the Hamiltonian, and
so does $\widehat{T}\left(  \mathbf{a}\right)  $.

Using then the identities:%
\begin{eqnarray}
&&\exp\left\{  A+B\right\}  =\exp\left\{  A\right\}  \exp\left\{  B\right\}
\exp\left(  -\frac{1}{2}\left[  A,B\right]  \right)\\
&& \exp\left\{  A\right\}  \exp\left\{  B\right\}  =\exp\left\{  B\right\}
\exp\left\{  A\right\}  \exp\left(  \left[  A,B\right]  \right)
\end{eqnarray}
valid whenever: $\left[  A,\left[  A,B\right]  \right]  =\left[  B,\left[
A,B\right]  \right]  =0$, and noting that:%
\begin{equation}
\left[  \pi_{i},\left(  \mathbf{B}\times\mathbf{r}-\mathbf{A}\right)
_{j}\right]  =i\hbar\frac{\partial A_{j}}{\partial x_{i}}%
\end{equation}
one finds for the action of $\widehat{T}\left(  \mathbf{a}\right)  $ on
wavefunctions:%
\begin{equation}
\left(  \widehat{T}\left(  \mathbf{a}\right)  \psi\right)  \left(
\mathbf{r}\right)  =\exp\left\{  -i\frac{q}{\hbar c}\mathbf{a}\cdot
\mathbf{A}\right\}  \psi\left(  \mathbf{r}+\mathbf{a}\right)
\end{equation}
in the symmetric gauge, and, e.g.:%
\begin{equation}
\left(  \widehat{T}\left(  \mathbf{a}\right)  \psi\right)  \left(
\mathbf{r}\right)  =\exp\left\{  -i\frac{q}{\hbar c}Ba_{1}\left(
y-a_{2}/2\right)  \right\}  \psi\left(  \mathbf{r}+\mathbf{a}\right)
\end{equation}
in the Landau gauge: $\mathbf{A}=B\left(  0,x,0\right)  $. Then it is easy to
prove that:
\begin{equation}
\widehat{T}\left(  \mathbf{a}\right)  \widehat{T}\left(  \mathbf{b}\right)
=\exp\left\{  \frac{iq}{2\hbar c}\mathbf{B}\cdot\mathbf{a}\times
\mathbf{b}\right\}  \widehat{T}\left(  \mathbf{a}+\mathbf{b}\right)
\end{equation}
and:%
\begin{equation}
\widehat{T}\left(  \mathbf{a}\right)  \widehat{T}\left(  \mathbf{b}\right)
=\widehat{T}\left(  \mathbf{b}\right)  \widehat{T}\left(  \mathbf{a}\right)
\exp\left\{  \frac{iq}{\hbar c}\mathbf{B}\cdot\mathbf{a}\times\mathbf{b}%
\right\}
\end{equation}

But:
\begin{equation}
\frac{q}{c}\mathbf{B}\cdot\mathbf{a}\times\mathbf{b}=\omega_{\mathcal{L}%
}\left(  \mathbf{a},\mathbf{b}\right)
\end{equation}
and hence:
\begin{equation}
\widehat{T}\left(  \mathbf{a}\right)  \widehat{T}\left(  \mathbf{b}\right)
=\exp\left\{  \frac{i}{2\hbar}\omega_{\mathcal{L}}\left(  \mathbf{a}%
,\mathbf{b}\right)  \right\}  \widehat{T}\left(  \mathbf{a}+\mathbf{b}\right)
\end{equation}
The magnetic translation operators are therefore an instance \cite{Zak3} of a
Weyl system on the configuration space.

\newpage

%% file: Chapt6rev.tex
\section{Quantum Mechanics in Phase Space}\label{ch:QMPS}

\subsection{The Weyl and Wigner Maps}\label{Weyl-Wig}
\bigskip

We will work in $S\approx\mathbb{R}^{2}$ for simplicity. Generalizations to
higher dimensions are easy to work out.

As a preliminary remark, let's observe that we have the identity
($f\in\mathcal{L}_{2}\left(  \mathbb{R}^{2}\right)  $):%
\begin{equation}%
{\displaystyle\iiiint}
\frac{d\xi d\eta dq^{\prime}dp^{\prime}}{\left(  2\pi\hbar\right)  ^{2}%
}f\left(  q^{\prime},p^{\prime}\right)  e^{-i\omega_{0}\left(  \left(
q^{\prime},p^{\prime}\right)  ,\left(  \xi,\eta\right)  \right)  /\hbar
}e^{i\left(  \xi p+\eta q\right)  /\hbar}\equiv f\left(  q,-p\right)
\label{id1}%
\end{equation}
This can also be rewritten as:
\begin{equation}%
{\displaystyle\iint}
\frac{d\xi d\eta}{2\pi\hbar}\left[  \frac{1}{\hbar}\mathcal{F}_{s}\left(
f\right)  \left(  \frac{\eta}{\hbar},\frac{\xi}{\hbar}\right)  \right]
e^{i\left(  \xi p+\eta q\right)  /\hbar}=f\left(  q,-p\right)  \label{id2}%
\end{equation}
where $\mathcal{F}_{s}\left(  f\right)  $ is the symplectic Fourier
transform\footnote{See Appendix $C$. The fact that we get a change in sign in
the second variable is precisely a byproduct of the use of the symplectic
Fourier transform. Had we used instead the ordinary Fourier transform we would
have obtained of course $f\left(  q,p\right)  $ instead of $f\left(
q,-p\right)  $ on the r.h.s. of (\ref{id2}).} \cite{Fo,Zam}:%
\begin{equation}
\mathcal{F}_{s}\left(  f\right)  \left(  \eta,\xi\right)  =%
{\displaystyle\iint}
\frac{dqdp}{2\pi}f\left(  q,p\right)  e^{-i\omega_{0}\left(  \left(
q,p\right)  ,\left(  \xi,\eta\right)  \right)  }%
\end{equation}
and, as usual: $\omega_{0}\left(  \left(  q,p\right)  ,\left(  \xi
,\eta\right)  \right)  =q\eta-p\xi$.

\bigskip

\textbf{Digression.}

Allowing also for distribution-valued transforms, we have, in
particular:%
\begin{equation}
\mathcal{F}_{s}\left(  q\right)  \left(  \eta,\xi\right)  =2\pi i\delta
^{\prime}\left(  \eta\right)  \delta\left(  \xi\right)  \label{sympq}%
\end{equation}
and:%
\begin{equation}
\mathcal{F}_{s}\left(  p\right)  \left(  \eta,\xi\right)  =-2\pi
i\delta\left(  \eta\right)  \delta^{\prime}\left(  \xi\right)  \label{sympp}%
\end{equation}

\bigskip

The \textit{Weyl map} \cite{We} amounts to the replacement, in Eq.(\ref{id2}):%
\begin{equation}
\exp\left\{  i\left(  \xi p+\eta q\right)  /\hbar\right\}  \rightarrow
\exp\left\{  i\left(  \xi\widehat{P}+\eta\widehat{Q}\right)  /\hbar\right\}
\equiv\widehat{W}\left(  \xi,\eta\right)
\end{equation}
whereby one obtains the map:%
\begin{equation}
\Omega:\mathcal{F}\left(  \mathbb{R}^{2}\right)  \rightarrow\mathcal{O}%
p\left(  \mathcal{H}\right)
\end{equation}
defined by:%
\begin{eqnarray}
\Omega\left(  f\right)  &=&
{\displaystyle\iint}
\frac{d\xi d\eta}{2\pi\hbar}\left[  \frac{1}{\hbar}\mathcal{F}_{s}\left(
f\right)  \left(  \frac{\eta}{\hbar},\frac{\xi}{\hbar}\right)  \right]
\widehat{W}\left(  \xi,\eta\right)  = \label{operator} \\
&\equiv&
{\displaystyle\iint}
\frac{d\xi d\eta}{2\pi}\mathcal{F}_{s}\left(  f\right)  \left(  \eta
,\xi\right)  \widehat{W}\left(  \hbar\xi,\hbar\eta\right) \nonumber
\end{eqnarray}

\bigskip It is simple to show that, if $f$ \ is real, then:%
\begin{equation}
\overline{\mathcal{F}_{s}\left(  f\right)  \left(  \eta,\xi\right)
}=\mathcal{F}_{s}\left(  f\right)  \left(  -\eta,-\xi\right)
\end{equation}
and this proves that $\Omega\left(  f\right)  $ is (at least) a
\textit{symmetric} \cite{RS} operator (more on this later on). Using then:%
\begin{equation}
\left(  \widehat{W}\left(  \xi,\eta\right)  \psi\right)  \left(  x\right)
=\exp\left\{  i\eta\left[  x+\xi/2\right]  /\hbar\right\}  \psi\left(
x+\xi\right)
\end{equation}
we obtain:%
\begin{equation}
\left(  \Omega\left(  f\right)  \psi\right)  \left(  x\right)  =%
{\displaystyle\iint}
\frac{d\xi d\eta}{2\pi}\mathcal{F}_{s}\left(  f\right)  \left(  \eta
,\xi\right)  \exp\left[  i\eta\left(  x+\hbar\xi/2\right)  \right]
\psi\left(  x+\hbar\xi\right)  \label{Omegaf}%
\end{equation}
In particular, using (\ref{sympq}) and (\ref{sympp}):
\begin{equation}
\left(  \Omega\left(  q\right)  \psi\right)  \left(  x\right)  =x\psi\left(
x\right)  ,\text{ \ \ }\left(  \Omega\left(  p\right)  \psi\right)  \left(
x\right)  =i\hbar\frac{d\psi}{dx}%
\end{equation}
In other words:%
\begin{equation}
\Omega\left(  q\right)  =\widehat{Q} \label{omegaq}%
\end{equation}
while (cfr. the discussion in the previous footnote):%
\begin{equation}
\Omega\left(  p\right)  =-\widehat{P} \label{omegap}%
\end{equation}

More generally, for arbitrary integers $n$ and $m$:%
\begin{equation}
\mathcal{F}_{s}\left(  q^{n}p^{m}\right)  \left(  \eta,\xi\right)
=2\pi\left(  -\right)  ^{m}i^{n+m}\delta^{\left(  n\right)  }\left(
\eta\right)  \delta^{\left(  m\right)  }\left(  \xi\right)
\end{equation}
which implies:%
\begin{equation}
\left(  \Omega\left(  q^{n}p^{m}\right)  \psi\right)  \left(  x\right)
=\left(  i\frac{d}{d\xi}\right)  ^{m}\left[  \left(  x+\hbar\xi/2\right)
^{n}\psi\left(  x+\hbar\xi\right)  \right]  |_{\xi=0}%
\end{equation}
which can be rearranged \cite{Zam} in the form:%
\begin{equation}
\left(  \Omega\left(  q^{n}p^{m}\right)  \psi\right)  \left(  x\right)
=\frac{1}{2^{n}}%
{\displaystyle\sum\limits_{k=0}^{n}}
\binom{n}{k}x^{k}\left(  i\hbar\frac{d}{dx}\right)  ^{m}\left[  x^{n-k}%
\psi\left(  x\right)  \right]
\end{equation}
Hence:%
\begin{equation}
\Omega\left(  q^{n}p^{m}\right)  =\frac{1}{2^{n}}%
{\displaystyle\sum\limits_{k=0}^{n}}
\binom{n}{k}\left[  \Omega\left(  q\right)  \right]  ^{k}\cdot\left[
\Omega\left(  p\right)  \right]  ^{m}\cdot\left[  \Omega\left(  q\right)
\right]  ^{n-k} \label{productnm}%
\end{equation}
In particular, for $n=m=1$:%
\begin{equation}
\Omega\left(  qp\right)  =\frac{1}{2}\left(  \Omega\left(  q\right)
\cdot\Omega\left(  p\right)  +\Omega\left(  p\right)  \cdot\Omega\left(
q\right)  \right)  \label{prod}%
\end{equation}
Notice that:%
\begin{equation}
\Omega\left(  qp\right)  =\Omega\left(  pq\right)
\end{equation}
but:%
\begin{equation}
\Omega\left(  qp\right)  \neq\Omega\left(  q\right)  \cdot\Omega\left(
p\right)
\end{equation}
Also, as can be shown on examples, in general:%
\begin{equation}
\Omega\left(  fg\right)  \neq\frac{1}{2}\left(  \Omega\left(  f\right)
\cdot\Omega\left(  g\right)  +\Omega\left(  g\right)  \cdot\Omega\left(
f\right)  \right)  \label{symm}%
\end{equation}
as can be seen already from Eq.(\ref{productnm}) when $m$ and/or $n\neq1$, i.e.
the "Weyl symmetrization procedure" (\ref{symm}) \cite{We} holds only in very
special cases.

\bigskip Using Eq.(\ref{Omegaf}) we obtain, for the matrix elements of the
Weyl operator $\Omega\left(  f\right)  $:%
\begin{equation}
\left\langle \phi|\Omega\left(  f\right)  |\psi\right\rangle =\int\frac{dxd\xi
d\eta}{2\pi}\mathcal{F}_{s}\left(  f\right)  \left(  \eta,\xi\right)
e^{i\eta\left(  x+\hbar\xi/2\right)  }\overline{\phi\left(  x\right)  }%
\psi\left(  x+\hbar\xi\right)
\end{equation}
In particular, in a plane-wave basis ($\psi\left(  x\right)  =\left(
1/\sqrt{2\pi}\right)  \exp\left\{  ikx\right\}  $ etc.):%
\begin{equation}
\left\langle k^{\prime}|\Omega\left(  f\right)  |k\right\rangle =\int
\frac{d\xi}{2\pi}\mathcal{F}_{s}\left(  f\right)  \left(  k^{\prime}%
-k,\xi\right)  \exp\left\{  i\hbar\xi\left(  k+k^{\prime}\right)  /2\right\}
\end{equation}
or:%
\begin{equation}
\left\langle K+k/2|\Omega\left(  f\right)  |K-k/2\right\rangle =\int\frac
{d\xi}{2\pi}\mathcal{F}_{s}\left(  f\right)  \left(  k,\xi\right)
\exp\left\{  i\hbar\xi K/2\right\}
\end{equation}
Inserting then the explicit form of the symplectic Fourier transform we find
eventually:%
\begin{equation}
\left\langle K+k/2|\Omega\left(  f\right)  |K-k/2\right\rangle =\int\frac
{dq}{2\pi}f\left(  q,-\hbar K\right)  \exp\left\{  -ikq\right\}
\end{equation}
For example, $f\left(  q,p\right)  =p$ yields:%
\begin{equation}
\left\langle K+k/2|\Omega\left(  p\right)  |K-k/2\right\rangle =-\hbar
K\delta\left(  k\right)
\end{equation}
which (cfr. Eq.(\ref{omegap})) is the correct result.

\bigskip

The Weyl map can be inverted, i.e there exists a map, called the
\textit{Wigner map}:%
\begin{equation}
\Omega^{-1}:\mathcal{O}p\left(  \mathcal{H}\right)  \rightarrow\mathcal{F}%
\left(  \mathbb{R}^{2}\right)
\end{equation}
such that:%
\begin{equation}
\Omega^{-1}\left(  \Omega\left(  f\right)  \right)  =f \label{inverse}%
\end{equation}

In general, given any operator $\widehat{O}$ such that $Tr\left[  \widehat
{O}\widehat{W}\left(  x,k\right)  \right]  $ exists\footnote{As $W$ is a
bounded operator, this will be granted, e.g., if $A$ is trace-class.}, the
Wigner map is defined as:%
\begin{equation}
\Omega^{-1}\left(  \widehat{O}\right)  \left(  q,p\right)  =:%
{\displaystyle\iint}
\frac{dxdk}{2\pi\hbar}\exp\left\{  -i\omega_{0}\left(  \left(  x,k\right)
,\left(  q,p\right)  \right)  /\hbar\right\}  Tr\left[  \widehat{O}\widehat
{W}^{\dag}\left(  x,k\right)  \right]  \label{inverse2}%
\end{equation}
In order to prove Eq.(\ref{inverse}), we need the trace:%
\begin{equation}
Tr[\widehat{W}\left(  x,k\right)  \widehat{W}^{\dag}\left(  \xi,\eta\right)
]=\int dhdh^{\prime}\left\langle h\left\vert \widehat{W}\left(  x,k\right)
\right\vert h^{\prime}\right\rangle \left\langle h^{\prime}\left\vert
\widehat{W}^{\dag}\mathcal{(\xi},\eta)\right\vert h\right\rangle
\end{equation}

Using Eq. (\ref{matelement}) we obtain:%
\begin{equation}
Tr\left[  \widehat{W}\left(  x,k\right)  \widehat{W}^{\dag}\left(  \xi
,\eta\right)  \right]  =2\pi\hbar\delta\left(  x-\xi\right)  \delta\left(
k-\eta\right)  \label{trace}%
\end{equation}
Inserting then (\ref{trace}) into (\ref{inverse2}) and using this result, we
obtain:
\begin{equation}
\Omega^{-1}\left(  \Omega\left(  f\right)  \right)  \left(  q,p\right)
=\int\frac{d\xi d\eta}{2\pi}\mathcal{F}_{s}\left(  \eta,\xi\right)
\exp\left\{  -i\omega\left(  \left(  \xi,\eta\right)  ,\left(  q,p\right)
\right)  \right\}  =f\left(  q,p\right)
\end{equation}
$\blacksquare$

\subsection{A Digression on: Phase-Point Operators}
\bigskip

Going back to Eq.(\ref{operator}), which reads:}%
\begin{equation}
\Omega\left(  f\right)  =%
{\displaystyle\iint}
\frac{d\xi d\eta}{(2\pi\hbar)^{2}}\widehat{W}\left(
\xi,\eta\right)  \iint dqdpe^{-i\left(  q\eta-p\xi\right)
/\hbar}f\left(  q,p\right)
\end{equation}
and, to the extent that it is legitimate to apply Fubini's
theorem, we
obtain:%
\begin{equation}
\Omega\left(  f\right)  =\iint\frac{dqdp}{2\pi\hbar}f\left(
q,p\right) \widehat{A}\left(  q,p\right)\label{phase1}
\end{equation}
where the symplectic Fourier transform of
$\widehat{W}\left(
\xi,\eta\right)  $\textit{, i.e.:}%
\begin{equation}
\widehat{A}\left(  q,p\right)  =:%
{\displaystyle\iint}
\frac{d\xi d\eta}{2\pi\hbar}e^{-i\left(  q\eta-p\xi\right)
/\hbar}\widehat {W}\left(  \xi,\eta\right)
\end{equation}
defines the so-called "phase-point operators"
\cite{CEMMMS1,CEMMMS2,EMMM,Leo,Woo1,Woo2}. It is not hard to
prove, using: $\widehat{W}^{\dag}\left(  \xi,\eta\right)
=\widehat{W}\left(  -\xi ,-\eta\right)  $ and
Eqns.(\ref{wtrace}) and (\ref{trace})\ that:

\begin{itemize}
\item The phase-point operators are of unit trace:%
\begin{equation}
Tr\widehat{A}\left(  q,p\right)  =1
\end{equation}

\item They are Hermitian: %
\begin{equation}
\widehat{A}^{\dag}\left(  q,p\right)  =\widehat{A}\left(
q,p\right)
\end{equation}
and:

\item They are trace-orthogonal, i.e.:%
\begin{equation}\label{phase3}
Tr\left(  \widehat{A}\left(  q,p\right)  \widehat{A}\left(
q^{\prime },p^{\prime}\right)  \right)  =2\pi\hbar\delta\left(
q-q^{\prime}\right) \delta\left(  p-p^{\prime}\right)
\end{equation}

\item Moreover, a simple calculation shows that:%
\begin{equation}
\iint\frac{dqdp}{2\pi\hbar}\widehat{A}\left(  q,p\right)
=\widehat{W}\left(
0,0\right)  =\widehat{\mathbb{I}}%
\end{equation}
with $\widehat{\mathbb{I}}$ the identity
operator.\
\end{itemize}

All this proves that the phase-point operators are a \textit{complete (trace)
orthonormal set of Hermitian operators.} In particular, substituting the
Wigner function $\Omega^{-1}\left(  \widehat{O}\right)  $ for the function $f$
in Eq.(\ref{phase1}) and as: $\Omega\left(  \Omega^{-1}\left(  \widehat
{O}\right)  \right)  =\widehat{O}$, we obtain at once the reconstruction:%
\begin{equation}
\widehat{O}=\iint\frac{dqdp}{2\pi\hbar}\Omega^{-1}\left(  \widehat{O}\right)
\left(  q,p\right)  \widehat{A}\left(  q,p\right)
\end{equation}
in terms of the Wigner function and the phase-point operators, as well as,
using Eq.(\ref{phase3}):%
\begin{equation}
\Omega^{-1}\left(  \widehat{O}\right)  \left(  q,p\right)  =Tr\left\{
\widehat{O}\widehat{A}\left(  q,p\right)  \right\}
\end{equation}

An explicit representation of phase-point operators
satisfying all of
the above properties is:%
\begin{equation}
\widehat{A}\left(  q,p\right)  =\int
dq^{\prime}|q+q^{\prime}/2\rangle \exp\left(
iq^{\prime}p/\hbar\right)  \langle q-q^{\prime}/2|
\end{equation}
with matrix elements:%
\begin{equation}
\left\langle x\left\vert \widehat{A}\left(  q,p\right)
\right\vert x^{\prime }\right\rangle =2\delta\left(
x+x^{\prime}-2q\right)  \exp\left\{  ip\left( x-x^{\prime}\right)
/\hbar\right\}
\end{equation}

\subsection{More on the Wigner Map}
\bigskip

It is useful to have an expression for the Wigner map directly in terms of the
matrix elements of the operators. Working again for simplicity in
$\mathbb{R}^{2}$ and introducing resolutions of the identity in terms of plane
waves: $\left\{  \left.  |m\right\rangle \right\}  $\footnote{$\left\langle
x|m\right\rangle =\frac{1}{\sqrt{2\pi}}\exp\left\{  imx\right\}  $, and:$%
{\displaystyle\int}
\left.  dm|m\right\rangle \left\langle m|\right.  =\mathbb{I}$.}:%
\begin{equation}
\Omega^{-1}\left(  \widehat{O}\right)  \left(  q,p\right)  =%
{\displaystyle\int}
\frac{dxd\pi dldm}{2\pi\hbar}\exp\left\{  -i\left(  xp-\pi q\right)
/\hbar\right\}  \left\langle l|\widehat{O}|m\right\rangle \left\langle
m|\widehat{W}^{\dag}\left(  x,\pi\right)  |l\right\rangle
\end{equation}
or ($\pi=\hbar k$):%
\begin{equation}
\Omega^{-1}\left(  \widehat{O}\right)  \left(  q,p\right)  =%
{\displaystyle\int}
\frac{dxdkdldm}{2\pi}\exp\left\{  -i\left(  xp/\hbar-kq\right)  \right\}
\left\langle l|\widehat{O}|m\right\rangle \left\langle m|\widehat{W}^{\dag
}\left(  x,\hbar k\right)  |l\right\rangle
\end{equation}
and using:%
\begin{equation}
\left\langle m|\widehat{W}^{\dag}\left(  x,\hbar k\right)  |l\right\rangle
=\exp\left\{  -ix\left(  m+l\right)  /2\right\}  \delta\left(  l-m-k\right)
\end{equation}
one finds eventually:%
\begin{equation}
\Omega^{-1}\left(  \widehat{O}\right)  \left(  q,p\right)  =\int
dke^{iqk}\left\langle -p/\hbar+k/2|\widehat{O}|-p/\hbar-k/2\right\rangle
\label{Wigner}%
\end{equation}
with obvious generalizations to higher dimensions. As an example, if:
$\widehat{A}=-\widehat{P}$, then, as: $\widehat{P}\left.  |m\right\rangle
=\hbar m\left.  |m\right\rangle $:%
\begin{eqnarray}
\left\langle -p/\hbar+k/2|(-\widehat{P})|-p/\hbar-k/2\right\rangle &=&\left(
p\hbar+k/2\right)  \left\langle -p/\hbar+k/2|-p/\hbar-k/2\right\rangle  \nonumber \\
&\equiv&
p\delta\left(  k\right)
\end{eqnarray}
and we find:%
\begin{equation}
\Omega^{-1}\left(  (-\widehat{P})\right)  \left(  q,p\right)  =p
\end{equation}
as expected.

Also, it is easy to prove that:%
\begin{equation}
\Omega^{-1}\left(  \widehat{W}\left(  q^{\prime},,p^{\prime}\right)  \right)
\left(  q,p\right)  =\exp\left\{  i\omega_{0}\left(  \left(  q,p\right)
,\left(  q^{\prime},p^{\prime}\right)  \right)  /\hbar\right\}
\end{equation}

Introducing now resolutions of the identity relative to the coordinates:
\begin{equation}
\Omega^{-1}\left(  \widehat{A}\right)  \left(  q,p\right)  =\int
dkdxdx^{\prime}e^{iqk}\left\langle -p/\hbar+k/2|\left.  x\right\rangle
\left\langle x\right.  |\widehat{A}|\left.  x^{\prime}\right\rangle
\left\langle x^{\prime}\right.  |-p/\hbar-k/2\right\rangle
\end{equation}
the integration over $k$ yields a delta-function, and we obtain, eventually,
the celebrated \textit{Wigner formula }\cite{Leo,Wig4}, or \textit{Wigner
transform }:%
\begin{equation}
\Omega^{-1}\left(  \widehat{O}\right)  \left(  q,p\right)  =\int d\xi
e^{ip\xi/\hbar}\left\langle q+\xi/2|\widehat{O}\left\vert q-\xi/2\right.
\right\rangle \label{transform}%
\end{equation}
Here too, setting: $\widehat{A}=\widehat{Q}$, we find at once: $\Omega
^{-1}\left(  \widehat{Q}\right)  \left(  q,p\right)  =q$, as expected. As
another example, consider, e.g.: $\widehat{A}=|\phi\rangle\langle\psi|$ (which
is a prototype of a finite-rank operator). Then it is immediate to see that:%
\begin{equation}
\Omega^{-1}\left(  |\phi\rangle\langle\psi|\right)  \left(  q,p\right)  =%
{\displaystyle\int\limits_{-\infty}^{\infty}}
d\xi e^{ip\xi/\hbar}\phi\left(  q+\xi/2\right)  \overline{\psi\left(
q-\xi/2\right)  } \label{transform2}%
\end{equation}


\begin{remark}

\textit{From Eq.(\ref{transform2})}\bigskip\ \textit{we obtain}:%
\begin{equation}
\left\vert \Omega^{-1}\left(  |\phi\rangle\langle\psi|\right)  \left(
q,p\right)  \right\vert \leq2%
{\displaystyle\int\limits_{-\infty}^{\infty}}
d\eta\left\vert \phi\left(  q+\eta\right)  \right\vert \left\vert \psi\left(
q-\eta\right)  \right\vert
\end{equation}
\textit{and, using Schwartz's inequality:}%
\begin{equation}
\left\vert \Omega^{-1}\left(  |\phi\rangle\langle\psi|\right)  \left(
q,p\right)  \right\vert \leq2\left\Vert \phi\right\Vert \left\Vert
\psi\right\Vert
\end{equation}
\textit{In particular, if }$\left\vert \psi\rangle=\right.  \left\vert
\phi\rangle\right.  $ \textit{and:} $\left\langle \phi|\phi\right\rangle =1$,
\textit{i.e. for a one-dimensional projector: }$P_{\phi}=|\phi\rangle
\langle\phi|$\textit{:}%
\begin{equation}
\left\vert \Omega^{-1}\left(  P_{\phi}\right)  \left(  q,p\right)  \right\vert
\leq2 \label{uniform1}%
\end{equation}
\textit{As every density matrix can be written as a convex linear combination
of one-dimensional projectors, we obtain eventually the uniform
bound\footnote{Note that we are using here a slightly different
normalization than that used in Ref.\cite{EMMM}.} \cite{EMMM}:}%
\begin{equation}
\left\vert \Omega^{-1}\left(  \widehat{\rho}\right)  \left(  q,p\right)
\right\vert \leq2 \label{uniform2}%
\end{equation}
\textit{if }$\widehat{\rho}$ \ \textit{is} \textit{a density matrix.}

\textit{Proceeding in a somewhat heuristic manner, let now }$\widehat{O}$
\textit{be a self-adjoint operator with a completely discrete spectrum:
}$\widehat{O}|\phi_{n}\rangle=\lambda_{n}|\phi_{n}\rangle,$ $\left\langle
\phi_{n}|\phi_{m}\right\rangle =\delta_{nm}$ \textit{and: }$\sum_{n}|\phi
_{n}\rangle\langle\phi_{n}|=\mathbb{I}$. \textit{Then:}%
\begin{equation}
\Omega^{-1}\left(  \widehat{O}\right)  \left(  q,p\right)  =\sum
\limits_{n}\lambda_{n}\int d\xi e^{ip\xi/\hbar}\phi_{n}\left(  q+\xi/2\right)
\overline{\phi_{n}\left(  q+\xi/2\right)  }%
\end{equation}
\textit{and hence, proceeding as before:}%
\begin{equation}
\left\vert \Omega^{-1}\left(  \widehat{O}\right)  \left(  q,p\right)
\right\vert \leq2\sum\limits_{n}\left\vert \lambda_{n}\right\vert
=2Tr\left\vert \widehat{O}\right\vert
\end{equation}
\textit{where \cite{RS}: }$\left\vert \widehat{O}\right\vert =:\sqrt
{\widehat{O}^{\dag}\widehat{O}}$. \textit{Trace-class operators\footnote{A
class of operators comprising, in particular, finite-rank projection operators
as well as density states.} are defined \cite{RS} by requiring finiteness of
}$Tr\left\vert \widehat{O}\right\vert $. \textit{Therefore:}

\textit{The Wigner function of any trace-class operator}
$\widehat{O}$ \textit{will be uniformly bounded by }$2Tr\left\vert \widehat
{O}\right\vert$ . 
\end{remark}


It is easy to check that the Wigner transform inverts to:%
\begin{equation}
\left\langle x|\widehat{O}|x^{\prime}\right\rangle =\int\frac{dp}{2\pi\hbar
}\exp\left\{  -ip\left(  x-x^{\prime}\right)  /\hbar\right\}  \Omega
^{-1}\left(  \widehat{O}\right)  \left(  \frac{x+x^{\prime}}{2},p\right)
\label{pretr}%
\end{equation}
\bigskip

As an example, let's consider the Wigner transform of $\widehat{O}%
=|\phi\rangle\langle\psi|$ as given by Eq.(\ref{transform2}). Then it is
immediate to check that, indeed:%
\begin{equation}%
{\displaystyle\int}
\frac{dp}{2\pi\hbar}e^{\{-ip\left(  x-x^{\prime}\right)  /\hbar\}}\Omega
^{-1}\left(  \widehat{O}\right)  \left(  \frac{x+x^{\prime}}{2},p\right)
=\phi\left(  x\right)  \overline{\psi\left(  x^{\prime}\right)  }=\left\langle
x|\phi\rangle\langle\psi|x^{\prime}\right\rangle
\end{equation}

\begin{example}

\textit{As a less simple example as compared to the previous ones, let us
consider a }$1D$\textit{ harmonic oscillator of mass }$m$\textit{ and proper
frequency }$\omega$\textit{. The corresponding Hamiltonian is: }%
\begin{equation}
\widehat{H}=\frac{\widehat{P}^{2}}{2m}+\frac{1}{2}m\omega^{2}\widehat{Q}^{2}%
\end{equation}
\textit{with eigenvalues: }$E_{n}=\left(  n+1/2\right)  \hbar\omega$\textit{ ,
}$n\geq0$\textit{ and eigenfunctions: }%
\begin{equation}
\psi_{n}\left(  x\right)  =\sqrt[4]{\frac{m\omega}{\pi\hbar}}\frac{1}%
{\sqrt{2^{n}n!}}\exp\left(  -\zeta^{2}/2\right)  H_{n}\left(  \zeta\right)
\label{Herm}%
\end{equation}
\textit{where }$\zeta$\textit{ is the dimensionless variable: }$\zeta
=x\sqrt{m\omega/\hbar}$\textit{ and the }$H_{n}$\textit{'s are the Hermite
polynomials \cite{EM}. We want to evaluate here the Wigner function (the
Wigner map) associated with the "Boltzmann factor" }$\widehat{A}=\exp\left(
-\beta\widehat{H}\right)  $\textit{, with }$\beta$\textit{ the inverse
temperature . Of course:}%
\begin{equation}
\left\langle x\left\vert e^{-\beta\widehat{H}}\right\vert x^{\prime
}\right\rangle =%
{\displaystyle\sum\limits_{n=0}^{\infty}}
e^{-\beta E_{n}}\psi_{n}\left(  x\right)  \psi_{n}\left(  x^{\prime}\right)
\label{har1}%
\end{equation}
\textit{Inserting the explicit form (\ref{Herm}) of the eigenfunctions:}%
\begin{equation}
\left\langle x\left\vert e^{-\beta\widehat{H}}\right\vert x^{\prime
}\right\rangle =\sqrt{\frac{m\omega z}{\pi\hbar}}%
{\displaystyle\sum\limits_{n=0}^{\infty}}
\frac{z^{n}}{2^{n}n!}e^{-\left(  \zeta^{2}+\zeta^{\prime2}\right)  /2}%
H_{n}\left(  \zeta\right)  H_{n}\left(  \zeta^{\prime}\right)  ,\text{
\ }z=\exp\left(  -\beta\hbar\omega\right)
\end{equation}
\textit{Now, it turns out that\footnote{This is known also as \textit{Mehler's
formula.}} \cite{Fo,LM}: }%
\begin{equation}%
{\displaystyle\sum\limits_{n=0}^{\infty}}
\frac{z^{n}}{2^{n}n!}H_{n}\left(  \zeta\right)  H_{n}\left(  \zeta^{\prime
}\right)  =\frac{1}{\sqrt{1-z^{2}}}\exp\left\{  \frac{2z\zeta\zeta^{\prime
}-z^{2}\left(  \zeta^{2}+\zeta^{\prime2}\right)  }{1-z^{2}}\right\}  ,\text{
\ }\left\vert z\right\vert <1
\end{equation}
\textit{and therefore the matrix element (\ref{har1})can be expressed in
closed form as: }%
\begin{equation}
\left\langle x\left\vert e^{-\beta\widehat{H}}\right\vert x^{\prime
}\right\rangle =\sqrt{\frac{m\omega}{\pi\hbar}}e^{-\left(  \zeta^{2}%
+\zeta^{\prime2}\right)  /2}\sqrt{\frac{z}{1-z^{2}}}\exp[\frac{2z\zeta
\zeta^{\prime}-z^{2}\left(  \zeta^{2}+\zeta^{\prime2}\right)  }{1-z^{2}}]
\end{equation}
\textit{Setting then: }$x=q+\xi/2,x^{\prime}=q-\xi/2$\textit{ and inserting
the result into Eq.(\ref{transform}) one finds eventually the Wigner
function:}%
\begin{equation}
\Omega^{-1}\left(  e^{-\beta\widehat{H}}\right)  \left(  q,p\right)  =\frac
{1}{\cosh\left(  \beta\hslash\omega/2\right)  }\exp\left\{  -\tanh\left(
\beta\hslash\omega/2\right)  \left[  \frac{m\omega}{\hslash}q^{2}+\frac{p^{2}%
}{m\hslash\omega}\right]  \right\}  \label{har2}%
\end{equation}
\end{example}

Coming back now to the main object of this Section, an interesting consequence
of Eq.(\ref{Wigner}) is the following. Let's calculate the $\mathcal{L}^{2}$
norm of $\Omega^{-1}\left(  \widehat{A}\right)  \left(  q,p\right)  $, i.e.:%
\begin{equation}
\left\Vert \Omega^{-1}\left(  \widehat{A}\right)  \right\Vert ^{2}=\int
\frac{dqdp}{2\pi\hbar}\left\vert \Omega^{-1}\left(  \widehat{A}\right)
\left(  q,p\right)  \right\vert ^{2}%
\end{equation}
Explicitly:%
\begin{eqnarray}
&&\left\Vert \Omega^{-1}\left(  \widehat{A}\right)  \right\Vert ^{2}= \\
&&=\int\frac{dqdp}{2\pi\hbar}dkdk^{\prime}e^{i\left(  k^{\prime}-k\right)
q}\left\langle p/\hbar-k/2|\widehat{A}^{\dag}|p/\hbar+k/2\right\rangle
\left\langle p/\hbar+k^{\prime}/2|\widehat{A}|p/\hbar-k^{\prime}%
/2\right\rangle \nonumber
\end{eqnarray}
Performing the integration over $q$ , which produces a delta-function, and
shifting variables: $p\rightarrow p+\hbar k/2$:%
\begin{equation}
\left\Vert \Omega^{-1}\left(  \widehat{A}\right)  \right\Vert ^{2}=\int
d\left(  p/\hbar\right)  dk\left\langle p/\hbar|\widehat{A}^{\dag}%
|p/\hbar+k\right\rangle \left\langle p/\hbar+k|\widehat{A}|p/\hbar
\right\rangle
\end{equation}
The integration over $k$ yields a resolution of the identity, and we end up
with:%
\begin{equation}
\left\Vert \Omega^{-1}\left(  \widehat{A}\right)  \right\Vert ^{2}=\int
d\left(  p/\hbar\right)  \left\langle p/\hbar|\widehat{A}^{\dag}\widehat
{A}|p/\hbar\right\rangle
\end{equation}
i.e., eventually:%
\begin{equation}
\left\Vert \Omega^{-1}\left(  \widehat{A}\right)  \right\Vert ^{2}=Tr\left\{
\widehat{A}^{\dag}\widehat{A}\right\}  \label{norm1}%
\end{equation}
and, if: $\widehat{A}=\Omega\left(  f\right)  $:%
\begin{equation}
\left\Vert f\right\Vert ^{2}=Tr\left\{  \Omega\left(  f\right)  ^{\dag}%
\Omega\left(  f\right)  \right\}  \label{norm2}%
\end{equation}
The condition of finiteness (positivity is obvious) of $Tr\left\{  A^{\dag
}A\right\}  $ characterizes $A$ as a \textit{Hilbert-Schmidt } \cite{RS}
\ operator. Therefore \cite{Poo}:
\begin{theorem}
$f$ \textit{\ will be square-integrable (}$f\in\mathcal{L}^{2}\left(
\mathbb{R}^{2}\right)  $\textit{) if and only if }$\Omega\left(  f\right)  $
\textit{is Hilbert-Schmidt}$.$\textit{Quite similarly}: \textit{ }$\Omega
^{-1}\left(  \widehat{A}\right)  $ \textit{will be square-integrable if and
only if }$\widehat{A}$ \textit{is Hilbert-Schmidt.}
\end{theorem}

\bigskip

The Weyl and Wigner maps establish therefore a \textit{bijection}
\cite{Fo,Goe} between Hilbert-Schmidt operators and square-integrable
functions on phase space. This is consistent with the fact that both spaces
are Hilbert spaces. Moreover, Eqs.(\ref{norm1}) and (\ref{norm2}) prove that
the bijection, being an isometry, is also (strongly) bicontinuous.

The fact that: $\overline{\mathcal{F}_{s}\left(  \eta,\xi\right)
}=\mathcal{F}_{s}\left(  -\eta,-\xi\right)  $ as well as that: $\widehat
{W}^{\dag}\left(  \xi,\eta\right)  =\allowbreak\widehat{W}\left(  -\xi
,-\eta\right)  $ allows also to prove at once that the Weyl and Wigner maps
\textit{"preserve conjugation"}, i.e. that:%
\begin{equation}
\Omega\left(  \overline{f}\right)  =\Omega\left(  f\right)  ^{\dag}%
\end{equation}
as well as:%
\begin{equation}
\Omega^{-1}\left(  \widehat{O}^{\dag}\right)  =\overline{\Omega^{-1}\left(
\widehat{O}\right)  }%
\end{equation}
Therefore, in particular, if $f$ is real, then, as already mentioned,
$\Omega\left(  f\right)  $ will be a \textit{symmetric} operator.

\bigskip

\bigskip As a final remark, we observe that Eq. (\ref{pretr}) implies also:%

\begin{equation}
Tr_{x}\left(  \widehat{O}\right)  =:%
{\displaystyle\int}
dx\left\langle x\left\vert O\right\vert x\right\rangle =%
{\displaystyle\int}
\frac{dqdp}{2\pi\hbar}\Omega^{-1}\left(  \widehat{O}\right)  \left(
q,p\right)  \label{tr1}%
\end{equation}
(with the same result for the similarly defined $Tr_{p}\left(  \widehat
{A}\right)  $) as well as, of course:%
\begin{equation}%
{\displaystyle\int}
\frac{dqdp}{2\pi\hbar}f\left(  q,p\right)  =Tr\left(  \Omega\left(  f\right)
\right)  \label{tr2}%
\end{equation}
and this defines formally a \textit{"trace" operation} on phase space:
\begin{equation}
Tr\left(  f\right)  =:\int\frac{dqdp}{2\pi\hbar}f\left(  q,p\right)
\label{tr3}%
\end{equation}
Of course, all these results will make sense when all the quantities in the
previous equations are finite. For example, if: $\widehat{A}=P_{\psi
}=\left\vert \psi\rangle\langle\psi\right\vert ,\left\langle \psi
|\psi\right\rangle =1$ is a one-dimensional projector, then:%
\begin{equation}
\Omega^{-1}\left(  P_{\psi}\right)  \left(  q,p\right)  =\int d\xi
e^{ip\xi/\hbar}\left\langle q+\xi/2|\psi\right\rangle \left\langle \psi
|q-\xi/2\right\rangle
\end{equation}
and:%
\begin{equation}
\int\frac{dqdp}{2\pi\hbar}\Omega^{-1}\left(  P_{\psi}\right)  \left(
q,p\right)  =\int dq\left\langle q|\psi\right\rangle \left\langle
\psi|q\right\rangle =\left\Vert \psi\right\Vert ^{2}=1
\end{equation}

As a less trivial example, in the case of the harmonic oscillator we find with
some long but elementary algebra using Eq.(\ref{har2}):%
\begin{equation}
Tr\left\{  \Omega^{-1}\left(  e^{-\beta\widehat{H}}\right)  \right\}
=\int\frac{dqdp}{2\pi\hbar}\Omega^{-1}\left(  e^{-\beta\widehat{H}}\right)
=\frac{1}{2\sinh\left(  \beta\hbar\omega/2\right)  }%
\end{equation}
which is the expected result \cite{Mes} for the canonical partition function
of a $1D$ harmonic oscillator.

\begin{remark}

\textit{The mere existence of the phase-space trace of }$\Omega^{-1}\left(
\widehat{O}\right)  $\textit{, i.e. finiteness of }$\int\left(  dqdp/2\pi
\hbar\right)  \Omega^{-1}\left(  \widehat{O}\right)  \left(  q,p\right)
$\textit{ does not however guarantee that }$\widehat{A}$\textit{ \ be
trace-class, as this requires, as already recalled \cite{RS}, the more
stringent condition that :}%

\begin{equation}
Tr\left(  \left\vert \widehat{O}\right\vert \right)  <\infty,\text{
\ }\left\vert \widehat{O}\right\vert =:\sqrt{O^{\dag}O}%
\end{equation}
\textit{and }$\left\vert \widehat{O}\right\vert $\textit{ is not connected to
the Wigner function }$\Omega^{-1}\left(  \widehat{O}\right)  $\textit{ in any
simple manner.}

\end{remark}

\subsection{The Moyal Product \label{Moy1}}
\bigskip

Working again for simplicity\footnote{We stress once again that extensions to
higher dimensions are essentially straightforward.} in $\mathcal{S}%
\approx\mathbb{R}^{2}$, the Wigner map allows for the definition of a new
algebra structure on the space of functions $\mathcal{F}\left(  \mathbb{R}%
^{2}\right)  $, the \textit{Moyal }$"\ast"$-\textit{product}
\cite{Gro,Moy,Wig4}, that is defined as:%
\begin{equation}
f\ast g=:\Omega^{-1}\left(  \widehat{\Omega}\left(  f\right)  \cdot
\widehat{\Omega}\left(  g\right)  \right)  \label{Moyal1}%
\end{equation}
(as, generically: \ \ $\widehat{\Omega}\left(  f\right)  \cdot\widehat{\Omega
}\left(  g\right)  \neq\widehat{\Omega}\left(  g\right)  \cdot\widehat{\Omega
}\left(  f\right)  $, it is clear that, again generically: $f\ast g\neq g\ast
f$).

This product is \textit{associative}\footnote{$f\ast\left(  g\ast h\right)
=\left(  f\ast g\right)  \ast h$} (as the algebra of operators is), it is
\textit{distributive} w.r.t. the sum\footnote{$f\ast\left(  g+h\right)  =f\ast
g+f\ast h$} (as $\widehat{\Omega}\left(  .\right)  $ is linear), but it is
\textit{non-local} and \textit{non-commutative. }Indeed:%
\begin{equation}
\left(  f\ast g\right)  \left(  q,p\right)  =%
{\displaystyle\iint}
\frac{dxdk}{2\pi\hbar}\exp\left\{  -i\omega_{0}\left(  \left(  x,k\right)
,\left(  q,p\right)  \right)  /\hbar\right\}  Tr\left[  \widehat{\Omega
}\left(  f\right)  \cdot\widehat{\Omega}\left(  g\right)  \widehat{W}^{\dag
}\left(  x,k\right)  \right]  \label{Moyal2}%
\end{equation}
and:%
\begin{equation}%
\begin{array}
[c]{c}%
Tr\left[  \widehat{\Omega}\left(  f\right)  \cdot\widehat{\Omega}\left(
g\right)  \widehat{W}^{\dag}\left(  x,k\right)  \right]  =\\
=%
{\displaystyle\int}
\frac{d\xi d\eta d\xi^{\prime}d\eta^{\prime}}{\left(  2\pi\right)  ^{2}%
}\mathcal{F}_{s}\left(  f\right)  \left(  \eta,\xi\right)  \mathcal{F}%
_{s}\left(  g\right)  \left(  \eta^{\prime},\xi^{\prime}\right)  Tr\left[
\widehat{W}\left(  \hbar\xi,\hbar\eta\right)  \widehat{W}\left(  \hbar
\xi^{\prime},\hbar\eta^{\prime}\right)  \widehat{W}^{\dag}\left(  x,k\right)
\right]
\end{array}
\end{equation}
Now:%
\begin{equation}%
\begin{array}
[c]{c}%
Tr\left[  \widehat{W}\left(  \alpha,\beta\right)  \widehat{W}\left(
\sigma,\tau\right)  \widehat{W}^{\dag}\left(  x,k\right)  \right]  =\\
=2\pi\delta\left(  \alpha+\sigma-x\right)  \delta\left(  \beta+\tau-k\right)
\exp\left\{  -i\left[  \beta\left(  \alpha+\sigma\right)  +k\left(
\sigma-x\right)  \right]  /2\hbar\right\}
\end{array}
\end{equation}
Hence:%
\begin{equation}%
\begin{array}
[c]{c}%
Tr\left[  \widehat{\Omega}\left(  f\right)  \cdot\widehat{\Omega}\left(
g\right)  \widehat{W}^{\dag}\left(  x,k\right)  \right]  =\\%
{\displaystyle\int}
\frac{d\xi d\xi^{\prime}d\eta d\eta^{\prime}}{2\pi\hbar}\mathcal{F}_{s}\left(
f\right)  \left(  \eta,\xi\right)  \mathcal{F}_{s}\left(  g\right)  \left(
\eta^{\prime},\xi^{\prime}\right)  e^{-i\left(  \eta x-k\xi\right)  /2}%
\delta\left(  \xi+\xi^{\prime}-x/\hbar\right)  \delta\left(  \eta+\eta
^{\prime}-k/\hbar\right)
\end{array}
\end{equation}
and, using the deltas to get rid of the $\xi^{\prime},\eta^{\prime}$
integrations and the explicit form of the symplectic Fourier transforms:%
\begin{equation}%
\begin{array}
[c]{c}%
Tr\left[  \widehat{\Omega}\left(  f\right)  \cdot\widehat{\Omega}\left(
g\right)  \widehat{W}^{\dag}\left(  x,k\right)  \right]  =\\
4%
{\displaystyle\int}
\frac{dadbdsdt}{2\pi\hbar}f\left(  a,b\right)  g\left(  s,t\right)
e^{-i\left(  sk-tx\right)  /\hbar}\delta\left(  k-2\left(  t-b\right)
\right)  \delta\left(  x-2\left(  s-a\right)  \right)
\end{array}
\end{equation}
Inserting this result into Eq.(\ref{Moyal2}) we eventually obtain:%
\begin{equation}
\left(  f\ast g\right)  \left(  q,p\right)  =4%
{\displaystyle\int}
\frac{dadbdsdt}{(2\pi\hbar)^{2}}f\left(  a,b\right)  g\left(  s,t\right)
\exp\left\{  -\frac{2i}{\hbar}\left[  \left(  a-q\right)  \left(  t-p\right)
+\left(  s-q\right)  \left(  p-b\right)  \right]  \right\}
\end{equation}
or:%
\begin{equation}
\left(  f\ast g\right)  \left(  q,p\right)  =4%
{\displaystyle\int}
\frac{dadbdsdt}{\left(  2\pi\hbar\right)  ^{2}}f\left(  a,b\right)  g\left(
s,t\right)  \exp\left\{  2i\omega_{0}\left(  \left(  q-a,p-b\right)  ,\left(
q-s,p-t\right)  \right)  /\hbar\right\}
\end{equation}

and this exhibits explicitly the non-locality of the Moyal product.

It can be shown\footnote{See,e.g., Ref.\cite{Zam} for details.} that:

\begin{itemize}
\item The Moyal product can be recast in the form:%
\begin{equation}
\left(  f\ast g\right)  \left(  q,p\right)  =%
{\displaystyle\sum\limits_{n,m=0}^{\infty}}
\left(  \frac{i\hbar}{2}\right)  ^{n+m}\frac{\left(  -1\right)  ^{n}}%
{n!m!}\left\{  \frac{\partial^{m+n}f\left(  a,b\right)  }{\partial
a^{m}\partial b^{n}}\frac{\partial^{m+n}g\left(  a,b\right)  }{\partial
a^{n}\partial b^{m}}\right\}  |_{a=q,b=p} \label{Moyal3}%
\end{equation}

and that:

\item Eq.(\ref{Moyal3}) can be rewritten in compact form as:%
\begin{equation}
\left(  f\ast g\right)  \left(  q,p\right)  =f\left(  q,p\right)  \exp\left\{
\frac{i\hbar}{2}\left[  \frac{\overleftarrow{\partial}}{\partial q}%
\frac{\overrightarrow{\partial}}{\partial p}-\frac{\overleftarrow{\partial}%
}{\partial p}\frac{\overrightarrow{\partial}}{\partial q}\right]  \right\}
g\left(  q,p\right)  \label{Moyal4}%
\end{equation}

\end{itemize}

Other equivalent forms of the Moyal product are:%
\begin{equation}
\left(  f\ast g\right)  \left(  q,p\right)  =f\left(  q+\frac{i\hbar}{2}%
\frac{\overrightarrow{\partial}}{\partial p},p-\frac{i\hbar}{2}\frac
{\overrightarrow{\partial}}{\partial q}\right)  g\left(  q,p\right)
\label{Moyal5}%
\end{equation}
or:%
\begin{equation}
\left(  f\ast g\right)  \left(  q,p\right)  =f\left(  q,p\right)  g\left(
q-\frac{i\hbar}{2}\frac{\overleftarrow{\partial}}{\partial p},p+\frac{i\hbar
}{2}\frac{\overleftarrow{\partial}}{\partial q}\right)
\end{equation}

\bigskip

\begin{remark}
All the above expressions for the Moyal product apply
of course to functions that are regular enough for the right-hand side of the
defining equations to make sense. In particular, they will hold when 
$f,g$ are "Schwartzian" functions \cite{Ric} in $S\left(
\mathbb{R}^{2}\right)  $, i.e. they are of class $C^{\infty}$
and of fast decrease at infinity.
\end{remark}

\bigskip

The form (\ref{Moyal3}) exhibits explicitly the Moyal product as a series
expansion in powers of $\hbar$. To lowest order:%
\begin{equation}
f\ast g=fg+\frac{i\hbar}{2}\left\{  f,g\right\}  +\mathcal{O}\left(  \hbar
^{2}\right)  \label{classic1}%
\end{equation}
where $\left\{  .,.\right\}  $ is the Poisson bracket. The Planck constant
$\hbar$ acts then as a "deformation parameter" of the usual associative
product structure on the algebra of functions, making the product
non-commutative. Indeed, it can be seen, e.g., from the expansion of the
exponential in Eq.(\ref{Moyal4}), that terms proportional to even powers of
$\hbar$ are symmetric under the interchange $f\leftrightarrow g$, but terms
proportional to odd powers are \textit{anti}symmetric, and this makes the
product non-commutative.

\bigskip

\begin{example} ~

\begin{itemize}
\item $f\equiv1$\textit{ or }$g\equiv1$\textit{. Then:}%
\begin{equation}
\left(  1\ast g\right)  \left(  q,p\right)  =g\left(  q,p\right)  ,\left(
f\ast1\right)  \left(  q,p\right)  =f\left(  q,p\right)
\end{equation}

\item $f=q$\textit{. Then, at least if }$g\in S^{\infty}\left(  \mathbb{R}%
^{2}\right)  $\textit{:}%
\begin{equation}%
\begin{array}
[c]{c}%
\left(  q\ast g\right)  \left(  q,p\right)  =4%
{\displaystyle\int}
\frac{dadbdsdt}{(2\pi\hbar)^{2}}ag\left(  s,t\right)  \exp\left\{  \frac
{2i}{\hbar}\left[  \left(  a-q\right)  \left(  t-p\right)  +\left(
s-q\right)  \left(  p-b\right)  \right]  \right\}  =\\
= 4%
{\displaystyle\int}
\frac{dadbdsdt}{(2\pi\hbar)^{2}}g\left(  s,t\right)  \left(  q+\frac{i\hbar
}{2}\frac{\partial}{\partial t}\right)  \exp\left\{  \frac{2i}{\hbar}\left[
\left(  a-q\right)  \left(  t-p\right)  +\left(  s-q\right)  \left(
p-b\right)  \right]  \right\}
\end{array}
\end{equation}
\textit{and, integrating by parts in the second integral and using the
previous result:}%
\begin{equation}
\left(  q\ast g\right)  \left(  q,p\right)  =\left(  q+\frac{i\hbar}{2}%
\frac{\partial}{\partial p}\right)  g\left(  q,p\right)  \label{qugi}%
\end{equation}
\textit{Then, in view of the symmetry properties of the various terms in the
expansion of the Moyal product in powers of }$\hbar$\textit{:}%
\begin{equation}
\left(  g\ast q\right)  \left(  q,p\right)  =\left(  q-\frac{i\hbar}{2}%
\frac{\partial}{\partial p}\right)  g\left(  q,p\right)
\end{equation}

\item \textit{In the same way, if }$f=p$\textit{:}%
\begin{equation}
\left(  p\ast g\right)  \left(  q,p\right)  =\left(  p-\frac{i\hbar}{2}%
\frac{\partial}{\partial q}\right)  g\left(  q,p\right)
\end{equation}
\textit{etc.}

\item \textit{If }$f=q,g=p$\textit{ (or viceversa), then, using, e.g., Eq.
(\ref{Moyal5}):}%
\begin{equation}
\left(  q\ast p\right)  \left(  q,p\right)  =qp+\frac{i\hbar}{2};\left(  p\ast
q\right)  \left(  q,p\right)  =qp-\frac{i\hbar}{2}%
\end{equation}

\end{itemize}
\textit{Notice that Eq.(\ref{qugi}) implies:}%
\begin{equation}
\widehat{\Omega}\left(  q\right)  \cdot\widehat{\Omega}\left(  g\right)
=\widehat{\Omega}\left(  qg\right)  +\frac{i\hbar}{2}\widehat{\Omega}\left(
\frac{\partial g}{\partial p}\right)  \label{qugi2}%
\end{equation}
\textit{and similarly for the others.}
\end{example}

The generalization of these results, as well as of those of the following
Subsections, to higher dimensions, i.e. to: $\mathcal{S}=\mathbb{R}^{2n}$ with
$n>1$, are straightforward, so we will omit details here.

\subsection{The Moyal Bracket(s), "Moyal" Quantum Mechanics and the
Quantum-Classical Transition\label{Moy3}}

\bigskip

\subsubsection{The Moyal Bracket\label{Moy4}}
\bigskip

Using \ the Moyal product we can define the \textit{Moyal Bracket} $\left\{
.,.\right\}  _{M}$ as:
\begin{equation}
\left\{  .,.\right\}  _{M}:\mathcal{F}\left(  \mathbb{R}^{2}\right)
\times\mathcal{F}\left(  \mathbb{R}^{2}\right)  \rightarrow\mathcal{F}\left(
\mathbb{R}^{2}\right)  ;\text{ }\left\{  f,g\right\}  _{M}=:\frac{1}{i\hbar
}\left(  f\ast g-g\ast f\right)  \label{MB1}%
\end{equation}
Hence, in particular:%
\begin{equation}
\left\{  f,g\right\}  _{M}=\left\{  f,g\right\}  +\mathcal{O}\left(  \hbar
^{2}\right)  \label{MB2}%
\end{equation}
where $\left\{  .,.\right\}  $ is the standard Poisson bracket\footnote{The
difference between the Moyal and Poisson brackets is $\mathcal{O(}\hbar^{2})$,
and not $\mathcal{O}\left(  \hbar\right)  $ as one could expect, and that
because the difference $f\ast g-g\ast f$ contains only \textit{odd} powers of
$\hbar$.}.

Being defined in terms of an associative product, the Moyal bracket fulfills
all the properties of a Poisson bracket (linearity, anti-symmetry and the
Jacobi identity), and defines a new Poisson structure on the (non-commutative)
algebra of functions with the Moyal product. In particular, just as for the
ordinary Poisson brackets, the Jacobi identity implies:%
\begin{equation}
\left\{  f,g\ast h\right\}  _{M}=\left\{  f,g\right\}  _{M}\ast h+g\ast
\left\{  f,h\right\}  _{M}%
\end{equation}
i.e. that \ $\left\{  f,.\right\}  $ is a \textit{derivation} (with respect to
the $\ast$-product) on the algebra of functions. Writing down explicitly the
second term in $\ $(\ref{MB2}): $\left\{  f,g\right\}  _{M}=\left\{
f,g\right\}  +\hbar^{2}\left\{  f,g\right\}  _{2}+...$, we obtain:%
\begin{equation}
\left\{  f,g\right\}  _{2}\left(  q,p\right)  =\frac{1}{24}\left\{
\frac{\partial^{3}f}{\partial q^{3}}\frac{\partial^{3}g}{\partial p^{3}%
}-3\frac{\partial^{3}f}{\partial p\partial q^{2}}\frac{\partial^{3}g}{\partial
q\partial p^{2}}+3\frac{\partial^{3}f}{\partial p^{2}\partial q}\frac
{\partial^{3}g}{\partial q\partial q^{2}}-\frac{\partial^{3}f}{\partial p^{3}%
}\frac{\partial^{3}g}{\partial q^{3}}\right\}
\end{equation}
Therefore, $\left\{  f,g\right\}  _{M}$ \ \ contains, besides first-order
derivatives, third and higher-order derivatives, and, although it is a
derivation on the algebra of functions with the $"\ast"$ product, it is
\textit{not} a vector field (while $\left\{  f,.\right\}  $ \textit{is} a
vector field). The reason for that is precisely that the Moyal bracket is
non-local, and hence Willmore's theorem \cite{Wil} connecting (inner)
derivations with vector fields does not apply. It is only when $f$ is
\textit{at most a quadratic polynomial} that $\left\{  f,.\right\}  _{M}$
becomes a derivation on the usual pointwise product. Indeed, if this is the
case, the Moyal and Poisson brackets of $f$ with other functions coincide. As
a check, we see that, in simple cases, we obtain:%
\begin{equation}
\left\{  q,p\right\}  _{M}=1,\text{ }\left\{  q,g\right\}  _{M}=\frac{\partial
g}{\partial p},\text{ }\left\{  p,g\right\}  _{M}=-\frac{\partial g}{\partial
q} \label{MB3}%
\end{equation}

Using the definitions of the Weyl and Wigner maps we have, in general:%
\begin{equation}
\left\{  f,g\right\}  _{M}=i\Omega^{-1}\left(  \widehat{\Omega}\left(
f\right)  \cdot\widehat{\Omega}\left(  g\right)  -\widehat{\Omega}\left(
g\right)  \cdot\widehat{\Omega}\left(  f\right)  \right)  /\hbar
\end{equation}
i.e.:%
\begin{equation}
\left[  \widehat{\Omega}\left(  f\right)  ,\widehat{\Omega}\left(  g\right)
\right]  =-i\hbar\widehat{\Omega}\left(  \left\{  f,g\right\}  _{M}\right)
\end{equation}

In particular, using (\ref{MB3}) (and: $\widehat{\Omega}\left(  1\right)
=\mathbb{I}$)\footnote{The minus sign in the first commutator stems from the
fact that $\Omega\left(  p\right)  =-\widehat{P}$, i.e. ultimately from the
fact that we are using the symplectic and not the ordinary Fourier
transform.}:%
\begin{align}
& \left[  \widehat{\Omega}\left(  q\right)  ,\widehat{\Omega}\left(  p\right)
\right]     =-i\hbar\mathbb{I},\text{ }\left[  \widehat{\Omega}\left(
q\right)  ,\widehat{\Omega}\left(  \mathcal{H}\right)  \right]  =-i\hbar
\widehat{\Omega}\left(  \partial\mathcal{H}/\partial p\right) \\
& \left[  \widehat{\Omega}\left(  p\right)  ,\widehat{\Omega}\left(
\mathcal{H}\right)  \right]     =i\hbar\widehat{\Omega}\left(  \partial
\mathcal{H}/\partial q\right)
\end{align}

\bigskip

Unless $f$ and/or $g$ are \textit{at most} quadratic, $\left\{  f,g\right\}
_{M}\neq\left\{  f,g\right\}  $. Therefore, the commutator of the quantum
operators associated with observables on phases space is \textit{not}
(\textit{"modulo"} a multiplicative constant) the quantum operator associated
with the Poisson bracket \cite{Dir3}. Generically, it becomes so only to
lowest order in $\hbar$, and reproduces the Ehrenfest theorem \cite{EM}.

\subsubsection{Quantum Mechanics in Phase Space\label{Moy5}}
\bigskip

First of all, it is of some interest, in view of the relevant r\^{o}le they
play in Quantum Mechanics, to see here which phase-space functions correspond
to projection operators on the Hilbert space. The latter, that we will denote as $\widehat{\mathcal{P}}$, are completely
characterized by:

\begin{itemize}
\item
\begin{equation}
\widehat{\mathcal{P}}^{2}=\widehat{\mathcal{P}},\text{ \ }idempotency \label{proj1}%
\end{equation}

\item
\begin{equation}
\widehat{\mathcal{P}}^{\dag}=\widehat{\mathcal{P}},\text{ \ }self-adjointness \label{proj2}%
\end{equation}

\end{itemize}

As to (\ref{proj2}), this requires the associated Wigner function $\Omega
^{-1}\left(  \widehat{\mathcal{P}}\right)  $ to be \textit{real}. As to (\ref{proj1}),
this implies, in terms of the Moyal product (cfr. Eq. (\ref{Moyal1})):%
\begin{equation}
\Omega^{-1}\left(  \widehat{\mathcal{P}}^{2}\right)  =\Omega^{-1}\left(  \widehat
{\mathcal{P}}\right)  =\Omega^{-1}\left(  \widehat{\mathcal{P}}\right)  \ast\Omega^{-1}\left(
\widehat{\mathcal{P}}\right)
\end{equation}
Moreover:
\begin{equation}
Tr\left(  \Omega^{-1}\left(  \widehat{\mathcal{P}}\right)  \right)  =Tr(\widehat{\mathcal{P}})
\end{equation}
and: $Tr\left(  \Omega^{-1}\left(  \widehat{\mathcal{P}}\right)  \right)  $ will be
finite iff $\widehat{\mathcal{P}}$ is a finite-rank projection operator.

Therefore:

\textit{\underline{\textit{Projection operators}} are represented in phase
space by real, uniformly-bounded (cfr. Eq.(\ref{uniform2})) functions
satisfying:}%
\begin{equation}
f\ast f=f
\end{equation}
\textit{and}:
\begin{equation}
Tr\left(  f\right)  <+\infty
\end{equation}
\textit{iff \ the associated projector is of finite rank. Density
states will be represented in turn by real, again
uniformly-bounded, phase-space functions }$f\left(  q,p\right)  $
\textit{satisfying: }$Tr\left(  f\right)  =1$ and:%
\begin{equation}
Tr\left(  f\ast f\right)  \leq1
\end{equation}

\bigskip

As discussed in Chapt.\ref{QM},\ Quantum Mechanics can (and should) be
consistently described in the framework of the projective Hilbert space
$P\mathcal{H}$. Once this is identified (via the Hermitian structure, see the
discussion in Chapt.\ref{QM}) with the space of rank-one projectors, it is
natural to pose eigenvalue problems not for vectors in the Hilbert space but
for the associate rank-one projectors, i.e. in the form:%
\begin{equation}
\widehat{O}\widehat{P}=\lambda\widehat{P};\text{ }\widehat{P}^{\dag}%
=\widehat{P},\text{ }\widehat{P}^{2}=\widehat{P},\text{ }Tr\widehat{P}=1
\label{eigen1}%
\end{equation}
with $\widehat{O}$ an observable and $\lambda\in\mathbb{R}$ the corresponding
eigenvalue\footnote{To avoid unnecessary technical complications, we pose here
the problem in the discrete spectrum. Also, the last condition in
Eq.(\ref{eigen1}) can be relaxed in favor of $P$ becoming then a not
necessarily one-dimensional eigenprojector onto the subspace spanned by the
eigenvalue $\lambda$.}. Put in this form, the eigenvalue problem can be easily
formulated on phase space. Indeed, denoting by simplicity as $f_{\widehat{O}%
}=$ $\Omega^{-1}\left(  \widehat{O}\right)  $ the Wigner function associated
with $\widehat{O}$, the equivalent phase-space formulation will be:%
\begin{equation}
f_{\widehat{O}}\ast f=\lambda f;\text{ }f\ast f=f,\text{ }f\in L_{2}\left(
T^{\ast}Q\right)
\end{equation}
for a real (and uniformly-bounded) function $f$. This will qualify $f$ as the
Wigner function associated with a projection operator: $f=$ $\Omega
^{-1}\left(  \widehat{P}\right)  $, with: $Trf=1$ if it corresponds to a pure state.

A superposition rule capturing also interference phenomena can be formulated
in terms of Wigner functions \cite{MMSZ4,MMSZ6,MMSZ5,MMSZ1,MMSZ2} following
the lines of the discussion of Sect.\ref{introd:geometry}. If we denote as
$f_{0}$ the Wigner function associated with a reference (pure) state (see
Sect.\ref{introd:geometry} for more details) and as $f_{1},f_{2}$ those
associated with two orthogonal (i.e.: $f_{1}\ast f_{2}=0$) pure states, then
to the linear superposition with coefficients $c_{1}$ and $c_{2}$, $\left\vert
c_{1}\right\vert ^{2}+$ $\left\vert c_{2}\right\vert ^{2}=1$, there
corresponds the Wigner function associated with Eq.(\ref{purif2}), namely:%
\begin{equation}
f=\sum\limits_{i,j=1}^{2}c_{i}c_{j}^{\ast}\frac{f_{i}\ast f_{0}\ast f_{j}%
}{\sqrt{Tr\left(  f_{i}\ast f_{0}\ast f_{j}\ast f_{0}\right)  }}%
\end{equation}
where the phase-space trace has been defined in Eq.(\ref{tr3}).

Coming now to quantum evolution, an observable (a self-adjoint operator)
$\widehat{O}$ \ will evolve in time as:%
\begin{equation}
\widehat{O}\left(  t\right)  =\widehat{U}^{\dag}\left(  t\right)
\cdot\widehat{O}\cdot\widehat{U}\left(  t\right)
\end{equation}
where the evolution operator is given by:%
\begin{equation}
\widehat{U}\left(  t\right)  =\exp\left(  -it\widehat{H}/\hbar\right)
\label{evol1}%
\end{equation}
$\widehat{H}$ \ being the Hamiltonian operator. Denoting again the Wigner
function associated with $\widehat{O}$ as $f_{\widehat{O}}$, and from the very
definition of the Moyal product:%
\begin{equation}
f_{\widehat{O}\left(  t\right)  }=f_{\widehat{U}^{\dag}\left(  t\right)
\cdot\widehat{O}\cdot\widehat{U}\left(  t\right)  }=f_{\widehat{U}^{\dag
}\left(  t\right)  }\ast f_{\widehat{O}}\ast f_{\widehat{U}\left(  t\right)  }%
\end{equation}

Using the (formal) series expansion of the evolution operator (\ref{evol1}) we
can also write explicitly the evolution operator in phase space $f_{\widehat
{U}\left(  t\right)  }$ as \cite{Bas1}:%
\begin{equation}
f_{\widehat{U}\left(  t\right)  }=\exp_{\ast}\left(  -itf_{\widehat{H}}%
/\hbar\right)  =:%
{\displaystyle\sum\limits_{n=0}^{\infty}}
\frac{\left(  -it/\hbar\right)  ^{n}}{n!}\left(  f_{\widehat{H}}\right)
_{\ast}^{n} \label{evol2}%
\end{equation}
where $\left(  .\right)  _{\ast}^{n}$ stands for an $n$-fold star-product.

Now, to lowest order in $t$: $f_{\widehat{U}\left(  t\right)  }\approx
1-\left(  it/\hbar\right)  f_{\widehat{H}}$ etc., and we obtain easily:%

\begin{equation}
\frac{d}{dt}f_{\widehat{O}\left(  t\right)  }=\left\{  f_{\widehat{O}\left(
t\right)  },f_{\widehat{H}}\right\}  _{M} \label{Q-dyn}%
\end{equation}
or, more generally:%
\begin{equation}
\frac{d}{dt}f\left(  t\right)  =\left\{  f\left(  t\right)  ,f_{\widehat{H}%
}\right\}  _{M};\text{ \ }f\left(  0\right)  =f \label{QM2}%
\end{equation}
with $f$ \ any suitable function (e.g., a square-integrable function) on phase
space, leading to:%
\begin{equation}
f\left(  t\right)  =\exp_{\ast}\left(  itf_{\widehat{H}}/\hbar\right)  \ast
f\ast\exp_{\ast}\left(  -itf_{\widehat{H}}/\hbar\right)  \label{QM3}%
\end{equation}
and this is the phase-space description of quantum dynamics. As the classical
($\hbar\rightarrow0$) limit of the Moyal bracket is the Poisson bracket,
Eqs.(\ref{Q-dyn}) and/or (\ref{QM2}) reduce, in the classical limit, to the
description of the dynamics in terms of Poisson brackets.

\subsection{"Alternative" Quantum Mechanics and Their Classical
Counterparts\label{Moy6}}
\bigskip

We can begin by recalling a theorem due to Dirac (see \cite{Dir3} and
\cite{GM} for a more general discussion) which states that, given an
associative, non-Abelian and maximally non-commutative\footnote{That is, such
that \cite{GM} the derived algebra: $\mathcal{A}^{\prime}=Span\{\left[
a,b\right]  \};a,b\in\mathcal{A}$, together with the identity, spans the whole
of $\mathcal{A}$.} algebra $\mathcal{A}$ with identity over $\mathbb{R}$ or
$\mathbb{C}$, and defining a "Poisson bracket"\footnote{Having in mind the
algebra of operators on a Hilbert space, Dirac \cite{Dir3} calls it a "Quantum
Poisson bracket".} on $\mathcal{A}$ as a map:%
\begin{equation}
\left\{  .,.\right\}  :\mathcal{A}\times\mathcal{A}\longrightarrow\mathcal{A}%
\end{equation}
that is bilinear, antisymmetric, satisfies the Jacobi identity:%
\begin{equation}
\left\{  a,\left\{  b,c\right\}  \right\}  +\left\{  b,\left\{  c,a\right\}
\right\}  +\left\{  c,\left\{  a,b\right\}  \right\}  =0\text{ }\forall
a,b,c\in\mathcal{A}%
\end{equation}
and acts as a derivation on the product on the algebra, i.e.:%
\begin{equation}
\left\{  a,bc\right\}  =\left\{  a,b\right\}  c+b\left\{  a,c\right\}  \text{
}\forall a,b,c\in\mathcal{A}%
\end{equation}
then\footnote{See, e.g., Ref.\cite{Zam} for details of the proof.} the Poisson
bracket $\left\{  a,b\right\}  $ is necessarily proportional to the "standard"
commutator $ab-ba$.

This theorem was actually one of the main motivations why, in
Chapt.\ref{introduc}, we discussed alternative approaches to Quantum Mechanics
involving modifications of the Hermitian product or, equivalently, of the
associative product between operators.

Sticking to this last approach,\ we consider now a "deformed" associative
product between operators defined as:%
\begin{equation}
\widehat{A}\underset{(\widehat{K})}{\mathbf{\cdot}}\widehat{B}=:\widehat
{A}\cdot\widehat{K}\cdot\widehat{B} \label{defprod1}%
\end{equation}
where $\widehat{A},\widehat{B}$ are linear operators and $\widehat{K}$ is a
fixed, positive operator which is also a constant of the motion. This leads to
the definition of the "deformed" commutator:%
\begin{equation}
\left[  \widehat{A},\widehat{B}\right]  _{(\widehat{K})}=:\widehat{A}%
\underset{(\widehat{K})}{\mathbf{\cdot}}\widehat{B}-\widehat{B}\underset
{(\widehat{K})}{\mathbf{\cdot}}\widehat{A} \label{defcomm}%
\end{equation}
which satisfies again the Jacobi identity\footnote{See Sect.\ref{Alt1} for
further details.}.

Given then two phase-space functions $f$ and $g$, Eq.(\ref{defprod1}) leads to
the "deformed" Moyal product:%
\begin{equation}
f\underset{(k)}{\ast}g=f\ast k\ast g \label{defprod2}%
\end{equation}
where:%
\begin{equation}
k=:\Omega^{-1}\left(  \widehat{K}\right)
\end{equation}
is the Wigner function associated with the operator $\widehat{K}$, and to the
"deformed" Moyal bracket:%
\begin{equation}
\left\{  f,g\right\}  _{M,k}=:\frac{1}{i\hbar}(f\underset{(k)}{\ast
}g-g\underset{(k)}{\ast}f)\equiv\frac{1}{i\hbar}(f\ast k\ast g-g\ast k\ast f)
\label{defMoy}%
\end{equation}
and, of course:%
\begin{equation}
\left\{  f,g\right\}  _{M,1}\equiv\left\{  f,g\right\}  _{M}%
\end{equation}

\begin{remark}
 \textit{Requiring the operator }$\widehat{K}$ \textit{to be
strictly}\textbf{ }\textit{positive is a necessary condition\footnote{See
again Sect.\ref{Alt1}.} for the definition of a sensible "deformed" Hermitian
product on the Hilbert space. If this is the case, then \ }$\widehat{K}%
$\textit{ \ is invertible and the new (associative) algebra structure defined
by Eq.(\ref{defprod2}) will have an identity }$e$\textit{, given now by the
"}$\ast$-\textit{inverse" of }$k$\textit{: }$e=k^{\ast-1}$\textit{, where:
}$k^{\ast-1}=:$\textit{ }$\Omega^{-1}\left(  \widehat{K}^{-1}\right)  $
\textit{(i.e.: }$k\ast k^{\ast-1}=k^{\ast-1}\ast k=1$\textit{). This is of
course the counterpart of the fact that the inverse of }$\widehat{K}$,
$\widehat{K}^{-1}$, \textit{plays the r\^{o}le of the identity for the
deformed associative product (\ref{defprod1}) on the algebra of operators.}
\end{remark}

Again with reference to the discussion in Sect.\ref{Alt1}, and in particular
to Eq.(\ref{Heis}), we see that now the dynamics will be described, in phase
space, by the equation:%
\begin{equation}
\frac{d}{dt}f=\left\{  f,f_{\widehat{H}^{\prime}}\right\}  _{M,k}
\label{Q-dyn2}%
\end{equation}
in such a way that (cfr. Eq.(\ref{QM2})):%
\begin{equation}
\left\{  f,f_{\widehat{H}^{\prime}}\right\}  _{M,k}=\left\{  f,f_{\widehat{H}%
}\right\}  _{M}%
\end{equation}
where the new Hamiltonian function will be given by:%
\begin{equation}
f_{\widehat{H}^{\prime}}=\Omega^{-1}\left(  \widehat{H}\cdot\widehat{K}%
^{-1}\right)  =\Omega^{-1}\left(  \widehat{K}^{-1}\cdot\widehat{H}\right)
=f_{\widehat{H}}\ast f_{\widehat{K}^{-1}}%
\end{equation}
Moreover (cfr. Eq.(\ref{Kder})), time evolution will act again as a derivation
on the deformed algebra of functions, i.e.:%
\begin{equation}
\frac{d}{dt}\left(  f\underset{(k)}{\ast}g\right)  =\frac{df}{dt}%
\underset{(k)}{\ast}g+f\underset{\left(  k\right)  }{\ast}\frac{dg}{dt},\text{
\ }\forall f,g
\end{equation}

\bigskip

Turning now to the classical limit and using Eq.(\ref{classic1}), a simple
computation shows that, for $\hbar\rightarrow0$, Eq.(\ref{defprod2}) becomes:%
\begin{equation}
f\underset{(k)}{\ast}g\simeq fkg+\frac{i\hbar}{2}\left\{  f,g\right\}
_{k}+\mathcal{O}\left(  \hbar^{2}\right)
\end{equation}
with a "deformed" bracket is given now by:%
\begin{equation}
\left\{  f,g\right\}  _{k}=\underset{\hbar\rightarrow0}{\lim}\left\{
f,g\right\}  _{M,k}%
\end{equation}
and, explicitly:%
\begin{equation}
\left\{  f,g\right\}  _{k}=k\left\{  f,g\right\}  +f\left\{  k,g\right\}
-g\left\{  k,f\right\}  \label{defbracket}%
\end{equation}
(once again: $\left\{  f,g\right\}  _{1}\equiv\left\{  f,g\right\}  $).\ Being
defined in terms of an associative product, this new bracket\footnote{Also
called \cite{Li} a \textit{Jacobi bracket}.} satisfies the Jacobi identity,
but, at variance with the Poisson bracket and as it is clear from
Eq.(\ref{defbracket}), $\left\{  f,.\right\}  _{k}$ fails to be (for fixed
$f$) a derivation on the algebra of functions (it is not even zero on constant functions).

\subsubsection{Alternative Moyal-like brackets}
\bigskip

In Section we go back to the GNS construction for the finite-dimensional ${\mathbb C}^*$-algebra ${\cal B}( {\mathbb C}^n)$ we have discussed  in \ref{se:GNS}.
Recall that different states over ${\cal B}( {\mathbb C}^n)$ give rise to different representations and hence to different realizations of the corresponding Hilbert space.
We have already noticed that any such state is represented by a positive $n \times n$ matrix $K$ which can be used to define an alternative scalar product on ${\mathbb C}^n$ of the form
\begin{equation}
z \cdot_K w := \sum_{j,k=1}^{n} \bar{z}_j K_{jk} w_k
\end{equation}
for any $z,w \in {\mathbb C}^n$. In turn, we can define  a different multiplication rule in ${\cal B}( {\mathbb C}^n)$ by means of:
\begin{equation}
A \cdot_K B = A \cdot K \cdot B \label{prok}
\end {equation}
for any $A,B \in {\cal B}( {\mathbb C}^n)$. This product is associative, so that $({\cal B}( {\mathbb C}^n), \cdot_K)$ is a ${\mathbb C}^*$-algebra. Accordingly, we can define alternative Lie algebra and Jordan algebra structures via:
\begin{eqnarray}
&& [A,B]_K := \frac{i}{2} (A \cdot_K B - B \cdot_K A ) \\
&& A \circ_K B := \frac{1}{2} (A \cdot_K B + B \cdot_K A )
\end{eqnarray}
Let us consider now a quantum system whose dynamics is specified by a Hamiltonian $H$, yielding the standard Heisenberg equation:
\begin{equation}
i \hbar \dot{A} = [A,H]
\end{equation}
Suppose that $[H,K]=H\cdot K - K \dot H =0$.
By setting $H_K = K^{-1} \cdot H$, one can easily verify that, for any for any $A \in {\cal B}( {\mathbb C}^n)$:
\begin{equation}
[A,H] = A  \cdot_K H_K - H_K \cdot A = [A,H_K]_K
\end{equation}
Hence we have an alternative Hesienberg-like description which makes use of the alternative product (\ref{prok}):
\begin{equation}
i \hbar \dot{A} = [A,H_K]_K
\end{equation}
These alternative structures are therefore analogue to those we have examined in classical dynamics when we have studied bi-Hamiltonian systems.

We can analyze these structures also in terms of the Wigner-Weyl formalism introduced in the previous paragraphs. We already know (see Sect. \ref{se:cps}) that  on the space of K\"{a}hler functions  on the projective space, ${\cal F}^{\mathbb C}( P{\cal H})$, we can define a star-product, that of formula (\ref{spc}), such that:
\begin{equation}
f_A \star f_B = f_{AB}
\end{equation}
We can then define an antisymmetric star-bracket according to:
\begin{equation}
\{ f , g \}_\star := \frac{1}{2i}  (f \star g - g \star f)
\end{equation}
for any $f, g \in {\cal F}^{\mathbb C}( P{\cal H})$, which yields the standard Poisson bracket in the classical limit. Now, it is known \cite{Ru} that any associative local product in  ${\cal F}^{\mathbb C}( P{\cal H})$ is of the form:
\begin{equation}
f\cdot_k g := f\cdot k \cdot g
\end{equation}
for some $k\in {\cal F}^{\mathbb C}( P{\cal H})$, $k>0$. With this product, we can now define an alternative $\star_k$-product and $\star_k$ Lie and Jordan brackets:
\begin{eqnarray}
&& f_A \star_k f_B = f_A \star k \star f_B \\
&& \{f_A,f_B\}_{\star k} = \frac{1}{2i}  (f \star_k g - g \star_k f)\\
&& f_A \circ_k f_B = \frac{1}{2}  (f \star_k g + g \star_k f)
\end{eqnarray}
We are back here to the construction of "deformed" Moyal brackets we have discussed in the previous paragraph. We have already seen that, in the classical limit, we get:
\begin{equation}
\lim_{\hbar \rightarrow 0} \frac{1}{\hbar} \{f_A,f_B\}_{\star_k} = \{f,g\} + f \{k,g\} - g \{k,f\} := \{f,g\}_k
\end{equation}
obtaining the standard Poisson bracket only if $k=1$. In a similar way, we see that:
\begin{equation}
\lim_{\hbar \rightarrow 0} f \circ_k g = f\cdot k \cdot := f \cdot_k g
\end{equation}
This shows that the alternative quantization schemes we have introduced in the previous paragraph depend on the associative products $\star_k$ one can define on the originally commutative algebra $k\in {\cal F}^{\mathbb C}( P{\cal H})$.

\subsubsection{"Conformal" Poisson Tensors Associated with Deformed Moyal Products}
\bigskip

From now on we will consider the case $\mathcal{S}=\mathbb{R}^{2n}$ for
generic $n>1$, the main reason being that most of what will be said becomes
trivial for $n=1$.

As discussed in previous Sections, assigning a Poisson bracket is equivalent
to assigning a \textit{bi-vector field,} i.e. a totally antisymmetric tensor of type
$\left(  2,0\right)  $, the \textit{Poisson tensor}, given, in local
collective coordinates, as:%
\begin{equation}
\Lambda=\frac{1}{2}\Lambda^{ij}\frac{\partial}{\partial\xi^{i}}\wedge
\frac{\partial}{\partial\xi^{j}};\text{ }\Lambda^{ij}+\Lambda^{ji}=0
\end{equation}
and such that:%
\begin{equation}
\left\{  f,g\right\}  =\Lambda\left(  df,dg\right)
\end{equation}

In general, on can define, on multivectors, a bracket, the \textit{Schouten
bracket} \cite{Ni,Schouten}, that associates to every pair $X,Y$ of
multivectors of ranks $n$ and $m$ respectively a multivector $\left[
X,Y\right]  _{S}$ of rank $n+m-1$. Limiting ourselves to bi-vectors, if $X$
and $Y$ are monomials:%
\begin{equation}
X=\chi_{1}\wedge\chi_{2},\text{ \ }Y=\eta_{1}\wedge\eta_{2}%
\end{equation}
(with the $\chi$'s and $\eta$'s vector fields), then:%
\begin{equation}
\left[  X,Y\right]  _{S}=\left[  \chi_{1},\eta_{1}\right]  \wedge\chi
_{2}\wedge\eta_{2}-\left[  \chi_{1},\eta_{2}\right]  \wedge\chi_{2}\wedge
\eta_{1}-\left[  \chi_{2},\eta_{1}\right]  \wedge\chi_{1}\wedge\eta
_{2}+\left[  \chi_{2},\eta_{2}\right]  \wedge\chi_{1}\wedge\eta_{1}%
\end{equation}
It follows that, if $f,g$ are functions:%
\begin{eqnarray}
\left[  fX,gY\right]  _{S}&=&fg\left[  X,Y\right]  _{S}+\\
&+& f\left(  L_{\chi_{2}}g\right)  \chi_{1}\wedge\eta_{1}\wedge\eta_{2}-f\left(
L_{\chi_{1}}g\right)  \chi_{2}\wedge\eta_{1}\wedge\eta_{2}+\nonumber \\
&+& g\left(  L_{\eta_{2}}f\right)  \chi_{1}\wedge\chi_{2}\wedge\eta_{1}-g\left(
L_{\eta_{1}}f\right)  \chi_{1}\wedge\chi_{2}\wedge\eta_{2} \nonumber
\end{eqnarray}
and then the Schouten bracket can be extended by linearity to arbitrary bi-vectors.

The Jacobi identity can be expressed in terms of the Poisson tensor as:%
\begin{equation}
\left[  \Lambda,\Lambda\right]  _{S}=0
\end{equation}
and this is equivalent, whenever the Poisson tensor is not degenerate and
allows then for the definition of a symplectic two-form $\omega$, to the
closure of the latter.

\begin{remark}
As, in dimension two, there are no non-vanishing
tri-vector fields (and all two-forms are closed), it is clear why what we are saying
here becomes essentially void in dimension two. There, every pair of
bi-vector fields has \ a vanishing Schouten bracket.
\end{remark}

The "deformed" bracket (\ref{defbracket}) can be rewritten as:%
\begin{equation}
\left\{  f,g\right\}  _{k}=\Lambda^{\prime}\left(  df,dg\right)  +fL_{X_{k}%
}g-gL_{X_{k}}f \label{defbracket2}%
\end{equation}
where: $X_{k}=:\left\{  k,.\right\}  $ is the Hamiltonian vector field
associated with the function $k$, and:%
\begin{equation}
\Lambda^{\prime}=k\Lambda
\end{equation}
is what is called \cite{Bas1,Bas2} a \textit{conformal Poisson tensor} with
\textit{conformal factor} $k$. Equivalently:%
\begin{equation}
\left\{  f,g\right\}  _{k}=\Lambda^{\prime}\left(  df,dg\right)  +f\left\{
k,g\right\}  -g\left\{  k,f\right\}  \label{defbracket4}%
\end{equation}
Due to the presence of the conformal factor, the Schouten bracket of the
conformal Poisson tensor with itself does not vanish anymore. Instead
\cite{Bas1}:%
\begin{equation}
\left[  \Lambda^{\prime},\Lambda^{\prime}\right]  _{S}=-2X_{k}\wedge
\Lambda^{\prime}%
\end{equation}
and also, as $X_{k}$ is a Hamiltonian vector field:%
\begin{equation}
L_{X_{k}}\Lambda^{\prime}\equiv kL_{X_{k}}\Lambda=0
\end{equation}
\bigskip

\begin{remark}
\textit{The bracket (\ref{defbracket}) is $\mathbb{R}$-linear homogeneous in the
conformal factor $k$. So, any two such brackets with conformal factors, say,
$k_{1}$ and $k_{2}$, will give rise to a bracket of the same form ( a
"compatible" bracket, in this sense) with conformal factor: $k=k_{1}+k_{2}$.
This seems to imply that, in order to obtain non-compatible classical limits,
one should introduce some amount of non-linearity. This can be done by using
\ non-linearly related Poisson structures.}
\end{remark}

\begin{remark}
\textit{Extrapolating now the Jacobi bracket (\ref{defbracket2}) to dimension
\textit{one}, one finds nonetheless an interesting consequence. In this
case,\ and "a fortiori"',} $\Lambda=\Lambda^{\prime}\equiv0$, \textit{and hence:}%
\begin{equation}
\left\{  f,g\right\}  _{k}=fL_{X_{k}}g-gL_{X_{k}}f \label{defbracket3}%
\end{equation}
\textit{If we consider a circle $S^{1}$ with angular coordinate} $\varphi\in\left[
0,2\pi\right]  $ \textit{and measure} $d\varphi/2\pi$,\textit{consider periodic functions that
can be expanded in Fourier series on the} $O.N.$ \textit{basis:}%
\begin{equation}
f_{n}=e^{in\varphi},\text{ \ }n\in\mathbb{Z}%
\end{equation}
\textit{and take:}
\begin{equation}
X_{k}=i\frac{\partial}{\partial\varphi}%
\end{equation}
\textit{then Eq.(\ref{defbracket3}) yields at once:}%
\begin{equation}
\left\{  f_{n},f_{m}\right\}  _{k}=\left(  n-m\right)  f_{n+m}%
\end{equation}
\textit{which is nothing but the classical conformal algebra \cite{DFMS} (i.e. the
Virasoro algebra without central charge).}
\end{remark}

\subsubsection{Conformal Poisson Brackets and the \textit{KMS} Condition in Phase
Space}
\bigskip

We will consider here the algebra $\mathcal{A}$ of functions on
phase space equipped with the $\ast$-product (the Moyal product for
the time being) and with the associated bracket.

Evolution in time on this algebra is an automorphism of
$\mathcal{A}$ described by Eqs.(\ref{QM2}) and (\ref{QM3}). In
particular, the latter states
that:%
\begin{equation}
\mathcal{A}\ni f\rightarrow f\left(  t\right)  =\exp_{\ast}\left(
itf_{\widehat{H}}/\hbar\right)  \ast f\ast\exp_{\ast}\left(
-itf_{\widehat
{H}}/\hbar\right)  \label{QM4}%
\end{equation}

Let now $\omega$ be a state\footnote{i.e. \cite{Ha} a linear
functional that is \textit{real, positive }and \textit{normalized},
the latter condition being equivalent \cite{Ha} to: $\omega\left(
1\right)  =1$.} on the algebra. \textit{Correlation functions} will
be in general of the form: $\omega\left( f\left(  t\right)  \ast
g\left(  t^{\prime}\right)  \right)  $, $f,g\in \mathcal{A}$.
Time-translational invariance will be assumed \cite{MNE} for
equilibrium states \cite{Ha}. Hence:
\begin{equation}
\omega\left(  f\left(  t\right)  \ast g\left(  t^{\prime}\right)
\right) =\omega\left(  f\left(  t-t^{\prime}\right)  \ast g\right)
=\omega\left(
f\ast g\left(  t^{\prime}-t\right)  \right)  \label{time-trans1}%
\end{equation}
will be assumed throughout. In particular, setting $g=1$ in
Eq.(\ref{time-trans1}), we obtain:%
\begin{equation}
\omega\left(  f\left(  t\right)  \right)  \equiv\omega\left(
f\right)
\forall f,t\label{time-trans2}%
\end{equation}

With any pair $f,g\in\mathcal{A}$ we can associate the correlation
functions
\cite{MNE}:%
\begin{equation}
\mathcal{G}_{fg}\left(  t\right)  =\omega\left(  f\left(  t\right)
\ast g\right)
\end{equation}
and:%
\begin{equation}
\mathcal{F}_{fg}\left(  t\right)  =\omega\left(  g\ast f\left(
t\right) \right)
\end{equation}
Making $t$ into a complex variable, the state $\omega$ will be said
to be a (Kubo, Martin, Schwinger) $\mathit{KMS}$ \textit{state} at
(inverse) temperature $\beta$
\cite{AGGL,Ha,HHW,Hu,Kubo,MarSchwi,MNE,NT} if:

\begin{itemize}
\item $\mathcal{G}_{fg}\left(  t\right)  $ is bounded and continuous in the
strip: $-\hbar\beta\leq\operatorname{Im}t\leq0$ and analytic inside
the strip.

\item The same for $\mathcal{F}_{fg}\left(  t\right)  $ but in the strip
$0\leq\operatorname{Im}t\leq\hbar\beta$ and:

\item The two are connected by:%
\begin{equation}
\mathcal{G}_{fg}\left(  t\right)  =\mathcal{F}_{fg}\left(  t+i\hbar
\beta\right)  ,\text{ }-\hbar\beta<\operatorname{Im}t<0\label{KMS0}%
\end{equation}

\end{itemize}

Taking then boundary values on the real axis, we obtain the
$\mathit{KMS}$
\textit{condition:}%
\begin{equation}
\omega\left(  f\left(  t\right)  \ast g\right)  =\omega\left(  g\ast
f\left(
t+i\hbar\beta\right)  \right)  \ \label{KMS1}%
\end{equation}

\bigskip

\begin{remark}
\textit{In the operator language, the }$\mathit{KMS}$\textit{condition
is usually proved  (at least for bounded operators), using the
cyclic invariance of the trace \cite{KB,MNE} for systems whose
(thermodynamic) equilibrium states are described by the canonical
ensemble or (with minor modifications) by the grand-canonical
ensemble.}
\end{remark}

\begin{remark}
\textit{\ Although the }$\mathit{KMS}$ \textit{condition is usually
stated for equilibrium states at non-zero temperature, there is a
similar condition \cite{Hu} characterizing the ground state(s) at
zero temperature, namely that  }$\mathcal{G}_{fg}\left(  t\right)  $
\textit{be, for real times, the boundary value on the real axis of
an entire function that is uniformly bounded for
}$\operatorname{Im}t\leq0$\textit{.  \ }
\end{remark}

Noticing that:%
\begin{eqnarray}
f\left(  t+i\hbar\beta\right) 
&=& \exp_{\ast}\left(  i(t+i\hbar\beta)f_{\widehat{H}}/\hbar\right)
\ast
f\ast\exp_{\ast}\left(  -i(t+i\hbar\beta)f_{\widehat{H}}/\hbar\right)  = \nonumber \\
&=& \exp_{\ast}\left(  -\beta f_{\widehat{H}}\right)  \ast f\left(
t\right) \ast\exp_{\ast}\left(  \beta f_{\widehat{H}}\right)
\end{eqnarray}
and expanding the exponentials in the last expression::%
\begin{equation}
f\left(  t+i\hbar\beta\right)  \simeq f\left(  t\right)
+i\hbar\beta\left\{ f\left(  t\right)  ,f_{\widehat{H}}\right\}
_{M}+\mathcal{O}\left(  \hbar
^{2}\right)  \label{exp}%
\end{equation}
and, as: $\left\{  .,.\right\}  _{M}=\left\{  .,.\right\}  $ (the
classical Poisson bracket) to lowest order in $\hbar$, we obtain the
(correct)
expansion:%
\begin{equation}
f\left(  t+i\hbar\beta\right)  \simeq f\left(  t\right)
+i\hbar\beta\left\{ f\left(  t\right)  ,f_{\widehat{H}}\right\}
+\mathcal{O}\left(  \hbar ^{2}\right)
\end{equation}
and hence the \textit{classical }$\mathit{KMS}$ \textit{condition}
\cite{AGGL,Bas1,Bas2}:%
\begin{equation}
\omega\left(  \left\{  f\left(  t\right)  ,g\right\}  \right)
=\beta \omega\left(  g\left\{  f\left(  t\right)
,f_{\widehat{H}}\right\}  \right)
\label{KMS2}%
\end{equation}
Interchanging the r\^{o}les of $f$ and $g$ and taking differences,
we obtain
also:%
\begin{equation}
\omega\left(  \left\{  f\left(  t\right)  ,g\right\}  \right)  =\frac{1}%
{2}\beta\omega\left(  f\left(  t\right)  \left\{
f_{\widehat{H}},g\right\}
-g\left\{  f_{\widehat{H}},f\left(  t\right)  \right\}  \right)  \label{KMS3}%
\end{equation}

\begin{remark}
\textit{Setting }$g=1$ \textit{in Eq.(\ref{KMS3})
we obtain: }$\omega\left(  \left\{  f_{\widehat{H}},f\left(
t\right)  \right\}  \right) =0$ $\forall f\in\mathcal{A}$.
\textit{Adding then }$\left(  -1/2\right) \beta\omega\left(  \left\{
f_{\widehat{H}},f\left(  t\right)  g\right\} \right)  =0$ \textit{to
the r.h.s. of Eq.(\ref{KMS3}) we re-obtain Eq.(\ref{KMS2}), and the
two are therefore equivalent.}
\end{remark}

Noticing further that:%
\begin{equation}
\frac{1}{2}\beta\left\{  f_{\widehat{H}},.\right\}  \equiv-e^{\left(
1/2\right)  \beta f_{\widehat{H}}}\left\{  e^{-\left(  1/2\right)
\beta f_{\widehat{H}}},.\right\}
\end{equation}
we can rewrite Eq.(\ref{KMS3}) in the form:%
\begin{eqnarray}
\omega\left(  e^{\left(  1/2\right)  \beta f_{\widehat{H}}}\left[
e^{-\left( 1/2\right)  \beta f_{\widehat{H}}}\left\{  f\left(
t\right)  ,g \right\} \right. \right. 
&+& f\left(  t\right)  \left\{  e^{-\left(  1/2\right)  \beta f_{\widehat{H}}
},g\right\}\label{KMS4}  \\
&-&\left. \left. g\left\{  e^{-\left(  1/2\right)  \beta f_{\widehat{H}}%
},f\left(  t\right)  \right\}  \right]  \right)  =0 \nonumber
\end{eqnarray}

Comparison with Eq.(\ref{defbracket4}) shows then that:

\bigskip

\ \textit{The classical }$KMS$ \textit{condition (\ref{KMS2}) is
equivalent to
the condition}%
\begin{equation}
\omega\left(  e^{\left(  1/2\right)  \beta f_{\widehat{H}}}\left\{
f\left( t\right)  ,g\right\}  _{k}\right)  =0\text{ \ }\forall
f,g\in\mathcal{A}
\label{KMS5}%
\end{equation}
\ \textit{where the bracket on the l.h.s. of Eq.(\ref{KMS5})
i\textit{s the
conformal bracket} (\ref{defbracket4}) \textit{with conformal factor}}%
\begin{equation}
k=\exp\left(  -\left(  1/2\right)  \beta f_{\widehat{H}}\right)  \label{cf1}%
\end{equation}

\bigskip

We turn now to the full quantum case (i.e. away from the limit
$\hbar
\rightarrow0$). Define (cfr. Eqs.(\ref{cf1})and (\ref{QM3})):%
\begin{equation}
k_{\beta}=:\exp_{\ast}\left(  -\left(  1/2\right)  \beta f_{\widehat{H}%
}\right)  \label{cf2}%
\end{equation}
where:%
\begin{equation}
\exp_{\ast}f=:1+%
{\displaystyle\sum\limits_{n=1}^{\infty}}
\frac{1}{n!}\underset{n\text{ }times}{\underbrace{f\ast f\ast...\ast f}}%
\end{equation}
whose $\ast$-inverse is $k_{-\beta}$. This defines the automorphism:%
\begin{equation}
\sigma:f\rightarrow\sigma\left(  f\right)  =f\left(
i\hbar\beta/2\right)
=k_{\beta}\ast f\ast k_{-\beta}%
\end{equation}
(notice that: $\sigma\left(  f\ast g\right)  =\sigma\left(  f\right)
\ast\sigma\left(  g\right)  \forall f,g$) and the $KMS$ condition
(\ref{KMS1}) can be written as:%
\begin{equation}
\omega\left(  f\left(  t\right)  \ast g\right)  =\omega\left(
g\ast\sigma
^{2}\left(  f\left(  t\right)  \right)  \right)  \label{KMS6}%
\end{equation}

\bigskip

Substituting now $\sigma\left(  g\right)  $ for $g$ in
Eq.(\ref{KMS6}) we find:
\begin{equation}
\omega\left(  \sigma\left(  g\right)  \ast\sigma^{2}\left(  f\left(
t\right) \right)  \right)  =\omega\left(  \sigma(g\ast\sigma\left(
f\left(  t\right) \right)  )\right)  =\omega\left(
g\ast\sigma\left(  f\left(  t\right) \right)  \right)
\end{equation}
the last passage following from time-translational
invariance\footnote{If time-translational invariance is not assumed,
then Eq.(\ref{KMS6}) leads, setting $g=1$, to:
$\omega((\sigma^{2}-1)f)=0$. As what is needed to complete the
argument is instead the condition (see below): $\omega\left( \left(
\sigma-1\right)  f\right)  =0$, one has then to assume \cite{Bas1}
the mapping $\sigma+1$ to be invertible.} (Eq.(\ref{time-trans2}))
and, eventually:
\begin{equation}
\omega\left(  f\left(  t\right)  \ast\sigma\left(  g\right)  \right)
=\omega\left(  g\ast\sigma\left(  f\left(  t\right)  \right)
\right)
\end{equation}
or, in terms of the deformed Moyal bracket (\ref{defMoy}) with
deformation
factor $k=k_{\beta}$:%
\begin{equation}
\omega\left(  \left\{  f\left(  t\right)  ,g\right\}
_{M,k_{\beta}}\ast k_{\beta}^{-1}\right)  =0
\end{equation}
But:%
\begin{equation}
\left\{  f\left(  t\right)  ,g\right\}  _{M,k_{\beta}}\ast k_{\beta}%
^{-1}=\sigma\left[  k_{-\beta}\ast\left\{  f\left(  t\right)
,g\right\} _{M,k_{\beta}}\right]
\end{equation}
and, using again Eq.(\ref{time-trans2}) , we obtain eventually
\cite{Bas1,Bas2}:
\begin{equation}
\omega\left(  k_{-\beta}\ast\left\{  f\left(  t\right)  ,g\right\}
_{M,k_{\beta}}\right)  \equiv\omega\left(  \exp_{\ast}\left(  \left(
1/2\right)  \beta f_{\widehat{H}}\right)  \ast\left\{  f\left(
t\right)
,g\right\}  _{M,k_{\beta}}\right)  =0\label{KMS7}%
\end{equation}
which is the quantum version of the classical $KMS$ condition, with
exponentials replaced by "$\ast$-exponentials" and (deformed)
Poisson brackets replaced by (deformed) Moyal brackets.
\newpage

%% file: Chapt7rev.tex
\section{Additional Topics and Concluding Remarks}\label{ch:Add}

\subsection{Some Generalizations}
\bigskip

Weyl systems, the way we have presented them, have been built with the use of
a specific prescription (whose basic ingredients (see Chapt.\ref{Weyl0}) are a
vector spaces $E$ and a symplectic structure over $E$) to deal with a specific
prescription for the ordering problem that arises in the quantization
procedure, one that is known as the "Weyl ordering" prescription
(see Sect.\ref{Weyl-Wig}).

To deal with other ordering prescriptions that are available in the literature
(say, normal, antinormal or other "$s$-ordering" prescriptions (see, e.g. Ref.\cite{LRT}) one has to enlarge slightly the setting of Weyl systems.

Consider then a symplectic vector space with symplectic form $\omega\left(
.,.\right)  $, equipped however with an additional complex structure and
therefore (see Chapt.\ref{QM}) with an Hermitian structure $\left\langle
.|.\right\rangle $. In this way, having the Hermitian structure at hand, one
can replace the (conventional) Weyl map, i.e.:%
\begin{equation}
\widehat{W}\left(  \mathbf{v}_{1}\right)  \widehat{W}\left(  \mathbf{v}%
_{2}\right)  \widehat{W}^{-1}\left(  \mathbf{v}_{1}\right)  \widehat{W}%
^{-1}\left(  \mathbf{v}_{2}\right)  =e^{-i\omega\left(  \mathbf{v}%
_{1},\mathbf{v}_{2}\right)  }\mathbb{I}%
\end{equation}
with the following one:%
\begin{equation}
\widehat{W}\left(  \mathbf{v}_{1}\right)  \widehat{W}\left(  \mathbf{v}%
_{2}\right)  \widehat{W}^{-1}\left(  \mathbf{v}_{1}\right)  \widehat{W}%
^{-1}\left(  \mathbf{v}_{2}\right)  =e^{-\left\langle \mathbf{v}%
_{1}|\mathbf{v}_{2}\right\rangle }\mathbb{I}%
\end{equation}

Here the r.h.s. is no more a unitary transformation, i.e. an element of
$U\left(  1\right)  $, but it is instead an element of $\mathbb{C}_{0}\equiv
U\left(  1\right)  \times\mathbb{R}_{+}$.

More generally, by splitting, as we have done repeatedly, the Hermitian
structure into its real and imaginary parts: $\left\langle .|.\right\rangle
=g\left(  .,.\right)  +i\omega\left(  .,.\right)  $, it is possible to
consider a further generalization by setting:%
\begin{equation}
\widehat{W}\left(  \mathbf{v}_{1}\right)  \widehat{W}\left(  \mathbf{v}%
_{2}\right)  \widehat{W}^{-1}\left(  \mathbf{v}_{1}\right)  \widehat{W}%
^{-1}\left(  \mathbf{v}_{2}\right)  =e^{-sg\left(  \mathbf{v}_{1}%
,\mathbf{v}_{2}\right)  -i\omega\left(
\mathbf{v}_{1},\mathbf{v}_{2}\right) }\mathbb{I}%
\end{equation}
with the "deformation parameter" $s$ taking values in $\left[  -1,1\right]  $.
This kind of generalization can become quite useful in dealing with the
problem of second quantization (see below Sect.\ref{s:SecQuant}).

\begin{remark}
 \textit{Notice that, the metric tensor $g$ having
been replaced by $sg$, the link between the real and imaginary parts
of the Hermitian structure and the complex structure gets lost here
for all $s\neq\pm1$.}
\end{remark}

From our point of view, this kind of generalization raises a new problem
concerning the Moyal product. Namely, besides the bi-differential operator%
\begin{equation}
\exp\left[  i\left(  \overleftarrow{\frac{\partial}{\partial x^{\mu}}}%
\wedge\overrightarrow{\frac{\partial}{\partial p_{\mu}}}\right)  \right]
\label{bidiff1}%
\end{equation}
we will be forced to consider in addition also the bi-differential operator%

\begin{equation}
\exp\left[  s\left(  \delta^{\mu\nu}\overleftarrow{\frac{\partial}{\partial
x^{\mu}}}\otimes\overrightarrow{\frac{\partial}{\partial x^{\nu}}}+\delta
_{\mu\nu}\overleftarrow{\frac{\partial}{\partial p_{\mu}}}\otimes
\overrightarrow{\frac{\partial}{\partial p_{\nu}}}\right)  \right]
\label{bidiff2}%
\end{equation}
In the framework of our "deformation" construction, and with reference to the
discussion of Nijenhuis operators and of the Hochschild cohomology that is
summarized in App.$A$, it is possible however to show that these additional
terms do not change the cohomology class of the algebra we obtain by using
only the Poisson tensor, i.e. the bi-differential operator (\ref{bidiff1}), as
the following example shows.

\begin{example}
\textit{To illustrate the situation, it will be enough to
consider the new product on functions defined on }$R$\textit{ along with the
deformation of the usual pointwise product. We can consider then the bilinear
map:}%
\begin{equation}
\left(  f,g\right)  \rightarrow"f\ast g":=f\exp\left\{  -s\overleftarrow
{\frac{\partial}{\partial x}}\otimes\overrightarrow{\frac{\partial}{\partial
x}}\right\}  g\label{bidiff3}%
\end{equation}
\textit{Now, it is possible to show that the linear map }$T$\textit{ defined
by:}%
\begin{equation}
T=\exp\left\{  -\frac{s}{2}\frac{\partial^{2}}{\partial x^{2}}\right\}
\label{bidiff4}%
\end{equation}
\textit{is such that: }%
\begin{equation}
"f\ast g"=T\left(  f\cdot g\right)  -T\left(  f\right)  \cdot g-f\cdot
T\left(  g\right)
\end{equation}
\textit{(with the dot denoting the usual pointwise product), thus
proving (see again App.}$A$\textit{) that the bilinear map
(\ref{bidiff3}) is indeed a coboundary in the Hochschild cohomology
of the algebra of functions with the pointwise product.}
\end{example}

\subsection{Pseudo-Hermitian Quantum Mechanics}\label{s:PHQM}
\bigskip

It is appropriate at this point of our exposition to mention that
many aspects of our mathematical considerations have also appeared
in a setting that has a completely different origin, namely the
field of pseudo-Hermitian Quantum Mechanics. Pseudo-Hermitian Quantum
Mechanics (PHQM) is an attempt to generalize Quantum Mechanics due
mainly to C.M.Bender and collaborators (see, e.g., \cite{Bend} and
references therein). One starts with a Hilbert space equipped with
an Hermitian product $\langle \cdot, , \cdot \rangle$ and a
Hamiltonian $H$ which is diagonalizable but is not Hermitian, i.e.,
in general:
\begin{equation}
\langle \psi , H \phi \rangle \neq \langle H\psi, \phi \rangle
\end{equation}
We shall assume for simplicity the spectrum of H to be entirely
discrete, this meaning that the eigenvalue equation
\begin{equation}
H |\psi_n\rangle = \lambda_n |\psi_n\rangle
\end{equation}
admits of a complete set $\{ |\psi_n\rangle\}_n$ of eigenfunctions
which cannot, in general, be chosen to be orthonormal. Suppose in
addition that $\{ |\psi_n\rangle\}_n$ admits of a bi-orthonormal
extension $\{ |\psi_n\rangle\}_n,  |\phi_n\rangle\}_n$, i.e. that
there exists another complete set $\{ |\phi_n\rangle\}_n$ such
that\footnote{Such a set always exists provided $\{
|\psi_n\rangle\}_n$ is a Riesz basis, i.e. provided one can find a
bounded invertible operator $A$ and an orthonormal basis $\{
|\chi_n\rangle\}_n$ such that $|\psi_n\rangle= A  |\chi_n\rangle$.
Indeed in this case one has: $ |\psi_n\rangle= \sum_{m} A_{mn}
|\chi_n\rangle$ with $A_{mn} = \langle \chi_m|A\chi_n\rangle$ and
can set: $|\phi_m\rangle = \sum_j (A^{-1})^*_{jm} |\chi_j\rangle$.}:
\begin{equation}
\langle \phi_m  |\psi_n\rangle = \delta_{mn}
\end{equation}
Notice that this implies:%
\begin{equation}
\left\langle \phi_{m}|H\psi_{n}\right\rangle =\lambda_{n}\delta_{mn}%
\end{equation}
which implies in turn:%
\begin{equation}
\left(  \lambda_{m}\langle\phi_{m}|-\langle\phi_{m}|H\right)  |\psi_{n}%
\rangle=0\text{ }\forall n
\end{equation}
and hence:%
\begin{equation}
\langle\phi_{m}|H=\lambda_{m}\langle\phi_{m}|
\end{equation}
i.e. that the $\langle\phi_{m}|$'s are (a complete set of) \textit{left}
eigenvectors of $H$.

Then one has a resolution of identity:
\begin{equation}
\mathbb I = \sum_n |\psi_n \rangle \langle \phi_n| = \sum_n
|\phi_n\rangle \langle \psi_n |
\end{equation}
Now one  defines a new operator $\eta$:
\begin{equation}
\eta = \sum |\phi_n\rangle \langle \phi_n|
\end{equation}
which can be easily shown \cite{Mosta} to be invertible, with
inverse
\begin{equation}
\eta^{-1} = \sum |\psi_n\rangle \langle \psi_n|
\end{equation}
and positive. Thus one can define a new new Hermitian product that
will be related to the original one by:
\begin{equation}
h(\cdot,\cdot) = \langle \cdot, , \eta \cdot \rangle
\end{equation}
In other words, $\eta$ is a positive operator that behaves as a (1,
1)-type tensor connecting the new and the old metrics. The latter is
then used only to identify the topology of the vector space of
states, which turns out to be equivalent \cite{Mosta} to the one
defined by the new scalar product.

It is immediate to see that: $(i)$ the complete set of
eigenfunctions $\{ |\psi_n\rangle\}_n$ becomes orthonormal w.r.t.
$h(\cdot,\cdot)$, $h(\psi_n,\psi_m)=\delta_{nm}$ and: $(ii)$ the
Hamiltonian $H$ becomes Hermitian, i.e.:
\begin{equation}
h(\psi,H\phi) = h(H\psi,\phi)
\end{equation}
provided that
\begin{equation}
H^\dagger = \eta H \eta^{-1}
\end{equation}
which is true iff the spectrum of $H$ is real, as one can easily
find after checking that: $H = \sum_m \lambda_m |\psi_m\rangle
\langle \phi_m|$, while: $ \eta H \eta^{-1} = \sum_m \lambda_m
|\phi_m\rangle \langle \psi_m|$ and: $H^\dagger = \sum_m \lambda^*_m
|\phi_m\rangle \langle \psi_m|$. Hermiticity of $H$ w.r.t. to the
new Hermitian product implies of course that $h(\cdot,\cdot)$ is
preserved by the dynamical evolution (while $\langle \cdot, , \cdot
\rangle$ is not). It is clear that, from our point of view, the
problem appears as a sort of Óinverse problemÓ, i.e. the problem of
determining all Hermitian products which are preserved by the flow
defined by the Hamiltonian $H$. Clearly, once a solution has been
found, there exist many others that can be found by using
appropriate operators in the commutant of $H$. Indeed, if $A$ is
such that $[A,H] = 0$, then:
\begin{equation}
(\eta A)H (\eta A)^{-1} \equiv \eta H \eta^{-1} = H^\dagger
\end{equation}
and this defines the new Hermitian product:
\begin{equation}
h_A(\cdot,\cdot) = h(\cdot,A \cdot) = \langle \cdot, , \eta A  \cdot
\rangle
\end{equation}
The appropriate conditions on $A$ will be that it be invertible and
that $\eta A$ be still a positive operator, and the conditions on
$H$, namely that it be diagonalizable with a real and discrete
spectrum, appear simply as conditions for the inverse problem to
have a solution (and hence in general many others). Thus, while in
the usual approach one fixes a Hilbert space (and hence an Hermitian
product) and looks for observables and unitary evolution, in PHQM it
is the dynamical evolution that is given, and one looks for the
Hermitian products that are preserved by the evolution. Recalling
our discussion of Sect. 1.2, we notice also that the new scalar
product $h(\cdot,\cdot) $ induces a new  associative product between
operators:
\begin{equation}
A \cdot_\eta B= A \eta B
\end{equation}
It is clear that even if $[A,B]=0$ then $[A,B]_{\eta}= A\eta B-B
\eta A \neq 0$ in general. For example, if both $A$ and $B$ admits
the following decomposition in term of the bi-orthonormal system:
\begin{equation}
A = \sum _n a_n |\psi_n\rangle \langle \phi_n| \;\; , \;\; B = \sum
_n b_n |\psi_n\rangle \langle \phi_n|
\end{equation}
so that $[A,B]=0$, one has:
\begin{equation}
[A,B]_{\eta} = \sum_{mn} (a_m b_n-a_n b_m) \langle \phi_m | \phi_n
\rangle |\psi_m \rangle \langle \phi_n|
\end{equation}
which is not zero since not all $\langle \phi_m | \phi_n \rangle $
are necessarily zero. When operators with continuous spectra are
involved, it may be the case that the Hermitian products rendering
the Hamiltonian Hermitian need not induce commutation relations for
which the operator is localizable. By this we mean that the position operators need not commute w.r.t. the new associative product that has been induced on the operators.

Let us end this section by giving a simple example of a
pseudo-hermitian operator \cite{KBZ} . We consider the Hilbert space
$L^2 ([0,d])$ with the standard scalar product $\langle \cdot, \cdot
\rangle$ and an operator $H_\alpha$ defined on twice (weakly)
differentiable functions in $L^2 ([0,d])$  given by the quadratic
form:
\begin{equation}
h_\alpha (\phi,\psi) = \langle \phi' , \psi' \rangle + i \alpha
\phi(d)^* \psi(d) - i \alpha \phi(0)^* \psi(0)
\end{equation}
where $\alpha$ is any real number. Some straightforward algebra
shows that the eigenvalue problem admits the following solutions:
\begin{eqnarray}
\psi_0 (x) = &A_0 \exp(-i\alpha x ) &  \lambda_0 = \alpha^2 \\
\psi_j (x) = &A_j \left[ \cos(k_j x) - i \frac{\alpha}{k_j} \sin(k_j x)\right]  &  \lambda_j = k_j^2 \\
 && \lambda_j = k_j^2 \; , \; k_j = j \frac{\pi}{d} \; , \; j=1,2,\cdots \nonumber
\end{eqnarray}
provided that $\alpha d/\pi \notin \mathbb Z - \{0\}$. It also easy
to see that $H_\alpha ^\dagger = H_{-\alpha}$ and its eigenfunctions
and eigenvalues are given by:
\begin{eqnarray}
\phi_0 (x) = &B_0 \exp(i\alpha x ) &  \lambda_0 = \alpha^2 \\
\phi_j (x) = &B_j \left[ \cos(k_j x) + i \frac{\alpha}{k_j} \sin(k_j x)\right]  &  \lambda_j = k_j^2 \\
 && \lambda_j = k_j^2 \; , \; k_j = j \frac{\pi}{d} \; , \; j=1,2,\cdots \nonumber
\end{eqnarray}
Both the sets $\{ |\psi_n\rangle\}_{n=0}^\infty$ and $\{
|\phi_n\rangle\}_{n=0}^\infty$ are complete \cite{KBZ} and the
coefficients $A_n, B_n$ can be chosen so that
\begin{equation}
\langle \phi_j | \psi_k \rangle =\delta_{jk}
\end{equation}
which shows that $\{ |\psi_n\rangle,  |\phi_n\rangle\}_{n=0}^\infty$
is a bi-orthonormal basis. Thus the invertible positive operator
$\eta_\alpha$, that can now be used to define a new scalar product
w.r.t. which $H_\alpha$ becomes hermitian, assumes the form:
\begin{equation}
\eta_\alpha = \sum_{j=0}^\infty \langle \phi_j , \cdot \rangle
\phi_j
\end{equation}
In ref. \cite{KBZ} it is shown that it can be recast in the
following form:
\begin{equation}
\eta_\alpha = \mathbb I + \langle \phi_0 , \cdot \rangle \phi_0 +
\theta_0 + i \alpha \theta_1 + \alpha^2 \theta_2
\end{equation}
where, for any $\psi(x) \in L^2 ([0,d])$:
\begin{eqnarray*}
&& (\theta_0 \psi ) (x) := -\frac{1}{d} (J\psi)(d) \\
&& (\theta_1 \psi ) (x) := 2 (J\psi)(x) - \frac{x}{d} (J\psi)(d)- \frac{1}{d} (J^2\psi)(d)\\
&& (\theta_2 \psi ) (x) := -(J^2\psi)(x) + \frac{x}{d} (J^2\psi)(d)
\end{eqnarray*}
with
\begin{equation}
(J\psi)(x) := \int_0^x dx \psi(x)
\end{equation}
which allows to prove  explicitly that indeed $\eta_\alpha$ is
bounded, invertible and positive.

\subsection{The R\^{o}le of Linear Structures in Statistical and Quantum
Mechanics}
\bigskip
\subsubsection{"Reformulating" the Von Neumann Theorem\label{evade}}
\bigskip

In \ Sects.\ref{se:linstr} and \ref{se:adapt} we have examined the situation
in which it is possible to define alternative linear structures at the
classical level. We will examine now the quantum case.

In general, if two non-linearly related linear structures (and associated
symplectic forms) are available for a classical system, then one can set up
two different Weyl systems realized on two different Hilbert space structures
made of functions defined on the same Lagrangian subspace (see the example
below) but anyhow with different Lebesgue measures. These two Lebesgue
measures, call them $d\mu$ and $d\mu^{\prime}$, will be associated with
different actions of the Abelian vector group of translations that are not
linearly related. When compared by writing both in the same coordinate system
they will not be simply proportional with a constant proportionality factor.
Functions that are square-integrable in one setting need not be such in the
other. Moreover, a necessary ingredient in the Weyl quantization program is
the use of the (standard or symplectic) Fourier transform. For the same
reasons as outlined above, it is clear that the two different linear
structures will define genuinely different Fourier transforms.

In this way one can "evade" the uniqueness part of von Neumann's theorem. What
the present discussion is actually meant at showing is that there are
assumptions, namely that the linear structure (and symplectic form) are given
once and for all and are unique, that are implicitly assumed but not
explicitly stated in the usual formulations of the theorem, and that, whenever
more structures are available, the situation can be much richer and lead to
genuinely and non-equivalent (in the unitary sense) formulations of Quantum Mechanics.

Let us illustrate these considerations by going back to the example of the
$1D$ harmonic oscillator that has been discussed in Sect.\ref{se:linstr}. To
quantize this system according to the Weyl scheme we have first of all to
select a Lagrangian subspace $\mathcal{L}$ of $\mathbb{R}^{2}$ and a Lebesgue
measure $d\mu$ on it defining then $L^{2}(\mathcal{L},d\mu)$. When we endow
$\mathbb{R}^{2}$ with the standard linear structure $\Delta=q\partial/\partial
q+p\partial/\partial p$, we can choose $\mathcal{L}=\{(q,0)\}$ and $d\mu=dq$.
Consider now, e.g., the change of coordinates: $\phi:\left(  q,p\right)
\leftrightarrow\left(  Q,P\right)  $ defined by \cite{EIMM}:%
\begin{equation}
q=Q\left(  1+\lambda R^{2}\right)  ,\text{ }p=P\left(  1+\lambda R^{2}\right)
\label{change1}%
\end{equation}
parametrized by: $\lambda\geq0$ and where: $R^{2}=Q^{2}+P^{2}$.
Eqs.(\ref{change1}) invert to $\left(  r=\sqrt{q^{2}+p^{2}}\right)  $:%
\begin{equation}
Q=qK\left(  r\right)  ,\text{ }P=pK\left(  r\right)  \label{change2}%
\end{equation}
where $K$ is a positive function, the (unique) real solution of the
equation\footnote{Eq.(\ref{change3}) below shows that, actually: $K=K\left(
\lambda r^{2}\right)  $. $K$ \ is monotonically decreasing for $\lambda\geq0$
and: $\lambda=0\leftrightarrow K\equiv1$, while: $K\underset{\lambda
\rightarrow\infty}{\approx}\left(  \lambda r^{2}\right)  ^{-1/3}$.}:%
\begin{equation}
\lambda r^{2}K^{3}+K-1=0 \label{change3}%
\end{equation}

Now we can consider the linear structure defined by: $\Delta^{\prime
}=Q\partial/\partial Q+P\partial/\partial P$ and take: $\mathcal{L}^{\prime
}=\left\{  \left(  Q,0\right)  \right\}  $ and: $d\mu^{\prime}=dQ$.

Notice that $\mathcal{L}$ and $\mathcal{L}^{\prime}$ are the same subset of
$\mathbb{R}^{2}$, defined by the conditions $P=p=0$ and with the coordinates
related by the relation $Q=qK(r=|q|)$. Nevertheless the two Hilbert spaces
$L^{2}(\mathcal{L},d\mu)$ and $L^{2}(\mathcal{L}^{\prime},d\mu^{\prime})$ are
not related via a unitary map since the Jacobian of the coordinate
transformations is not constant\footnote{In fact: $d\mu=(1+3\lambda Q^{2})d\mu^{\prime}$.}

As a second step in the Weyl scheme, we construct in $L^{2}(\mathcal{L},d\mu)$
the operator $\hat{U}(\alpha)$:
\begin{equation}
\left(  \hat{U}(\alpha)\psi\right)  (q)=e^{i\alpha q/\hbar}\psi(q)\;,\;\psi
(q)\in L^{2}(\mathcal{L},d\mu),
\end{equation}
whose generator is $\hat{x}=q$, and the operator $\hat{V}(h)$:
\begin{equation}
\left(  \hat{V}(h)\psi\right)  (q)=\psi(q+h)\;\psi(q)\in L^{2}(\mathcal{L}%
,d\mu),
\end{equation}
which is generated by $\hat{\pi}=-i\hbar\partial/\partial q$. The quantum
Hamiltonian can be written as $H=\hbar\left(  a^{\dagger}a+\frac{1}{2}\right)
$ where $a=(\hat{x}+i\hat{\pi})/\sqrt{2}\hbar$ (here the adjoint is taken with
respect to the complex structure compatible with the Lebesgue measure $d\mu$).
\newline Similar expressions hold in $L^{2}(\mathcal{L}^{\prime},d\mu^{\prime
})$, and we will obtain unitary operators $\hat{U}^{\prime}(\alpha)$, $\hat
{V}^{\prime}(h)$ \ with infinitesimal generators: $\widehat{X}=Q$ and:
$\widehat{\Pi}=-i\hbar\partial/\partial Q$. Notice that, when seen as an
operator in the previous Hilbert space, $\hat{V}^{\prime}(h)$ implements
\cite{EIMM} translations with respect to the linear structure defined, in the
notation of Sect.\ref{se:linstr} by:
\begin{equation}
(\hat{V}^{\prime}(h)\psi)(q)=\psi(q+_{(\phi)}h).
\end{equation}

Denoting as usual with a dagger but also with an additional prime the adjoints
taken with respect to the complex structure compatible with the Lebesgue
measure $d\mu^{\prime}$, the quantum Hamiltonian will be now: $H^{\prime
}=\hbar\left(  A{^{\dagger\prime}}A+\frac{1}{2}\right)  $ with $A=(\widehat
{X}+i\widehat{\Pi})/\sqrt{2}\hbar$.

It is interesting to notice that, in the respective Hilbert spaces:
$[a,a^{\dagger}]=\mathbb{I}$ as well as: $[A,A^{\dagger\prime}]=\mathbb{I}$,
so that we obtain two different and not linearly related realizations of the
Heisenberg algebra.

In terms of the "uppercase" variables, we obtain \cite{EIMM} with some
algebra:%
\begin{equation}
\widehat{x}=(1+\lambda\widehat{X}^{2})\widehat{X}%
\end{equation}
and:%
\begin{equation}
\widehat{\pi}=(1+3\lambda\widehat{X}^{2})^{-1}\widehat{\Pi}%
\end{equation}
so, while the position operator \ $\widehat{x}$ will be self-adjoint with
respect to both measures, the conjugate momentum operator will be not, and
indeed, while: $\widehat{x}^{\dag}=\widehat{x}^{\dag\prime}=\widehat{x}$ and:
$\widehat{\pi}^{\dag}=\widehat{\pi}$, we obtain instead \cite{EIMM}:%
\begin{equation}
\widehat{\pi}^{\dag\prime}=\widehat{\pi}-6i\lambda\widehat{X}(1+3\lambda
\widehat{X}^{2})^{-2}%
\end{equation}

Thus, the $C^{\ast}$-algebra generated by $\hat{x},\hat{\pi},\mathbf{I}$
seen as operators acting on $L^{2}(\mathcal{L},d\mu)$ is closed, whereas the
one generated by $\hat{x},\hat{\pi},\mathbf{I}$ and their adjoints $\hat
{x}^{\dagger\prime},\hat{\pi}^{\dagger\prime},\mathbf{I}^{\dagger\prime}$
acting on $L^{2}(\mathcal{L}^{\prime},d\mu^{\prime})$ does not close because
we generate new operators whenever we consider the commutator between
$\hat{\pi}$ and $\hat{\pi}^{\dagger\prime}$. As a consequence, the operators
$\hat{x},\hat{\pi}$ and $\hat{x}^{\prime},\hat{\pi}^{\prime}$ close on the
Heisenberg algebra only if we let them act on two different Hilbert spaces
generated, respectively, by the sets of the Fock states
\begin{align}
|n\rangle &  =\frac{1}{\sqrt{n!}}(a^{\dagger})^{n}|0\rangle,\\
|N\rangle &  =\frac{1}{\sqrt{N!}}(A^{\dagger\prime})^{N}|0\rangle.
\end{align}

\subsubsection{Alternative Descriptions and Statistical Mechanics}
\bigskip

By further considering the example of the $1D$ harmonic oscillator, we would
like to examine whether alternative Hamiltonian descriptions do lead to the
same thermodynamical description of a given system.

Let us start from the classical case, when the symplectic form can be
rewritten on $\mathbf{{R}^{2}-\{\mathbf{0}\}}$ as:
\begin{equation}
\omega=dp\wedge dq=dH\wedge\xi
\end{equation}
with:
\begin{equation}
\xi=dt=\frac{pdq-qdp}{2H}%
\end{equation}
and the \textquotedblright time function\textquotedblright\ $t$ will be given
by: $t=(1/\omega)\tan^{-1}\{m\omega q/p\}$, which emphasizes its local
character. Thus $\mathbf{{R}^{2}-\{\mathbf{0}\}}$ can be identified with
$\mathbf{{S}^{1}{{\times}}{R}^{+}}$ parametrized by $dH$ and $dt$. The
associated canonical\footnote{We will restrict here to the canonical ensemble
of (both classical and quantum) Statistical Mechanics.} partition function is
easily evaluated, and the well-known result \cite{MNE} is:
\begin{equation}
\mathcal{Z}=h^{-1}\int\limits_{\mathbf{{R}^{2}}}\exp\{-\beta H\}\omega
=h^{-1}\int\limits_{0}^{\infty}dE\exp\{-\beta E)\int\limits_{\Sigma
(E)}dt=\frac{1}{\beta\hbar\omega}%
\end{equation}
Here $\Sigma(E)$ denotes the one-dimensional \textquotedblright
surface\textquotedblright\ of constant energy $E$, $\beta=1/k_{B}T$ with $T$
the (absolute) temperature and $k_{B}$ the Boltzmann constant, while $h$ (and:
$\hbar=h/2\pi$) is a numerically undetermined constant with the dimension of
an action{\footnote{It is well known that one is forced \cite{MNE} to
introduce it in the context of classical Statistical Mechanics in order to
obtain a dimensionless expression for the partition function, so as to make
sense of expressions such as : $\mathcal{F}=-\beta^{-1}\ln\mathcal{Z}$ for the
(Helmoltz) free energy. The value of $h$ is fixed unambiguously at that of
Planck's constant at the quantum level of Statistical Mechanics.}}.

In order to keep track of the correct dimensions of the various physical
quantities involved, let's consider a new Hamiltonian of the form:
\begin{equation}
H_{f}=\beta_{0}^{-1}f(\beta_{0}H)
\end{equation}
where $\beta_{0}$ is a \textquotedblright fiducial\textquotedblright%
\ quantity, fixed once and for all and having dimension $[energy]^{-1}$, and
$f(.)$ is a real function\footnote{We will assume $f^{\prime}>0$ throughout,
and that in order: $i)$ to give a sensible meaning to integrals (see below)
over phase space and: $ii)$ not to change the number of critical points. The
original Hamiltonian will correspond of course to $f(x)=x$.}. It is easy to
prove that if $\Gamma$ is Hamiltonian w.r.t. $(H,\omega)$, then it will be
Hamiltonian as well w.r.t. $(H_{f},\omega_{f})$, where $\omega_{f}$ is defined
as:
\begin{equation}
\omega_{f}=dH_{f}\wedge dt
\end{equation}
Having redefined (through the new symplectic form) the volume element in phase
space, it is natural to redefine the partition function as:
\begin{equation}
\mathcal{Z}_{f}=h^{-1}\int\limits_{\mathbf{{R}^{2}}}\exp\{-\beta H_{f}%
\}\omega_{f}%
\end{equation}
But then:
\begin{equation}
\mathcal{Z}_{f}=h^{-1}\int dE_{f}\exp\{-\beta E_{f}\}\int\limits_{\Sigma
(E_{f})}dt
\end{equation}
We notice that the nonlinear change of coordinates (\ref{change}) defines such
a transformation on the Hamiltonian if we set: $f(\beta_{0}H)\equiv\phi(H)$.

We come now to the analogous problem in the context of Quantum Mechanics. In
terms of the creation and annihilation operators $a$ and $a^{\dagger}$, with
the standard commutation relations:%

\begin{equation}
\lbrack a,a^{\dagger}]=1
\end{equation}
one constructs a basis in the Fock space as:%

\begin{equation}
|n\rangle_{1}=\frac{(a^{\dagger})^{n}}{\sqrt{n!}}|0\rangle
\end{equation}
with $|0\rangle$ the Fock vacuum and the standard scalar product, that we will
denote as $\langle.|.\rangle_{1}$:%

\begin{equation}
\langle n|m\rangle _{1}=\delta_{nm}%
\end{equation}
We need to define for any (trace-class) linear operators the trace as:
\begin{equation}
Tr_{1}\hat{O}=\sum\limits_{n=0}^{\infty}\left\langle n|\hat{O}|n\right\rangle
_{1}%
\end{equation}
in order to be able to calculate the partition function at the quantum level
as:
\begin{equation}
Z \equiv Tr\exp\{-\beta H\}=\sum_{n}\langle n |\exp\{-\beta H\}| n \rangle_{1}%
\end{equation}

Now, we perform a ''nonlinear change of variables'' by defining \cite{MMSZ,
EMM} new operators as:
\begin{equation}
A=f(\widehat{n})a
\end{equation}
with $\ f(\widehat{n})$ a \ positive, monotonically increasing and nowhere
vanishing function of the number operator $\widehat{n}=a^{\dagger}a$.

At this point, a little care is required when defining the adjoint of any
operator: with the scalar product $\left\langle .|.\right\rangle _{1}$, with
which $a^{\dagger}$ is the adjoint of $a$, the adjoint of $A$ is of course:
$A^{\dagger}=a^{\dagger}f(\hat{n})$.

It is pretty clear that, \ $\widehat{n}$ \ being a constant of the motion, the
equations of motion for $A$ and $A^{\dagger}$ will be the \ same as before. We
can however reconstruct a different Fock space by \ assuming the same vacuum
and defining new states\footnote{Note that, with this definition:
$|n\rangle_{2}=\{\prod_{k=0}^{N-1}f(k)\}|n\rangle_{1}$} as:
\begin{equation}
|n\rangle_{2}=\frac{(A^{\dagger})^{n}}{\sqrt{n!}}|0\rangle
\end{equation}
with a new scalar product defined as:
\begin{equation}
\langle n|m\rangle _{2}=\delta_{nm}%
\end{equation}

The nonlinearity of the transformation reflects itself in the fact that,
despite the fact that \ $|n\rangle_{1}$ and \ $|n\rangle_{2}$ are proportional,
the linear structure in the Fock space labeled by $"1"$ does not carry over
to the linear structure of space $"2"$. This has to do with the fact that the
proportionality factors between the $|n\rangle_{1}$'s and the $|n\rangle_{2}%
$'s depend on $n$. In other words, if we try to induce on space $"2"$ a linear
structure modeled on that of $\ $space $"1",$ the latter will not be
compatible with the bilinearity of the scalar product $\left\langle
.|.\right\rangle _{2}$ that we have just defined.

Now, $A^{\dagger}$ is no more the adjoint of $A$ w.r.t. the new Hermitian
structure we have introduced. If \ we denote by $(.)_{2}^{\dagger}$ the
adjoint of any operator w.r.t. the second Hermitian structure, then we find:
\begin{equation}
(A^{\dagger})_{2}^{\dagger}=\frac{1}{f(\hat{n})}a
\end{equation}
which is quite different from $A$. The pair $\{(A^{\dagger})_{2}^{\dagger
},A^{\dagger}\}$ will yield a new (''nonlinear'') realization of the
Heisenberg algebra, and indeed it is immediate to see that:
\begin{equation}
\lbrack(A^{\dagger})_{2}^{\dagger},A^{\dagger}]=1
\end{equation}
Now, $(A^{\dagger})_{2}^{\dagger}$ and $A^{\dagger}$ will obey the same
equations of motion as $a$ and $a^{\dagger}$, that can be derived from the
previous commutation relations and from the Hamiltonian: $\widetilde
{H}=A^{\dagger}(A^{\dagger})_{2}^{\dagger}+1/2$ (which turns out actually to
coincide with the old one when written in terms of the original creation and
annihilation operators) and that will have therefore the same spectrum.
Defining then consistently the trace of any operator $\hat{O}$ as:
\begin{equation}
Tr_{2}\hat{O}=\sum\limits_{n=0}^{\infty}\left\langle n|\hat{O}|n\right\rangle
_{2}%
\end{equation}
will lead to the same partition function.

\subsection{Weyl Systems and Second Quantization}\label{s:SecQuant}
\bigskip

\subsubsection{Some Preliminaries}
\bigskip

We recall here, mainly to fix the notation, what are the main ingredients for
the construction of \ a Weyl system that were discussed at the beginning of this Chapter.
What we need is:

\begin{itemize}
\item A real, symplectic vector space $\mathcal{S}$ whose symplectic form
(skew-symmetric and non-degenerate) will be denotes as $\omega\left(
.,.\right)  $. If $\mathcal{S}$ is finite-dimensional, then: $\dim
\mathcal{S}=2n$ for some integer $n$. $\mathcal{S}$ will be required (see
Sect.\ref{s:prelim} for more details) to possess also a complex structure $J$,
i.e. a $\left(  1,1\right)  $-tensor satisfying: $J^{2}=-\mathbb{I}%
_{2n\times2n}$ and compatible with $\omega$, which means:%
\begin{equation}
\omega\left(  z,Jz^{\prime}\right)  +\omega\left(  Jz,z^{\prime}\right)
=0\text{ }\forall z,z^{\prime}\in\mathcal{S} \label{compat}%
\end{equation}
and implies that:
\begin{equation}
g\left(  .,.\right)  =:\omega\left(  .,J\left(  .\right)  \right)
\label{metric1}%
\end{equation}
$\left(  g\left(  z,z^{\prime}\right)  =\omega\left(  z,Jz^{\prime}\right)
\right)  $ will be symmetric and nondegenerate, hence a metric and a positive
one iff:%
\begin{equation}
\omega\left(  z,J.z\right)  >0,\forall z\neq0 \label{metric2}%
\end{equation}
\ \ It is always possible to decompose\textit{ }$S$\textit{ }into the direct
sum of two Lagrangian subspaces $S_{1}$ and $S_{2}$, $S=S_{1}\oplus S_{2}%
$, in such a way that, writing (in an unique way): $z=\left(  z_{1}%
,z_{2}\right)  =\left(  z_{1},0\right)  +\left(  z_{2},0\right)  ,z_{1}\in S$
$_{1},z_{2}\in S$ $_{2}$,\ $\omega$ can be written "in Darboux form", being
represented by the matrix:%
\begin{equation}
\left\Vert \omega_{ij}\right\Vert =\left\vert
\begin{array}
[c]{cc}%
\mathbf{0}_{n\times n} & \mathbb{I}_{n\times n}\\
-\mathbb{I}_{n\times n} & \mathbf{0}_{n\times n}%
\end{array}
\right\vert
\end{equation}
i.e.:%
\begin{equation}
\omega\left(  z,z^{\prime}\right)  =z_{1}\cdot z_{2}^{\prime}-z_{2}\cdot
z_{1}^{\prime}%
\end{equation}
the dot denoting the standard Euclidean scalar product. The (compatible)
complex structure $J$ will act as\footnote{Notice that $J$ is not unique. For
example \cite{Bong}, if $J$ is a complex structure, then also: $J^{\prime
}=S^{-1}JS$ will be such if $S$ is any symplectic transformation.}:%
\begin{equation}
J:\left(  z_{1},z_{2}\right)  \mapsto\left(  -z_{2},z_{1}\right)
\end{equation}
The vector space $\mathcal{S}$ can be viewed either as the cotangent space of
either $\mathcal{S}_{1}$ or $\mathcal{S}_{2}$ or, alternatively, as the
realification \cite{Ar1} of a complex vector space of complex dimension $n$,
in which case, writing, e.g.: $z=z_{1}+iz_{2}$, the complex structure will act
as multiplication by the imaginary unit $i$. \ A Weyl system will consist then of:

\item A map: $W:$ $\mathcal{S}\rightarrow\mathcal{U}\left(  \mathcal{H}%
\right)  ;\mathcal{S}\ni z\mapsto \widehat{W}(z)\in\mathcal{U}\left(  \mathcal{H}\right)  $
into the set $\mathcal{U}\left(  \mathcal{H}\right)  $ of the unitary
operators over a Hilbert space $\mathcal{H}$ which is strongly continuous and
satisfies:%
\begin{equation}
\widehat{W}\left(  z\right)  \widehat{W}\left(  z^{\prime}\right)
=\widehat{W}\left(  z+z^{\prime}\right)  \exp\left\{  i\omega\left(
z,z^{\prime}\right)  /2\right\}  \text{ },\forall z,z^{\prime}\in\mathcal{S}
\label{Weyl22b}%
\end{equation}
where (here and in the following) we have set for simplicity $\hbar=1$. We
have already discusses how, using Stone's theorem \cite{RS}, one can represent
$\widehat{W}\left(  z\right)  $ as:%
\begin{equation}
\widehat{W}\left(  z\right)  =\exp\left\{  i\widehat{G}\left(  z\right)
\right\}  \label{Weyl23}%
\end{equation}
with $\widehat{G}\left(  z\right)  $ (essentially) self-adjoint, $\widehat
{G}\left(  tz\right)  =t\widehat{G}\left(  z\right)  $ and:
\begin{equation}
\left[  \widehat{G}\left(  z\right)  ,\widehat{G}\left(  z^{\prime}\right)
\right]  =-i\omega\left(  z,z^{\prime}\right)  \label{Weyl24}%
\end{equation}

\end{itemize}

\begin{remark}\textit{Using the truncated Baker-Campbell-Hausdorff
formula\footnote{$e^{A}e^{B}=e^{A+B}e^{\frac{1}{2}\left[  A,B\right]  }$
whenever: $\left[  A,\left[  A,B\right]  \right]  =\left[  B\left[
A,B\right]  \right]  =0$.} one can also write:}
\begin{eqnarray}
\exp\left\{  it\widehat{G}\left(  z\right)  \right\}  \cdot\exp\left\{
i\widehat{tG}\left(  z^{\prime}\right)  \right\}  &=& \exp\left\{  it\left[
\widehat{G}\left(  z\right)  +\widehat{G}\left(  z^{\prime}\right)  \right]
\right\}  \cdot \\
&\cdot  & \exp\left\{  -\frac{1}{2}t^{2}\left[  \widehat{G}\left(  z\right)
,\widehat{G}\left(  z^{\prime}\right)  \right]  \right\} \nonumber 
\end{eqnarray}
\textit{whence, comparing with Eqs.(\ref{Weyl22b}) and (\ref{Weyl24}) and
expanding in }$t$:%
\begin{equation}
\widehat{G}\left(  z\right)  +\widehat{G}\left(  z^{\prime}\right)
=\widehat{G}\left(  z+z^{\prime}\right), \forall z,z' \label{Weyl25}%
\end{equation}
\end{remark}

\begin{remark}
\textit{To be more precise, the l.h.s.'s of both
Eqs.(\ref{Weyl24}) and (\ref{Weyl25}) should be properly understood
\cite{Bong} as the closures of the commutator and of the sum respectively.}
\end{remark}

We know also from Sect.\ref{sec:Neumann} that, via the von Neumann theorem
\cite{Neu2}, one can realize concretely $\mathcal{H}$ as the Hilbert space of
square-integrable functions over a Lagrangian submanifold $Q\subset
\mathcal{S}$, and how\footnote{As long as we do not alter (see
Sect.\ref{evade}) the linear structure in a non-linear way.} different
realizations of $\mathcal{H}$ are mutually unitarily related.

\subsubsection{Weyl Systems over a Hilbert Space. Second Quantization}
\bigskip

Following the scheme set up in Sect.\ref{sec:Neumann}, assume that we have
realized the Hilbert space $\mathcal{H}$ as the (complete) Hilbert space
$L_{2}\left(  Q\right)  $, with $Q$ a Lagrangian submanifold of the original
(real) vector space $\mathcal{S}$. To fix the ideas, and in the notation of
the previous Subsection, we can take, e.g.: $Q=\mathcal{S}_{1}$ and, writing
now: $z=\left(  \mathbf{q},\mathbf{p}\right)  $ and: $\widehat{W}\left(
z\right)  =\widehat{W}\left(  \mathbf{q},\mathbf{p}\right)  $ , we have then,
with: $\psi\in L_{2}\left(  \mathcal{S}_{1}\right)  $ and: $\mathbf{x}%
\in\mathcal{S}_{1}$:%
\begin{equation}
\left(  \widehat{W}\left(  \mathbf{q},0\right)  \psi\right)  \left(
\mathbf{x}\right)  =:\left(  \widehat{U}\left(  \mathbf{q}\right)
\psi\right)  \left(  \mathbf{x}\right)  =\psi\left(  \mathbf{x}+\mathbf{q}%
\right)
\end{equation}
and:%
\begin{equation}
\left(  \widehat{W}\left(  0,\mathbf{p}\right)  \psi\right)  \left(
\mathbf{x}\right)  =:\left(  \widehat{V}\left(  \mathbf{p}\right)
\psi\right)  \left(  \mathbf{x}\right)  =\exp\left\{  i\mathbf{p\cdot
x}\right\}  \psi\left(  \mathbf{x}\right) 
\end{equation}
(here too we are setting: $\hbar=1$).

We will consider here $\mathcal{H\simeq}L_{2}\left(  Q\right)  $ as a
"single-particle Hilbert space", and we will proceed to setting up a
description of an assembly of identical particles, fixing our attention, for
the sake of illustration, on the case of particles obeying Bose
statistics.

We turn now explicitly to the Hilbert space $L_{2}\left(  Q\right)  $, which
is endowed with the Hermitian (linear in the second factor) scalar product
($d\mathbf{x}$ standing for the Lebesgue measure):%
\begin{equation}
h\left(  \psi,\psi^{\prime}\right)  =:\int d\mathbf{x}\overline{\psi}\left(
\mathbf{x}\right)  \psi^{\prime}\left(  \mathbf{x}\right)
\end{equation}

Writing: $\psi=u+iv$ for every $\psi\in\mathcal{H}$, the complex Hilbert space
$\mathcal{H}$ can be realified \cite{Ar1} into the real linear vector space of
pairs $\left(  u,v\right)  $, equipped with both a (positive) metric:%
\begin{equation}
g\left(  \left(  u,v\right)  ,\left(  u^{\prime},v^{\prime}\right)  \right)
=\int d\mathbf{x}\left[  uu^{\prime}+vv^{\prime}\right]  =\operatorname{Re}%
h\left(  \psi,\psi^{\prime}\right)
\end{equation}
and a symplectic form:%
\begin{equation}
\omega\left(  \left(  u,v\right)  ,\left(  u^{\prime},v^{\prime}\right)
\right)  =\int d\mathbf{x}\left[  uv^{\prime}-vu^{\prime}\right]
=\operatorname{Im}h\left(  \psi,\psi^{\prime}\right)
\end{equation}
i.e.:%
\begin{equation}
h\left(  .,.\right)  =g\left(  .,.\right)  +i\omega\left(  .,.\right)
\end{equation}
with the complex structure (see the previous Subsection) acting as:%
\begin{equation}
J:\left(  u,v\right)  \mapsto\left(  -v,u\right)
\end{equation}
(and hence: $g\left(  \left(  u,v\right)  ,\left(  u^{\prime},v^{\prime
}\right)  \right)  \equiv\omega\left(  \left(  u,v\right)  ,J\left(
u^{\prime},v^{\prime}\right)  \right)  $).

One can set up now a Weyl system in the form:%
\begin{eqnarray}
&& \left(  u,v\right)  \mapsto\widehat{W}\left(  u,v\right) \\
&& \widehat {W}\left(  u,v\right)  \widehat{W}\left(  u^{\prime},v^{\prime}\right)
=\widehat{W}\left(  u+u^{\prime},v+v^{\prime}\right)  \exp\left\{
i\omega\left(  \left(  u,v\right)  ,\left(  u^{\prime},v^{\prime}\right)
\right)  /2\right\} \nonumber
\end{eqnarray}

Representing $\widehat{W}\left(  u,v\right)  $ as:%
\begin{equation}
\widehat{W}\left(  u,v\right)  =\exp\left\{  i\widehat{G}\left(  u,v\right)
\right\}
\end{equation}
with a self-adjoint generator $\widehat{G}$, we have (cfr. Eqs.(\ref{Weyl24})
and (\ref{Weyl25})):%
\begin{equation}
\left[  \widehat{G}\left(  u,v\right)  ,\widehat{G}\left(  u^{\prime
},v^{\prime}\right)  \right]  =-i\omega\left(  \left(  u,v\right)  ,\left(
u^{\prime},v^{\prime}\right)  \right)  \label{Weyl26}%
\end{equation}
as well as:%
\begin{equation}
\widehat{G}\left(  u,v\right)  =\widehat{\Pi}\left(  u\right)  +\widehat{\Psi
}\left(  v\right)  \text{ };\widehat{\Pi}\left(  u\right)  =:\widehat
{G}\left(  u,0\right)  ,\widehat{\Psi}\left(  v\right)  =:\widehat{G}\left(
0,v\right)
\end{equation}
with the commutation relations\footnote{See however the Remark following
Eq.(\ref{Weyl25}).}:%
\begin{equation}
\left[  \widehat{\Psi}\left(  v\right)  ,\widehat{\Pi}\left(  u\right)
\right]  =i\int d\mathbf{x} \, u\left(  \mathbf{x}\right)  v\left(  \mathbf{x}%
\right)  \label{Weyl27}%
\end{equation}
as well as:
\begin{equation}
\left[  \widehat{\Psi}\left(  v\right)  ,\widehat{\Psi}\left(  v^{\prime
}\right)  \right]  =\left[  \widehat{\Pi}\left(  u\right)  ,\widehat{\Pi
}\left(  u^{\prime}\right)  \right]  =0 \label{Weyl28}%
\end{equation}

Being $\mathbb{R}$-linear in their arguments, it is customary to represent
both operators $\widehat{\Psi}$ and $\widehat{\Pi}$ in the form \cite{Bong,S3}%
:%
\begin{equation}
\widehat{\Psi}\left(  v\right)  =\int d\mathbf{x}\widehat{\Psi}\left(
\mathbf{x}\right)  v\left(  \mathbf{x}\right)  ;\text{ }\widehat{\Pi}\left(
u\right)  =\int d\mathbf{x}\widehat{\Pi}\left(  \mathbf{x}\right)  u\left(
\mathbf{x}\right)
\end{equation}
i.e. in terms of the distribution-valued (Hermitian) \textit{field operator}
$\widehat{\Psi}\left(  \mathbf{x}\right)  $ and of its \textit{conjugate
momentum }$\widehat{\Pi}\left(  \mathbf{x}\right)  $ obeying, as a consequence
of Eqs.(\ref{Weyl27}) and (\ref{Weyl28}), the (equal-time) commutation
relations:%
\begin{equation}
\left[  \widehat{\Psi}\left(  \mathbf{x}\right)  ,\widehat{\Pi}\left(
\mathbf{x}^{\prime}\right)  \right]  =i\delta\left(  \mathbf{x}-\mathbf{x}%
^{\prime}\right)
\end{equation}
as well as:
\begin{equation}
\left[  \widehat{\Psi}\left(  \mathbf{x}\right)  ,\widehat{\Psi}\left(
\mathbf{x}^{\prime}\right)  \right]  =\left[  \widehat{\Pi}\left(
\mathbf{x}\right)  ,\widehat{\Pi}\left(  \mathbf{x}^{\prime}\right)  \right]
=0,  \forall\textbf{x,x'}
\end{equation}

These operators are easily recognized to be  appropriate for the description
\cite{S3} of a bosonic field. Having constructed (admittedly in a partly
heuristic way) the algebra of field operators, one should then proceed to
construct the physical vacuum\footnote{We do not discuss here problems of
uniqueness of the vacuum state.} and of the associated Hilbert space on which
this algebra of operators acts via, e.g., the $GNS$ construction
\cite{BSZ,Ha} or defining \cite{Bong}, in terms of the $\widehat{W}$'s, a generating functional for the Wightman functions \cite{Ha}, using the "reconstruction theorem" of Axiomatic Field Theory \cite{SW}. We shall outline here however a slightly different route that
leads more directly to the usual Fock space description of (bosonic) quantum fields.

Reinstating for brevity the notation: $\psi=(u,v)$ for the (real) pair $\left(
u,v\right)  $, we can use the generators: $\widehat{G}\left(  \psi\right)
=:\widehat{G}\left(  u,v\right)  $ to define annihilation and creation
operators $\widehat{a}\left(  \psi\right)  $ and $\widehat{a}^{\dag}\left(
\psi\right)  $ associated with the state $\psi$ as:%
\begin{equation}
\widehat{a}\left(  \psi\right)  =\frac{1}{\sqrt{2}}\left[  \widehat{G}\left(
\psi\right)  +i\widehat{G}\left(  J\psi\right)  \right]  ;\widehat{a}^{\dag
}\left(  \psi\right)  =\frac{1}{\sqrt{2}}\left[  \widehat{G}\left(
\psi\right)  -i\widehat{G}\left(  J\psi\right)  \right]  \text{ }\label{ann1}%
\end{equation}
A little algebra shows then that:%
\begin{equation}
\left[  \widehat{a}\left(  \psi\right)  ,\widehat{a}\left(  \psi^{\prime
}\right)  \right]  =\left[  \widehat{a}^{\dag}\left(  \psi\right)
,\widehat{a}^{\dag}\left(  \psi^{\prime}\right)  \right]  =0\text{ }%
\forall\psi,\psi^{\prime}%
\end{equation}
while:%
\begin{equation}
\left[  \widehat{a}\left(  \psi\right)  ,\widehat{a}^\dag\left(  \psi^{\prime
}\right)  \right]  =h\left(  \psi,\psi^{\prime}\right)
\end{equation}

If we consider in particular an $ON$ basis\footnote{For example, if
$\mathcal{H}=L_{2}\left(  \mathbb{R}\right)  $, we could choose \cite{BSZ} the
basis of the eigenfunctions of the $1D$ harmonic oscillator (the Hermite
functions \cite{Fo}).} $\left\{  \psi_{n}\right\}  _{0}^{\infty}$ in the
"single-particle" Hilbert space $\mathcal{H}$ ($h\left(  \psi_{n},\psi
_{m}\right)  =\delta_{nm}$) and define:%
\begin{equation}
\widehat{a}_{n}=:\widehat{a}\left(  \psi_{n}\right)
\end{equation}
then:%
\begin{equation}
\left[  \widehat{a}_{n},\widehat{a}_{m}^{\dag}\right]  =\delta_{nm}%
\end{equation}
and all the other commutators vanish. With these operators at hand, one can
then proceed to the construction of the Fock space following, e.g., the approach
discussed by J.M.Cook \cite{Co} already in the early Fifties.

Of course, one can also work directly with the exponential form (\ref{Weyl22b}) of a Weyl system, as we will see now.
The possibility to do so relies on the following observations that can be easily verified if we work on a finite $n$-dimensional Hilbert space ${\cal H}$. We will denote with $\mathbb K$ the space of (complex) functions $f(z)=f(z_1,z_2, \cdots, z_n)$, $z_j\in \mathbb C$, on ${\cal H}$ which are square-integrable according to the (Gaussian) measure:
\begin{equation}
\| f\|^2 =: \int \left( \prod_{j=1}^n \frac{d\mbox{Re} z_j\,  d\mbox{Im} z_j}{\pi} \right) e^{-\langle z,z \rangle } |f(z)|^2  < \infty
\end{equation}
On such space, for any $z \in {\cal H}$ let us consider the operator:
\begin{equation}
W(z): f(w) \mapsto f_z(w) =: f(w-z) \exp \left( \frac{\langle w,z\rangle}{2}- \frac{\langle z,z\rangle}{4} \right)  \label{www}
\end{equation}
which: $i)$ conserves the norm: $\| f\|^2 = \| f_z \|^2$ and $ii)$ satisfies the relation
\begin{equation}
W(z) W(z') = W(z+z') \exp \left( \frac{i  \mbox{Im}\langle z, z'\rangle}{2} \right)
\end{equation}
and hence allow for the definition of a Weyl system which is irreducible on the subspace of $\mathbb K$ of antiholomorphic functions, ${\cal F}_{\mathbb K}$, which can be seen \cite{BSZ}  as the closure w.r.t. the norm defined above of the space of antiholomorphic polynomials in the $n$ variables $z_1,z_2, \cdots, z_n$. A straightforward calculation shows that setting $u^{(j)} = (0,\cdots, 0, u_j,0, \cdots, 0)$ and $v^{(j)} = (0,\cdots, 0, i v_j,0, \cdots, 0)$, with $u_j,v_j \in \mathbb R$,
one has:
\begin{eqnarray*}
&&i G(u^{(j)}) = - \frac{\partial}{\partial\mbox{Re} w_j} + \frac{\bar{w_j}}{2} \\
&&i G(v^{(j)}) = - \frac{\partial}{\partial\mbox{Im} w_j} -i \frac{\bar{w_j}}{2} 
\end{eqnarray*}
so that the annihilation/creation operators are given by
\begin{eqnarray*}
\hat{a}_j & =: &\frac{G(u^{(j)})- i G(v^{(j)})}{\sqrt{2}} = \sqrt{2} i \partial_{w_j} - \frac{i}{\sqrt{2}} \bar{ w_j} \\
& =&   - \frac{i}{\sqrt{2}}  \bar{w_j} \mbox{ on }  {\cal F}_{\mathbb K} \\
\hat{a}^\dagger_j &=:& \frac{G(u^{(j)})+ i G(v^{(j)})}{\sqrt{2}} =  \sqrt{2} i \partial_{\bar{w}_j} 
\end{eqnarray*}
and clearly satisfy bosonic-like  commutation relations. Then  the vacuum (or cyclic vector) is given by the constant unit monomial $P_0(z) = 1$. Notice also that for any unitary operator $U\in {\cal U}({\cal H})$ we may construct a unitary operator $\Gamma (U) \in {\cal U}({\cal F}_{\mathbb K})$ via the map:
\begin{equation}
\Gamma (U) : W(z) \mapsto W(U^{-1}z)
\end{equation}

A generalization of such results to an infinite dimensional Hilbert space ${\cal H}$ requires of course caution in the definition of domains of operators as well in the definition of the spaces  $\mathbb K$ and ${\cal F}_{\mathbb K}$. This can be done by introducing the so called isonormal \cite{BSZ} distribution $g$, which determines a measure $dg$ on the Hilbert space which, when restricted to finite dimensional subspaces looks like a Gaussian measure with variance $\sigma$,  and defining the space $\mathbb K$ as the completion of the space of polynomials on ${\cal H}$ w.r.t. the inner product
\begin{equation}
\int_L \overline{P'(\psi)} P(\psi) dg(\psi)
\end{equation}
$L$ being any finite-dimensional subspace of ${\cal H}$ on which the polynomials $P,P'$ have support. The space ${\cal F}_{\mathbb K}$ is now the subspace of those functions $F$ on ${\cal H}$ such that their restrictions $F|_L$ on any finite-dimensional subspace $L$ are antiholomorphic in the usual sense. Thus one gets a complex representation for the  bosonic field in which the Weyl system is given by the operators \cite{BSZ}:
\begin{equation}
W(\psi) : F(\phi) \mapsto F(\phi - \psi)  \exp \left( \frac{\langle \psi,\phi\rangle}{2\sigma}- \frac{\langle \phi,\phi\rangle}{4} \right) \;, \; \forall \psi \in {\cal H}
\end{equation}
For such representation, the cyclic vector is the function on ${\cal H}$ identically equal to one. Also, for any $U\in {\cal U}({\cal H})$ we have a unitary operator $\Gamma (U) \in {\cal U}({\cal F}_{\mathbb K})$ such that $W(\psi) \mapsto W(U^{-1}\psi)$

This completes the discussion of how Weyl's approach can lead, in a rather
natural and elegant way, to the formalism of second quantization and
hence of Field Theory. We have done that for bosons, and we refer to the literature
(see in particular Refs. \cite{BSZ},\cite{Bong} and \cite{Co}) for the parent
construction for the case of fermions.
Alternative Hilbert space structures will give rise also to additional ambiguities in the commutation relations for the fields.

\subsection{Concluding Remarks}
\bigskip

By using the geometrical formulation of Quantum Mechanics we have bee able to
"export" \ from the classical to the quantum framework many problems that
arise in the classical setting, and we have constructed a more direct "bridge"
which realizes Dirac's demand \cite{Dir3}that problems arising in Classical Mechanics
must be a suitable limit of analogous problems arising in Quantum Mechanics.

In particular, we have addressed the problem of the quantum interpretation of
the bi-Hamiltonian description of completely integrable systems in the
classical setting.

Alternative quantum Hamiltonian descriptions have been provided in various pictures of Quantum Mechanics, the Schr\"{o}dinger, Heisenberg and Weyl-Wigner-Moyal pictures.

We have also shown that it is possible to deal with nonlinear transformations
in Quantum Mechanics without giving up the superposition principle which is
associated with quantum interference phenomena.

The r\^{o}le of dynamically determined structures versus pre-assigned
mathematical structures \ in the formalization of Quantum Mechanics has been
further elucidated. 

One may wonder if, in analogy with what happens in General Relativity, where the metric is determined by solving the Einstein equations, one can conceive of some field equations whose solutions would provide the Hermitian tensor to be used in the description of quantum systems.

By mentioning how to deal with Second Quantization and Quantum Field Theories
in this framework we have hinted at the idea that this approach may provide
suggestions for the introduction of interactions in a pure quantum
field-theoretic setting.

At the end of this journey, we believe it to be rewarding to know that many
sophisticated methods of Classical Physics may find their way into the
formalism of Quantum Physics.
\newpage

%% file: AppArev.tex
\section[Nijenhuis torsions and Nijenhuis Tensors]{Nijenhuis torsions and Nijenhuis Tensors}
\bigskip
\subsection*{Nijenhuis Torsions and Tensors on Smooth Manifolds}
\addcontentsline{toc}{subsection}{Nijenhuis Torsions and Tensors on Smooth Manifolds}
\bigskip
 Let us consider, to begin with, the set $\mathfrak{X}\left(
\mathcal{M}\right)  $ of vector fields over some (smooth) manifold
$\mathcal{M}.$ $\mathfrak{X}\left(  \mathcal{M}\right)  $ has, as is well
known, the structure of a (actually an infinite-dimensional) Lie algebra
defined by the Lie bracket:
\begin{equation}
\left[  .,.\right]  :\mathfrak{X}\left(  \mathcal{M}\right)  \rightarrow
\mathfrak{X}\left(  \mathcal{M}\right)  ;\text{ }\left(  X,Y\right)
\mapsto\left[  X,Y\right]  =:\mathcal{L}_{X}Y=-\mathcal{L}_{Y}X;\text{ }%
X,Y\in\mathfrak{X}\left(  \mathcal{M}\right)
\end{equation}
with $\mathcal{L}_{\cdot}$ the Lie derivative. Let then $T$ be a $\left(
1-1\right)  $ tensor \ viewed as a map: $T:\mathfrak{X}\left(  \mathcal{M}%
\right)  \rightarrow\mathfrak{X}\left(  \mathcal{M}\right)  $. One
can associate\footnote{See Ref. \cite{MFVMR} for more details} with
$T$ an antiderivation $d_{T}$ of degree one whose actions on zero-
and one-forms is
given by:%
\begin{equation}
d_{T}f\left(  X\right)  =df\left(  TX\right)
\end{equation}
on functions, and:%
\begin{equation}
d_{T}\theta\left(  X,Y\right)  =\left(
\mathcal{L}_{TX}\theta\right)  \left( Y\right)  -\left(
\mathcal{L}_{TY}\theta\right)  \left(  X\right) +\theta\left(
T\left[  X,Y\right]  \right)
\end{equation}
on one-forms (recall that a (anti)derivation is entirely defined \ \cite{CM}
by its action on zero- and one-forms). One proves that $d_{T}^{2}$ is a
derivation (of degree two) commuting with $d$: $d\circ d_{T}^{2}=d_{T}%
^{2}\circ d$. As such, its action is entirely defined \cite{CM}
 by that on zero-forms (functions), and one finds:%
\begin{equation}
\left(  d_{T}^{2}f\right)  \left(  X,Y\right)  =-df\left(  N_{T}\left(
X,Y\right)  \right)
\end{equation}
where \cite{FroNi,Mar0,MFVMR,Ni} \ the\textit{Nijenhuis torsion }$N_{T}$ of
$T$ is the $\left(  1-2\right)  $-type tensor defined by\footnote{Note that
what we call here, following the literature, the "Nijenhuis torsion" was
called the "Nijenhuis tensor" in Ref. \cite{MFVMR}.}:%
\begin{equation}
N_{T}\left(  X,Y\right)  =\left\{  \left(  T\circ\mathcal{L}_{X}\left(
T\right)  \right)  -\left(  \mathcal{L}_{TX}\left(  T\right)  \right)
\right\}  \left(  Y\right)  \label{Nij1}%
\end{equation}
or, more explicitly:%
\begin{equation}
N_{T}\left(  X,Y\right)  =T\left[  TX,Y\right]  +T\left[  X,TY\right]
-T^{2}\left[  X,Y\right]  -\left[  TX,TY\right]  \label{Nij2}%
\end{equation}
$T$ \ will be said to be a \textit{Nijenhuis tensor} if its Nijenhuis torsion
vanishes, i.e. if:%
\begin{equation}
N_{T}=0
\end{equation}

\begin{remark}
In local coordinates $x^{i}$, if:
\begin{equation}
T=T^{i}\text{ }_{j}\frac{\partial}{\partial x^{i}}\otimes dx^{j}%
\end{equation}
then:
\begin{equation}
N_{T}=\frac{1}{2}\left(  N_{T}\right)  ^{i}\text{ }_{km}\frac{\partial
}{\partial x^{i}}\otimes dx^{k}\wedge dx^{m}%
\end{equation}
where:%
\begin{equation}
\left(  N_{T}\right)  ^{i}\text{ }_{km}=\frac{\partial T^{i}\text{ }_{k}%
}{\partial x^{j}}T^{j}\text{ }_{m}+T^{i}\text{ }_{j}\frac{\partial T^{j}\text{
}_{m}}{\partial xk}-\left(  k\longleftrightarrow m\right)
\end{equation}
and, obviously: $N_{T}=0$ whenever the representative matrix of $T$ \ is a
matrix with constant entries.
\end{remark}
\bigskip
\subsection*{Nijenhuis Torsions and Tensors on Associative Algebras}
\addcontentsline{toc}{subsection}{Nijenhuis Torsions and Tensors on Associative Algebras}
\bigskip
Eqn.(\ref{Nij1}) defines the Nijenhuis torsion on a Lie algebra. Nijenhuis-type
tensors and torsions can be given however a more general setting
\cite{CGM1,CGM2}in the framework of associative algebras. We
recall\footnote{It goes without saying that the "star-product" $\ast$ we are
talking about here has nothing to do with the Moyal product.} that an
associative algebra $\left(  \mathcal{A},\ast\right)  $ becomes also a Lie
algebra under commutation, \ i.e. with a bracket defined as:%
\begin{equation}
\left[  A,B\right]  =:A\ast B-B\ast A;\text{ \ \ }A,B\in\mathcal{A}
\label{comm}%
\end{equation}
and associativity of the algebra guarantees that the bracket does satisfy the
Jacobi identity, so it is indeed a Lie bracket.

Let then $\left(  \mathcal{A},\ast\right)  $ be an associative algebra over a
field $\mathbb{K}$ $\ $($\mathbb{K=R}$ or $\mathbb{K=C}$ for our purposes), an
let: $T:\mathcal{A}\rightarrow\mathcal{A}$ be a linear map. $T$ will be a
\textit{derivation} of the algebra $\left(  \mathcal{A},\ast\right)  $ if (and
only if):%
\begin{equation}
T\left(  A\ast B\right)  =T\left(  A\right)  \ast B+A\ast T\left(  B\right)
\text{ }\forall A,B\in\mathcal{A}%
\end{equation}
Be it as it may, given $T$ one can define in general the bilinear map:
\begin{equation}
\ast_{T}:\left(  A,B\right)  \rightarrow A\ast_{T}B=T\left(  A\right)  \ast
B+A\ast T\left(  B\right)  -T\left(  A\ast B\right)  \label{starproduct}%
\end{equation}%

and $\ast_{T}$ will be trivial if (and only if) $T$ is a derivation. In
general (with $T$ not a derivation), $\ast_{T}$ will define a (non-trivial)
new algebra structure $\left(  \mathcal{A},\ast_{T}\right)  $.

\bigskip

As a simple  example, let's take $T\in\mathcal{A}$, an hence: $T\left(
A\right)  =T\ast A$. Then, a simple calculation shows that:%
\begin{equation}
A\ast_{T}B=A\ast T\ast B\label{Kprod}%
\end{equation}
Products of this sort will be employed in the text in the discussion of
alternative commutation relations in Quantum Mechanics.

\bigskip

\subsection*{A Digression on: Hochschild Cohomologies}
\addcontentsline{toc}{subsection}{A Digression on: Hochschild Cohomologies}
\bigskip
Given an associative algebra $\left(  \mathcal{A},\ast\right)  $ and
an $\mathcal{A}$-bimodule $V$ (what we will have in mind will be the
case in which $V$ is the additive group of $\mathcal{A}$ and the
bimodule structure is given by left and right multiplication), \ an
\textit{n-cochain} will be an
$n$-linear mapping:%
\begin{equation}
\alpha:\underset{n\text{ \ }times}{\underbrace{\mathcal{A\times A\times
}...\mathcal{A}}}\text{ }\rightarrow V
\end{equation}
The space $C^{n}\left(  \mathcal{A},V\right)  $ of $n$-cochains has
a group structure under addition. Then, for every \thinspace$n$, the
\textit{Hochschild coboundary operator}\footnote{The suffix serves
here to stress that the operators and the ensuing properties are all
relative to the binary product ("star-product") in the algebra.}:
$\delta_{\ast}:$ $C^{n}\left( \mathcal{A},V\right) \rightarrow
C^{n+1}\left( \mathcal{A},V\right)  $ is
defined ($\alpha\in C^{n}\left(  \mathcal{A},V\right)  ,$ $a_{1}%
,...,a_{n+1}\in\mathcal{A}$) via \cite{Hoch}:%
\begin{equation}%
\begin{array}
[c]{c}%
\left(  \delta_{\ast}\alpha\right)  (a_{1},...,a_{n+1})=a_{1}\alpha\left(
a_{2},...,a_{n+1}\right)  +\\
+%
{\displaystyle\sum\limits_{n=1}^{n}}
\left(  -1\right)  ^{i}\alpha\left(  a_{1},..,a_{i}\ast a_{i+1},..,a_{n+1}%
\right)  +\\
+\left(  -1\right)  ^{n+1}\alpha\left(  a_{1},...,a_{n}\right)  a_{n+1}%
\end{array}
\end{equation}
where $a\alpha\left(  ..\right)  $ and $\alpha\left(  ..\right)  a$ denote the
left and right actions of $\mathcal{A}$ on $V$ respectively. One can check
directly that:%
\begin{equation}
\delta_{\ast}\circ\delta_{\ast}=0 \label{cob}%
\end{equation}
As an example, for $n=1$:%
\begin{equation}
\left(  \delta_{\ast}\alpha\right)  (a_{1},a_{2})=a_{1}\alpha\left(
a_{2}\right)  +\alpha\left(  a_{1}\right)  a_{2}-\alpha\left(  a_{1}\ast
a_{2}\right)
\end{equation}

An $n$-cochain $\alpha$ is called an $\mathit{\ }n-$\textit{cocycle} if
$\delta_{\ast}\alpha=0$, an $n-$\textit{coboundary} if $\alpha=\delta_{\ast
}\beta$ for some $\left(  n-1\right)  -$cochain $\beta$. $n-$cocycles form an
additive group usually denoted as $\mathbb{Z}^{n}\left(  \mathcal{A},V\right)
$, and (in view of (\ref{cob})) $n-$coboundaries form an subgroup
$\mathbb{B}^{n}\left(  \mathcal{A},V\right)  $ of $\mathbb{Z}^{n}\left(
\mathcal{A},V\right)  $. The $n-$\textit{(Hochschild) cohomology group}
$\mathbb{H}^{n}\left(  \mathcal{A},V\right)  $ is defined then as the
quotient:%
\begin{equation}
\mathbb{H}^{n}\left(  \mathcal{A},V\right)  =\mathbb{Z}^{n}\left(
\mathcal{A},V\right)  /\mathbb{B}^{n}\left(  \mathcal{A},V\right)
\end{equation}
The linear mapping $T$ can be considered as a one-cochain and, looking then at
Eqn.(\ref{starproduct})we can conclude that:%
\begin{equation}
A\ast_{T}B=\delta_{\ast}T\left(  A,B\right)
\end{equation}
and hence we can rephrase what has been said previously by saying that $T$
\ \ will be a derivation if and only if it is a one-cocycle in the Hochschild
cohomology associated with the "star-product".

\bigskip

The $\ast-$\textit{Nijenhuis torsion} of $T$ \ is defined as:%
\begin{equation}
N_{T}\left(  A,B\right)  =T\left(  A\ast_{T}B\right)  -T\left(  A\right)  \ast
T\left(  B\right)  \label{Nij3}%
\end{equation}
or, more explicitly:%
\begin{equation}
N_{T}\left(  A,B\right)  =T\left(  T\left(  A\right)  \ast B\right)  +T\left(
A\ast T\left(  B\right)  \right)  -T^{2}\left(  A\ast B\right)  -T\left(
A\right)  \ast T\left(  B\right)  \label{Nij4}%
\end{equation}
It is clear from Eqn.(\ref{Nij3}) that the Nijenhuis torsion of $T$ measures
the obstruction for the linear map $T$ to be a homomorphism of the two
products.

Here too it will be said that $T$ is a $\ast-$\textit{Nijenhuis tensor }if its
Nijenhuis torsion vanishes. For example, it is easy to see that $N_{T}=0$ if
$T\in\mathcal{A}$ and the associated product is given by Eqn.(\ref{Kprod}). Hence, $T$ is a Nijenhuis tensor.

\bigskip
\subsection*{Making Contacts}
\addcontentsline{toc}{subsection}{Making Contacts}
\bigskip
To make contact with the initial definition of the Nijenhuis torsion, we
recall what has already been said, i.e. that an associative algebra can be
made into a Lie algebra using the commutator (\ref{comm}). If we substitute
the "star-product" with the commutator, then Eqn.(\ref{Nij4}) becomes:%
\begin{equation}
N_{T}\left(  A,B\right)  =T\left[  T\left(  A\right)  ,B\right]  +T\left[
A,T\left(  B\right)  \right]  -T^{2}\left[  A,B\right]  -\left[  T\left(
A\right)  ,T\left(  B\right)  \right]  \label{Nij5}%
\end{equation}
which coincides with Eqn.(\ref{Nij2}) if we substitute for $A,B,..$
vector fields on a manifold and the commutator with the Lie bracket. This establishes the link between the two
definitions of the Nijenhuis torsion that have been given here. The Nijenhuis
torsion defined on an associative algebra will play a r\^{o}le in the
discussion, in the text, of alternative associative products on the algebra of
(bounded) operators on a Hilbert space. Completeness would require discussing
also how the (Lie) algebra of vector fields can be embedded into a larger
associative algebra (the enveloping algebra), but we will not insist on this
point not too lengthen too much the discussion.

\newpage

%% file: AppBrev.tex
\section[Recursion Operators]{Recursion Operators}
\bigskip
\subsection*{Some Preliminaries}
\addcontentsline{toc}{subsection}{Some Preliminaries}
\bigskip

Let $T$ be a $\left(  1,1\right)  $-type tensor field: $T\in\mathcal{F}%
_{1}^{1}\left(  \mathcal{M}\right)  $. As is already known, the
action of $T$ \ on vector fields (denoted with the same symbol)
and one-forms (defined as
$\widetilde{T}$ ) is defined uniquely by:%
\begin{equation}
\left\langle TX|\alpha\right\rangle =:\left\langle X|\widetilde{T}%
\alpha\right\rangle ,\text{ }X\in\mathcal{X}\left(
\mathcal{M}\right)
,\alpha\in\mathcal{X}^{\ast}\left(  \mathcal{M}\right)  \label{pairing}%
\end{equation}
where $\left\langle .|.\right\rangle $ denotes the usual pairing.
In coordinates, if:
\begin{equation}
T=T_{j}^{i}dx^{j}\otimes\frac{\partial}{\partial x^{i}}%
\end{equation}
is represented by the matrix\footnote{With some abuse of notation,
we will denote here with the same symbol $\left(  1,1\right)  $
tensors and their representative matrices.}: $T=\left\Vert
T^{i}\text{ }_{j}\right\Vert $ then $\widetilde{T}$ \ will be
represented by \ the matrix: $\widetilde {T}=:\left\Vert
\widetilde{T}_{j}\text{ }^{i}\right\Vert $ and
Eqn.(\ref{pairing}) implies:%
\begin{equation}
\widetilde{T}_{j}\text{ }^{i}=T^{i}\text{ }_{j} \label{transpose}%
\end{equation}
i.e. that $\widetilde{T}$ be the \textit{transpose} of $T$:%
\begin{equation}
\widetilde{T}=T^{t}%
\end{equation}
All this is well known and is repeated here only for completeness.

One can consider extending the action of the $\widetilde{T}$ on
forms oh higher rank, as well as that of $T$ on multivectors. We
will concentrate here only on the former, recollecting some
results that can be found in the literature (\cite{MFVMR}).

The extension under consideration is not unique. Let, e.g.,
$\omega$ be \ a
two-form. In particular, $\omega$ will be considered as the map:%
\begin{equation}%
\begin{array}
[c]{c}%
\omega:\mathcal{X}\left(  \mathcal{M}\right)
\rightarrow\mathcal{X}^{\ast }\left(  \mathcal{M}\right)  ;\text{
}\omega\text{\ }:Y\rightarrow
\omega\left(  .,Y\right)  =-i_{Y}\omega\\
\left\langle \omega\left(  .,Y\right)  |X\right\rangle =-i_{X}i_{Y}%
\omega=\omega\left(  X,Y\right)
\end{array}
\end{equation}
($\left(  \omega\left(  .,Y\right)  \right)
=\omega_{ij}Y^{j}dx^{i}$). Hence we can compose $\widetilde{T}$
with $\omega$ to obtain the $\left(
0,2\right)  $ tensor:%
\begin{equation}
\widetilde{T}\circ\omega:\text{ }\left(  X,Y\right)
\rightarrow\left\langle \widetilde{T}\circ\omega\left(  .,Y\right)
|X\right\rangle =\left\langle \omega\left(  .,Y\right)
|TX\right\rangle
\end{equation}
i.e.:%
\begin{equation}
\left(  \widetilde{T}\circ\omega\right)  \left(  X,Y\right)
=\omega\left(
TX,Y\right)  \label{extension1}%
\end{equation}
This is a linear extension. In terms of representative matrices \
$\widetilde {T}\circ\omega$ is represented by the matrix
$T^{t}\omega$, i.e. (cfr.
Eqn.(\ref{transpose})):%
\begin{equation}
\widetilde{T}\circ\omega=\left(  T^{t}\omega\right)
_{ij}dx^{i}\otimes
dx^{j}=T^{k}\text{ }_{i}\omega_{kj}dx^{i}\otimes dx^{j}%
\end{equation}

Another possible and more symmetric linear extension is provided by:%
\begin{equation}
\left(  \widetilde{T}\circ\omega\right)  \left(  X,Y\right)
=\omega\left(
TX,Y\right)  +\omega\left(  X,TY\right)  \label{extension2}%
\end{equation}
Also, a nonlinear extension such as:
\begin{equation}
\left(  \widetilde{T}\circ\omega\right)  \left(  X,Y\right)
=\omega\left(
TX,TY\right)  \label{extension3}%
\end{equation}
may be envisaged, with even more possibilities for forms of higher
rank.

Notice that, while the extensions (\ref{extension2}) and
(\ref{extension3}) map two-forms into two-forms, this is not true
in general for the extension (\ref{extension1}) which will yield
in general a $\left(  0,2\right)  $-type tensor but not a
two-form.

The linear extension (\ref{extension2}) allows for the association
with $T$ of an antiderivation of degree one usually denote as
$d_{T}$ that acts on zero-
and one-forms as:%
\begin{equation}
d_{T}f=\widetilde{T}df;\text{ }d_{T}f\left(  X\right)  =:df\left(
TX\right)
\end{equation}
and:%
\begin{equation}
\left(  d_{T}\theta\right)  \left(  X,Y\right)  =\left(  \mathcal{L}%
_{TX}\theta\right)  \left(  Y\right)  -\left(
\mathcal{L}_{TY}\theta\right) \left(  X\right)  +\theta\left(
T\left[  X,Y\right]  \right)
\end{equation}
$d_{T}$ can be shown to be nilpotent $\left(  d_{T}\circ d_{T}=:d_{T}%
^{2}=0\right)  $ like the ordinary exterior differential $d$ if
and only if $T$ has a vanishing Nijenhuis torsion, but we will not
insist on that.

Returning instead to the extension (\ref{extension1}), one can
prove the following:

\textit{The extension of the action of }$T$ \textit{on two-forms defined by:}%
\begin{equation}
\left(  \widetilde{T}\circ\omega\right)  \left(  X,Y\right)
=:\omega\left( TX,Y\right)
\end{equation}
\textit{will be a two-form (i.e. it will be skew-symmetric) if and only if:}%
\begin{equation}
\omega\left(  TX,Y\right)  =\omega\left(  X,TY\right)  \text{ \
}\forall X,Y
\label{skew0}%
\end{equation}

Indeed, if the condition (\ref{skew0}) holds, then:
\begin{eqnarray}
\left(  \widetilde{T}\circ\omega\right)  \left(  X,Y\right)
&=:&\omega\left(
TX,Y\right)  =-\omega\left(  Y,TX\right)  = \nonumber \\
&=& - \omega\left(  TY,X\right)  =-\left(
\widetilde{T}\circ\omega\right)  \left( Y,X\right)
\end{eqnarray}
and $\ \widetilde{T}\circ\omega$ is skew-symmetric. Viceversa, if
$\omega
_{1}=:\widetilde{T}\circ\omega$ is skew-symmetric, then:%
\begin{eqnarray}
\omega\left(  X,TY\right)  &=& -\omega\left(  TY,X\right)
=-\omega_{1}\left(
Y,X\right)  =\nonumber \\
&=& \omega_{1}\left(  X,Y\right)  =\omega\left(  TX,Y\right)
\end{eqnarray}
and (\ref{skew0}) holds.$\blacksquare$

Notice that, in this case:%
\begin{equation}
\omega\left(  TX,Y\right)  =\frac{1}{2}\left\{  \omega\left(
TX,Y\right) +\omega\left(  X,TY\right)  \right\}
\end{equation}
and there is no real difference between the two linear extensions.

\bigskip

\subsection*{$\mathcal{H}-$weak and $\omega$-weak Recursion Operators. Strong
Recursion Operators}
\addcontentsline{toc}{subsection}{$\mathcal{H}-$weak and $\omega$-weak Recursion Operators. Strong
Recursion Operators}

\bigskip

Let $\ \Gamma$ \ be a Hamiltonian vector field \ with Hamiltonian
\ $\mathcal{H}$ \ w.r.t. a given symplectic form $\omega$, i.e.:%
\begin{equation}
i_{\Gamma}\omega=d\mathcal{H}%
\end{equation}

Then \cite{DSVM1,LMV2,Mar17,ZK}, a $\left(  1,1\right)  $-type
tensor field $T$ \ compatible with the
dynamics, i.e. such that:%
\begin{equation}
\mathcal{L}_{\Gamma}T=0 \label{compatible}%
\end{equation}
is called:

\begin{itemize}
\item A \textit{\ }$\mathcal{H}$-\textit{weak recursion operator}
if it
"generates new Hamiltonians" in the sense that:%
\begin{equation}
d\left(  \widetilde{T}^{k}d\mathcal{H}\right)  =0,\text{
}k=1,2,3,...
\label{closure0}%
\end{equation}
i.e., locally at least:%
\begin{equation}
\widetilde{T}^{k}d\mathcal{H}=d\mathcal{H}_{k},\text{ }k\geq1 \label{newham}%
\end{equation}
for some $\mathcal{H}_{k}\in\mathcal{F}\left(  \mathcal{M}\right)
$. \ It is called instead:

\item A $\omega$-\textit{weak recursion operator} if it "generates
new
symplectic forms" in the sense that:%
\begin{equation}
\omega_{k}=:\underset{k\text{
}times}{\underbrace{\widetilde{T}\circ
\widetilde{T}\circ....\circ\widetilde{T}}}\omega=:\widetilde{T}^{k}\circ
\omega,\text{ }k=1,2,3... \label{newsymp}%
\end{equation}
is closed and skew-symmetric (and hence a symplectic form if $T$
is invertible). Finally, $T$ is called:

\item A \textit{strong recursion operator} if it is both
$\mathcal{H}$-weak and $\omega$-weak.
\end{itemize}

\bigskip

Before discussing the conditions under which a $\left(  1,1\right)
$ tensor is $\mathcal{H}$-weak and/or $\omega$-weak, let us
examine some consequences of these definitions.

First of all, if $T$ is $\mathcal{H}$-weak, it may well happen
that:
$d\mathcal{H}_{k}\wedge d\mathcal{H}=0$ for some $k$ (even for $k=1$%
\footnote{This seems to be the case for the Kepler problem
\cite{Mar17}}) , and the process of generating new Hamiltonian
functions will stop at this stage. Barring this case, one can
generate then a set of $\omega$-Hamiltonian vector fields
$\Gamma_{k\text{ }}$
via:%
\begin{equation}
i_{\Gamma_{k}}\omega=d\mathcal{H}_{k},\text{ \ }k\geq1\label{newham2}%
\end{equation}
Taking the Lie derivative w.r.t. $\Gamma$ of Eqn.(\ref{newham})
and taking
into account the invariance of $T$ one finds at once:%
\begin{equation}
d\left(  \mathcal{L}_{\Gamma}\mathcal{H}_{k}\right)  =0
\end{equation}
This implies only: $\mathcal{L}_{\Gamma}\mathcal{H}_{k}=const.$
and not that $\mathcal{H}_{k}$ is a constant of the motion for
$\Gamma$. This will require some additional assumptions that will
be discussed shortly below.

If instead $T$ is $\omega$-weak, taking again the Lie derivative
w.r.t.
$\Gamma$ of Eqn.(\ref{newsymp}), invariance of $T$ leads at once to:%
\begin{equation}
\mathcal{L}_{\Gamma}\omega_{k}=0,\text{ \ }k\geq1
\end{equation}
In other words, $\Gamma$ will be also locally
$\omega_{k}$-Hamiltonian. Then,
locally at least:%
\begin{equation}
i_{\Gamma}\omega_{k}=d\widetilde{\mathcal{H}}_{k}%
\end{equation}
for some $\widetilde{\mathcal{H}}_{k}\in\mathcal{F}\left(
\mathcal{M}\right) $, and this will provide \textit{alternative
Hamiltonian descriptions} for the same dynamics. Notice that the
$\widetilde{\mathcal{H}}_{k}$'s are not related
(at least not in a simple way) to the $\mathcal{H}_{k}$'s of Eqn.(\ref{newham}%
). Alternatively, one can define a new set of vector fields
$\widetilde
{\Gamma}_{k}$ via:%
\begin{equation}
i_{\widetilde{\Gamma}_{k}}\omega_{k}=d\mathcal{H}%
\end{equation}
and these will be all Hamiltonian vector fields associated with
different symplectic structure but with \textit{the same}
Hamiltonian function.

Some relevant results concerning $\mathcal{H}$-weak and/or
$\omega$-weak \ recursion operators have been proved in the
literature. The main results that we will summarize here
(referring to the literature for details of the proof) are:

\begin{enumerate}
\item If $T$ satisfies the condition (\ref{closure0}) for $k=1$, i.e.:%
\begin{equation}
d\left(  \widetilde{T}d\mathcal{H}\right)  =0
\end{equation}
and has vanishing Nijenhuis torsion:%
\begin{equation}
N_{T}=0
\end{equation}
then it is a $\mathcal{H}$-weak recursion operator (i.e.
Eqn.(\ref{closure0}) will hold for every $k$).$\blacksquare$

\item If, moreover, $\widetilde{T}\circ\omega$ is skew-symmetric,
which means, in terms of the representative matrices, $\omega$
being already
skew-symmetric:%
\begin{equation}
T^{t}\omega=\omega T \label{skew2}%
\end{equation}
then the $\mathcal{H}_{k}$'s defined by Eqn.(\ref{newham}) are all
constants
of the motion for $\Gamma$ pairwise in involution:%
\begin{equation}
\left\{  \mathcal{H}_{k},\mathcal{H}_{l}\right\}  =:\omega\left(
\Gamma _{l},\Gamma_{k}\right)  =0\text{ }\forall k,l\geq0
\end{equation}
where $\left\{  .,.\right\}  $ denotes the Poisson bracket
associated with the symplectic form $\omega$.$\blacksquare$
\end{enumerate}


\begin{remark}
This last result has the following implications:

\begin{itemize}
\item As $\omega$ is non-degenerate, there can be at most a set of
$k\leq n=\left(  1/2\right)  \dim\left(  \mathcal{M}\right)  $
(functionally) independent constants of the motion pairwise in
involution, and:

\item If the set is maximal (i.e. $k=n$), the dynamics is
\textit{completely integrable} in the Liouville sense.
\end{itemize}
\end{remark}

Concerning $\omega$-weak recursion operators, it has also been
proved in the literature that, if \ $T$ has a vanishing Nijenhuis
torsion
and, moreover, $\widetilde{T}\circ\omega$ is closed:%
\begin{equation}
d\left(  \widetilde{T}\circ\omega\right)  =0 \label{close1}%
\end{equation}
and is skew-symmetric (Eqn.(\ref{skew2})), then $T$ is a
$\omega$-weak recursion operator.$\blacksquare$

All \ this has the consequence that:

\begin{itemize}
\item \textit{If }$T$ \textit{has a vanishing Nijenhuis torsion:}%
\begin{equation}
N_{T}=0
\end{equation}
\textit{If :}

\item $\widetilde{T}\circ\omega$ \textit{is skew-symmetric, i.e.,
in terms of
the representative matrices:}%
\begin{equation}
T^{t}\omega=\omega T \label{skew3}%
\end{equation}
\textit{and if:}

\item \textit{both }$\widetilde{T}\circ\omega$ and
$\widetilde{T}d\mathcal{H}$
\textit{are closed:}%
\begin{equation}
d\left(  \widetilde{T}\circ\omega\right)  =d\left(  \widetilde{T}%
d\mathcal{H}\right)  =0 \label{close2}%
\end{equation}
\textit{then }$T$ \textit{is a strong recursion
operator.}$\blacksquare$
\end{itemize}

\bigskip

In the next Section we shall discuss a relevant class of recursion
operators that happen to satisfy almost all of the above
conditions.

\bigskip


\subsection*{Factorizable Recursion Operators}\label{recursion}
\addcontentsline{toc}{subsection}{Factorizable Recursion Operators}

\bigskip

We will consider here dynamical systems that are \textit{bi-Hamiltonian}%
\footnote{Or, for that matter, \textit{bi-Lagrangian}.}. A
dynamical vector field $\Gamma$ is bi-Hamiltonian if there exist
two pairs $\left(  \omega _{1},\mathcal{H}_{1}\right)  $ and
$\left( \omega_{2},\mathcal{H}_{2}\right) $ such that\footnote{In
the Lagrangian case the same r\^{o}le will be played
by the Lagrangian two-forms and the associated energy functions.}:%
\begin{equation}
i_{\Gamma}\omega_{1}=d\mathcal{H}_{1}%
\end{equation}
as well as:%
\begin{equation}
i_{\Gamma}\omega_{2}=d\mathcal{H}_{2}%
\end{equation}
At least one of the two closed two- forms, say $\omega_{1}$, will
be assumed to be non-degenerate, hence a symplectic form. As such,
it will have an inverse $\omega_{1}^{-1}$ which will be the
bivector (actually a $\left(
2,0\text{ }\right)  $ tensor, a Poisson tensor):%
\begin{equation}
\omega_{1}^{-1}=\frac{1}{2}(\omega_{1})^{ij}\frac{\partial}{\partial x^{i}%
}\wedge\frac{\partial}{\partial x^{j}};\text{ \ }(\omega_{1})^{ik}%
(\omega_{_{1}})_{kj}=\delta^{i}\text{ }_{j}%
\end{equation}

Out \ of the two symplectic forms we can then build up the $\left(
1,1\right)  $ tensor $T$ defined via:%
\begin{equation}
\left(  \widetilde{T}\circ\omega_{1}\right)  \left(  X,Y\right)
=:\omega
_{1}\left(  TX,Y\right)  =\omega_{2}\left(  X,Y\right)  \label{recursion1}%
\end{equation}
or, for short:%

\begin{equation}
T=\omega_{1}^{-1}\circ\omega_{2} \label{recursion2}%
\end{equation}
Explicitly:%
\begin{equation}
T=T^{i}\text{ }_{j}dx^{j}\otimes\frac{\partial}{\partial x^{i}};\text{ }%
T^{i}\text{ }_{j}=(\omega_{1})^{ik}(\omega_{2})_{kj} \label{recursion3}%
\end{equation}
$\left(  1,1\right)  $ tensors that can constructed via the
composition of a $\left(  2,0\right)  $ and of a $\left(
0,2\right) $ tensor will be called \textit{factorizable}.

From now on, $\omega_{1}$ and $\mathcal{H}_{1}$ will play the
r\^{o}le of the $\omega,\mathcal{H}$ of the previous Section.

\bigskip

\begin{remark}
 It is pretty obvious from the definition
(\ref{recursion1})
that:%
\begin{equation}
Ker\left(  T\right)  \equiv Ker\left(  \omega_{2}\right)
\end{equation}
As the kernel of a closed two-form is is a Lie subalgebra of
$\mathcal{X}\left( \mathcal{M}\right)  $, i.e. it is involutive, if: $\dim Ker\left(
T\right)  $ has constant dimension, it is also a
distribution. Moreover, $T$ will be invertible $\left(
\det\left\Vert T^{i}\text{ }_{j}\right\Vert \neq0\right) $ iff,
besides $\omega_{1}$, $\omega_{2}$ is also non-degenerate, and
hence symplectic as well.
\end{remark}

The $\left(  1,1\right)  $ tensor $T$ is a natural candidate for a
recursion operator. Indeed, let us prove first that the closure
condition for
$\widetilde{T}d\mathcal{H}_{1}$is satisfied. We have:%

\begin{equation}
Td\mathcal{H}_{1}=\frac{\partial\mathcal{H}_{1}}{\partial
x^{i}}T^{i}\text{
}_{j}dx^{j}\equiv\frac{\partial\mathcal{H}_{1}}{\partial
x^{i}}\left(
\omega_{1}\right)  ^{ik}\left(  \omega_{2}\right)  _{kj}dx^{j}%
\end{equation}
But: $i_{\Gamma}\omega_{1}=d\mathcal{H}_{1}$ implies:%
\begin{equation}
\frac{\partial\mathcal{H}_{1}}{\partial x^{i}}\left(
\omega_{1}\right)
^{ik}=\Gamma^{k}%
\end{equation}
and hence:%
\begin{equation}
Td\mathcal{H}_{1}=d\mathcal{H}_{2}%
\end{equation}
which proves that \ $Td\mathcal{H}_{1}$ is not only closed, but
also exact.$\blacksquare$

Moreover:%
\begin{eqnarray}
\omega_{1}\left(  TX,Y\right)  &=&\omega_{2}\left(  X,Y\right)  
= -\omega_{2}\left(  Y,X\right)  \\
&=&-\omega_{1}\left(  TY,X\right)
=\omega _{1}\left(  X,TY\right)
\nonumber
\end{eqnarray}
which proves (cfr. Eqn. (\ref{skew0})) that
$\widetilde{T}\circ\omega_{1}$ is skew-symmetric.

This result could have been inferred more directly from
Eqn.(\ref{recursion1})
which states that:%
\begin{equation}
\widetilde{T}\circ\omega_{1}=\omega_{2}%
\end{equation}
which allows us also to conclude that
$\widetilde{T}\circ\omega_{1}$ is a closed two-form.

Therefore we obtain the following result:

\textit{If the }$\left(  1,1\right)  $ \textit{tensor field (}\ref{recursion2}%
\textit{) satisfies the Nijenhuis condition, i.e. if:}%
\begin{equation}
N_{T}=0
\end{equation}
\textit{then }$T$ \textit{is a strong recursion
operator.}$\blacksquare$

\newpage

%% file: AppCrev.tex
\section[Symplectic Fourier Transform]{Symplectic Fourier Transform}

\subsection*{Introduction}
\addcontentsline{toc}{subsection}{Introduction}
\bigskip

 Let us consider, for simplicity \cite{Fo,Zam}, $\mathbb{R}^{2}\approx
\mathbb{T}^{\ast}\mathbb{R}$\ \ with coordinates $\left(
q,p\right)  $. The standard Fourier transform (e.g. in
$L_{2}\left(  \mathbb{R}^{2}\right)  $) of
a function $f=f\left(  q,p\right)  $ is defined as:%
\begin{equation}
\mathcal{F(}f)\left(  \eta,\xi\right)  =%
{\displaystyle\iint}
\frac{dqdp}{2\pi}\exp\{-i(q\eta+p\xi)\}f\left(  q,p\right)
\end{equation}
with the known inversion formula (again in the sense of \ $L_{2}\left(
\mathbb{R}^{2}\right)  $) :%
\begin{equation}
f\left(  q,p\right)  =%
{\displaystyle\iint}
\frac{d\eta d\xi}{2\pi}\exp\{i(q\eta+p\xi)\}\mathcal{F(}f)\left(  \eta
,\xi\right)
\end{equation}

Notice that, with the standard Euclidean metric in $\mathbb{R}^{2}$,
$g=diag\left(  1,1\right)  $, $q\eta+p\xi=g\left(  (q,p),(\eta,\xi\right)  $.
Introducing the canonical symplectic form $\omega_{D}=dq\wedge dp$, with
representative matrix:%
\begin{equation}
\Omega_{D}=\left\vert
\begin{array}
[c]{cc}%
0 & 1\\
-1 & 0
\end{array}
\right\vert
\end{equation}
the \textit{symplectic} Fourier transform $\mathcal{F}_{s}\left(  f\right)  $
is defined as:%
\begin{equation}
\mathcal{F}_{s}\left(  f\right)  \left(\eta,\xi\right)  =:%
{\displaystyle\iint}
\frac{dqdp}{2\pi}\exp\left\{  -i\omega_{D}\left(  \left(  q,p\right)
,(\xi,\eta\right)  \right\}  f\left(  q,p\right)
\end{equation}
where, explicitly:
\begin{equation}
\omega_{D}\left(  \left(  q,p\right)  ,\xi,\eta\right)  =\left\vert
\begin{array}
[c]{cc}%
q & p
\end{array}
\right\vert \left\vert
\begin{array}
[c]{cc}%
0 & 1\\
-1 & 0
\end{array}
\right\vert \left\vert
\begin{array}
[c]{c}%
\xi\\
\eta
\end{array}
\right\vert =q\eta-p\xi
\end{equation}

Therefore:
\begin{equation}
\mathcal{F}_{s}\left(  f\right)  \left(  \eta,\xi\right)  =\mathcal{F}\left(
f\right)  \left(  \eta,-\xi\right)
\end{equation}
and the transform can be inverted into:%
\begin{equation}
f\left(  q,p\right)  =%
{\displaystyle\iint}
\frac{d\eta d\xi}{2\pi}\exp\left\{  i\omega_{D}\left(  \left(  q.p\right)
,\left(  \xi,\eta\right)  \right)  \right\}  \mathcal{F}_{s}\left(  f\right)
\left(  \eta,\xi\right)
\end{equation}
or:%
\begin{equation}
f\left(  q,p\right)  =%
{\displaystyle\iint}
\frac{d\eta d\xi}{2\pi}\exp\left\{  -i\omega_{D}\left(  \left(  \xi
,\eta\right)  ,\left(  q,p\right)  \right)  \right\}  \mathcal{F}_{s}\left(
f\right)  \left(  \eta,\xi\right)
\end{equation}
where, explicitly: $\omega_{D}\left(  \left(  q,p\right)  ,\left(  \xi
,\eta\right)  \right)  =q\eta-p\xi$.

A generic constant symplectic structure\ $\omega$ in $\mathbb{R}^{2}$ is of
course associated with a (real) skew-symmetric matrix of the form:%
\begin{equation}
\Omega=\left\vert
\begin{array}
[c]{cc}%
0 & a\\
-a & 0
\end{array}
\right\vert ,\text{ \ }a\neq0
\end{equation}
and there exists a nonsingular matrix $T\in\mathcal{A}\mathit{ut}\left(
\mathbb{R}^{2}\right)  =GL\left(  2,\mathbb{R}\right)  $ (a $\left(
1,1\right)  $ tensor) such that:%
\begin{equation}
\Omega=\widetilde{T}\omega_{D}T
\end{equation}
i.e. (always remember that, by definition: $\left(  \widetilde{T}\right)  _{i}$
$^{j}=T^{j}$ $_{i}$):%
\begin{equation}
\omega(x,y)=\omega_{D}\left(  Tx,Ty\right)  ,\text{ }x,y\in\mathbb{R}^{2}%
\end{equation}
Indeed, if:
\begin{equation}
T=\left\vert
\begin{array}
[c]{cc}%
\lambda & \mu\\
\mathbb{\nu} & \rho
\end{array}
\right\vert
\end{equation}
then the previous condition only requires:
\begin{equation}
\det T=\lambda\rho-\mu\nu=a
\end{equation}
and $T$ will be actually defined "modulo" left multiplication by any matrix
$U$ with $\det U=1$, i.e.: $U\in Sp\left(  2,\mathbb{R}\right)  \approx
SL\left(  2,\mathbb{R}\right)  $: $\widetilde{U}\omega_{D}U=\omega_{D}$. In
this slightly more general setting, the symplectic Fourier transform is
defined as:
\begin{equation}
\mathcal{F}_{sT}\left(  f\right)  \left(  \eta,\xi\right)  =\frac{J}{2\pi}%
{\displaystyle\iint}
dqdp\exp\left\{  -i\omega\left(  \left(  q,p\right)  ,(\xi,\eta\right)
\right\}  f\left(  q,p\right)
\end{equation}
where: $J=:\det T$. Now, if: $T(q,p)=:(x,k)$, then:%
\begin{equation}
\frac{\partial\left(  q,p\right)  }{\partial\left(  x,k\right)  }=J^{-1}%
\end{equation}
Moreover, with: $X=:\left(  q,p\right)  ,Y=:\left(  x,k\right)  $, $TX=Y$ and:
$Z=(\xi,\eta)$, $\ $we have: $\ \omega\left(  \left(  q,p\right)  ,(\xi
,\eta\right)  =\omega\left(  T^{-1}Y,Z\right)  =\omega_{D}\left(  Y,TZ\right)
$. Hence, changing variables:%
\begin{equation}
\mathcal{F}_{sT}\left(  f\right)  \left(  \eta,\xi\right)  =%
{\displaystyle\iint}
\frac{dxdk}{2\pi}\left(  f\circ T^{-1}\right)  \left(  x,k\right)
\exp\left\{  -i\omega_{D}\left(  \left(  x,k\right)  ,T\left(  \xi
,\eta\right)  \right)  \right\}
\end{equation}
i.e., setting: $\left(  \xi_{T},\eta_{T}\right)  =:T\left(  \xi,\eta\right)
$:%
\begin{equation}
\mathcal{F}_{sT}\left(  f\right)  \left(  \eta,\xi\right)  =\mathcal{F}%
_{s}\left(  f\circ T^{-1}\right)  \left(  \eta_{T},\xi_{T}\right)
\end{equation}

Noticing that:
\begin{equation}
f\left(  q,p\right)  \equiv\left(  f\circ T^{-1}\right)  \left(  T\left(
q,p\right)  \right)
\end{equation}
we can write, using the inversion formula for the "canonical" symplectic
transform:%
\begin{equation}
f\left(  q,p\right)  =%
{\displaystyle\iint}
\frac{d\xi_{T}d\eta_{T}}{2\pi}\mathcal{F}_{s}\left(  f\circ T^{-1}\right)
\left(  \eta_{T},\xi_{T}\right)  \exp\left\{  -i\omega_{D}\left(  \left(
\xi_{T},\eta_{T}\right)  ,T\left(  q,p\right)  \right)  \right\}
\end{equation}
or:%
\begin{equation}
f\left(  q,p\right)  =%
{\displaystyle\iint}
\frac{d\xi_{T}d\eta_{T}}{2\pi}\mathcal{F}_{sT}\left(  f\right)  \left(
\eta,\xi\right)  \exp\left\{  -i\omega_{D}\left(  T\left(  \xi,\eta\right)
,T\left(  q,p\right)  \right)  \right\}
\end{equation}
and eventually ($\partial\left(  \xi_{T},\eta_{T}\right)  /\partial\left(
\xi,\eta\right)  =J$) we obtain the inversion formula:%
\begin{equation}
f\left(  q,p\right)  =\frac{J}{2\pi}%
{\displaystyle\iint}
d\xi d\eta\mathcal{F}_{sT}\left(  f\right)  \left(  \eta,\xi\right)
\exp\left\{  -i\omega\left(  \left(  \xi,\eta\right)  ,\left(  q,p\right)
\right)  \right\}
\end{equation}

\subsection*{Equivariance}
\addcontentsline{toc}{subsection}{Equivariance}

\bigskip

What remains to be discussed is the role of the ambiguity in the definition of
$T$ ($T$ and $UT$, $U\in Sp\left(  2,\mathbb{R}\right)  $ playing the same
role). The question is whether or not $\mathcal{F}_{sT}\left(  f\right)
\left(  \eta,\xi\right)  $ and $\mathcal{F}_{sUT}\left(  f\right)  \left(
\eta,\xi\right)  $, i.e. $\mathcal{F}_{s}\left(  f\circ T^{-1}\right)  \left(
\eta_{T},\xi_{T}\right)  $ and $\mathcal{F}_{s}\left(  f\circ(UT)^{-1}\right)
\left(  \eta_{UT},\xi_{UT}\right)  $ define the same symplectic Fourier
transform. From the definition:%
\begin{equation}%
\begin{array}
[c]{c}%
\mathcal{F}_{s}\left(  f\circ(UT)^{-1}\right)  \left(
\eta_{UT},\xi
_{UT}\right)  =\\
=%
{\displaystyle\iint}
\frac{dqdp}{2\pi}\left(  f\circ T^{-1}\circ U^{-1}\right)  \left(
q,p\right) \exp\left\{  -i\omega_{D}\left(  \left(  q,p\right)
,U\circ T\left(  \xi ,\eta\right)  \right)  \right\}
\end{array}
\end{equation}
Setting: $U^{-1}\left(  q,p\right)  =\left(  x,k\right)  $ ($\det U=1$):%
\begin{equation}%
\begin{array}
[c]{c}%
\mathcal{F}_{s}\left(  f\circ(UT)^{-1}\right)  \left(
\eta_{UT},\xi
_{UT}\right)  =\\
=%
{\displaystyle\iint}
\frac{dqdp}{2\pi}\left(  f\circ T^{-1}\right)  \left(  x,k\right)
\exp\left\{  -i\omega_{D}\left(  U\left(  x,k\right)  ,U\circ
T\left( \xi,\eta\right)  \right)  \right\}
\end{array}
\end{equation}
But: $\omega_{D}\left(  U(.),U(.)\right)  =\omega_{D}\left(
(.),(.)\right)
$, and hence:%
\begin{equation}%
\begin{array}
[c]{c}%
\mathcal{F}_{s}\left(  f\circ(UT)^{-1}\right)  \left(
\eta_{UT},\xi
_{UT}\right)  =\\
=%
{\displaystyle\iint}
\frac{dqdp}{2\pi}\left(  f\circ T^{-1}\right)  \left(  x,k\right)
\exp\left\{  -i\omega_{D}\left(  \left(  x,k\right)  ,T\left(  \xi
,\eta\right)  \right)  \right\}
\end{array}
\end{equation}
i.e.:%
\begin{equation}
\mathcal{F}_{s}\left(  f\circ(UT)^{-1}\right)  \left(
\eta_{UT},\xi _{UT}\right)  =\mathcal{F}_{s}\left(  f\circ
T^{-1}\right)  \left(  \eta _{T},\xi_{T}\right)
\end{equation}
$\mathcal{F}_{sT}$ depends then only on the right coset of $T$ in
$GL\left(
2,\mathbb{R}\right)  $ relative to the subgroup\ $Sp\left(  2,\mathbb{R}%
\right)  $ of the symplectic linear maps. This result can be
summarized by
writing (for $T=\mathbb{I}$, otherwise we substitute $f$ with $f\circ T^{-1}%
$):%
\begin{equation}
\mathcal{F}_{s}\left(  f\circ U^{-1}\right)  \circ
U=\mathcal{F}_{s}\left( f\right)
\end{equation}
or, according to the standard definition of "pull-back" of a map:%
\begin{equation}
\phi^{\ast}\mathcal{F}_{s}\left(  f\right)  =\mathcal{F}_{s}\left(
\phi ^{\ast}f\right)
\end{equation}
where: $\phi=U^{-1}\in Sp(2,\mathbb{R)}$, which can then be
rephrased by saying that the symplectic Fourier transform is
equivariant, or that it is "natural", w.r.t. the symplectic group.

\newpage

%% file: EQMCCRfinal.bbl
\begin{thebibliography}{999}



\bibitem{AM} R.  Abraham, J. E.  Marsden, \textit{Foundations of
Mechanics}, $2$nd Edition, Benjamin/Cummings, Reading, 1978. 

\bibitem{Ahl} L. V.  Ahlfors, \textit{Complex Analysis}, McGraw-Hill, New York, 
1953. 

\bibitem{AGGL} M. Aizenman, G.  Gallavotti, S.  Goldstein,
J. L.  Lebowitz,  \textit{Stability and Equilibrium States of Infinite
Classical Systems},  Comm. Math. Phys.  \textbf{48} (1976) 1. 

\bibitem{Ar} V. I.  Arnol'd,  \textit{Mathematical Methods of
Classical Mechanics}, Springer, Berlin and New York, 1989. 

\bibitem{Ar1} V. I.  Arnol'd, \textit{Ordinary Differential Equations},
Springer, Berlin and New York, 1991. 

\bibitem{AS} A.  Ashtekhar, T. A.  Schilling, \textit{Geometrical
Formulation of Quantum Mechanics}, in  \textit{On Einstein's Path},
Springer, Berlin and New York, 1999. 

\bibitem{Ba} H.  Bacry, \textit{Group-Theoretical Analysis of Elementary Particles in an External Electromagnetic
Field}, Nuovo Cim.  \textbf{70A} (1970) 289. 

\bibitem{BSZ} J. C.  Baez, I. E.  Segal, Z.  Zhou, \textit{Introduction to Algebraic and Constructive Quantum Field Theory}, Princeton University Press, Princeton, 1992. 

\bibitem{BMMNSSZ} A. P.  Balachandran, G.  Marmo, N.  Mukunda, J. S.  Nilsson, A.  Simoni, E. C. G.  Sudarshan, F.  Zaccaria,
\textit{Unified Geometrical Approach to Relativistic Particle Dynamics}, J. Math. Phys.  \textbf{25} (1984) 167. 

\bibitem{BMSL}  A. P.  Balachandran, G.  Marmo, A.  Stern, \textit{A Lagrangian Approach to the No-Interaction Theorem},
Nuovo Cim. \textbf{A11} (1982) 69. 

\bibitem{Barg} V.  Bargmann, \textit{On Unitary Ray Representations of Continuous
Groups}, Ann. Math.  \textbf{59} (1954) 1. 

\bibitem{Bas1} H.  Basart, M.  Flato, A.  Lichnerowicz, D.  Sternheimer,
\textit{Deformation Theory Applied to Quantization and Statistical
Mechanics},  Lett. Math. Phys.  \textbf{8} (1984) 483. 

\bibitem{Bas2} H.  Basart, A.  Lichnerowicz,\textit{Conformal Symplectic Geometry, Deformations, Rigidity and
Geometrical (KMS) Conditions}, Lett. Math. Phys. 
\textbf{10} (1985) 167. 

\bibitem{Bec} J.  Beckers, N.  Debergh, J. F.  Cari$\widetilde{n}$ena, G.  Marmo, \textit{Non-Hermitian Oscillator-Like Hamiltonians and $\lambda$-Coherent States Revisited}, Mod. Phys. Letters \textbf{A16} (2001) 91. 

\bibitem{Bend} C. M.  Bender,  \textit{Making Sense of Non-Hermitian Hamiltonians}, Rep. Prog. Phys.  \textbf{70} (2007) 947. 

\bibitem{BCR1}S.  Benenti, C.  Chanu, G.  Rastelli,  \textit{Rematks on the Connection Between the Additive Separation of the Hamilton-Jacobi Equation and the Multiplicative Separability of the Schr\"{o}dinger Equation.  $I$. The Completeness and Robertson Condition}, J. Math. Phys.  \textbf{43} (2002) 5183. 

\bibitem{BCR2}S.  Benenti, C.  Chanu, G.  Rastelli,  \textit{Rematks on the Connection Between the Additive Separation of the Hamilton-Jacobi Equation and the Multiplicative Separability of the Schr\"{o}dinger Equation.  $II$. First Integrals and Symmetry Operators}, J. Math. Phys.  \textbf{43} (2002) 5223. 

\bibitem{BCR3}S.  Benenti, C.  Chanu, G.  Rastelli,   \textit{Variable Separation Theory for the Null Hamilton-Jacobi Equation}, J. Math. Phys.  \textbf{46} (2005) 042901. 

\bibitem{Ben} I.  Bengsson, K.  Zyczkovski, \textit{ Geometry of quantum states}, Cambridge Univ.  Press, Cambridge, 2006. 

\bibitem{BSS}A.  Benvegnu', N.  Sansonetto, M. Spera, \textit{Remarks on Geometric Quantum
Mechanics}, J. Geom.  and Phys.  \textbf{51} (2004) 229. 

\bibitem{Ber} P.  Bergmann, \textit{Introduction to the Theory of
Relativity}, Dover, New York, 1975. 

\bibitem{BM} G.  Birkhoff, S.  MacLane, \textit{A Survey of Modern
Algebra}, McMillan, New York, 1965. 

\bibitem{BHK} P.  Blasiak, A.   Horzela, G.  Kapuscik, \textit{Alternative Hamiltonians and Weyl Quantization},
 J. Opt.  Quantum Semicl.  \textbf{5} (2003) S245. 

\bibitem{Bong} P. J. M.  Bongaarts, \textit{Linear Fields According to I. E. Segal}, in  R. F. Streater (Ed.), \textit{Mathematics of Contemporary Physics}, Ac. Press, New York, 1972. 

\bibitem{BDe} D. G.  Boulware, S.  Deser, \textit{"`Ambiguities" of Harmonic-Oscillator
Commutation Relations}, Nuovo Cim. $XXX$ (1963) 230. 

\bibitem{bratteli} O.  Bratteli, D. W.  Robinson, \textit{Operator Algebras and Quantum Statistical Mechanics}, Springer-Verlag, Berlin and New York, 1987. 

\bibitem{BH}D. C.  Brody, L. P.   Hughston, \textit{Geometric Quantum Mechanics},
J. Geom. Phys.  \textbf{38} (2001) 19. 

\bibitem {Br} E.  Brown, \textit{Bloch Electrons in a Uniform Magnetic Field}, Phys. Rev. 
\textbf{A133}  (1964) 1038. 

\bibitem{Br2}L. M.  Brown (Ed.), \textit{Feynman's Thesis}, World
Scientific, Singapore, 2005. 

\bibitem{CalDeG} F.  Calogero, A.  De Gasperis, \textit{On the Quantization of Newton-Equivalent Hamiltonians}, Am. J. Phys.  \textbf{9} (2004) 1202. 

\bibitem{CJM}J. F.  Cari$\widetilde{n}$ena, J.  Clemente-Gallardo, G. 
Marmo, \textit{Introduction to Quantum Mechanics and the
Quantum-Classical Transition}, quant-ph/0707. 3539 (2007). 

\bibitem{CJM2}J. F.  Cari$\widetilde{n}$ena, J.  Clemente-Gallardo, G. 
Marmo,  \textit{Geometrization of Quantum Mechanics}, Theoretical and Mathematical Physics \textbf{152} (2007) 894. 

\bibitem{CGM1}J. F.  Cari$\widetilde{n}$ena, J.  Grabowski, G.  Marmo,
\textit{Quantum Bi-Hamiltonian Systems}, Int. J. Mod. Phys. 
\textbf{A15} (2000) 4797. 

\bibitem{CGM2}J. F.  Cari$\widetilde{n}$ena, J.  Grabowski, G.  Marmo, 
\textit{Contractions  Nijenhuis Tensors for General Algebraic
Structures}, J. Phys. \textbf{A34} (2001) 3769. 

\bibitem {CIMS} J. F.  Cari$\widetilde{n}$ena, L. A.  Ibort, G.  Marmo, A.  Stern,
\textit{The Feynman Problem and the Inverse Problem for Poisson Dynamics. },Phys. Repts. 
\textbf{263} (1995)  153. 

\bibitem{CMR}J. F.  Cari$\widetilde{n}$ena, G.  Marmo,
M. F.  Ra$\widetilde{n}$ada, \textit{Non-Symplectic Symmetries and
Bi-Hamiltonian Structures for the Rational Harmonic Oscillator},
J. Phys.  \textbf{A35} (2002) L679. 

\bibitem{CMS} S.  Cavallaro, G.  Morchio, F.  Strocchi,
\textit{A Generalization of the Stone-von Neumann Theorem of Non-Regular Representation of the $CCR$ Algebra},
 Lett. Math. Phys.  \textbf{47} (1999)  307. 

\bibitem{CEMMMS1} S.  Chaturvedi, E.  Ercolessi, G.  Marmo, G.  Morandi, N. 
Mukunda, R.  Simon, \textit{Wigner Distributions for
Finite-Dimensional Quantum Systems  An Algebraic Approach},
Pramana-J. Phys.  \textbf{65} (2005) 981. 

\bibitem{CEMMMS2}S.  Chaturvedi, E.  Ercolessi, G.  Marmo, G.  Morandi, N. 
Mukunda, R.  Simon, \textit{Wigner-Weyl Correspondence in
Quantum Mechanics for Continuous and Discrete Systems.  A
Dirac-inspired View}, J. Phys.  \textbf{A39} (2006) 1405. 

\bibitem{Ch}S. S.  Chern, \textit{Complex Manifolds without Potential
Theory}, 2nd edition, Springer-Verlag, Berlin and New York, 1967. 

\bibitem{CM}Y.  Choquet-Bruhat, C.  Morette-deWitt, \textit{Analysis, Manifolds and Physics},
$2nd$ Edition,  North-Holland, Amsterdam, 1982. 

\bibitem{chru}D.  Chru\'shinski, G.  Marmo, \textit{Remarks on the GNS Representation and the Geometry of Quantum States}, Open Systems and Information Dynamics \textbf{16} (2009) 157. 

\bibitem{Cir}R.  Cirelli, A.  Mania', L.  Pizzocchero, \textit{A Functional Representation for Non-commutative ${\mathbb C}^*$
Algebras}, Rev. Math. Phys.  \textbf{6} (1994) 675. 

\bibitem{CGM}J.  Clemente-Gallardo, G.  Marmo, \textit{The Space of Density States in Geometrical
 Quantum Mechanics}, in   F.  Cantrijn, M.  Crampin and B. Langerock (Eds.), \textit{Differential Geometric Methods in Mechanics and Field Theory}, Gent Academia Press, Gent, 2007. 

\bibitem{Co}J. M.  Cook,J. M.  \textit{The Mathematics of Second
 Quantization}, Trans. Am. Math. Soc.  \textbf{74} (1953) 222. 

\bibitem{CJS}D. G.  Currie, T. F.  Jordan, E. C. G.  Sudarshan, \textit{Relativistic Invariance and Hamiltonian Theories of Interacting Particles}, Rev. Mod. Phys.  \textbf{35} (1963) 350. 

\bibitem{CS}D. G.  Currie, E. J.   Saletan, \textit{q-Equivalent Particle Hamiltonians.  I.  The Classical One-Dimensional Case},
J. Math. Phys.  \textbf{7} (1966) 967. 

\bibitem{CS2}D. G.  Currie, E. J.   Saletan, \textit{Canonical Transformations and Quadratic Hamiltonians}, Nuovo Cim. \textbf{B9} (1972) 143. 

\bibitem {DZ}I.  Dana, J.  Zak,  \textit{Adams Representation and Localization in a Magnetic
Field}, Phys. Rev.  \textbf{B28} (1983) 694. 

\bibitem{Das}A.  Das, \textit{Integrable Models}, World Scientific, Singapore,
1989. 

\bibitem{DM} A.  D'Avanzo, G.  Marmo,  \textit{Reduction and Unfolding  The Kepler
Problem}, Int. J. Geom. Meth. Mod. Phys.  \textbf{2} (2004) 83.

\bibitem{DLMV}S.  DeFilippo, G.  Landi, G.  Marmo, G.  Vilasi,  \textit{Tensor Fields Defining a Tangent Bundle
Structure}, Ann. Inst. H. Poincare' \textbf{50}  (1989) 205. 

\bibitem{DSVM1}S.  DeFilippo, M.  Salerno, G.  Vilasi, G.  Marmo, 
\textit{Phase Manifold Geometry of Burgers Hierarchy}, Lett.  Nuovo
Cim.  \textbf{37} (1983) 105. 

\bibitem{DSVM2}S.  DeFilippo, G.  Vilasi, G.  Marmo, M.  Salerno 
\textit{A New Characterization of Completely Integrable Systems},
Nuovo Cim.  \textbf{83B}  (1984) 97. 

\bibitem{DFMS}P.  Di Francesco, P.  Mathieu, D.  Senechal, \textit{Conformal Field Theory}, Springer, Berlin, 1997. 

\bibitem{Dir3}P. A. M.  Dirac, \textit{The Principles of Quantum
Mechanics}, Oxford University Press, Oxford 1958 and $4^{th}$
Edition, 1962. 

\bibitem{Dir1}P. A. M.  Dirac, \textit{The Lagrangian in Quantum
Mehanics}, Physikalische Zeitschrift der Sowjetunion,
Band\textbf{3}, Heft\textbf{1} (1933) 64. 

\bibitem{Doug}J.  Douglas,  \textit{Solution of the Inverse Problem of the Calculus of
Variations},Trans. Am. Math. Soc.  \textbf{50} (1941) 71. 

\bibitem{DHS}D. A.  Dubin, M. A.  Hennings, T. B.  Smith,  \textit{Mathematical Aspects of Weyl Quantization and Phase},
World Scientific, Singapore, 2000. 

\bibitem{DGMS}B. A.  Dubrovin, M.  Giordano, G.  Marmo, A.   Simoni, \textit{Poisson Brackets on Presymplectic Manifolds},
Int. J. Mod. Phys.  \textbf{A8}  (1993) 3747. 

\bibitem{DMS}B. A.  Dubrovin, G.  Marmo, A.   Simoni,
\textit{Alternative Hamiltonian Descriptions for Quantum Systems},
Mod. Phys. Letters \textbf{A5}  (1990) 1229. 

\bibitem{DF}B. A.  Dubrovin, S. P.  Novikov, \textit{Ground States of a Two-Dimensional Electron in a Periodic
Potential}, Sov. Phys. JETP \textbf{52(3)} (1980) 511. 

\bibitem{Dys}F. J.  Dyson,  \textit{Feynman's Proof of the Maxwell
Equations}, Am. J. Phys.  \textbf{58} (1990) 209. 

\bibitem{Ein}A. Einstein,  \textit{Zum Quantenzatz von Sommerfeld und
Epstein},Verhandlungen Physikalischen Gesellshaft \textbf{19}
(1917) 82. 

\bibitem{Em}G. G.  Emch, \textit{Mathematical and Conceptual Foundations of $20$-th Century Physics},
North-Holland, Amsterdam, 1984. 

\bibitem{EIMM}E.  Ercolessi, L. A. Ibort, G.  Marmo, G.  Morandi,
\textit{Alternative Linear Structures for Classical and Quantum
Systems}, Int. J. Mod. Phys.  \textbf{A22} (2007) 3039.

\bibitem{EMM}E.  Ercolessi, G.  Marmo, G.  Morandi, \textit{Alternative Hamiltonian Descriptions and Statistical
Mechanics}, Int. J. Mod. Phys.  \textbf{A17}  (2002) 3779. 

\bibitem{EMMM}E.  Ercolessi, G.  Marmo, G.  Morandi, N.  Mukunda,
\textit{Wigner Distributions in Quantum Mechanics},J. Phys.  Conf.  Series \textbf{87} (2007) 012010. 

\bibitem{EM}G.  Esposito, G.  Marmo, G.  Sudarshan, \textit{From Classical to Quantum
Mechanics}, Cambridge University Press, New York, 2004. 

\bibitem{Fa}P.  Facchi, V.  Gorini, G.  Marmo, S.  Pascazio,
E. C. G.  Sudarshan,  \textit{Quantum Zeno Dynamics}, Phys. Lett. 
\textbf{A275} (2000) 12. 

\bibitem{Fano} U.  Fano, \textit{Description of States in Quantum Mechanics by Density Matrix and Operator Techniques}, Revs. Mod. Phys.  \textbf{29} (1957) 74. 

\bibitem{Fe1}C.  Ferrario, G.  LoVecchio, G.  Marmo, G.  Morandi,
C.  Rubano,  \textit{Separability of Completely-Integrable Dynamical
Systems Admitting Alternative Lagrangian Descriptions},
Lett. Math. Phys. \textbf{9} (1985) 140. 

\bibitem{Fe2}C.  Ferrario, G.  LoVecchio,G.  Marmo, G.  Morandi,
C.  Rubano,  \textit{A Separability Theorem for Dynamical Systems
Admitting Alternative Lagrangian Descriptions},
J. Phys. \textbf{A20}  (1987) 3225. 

\bibitem{Fe}R. P.  Feynman,  \textit{Space-Time Approach to Non-Relativistic Quantum Mechanics}, Revs. Mod. Phys.  \textbf{20} (1948) 367. 

\bibitem{FH}R. P.  Feynman, A. R.  Hibbs, \textit{Quantum Mechancs and Path
Integrals}, McGraw-Hill, New York, 1965. 

\bibitem{Fl}W.  Florek, \textit{Magnetic Translation Groups in $n$
Dimensions},  Repts. Math. Phys.  \textbf{38}  (1996) 235. 

\bibitem{Fo}G. B.  Folland,  \textit{Harmonic Analysis in Phase Space},
Princeton Univ. Press, Princeton, 1989. 

\bibitem{FroNi}A.  Frolicher, A.  Nijenhuis,  \textit{Theory of Vector-Valued Differential Forms}, Indag. Math. 
\textbf{18}  (1956)  338. 

\bibitem{Ga}A.  Galindo, \textit{Some Myriotic Paraboson Fields}, Nuovo Cim. $XXX$ (1963) 235. 

\bibitem{GD}I. M.  Gel'fand, I. Ya.  Dorfman,  \textit{The Schouten Bracket and Hamiltonian
Operators}, Funct. Anal. and Appl.  \textbf{14}  (1981) 223. 

\bibitem{GZ}I. M.  Gel'fand, I.  Zakharevich,  \textit{On Local geometry of a Bihamiltonian
Structure}, in  L.  Corwin, I. M.  Gel'fand, J.  Lepowski (Eds.),  
\textit{Gel'fand Mathematical Seminars Series}, vol. I, Birkhauser, Boston, 
1993. 

\bibitem{Ger}R.  Geroch,  \textit{Mathematical Physics}, Univ.  of
Chicago Press, Chicago, 1985. 

\bibitem{GMR}M.  Giordano, G.  Marmo, C.  Rubano, 
\textit{The Inverse Problem in the Hamiltonian Formalism 
Integrability of Linear Vector Fields}, Inverse Problems \textbf{9}  (1993) 443. 

\bibitem{GMSV}M.  Giordano, G.  Marmo, A.  Simoni, F.  Ventriglia,
\textit{Integrable and Super-Integrable Systems in Classical and
Quantum Mechanics}, in M. J.  Ablowitz, M.  Boiti, F.  Pempinelli, B. Prinari
(Eds.),  \textit{Nonlinear Physics  Theory and
Experiment.  II},  World Scientific, Singapore, 2003. 

\bibitem{GJ}1.  Glimm, A.  Jaffe,  \textit{Quantum Physics.  A Functional Integral Point of
View}, Springer-Verlag, Berlin and New York, 1981. 

\bibitem{GKM}J.  Grabowski, M.  Kus, G.  Marmo,  \textit{Geometry of Quantum Systems  Density States and Entanglement},
J. Phys.  \textbf{A38}  (2005)  10127. 

\bibitem{GKM2}J.  Grabowski, M.  Kus, G.  Marmo, \textit{Wigner's Theorem and the Geometry of Extreme Positive Maps}, J. Phys.  \textbf{A42} 345301(2009). 

\bibitem{GLV}J.  Grabowski, G.  Landi, G.  Vilasi,  \textit{Generalized Reduction Procedure},
Fortschr.  der Physik \textbf{42}  (1994) 393.

\bibitem{GM}J.  Grabowski, G.  Marmo,  \textit{Binary Operations in Classical and Quantum
Mechanics}, in J. Grabowski, P. Urbanski (Eds.),  \textit{Classical and
Quantum Integrability}, 
Banach Center Publ.  \textbf{59} (2003) 163. 

\bibitem{GLMV}J. M.  Gracia-Bondia, F.  Lizzi, G.  Marmo, P.  Vitale, 
\textit{Infinitely Many Star Products to Play With},
 J. High Energy Phis.  \textbf{4} (2002) 26. 

\bibitem{Gre}H. S.  Green,  \textit{A Generalized Method of Field Quantization}, Phys. Rev.  \textbf{90} (1953) 270. 

\bibitem{GreMe}O. W.  Greenberg, A. M. L.   Messiah,  \textit{Selection Rules for Parafields and the Absence of Para Particles in Nature}, Phys. Rev.  \textbf{B138}  (1965) B1155. 

\bibitem{Goe}A.  Groenewold,  \textit{On the Principles of Elementary Quantum Mechanics}, Physica \textbf{12} (1946) 405. 

\bibitem{Gro}A.  Grossmann, G.  Loupias, E. M.   Stein,  \textit{An Algebra of Pseudo-Differential Operators and
Quantum Mechanics in Phase Space}, Ann. Inst. Fourier, Grenoble
\textbf{18} (1968) 2343. 

\bibitem{Ha}R.  Haag,  \textit{Local Quantum Physics. Fields, Particles, Algebras},
Springer-Verlag, Berlin and New York, 1992. 

\bibitem{HHW}R.  Haag, N. M.   Hugenholtz, M.  Winnink,  \textit{On the Equilibrium States in Quantum Statistical
Mechanics}, Comm. Math. Phys.  \textbf{5} (1967) 215. 

\bibitem{kastler}R.  Haag, D.  Kastler,  \textit{An Algebaric Approach to Quantum Field Theory}, J. Math. Phys.  \textbf{5} (1964) 884. 

\bibitem{Hav}O.  Havas,  \textit{The Range of Application of the Lagrangian Formalism}, Nuovo Cim.  Suppl.  \textbf{3}(1957) 363. 

\bibitem{Hel}H.  Helmoltz,  \textit{Ueber die Phisikalische Bedeutung des Prinzip der Klenisten
Wirkung}, Z. Reine Angew.  Math.  \textbf{100} (1887) 137. 

\bibitem{HS}M.  Henneaux, L. C.  Shepley,  \textit{Lagrangians for Spherically Symmetric
Potentials}, J. Math. Phys.  \textbf{23}  (1982) 2101. 

\bibitem{Hoch}G.  Hochschild, \textit{On the Cohomology Theory for Associative Algebras},
Ann. Math.  \textbf{47}  (1946)  568. 

\bibitem{Ho}D. R.  Hofstadter, 
\textit{Energy Levels and Wavefunctions of Bloch Electrons in
Rational and Irrational Magnetic Fields}, Phys. Rev.  \textbf{B14}
 (1976) 2239. 

\bibitem{Hu}N. M.  Hugenholtz,  \textit{States and Representations in Statistical Mechanics}, in  R. F. Streater (Ed.), \textit{Mathematics of Contemporary Physics},  Ac. Press, New York, 1972. 

\bibitem{Hus}D.  Husemoller, \textit{Fibre Bundles}, $3d$ Edition,
Springer, Berlin and New York, 1994. 

\bibitem{Huy} D.  Huybrechts, \textit{Complex Geometry}, Springer, Berlin and New York, 2005. 

\bibitem{Ib}L. A.  Ibort, M. deLeon, G.  Marmo,  \textit{Reduction of
Jacobi Manifolds},  J. Phys.  \textbf{A30}  (1997) 2783. 

\bibitem{IMM} L. A.  Ibort, F.  Magri, G.  Marmo, \textit{Bi-Hamiltonian Structures and St\"{a}ckel Separability}, J. Geom. Phys.  \textbf{33} (2000) 210. 

\bibitem{Jo}P.  Jordan, \textit{Uber die Multiplikation Quantenmechanischer
Grossen}, Zeitschrift f.  Physik \textbf{87} (1934) 505. 

\bibitem{JNW}P.  Jordan,  J.  von Neumann, E. P.  Wigner,
\textit{On an Algebraic Generalization of the Quantum Mechanical
Formalism}, Ann. Math.  \textbf{35} (1934) 29. 

\bibitem{KB}L. P.  Kadanoff, G.   Baym,  \textit{Quantum Statistical
Mechanics}, Benjamin Inc., New York, 1962. 

\bibitem{Ka}E. K.  Kasner,  \textit{Differential Geometric Aspects of Dynamics}, A. M. S. , New York, 1913. 

\bibitem{Kato}T.  Kato,  \textit{Perturbation Theory for Linear Operators}, Springer, Berlin and New York, 1995. 

\bibitem{Kir}A. A.  Kirillov,  \textit{Elements of the Theory of
Representations}, Springer-Verlag, Berlin, 1976. 

\bibitem{Kir2}A. A.  Kirillov,  \textit{Merits and Demerits of the Orbit
Method}, Bull. Am. Math. Soc. \textbf{36} (1999) 433. 

\bibitem{Ko}B.  Konstant,  \textit{Quantization and Unitary Representations Part I. 
Prequantization}, in  \textit{Lecture Notes in Mathematics 170},
Springer-Verlag, Berlin, 1970. 

\bibitem{KBZ}D.  Krejcirik, H.  Bila, M.   Znojil,  \textit{Closed Formula for the Metric in the Hilbert Space of a $PT$-Symmetric
Model}, J. Phys.  \textbf{A39} (2006) 10143. 

\bibitem{Kubo}R.  Kubo,  \textit{Statistical Mechanical Theory of Irreversible Processes.  I. 
General Theory and Simple Applications to Magnetic and Conduction
Problems}, J. Phys. Soc. Japan \textbf{12} (1957) 570. 

\bibitem{Lag}J. L.  Lagrange,   \textit{Memoire sur la Theorie de la Variation des
Elements des Planetes}, Mem.  Cl.  Sci.  Math.  Phys.  Ins.  France
(1808) 1-72. 

\bibitem{LMV1}G.  Landi, G.  Marmo, G.  Vilasi,  \textit{An Algebraic Approach to
Integrability}, J.  Group Theory in Physics \textbf{3} (1994) 1. 

\bibitem{LMV2}G.  Landi, G.  Marmo, G.  Vilasi,
\textit{Recursion Operators Meaning and Existence for Completely
Integrable Systems}, J. Math. Phys.  \textbf{35} (1994) 808. 

\bibitem{LRT}F.  Langouche, D.  Roekaerts, E.  Tirapegui,  \textit{Functional Integration and Semiclassical Expansions}, Reidel, Boston, 1982. 

\bibitem{Lax}P. D.   Lax, \textit{Integrals of Nonlinear Equations and Solitary Waves}, Comm. Pure Appl. Math. 
\textbf{XXI}  (1968) 467. 

\bibitem{Lax2}P. D.   Lax, \textit{Periodic Solutions of the KdV Equation}, Comm. Pure Appl.  Math.  \textbf{XXVIII}  (1975) 141. 

\bibitem{Lax3}P. D.   Lax, \textit{Almost Periodic Solutions of the KdV Equation}, Siam Review \textbf{18}  (1976) 351. 

\bibitem{Leo}U.  Leonhardt,  \textit{Measuring the Quantum State of
Light}, Cambridge  Univ.  Press, New York, 1997. 

\bibitem{LSU}B. I.  Lev, A. A.  Semenov, C. V.  Usenko,C. V.   \textit{Scalar Charged Particle in Wigner-Moyal Phase Space.  Constant Magnetic Field}, J. Russian Laser Research \textbf{23}  (2002) 347. 

\bibitem{TLC}T.  Levi-Civita,  \textit{Fondamenti di Meccanica Relativistica}, Zanichelli, Bologna, 1928, pp. 48-53. 

\bibitem{LM}J. Q.  Liang, G.  Morandi, \textit{On the Extended Feynman Formula for the Harmonic Oscillator},
 Phys. Letters \textbf{A160} (1991) 9. 

\bibitem{Li}A.  Lichnerowicz,  \textit{Les Vari\'{e}t\'{e}s de Jacobi et Leurs  Alg\'{e}bres de Lie
Associ\'{e}es}, J. Math. Pures Appl.  \textbf{57} (1978) 453. 

\bibitem{LMR}C.  Lopez, E.  Martinez, M. F.  Ra$\widetilde{n}$ada,  \textit{Dynamical Symmetries, Non-Cartan Symmetries and Superintegrability
of the $n$-Dimensional Harmonic Oscillator}, J. Phys. A Math. Gen. 
\textbf{32} (1999) 1241. 

\bibitem{Lo}R.  Lopez-Pe$\widetilde{n}$a, V. I.  Man'ko, G.  Marmo,  \textit{Wigner's
Problem for a Precessing Magnetic Dipole}, Phys. Rev.  \textbf{A56}
(1997) 1126. 

\bibitem{Mac1}G. W.  Mackey,  \textit{Induced Representations of Groups and Quantum Mechanics},
Benjamin, New York, 1968. 

\bibitem{Mac2}G. M.  Mackey, \textit{Mathematical Foundations of Quantum
Mechanics}, Benjamin, New York, 1963 and  Dover, New York,  2004. 


\bibitem{Ma}F.  Magri,  \textit{A Simple Model of the Integrable Hamiltonian
Equation}, J. Math. Phys.  \textbf{19}  (1978)  1156. 

\bibitem{Man}D.  Mancusi, \textit{Meccanica Quantistica sullo Spazio delle Fasi}, Thesis, Napoli, 2003 (Unpublished). 

\bibitem{MMM}O. V.  Man'ko, V. I.  Man'ko, G.  Marmo,  \textit{Alternative Commutation Relations, Star Products and
Tomography}, J. Phys. \textbf{A35}  (2002)  699. 

\bibitem{MMM2}O. V.  Man'ko, V. I.  Man'ko, G.  Marmo, 
\textit{Star-Product of Generalized Wigner-Weyl Symbols on $SU(2)$
Group, Deformations and Tomographic Probability Distribution},
Physica Scripta \textbf{62} (2000) 446. 

\bibitem{MM1}V. I.  Man'ko, G.  Marmo,  \textit{Probability
Distributions and Hilbert Spaces  Quantum and Classical Systems},
Physica Scripta \textbf{60} (1999) 111. 

\bibitem{MM2} V. I.  Man'ko, G.  Marmo,  \textit{Aspects of Nonlinear
and Noncanonical Transformations in Quantum Mechanics}, Physica
Scripta \textbf{58}  (1998) 224. 

\bibitem{MMSSV}V. I.  Man'ko, G.  Marmo, A.  Simoni, F.  Ventriglia,
\textit{Tomograms in the Quantum-Classical Transition}. 
Phys. Lett. \textbf{A343} (2005) 251. 

\bibitem{MMSo}V. I.  Man'ko, G.  Marmo, S.  Solimeno, F. Zaccaria,
\textit{Physical Nonlinear Aspects of Classical and Quantum
$Q$-Oscillators}, Int. J. Mod. Phys.  \textbf{A8}  (1993) 3577. 

\bibitem{MMSo2}V. I.  Man'ko, G.  Marmo, S.  Solimeno, F. Zaccaria,
\textit{Correlation Functions of Quantum $Q$-Oscillators}, Phys. 
Letters \textbf{A176}  (1993)  173. 

\bibitem{MMSZ3}V. I.  Man'ko, G.  Marmo, E. C. G.  Sudarshan,
F.  Zaccaria,  \textit{The Geometry of Density States, Positive Maps
and Tomograms}, in   B.  Gruber,
G.  Marmo and N.  Yoshinaga (Eds.), \textit{Symmetries in Science XI}, Kluwer, New York, 2004. 

\bibitem{MMSZ4}V. I.  Man'ko, G.  Marmo, E. C. G.  Sudarshan,
F.  Zaccaria, \textit{Purification of Impure Density Operators and
the Recovery of Entanglement}, quant-ph/9910080 (1999).  

\bibitem{MMSZ6}V. I.  Man'ko, G.  Marmo, E. C. G.  Sudarshan,
F.  Zaccaria, \textit{Inner Composition Law for Pure States as a
Purificatin of Impure States}, Phys. Lett.  \textbf{A273} (2000) 31. 

\bibitem{MMSZ5} V. I.  Man'ko, G.  Marmo, E. C. G.  Sudarshan,
F.  Zaccaria,\textit{Interference and Entanglement  an Intrisic
Approach}, Int. J. Theor. Phys. \textbf{40} (2002) 1525. 

\bibitem{MMSZ1}V. I.  Man'ko, G.  Marmo, E. C. G.  Sudarshan, F.  Zaccaria,
\textit{Entanglement in Probability Representation of Quantum States
and Tomographic Criterion of Separability}, J. Opt. \textbf{B} 
Quantum Semiclass.  Opt.  \textbf{6} (2004) 172. 

\bibitem{MMSZ2}V. I.  Man'ko, G.  Marmo, E. C. G.  Sudarshan, F.  Zaccaria,
\textit{Differential geometry of Density States}, Repts. Math. Phys. 
\textbf{55}, 405 (2005). 

\bibitem{JS}V. I.  Man'ko, G.  Marmo, P.  Vitale,, F.  Zaccaria, \textit{A Generalization of the Jordan-Wigner Map  Classical Versions and its q-Deformations}, Int. J. Mod. Phys. \textbf{A9} (1994) 5541. 

\bibitem{MMSZ}V. I.  Man'ko, G.  Marmo, F.  Zaccaria, E. C. G.  Sudarshan, 
\textit{Wigner's Problem and Alternative Commutation Relations for
Quantum Mechanics}, Int. J. Mod. Phys.  \textbf{B11}  (1996) 1281. 

\bibitem{MarG}G.  Marmo,  \textit{Equivalent Lagrangians and Quasi-Canonical Transformations}, in  A.  James, T.  Janssen, M. Boon (Eds.),  \textit{Group Theoretical Methods in Physics}, Springer-Verlag, 1976. 

\bibitem{Mar0}G.  Marmo, \textit{Nijenhuis Operators in Classical
Dynamics}, in  \textit{Seminar on Group Theoretical Methods in
Physics}, USSR Academy of Sciences, Yurmala, Latvian SSR, 1985. 

\bibitem{Mar1}G.  Marmo,  \textit{The Quantum-Classical Transition for Systems
with Alternative Hamiltonian Descriptions}, in \textit{"Proceedings
of the IX Fall Workshop on Geometry and Physics"}, Real Sociedad
Matematica Espa\~{n}ola, Madrid, 2001. 

\bibitem{Mar2}G.  Marmo,  \textit{Alternative Commutation Relations and Quantum Bi-Hamiltonian
Systems}, Acta Applicandae Mathematicae \textbf{70}  (2002) 161. 

\bibitem{Mar3}G.  Marmo, \textit{The Inverse Problem for Quantum
Systems}, in  
W.  Sarlet and F.  Cantrijn (Eds.), \textit{Applied Differential Geometry and Mechanics}, Gent Academia Press, Gent, 2003. 


\bibitem{MarMor}G.  Marmo, G.  Morandi, \textit{The Inverse Problem with Symmetries and the Appearance
of Cohomologies in Classical Lagrangian Dynamics}, Reports on
Math. Phys. \textbf{28}  (1989) 389. 

\bibitem{MarMor2}G. Marmo, G. Morandi  \textit{Some Geometry and Topology}, in
 S.  Lundqvist, G.  Morandi, Yu Lu (Eds.), \textit{Low-Dimensional Quantum Field Theories for Condensed-Matter
Physicists}, World Scientific, Singapore, 1995. 

\bibitem{Muk0}G.  Marmo, G.  Morandi, N.  Mukunda,  \textit{A Geometrical Approach to the Hamilton-Jacobi Form of Dynamics and its Generalizations}, Riv.  Nuovo Cim.  \textbf{13} (1990) 1. 

\bibitem{Mar6}G.  Marmo, G.  Morandi, C.  Rubano,  \textit{Symmetries in the Lagrangian and Hamiltonian Formalism. 
The Equivariant Inverse Problem}, in   B.  Gruber, E. Iachello (Eds.), \textit{Symmetries in
Science III}, Plenum Press, New York, 1983. 

\bibitem{Mar7}G.  Marmo, G.  Morandi, A.  Simoni, F.  Ventriglia,\textit{Alternative Structures and Bi-Hamiltonian
Systems}, J. Phys.  \textbf{A35}  (2002) 8393. 

\bibitem{MMSS}G.  Marmo, G.  Morandi, A.  Simoni, E. C. G.  Sudarshan, 
\textit{Quasi-Invariance and Central Extensions}, Phys. Rev. 
\textbf{D37} (1988) 2196. 

\bibitem{MMSL}G.  Marmo, N.  Mukunda, E. C. G.  Sudarshan,  \textit{Relativistic Particle Dynamics-Lagrangian Proof of the No-Interaction Theorem}, Phys. Rev.  \textbf{D30} (1984) 2110. 

\bibitem{Mar8}G.  Marmo, C.  Rubano,  \textit{Alternative Lagrangians for a Charged Particle in a Magnetic Field},
Phys. Lett.  \textbf{A119}  (1987) 321. 

\bibitem{MarSal}G.  Marmo, E. J.   Saletan, \textit{Ambiguities in the Lagrangian and Hamiltonian Formalisms  Transformation
Properties}, Nuovo Cim.  \textbf{40B} (1977) 67. 

\bibitem{MarSal2}G.  Marmo, E. J.   Saletan, \textit{q-Equivalent Particle Hamiltonians.  III.  The Two-Dimensional
Quantum Oscillator}, Hadronic J.  \textbf{3} (1980) 1644. 

\bibitem{Mar9} G.  Marmo, E. J.  Saletan, R.  Schmid, A.   Simoni,
\textit{Bi-Hamiltonian Dynamical Systems and the
Quadratic-Hamiltonian Theorem}, Nuovo Cim.  \textbf{100B}
(1987) 297. 

\bibitem{Mar10}G.  Marmo, E. J.  Saletan, A.  Simoni, B.  Vitale,  \textit{Dynamical Systems},
J. Wiley\&Sons, New York, 1985. 

\bibitem{Mar11}G.  Marmo, G.  Scolarici, A.  Simoni,
F.  Ventriglia, \textit{Alternative Hamiltonian Descriptions for
Quantum Systems}, in  
R. F.  Alvarez-Estrada, A.  Dobado, L. A.  Fernandez, M. A.  Martin-Delgado,
A.  Munoz Sudupe (Eds.), \textit{Encuentro de Fisica Fundamental}, Aula Documental de Investigacion, Madrid, 2005. 

\bibitem{Mar112}G.  Marmo, G.  Scolarici, A.  Simoni, F.  Ventriglia,
\textit{Quantum Bi-Hamiltonian Systems, Alternative Structures and Bi-Unitary Transformations},
Note di Matematica  \textbf{23}  (2004) 173. 

\bibitem{Mar12}G.  Marmo, G.  Scolarici, A.  Simoni, F.  Ventriglia,
\textit{The Quantum-Classical Transition  the Fate of the Complex
Structure}, Int. J. Geom. Meth.   Mod.  Phys.  \textbf{2} (2005) 1. 

\bibitem{Mar13}G.  Marmo, G.  Scolarici, A.  Simoni, F.  Ventriglia,
\textit{Alternative Structures and Bi-Hamiltonian Systems on a
Hilbert Space}, J. Phys. \textbf{A38} (2005) 3813. 

\bibitem{Mar131}G.  Marmo, G.  Scolarici, A.  Simoni, F.  Ventriglia,
\textit{Alternative Algebraic Structures from Bi-Hamiltonian Quantum
Systems}, Int. J. Geom. Meth.  Mod.  Phys.  \textbf{2}  (2005)  919. 

\bibitem{Mar132}Marmo. G. , G.  Scolarici, A.  Simoni, F.  Ventriglia,
\textit{Classical and Quantum Systems  Alternative Hamiltonian
Descriptions}, Theor.  and Math. Phys.  \textbf{144}  (2005) 1190. 

\bibitem{Mar14}G.  Marmo, A.  Simoni, F.  Ventriglia, \textit{Bi-Hamiltonian Quantum Systems and Weyl Quantization},
Repts. Math. Phys.  \textbf{48}  (2001) 149. 

\bibitem{Mar141}G.  Marmo, A.  Simoni, F.  Ventriglia,
\textit{Quantum Systems  Real Spectra and Non-Hermitian
(Hamiltonian) Operators}, Repts. Math. Phys.  \textbf{51}
(2003) 275. 

\bibitem{Mar15}G.  Marmo, A.  Simoni, F.  Ventriglia, \textit{Bi-Hamiltonian Systems in the Quantum-Classical
Transition}, Rendic.  Circolo Mat.  Palermo, Serie \textbf{II}, Suppl. 
\textbf{69}  (2002) 19. 

\bibitem{Mar16}G.  Marmo, A.  Simoni, F.  Ventriglia,\textit{Quantum Systems and Alternative Unitary Descriptions},
Int. J. Mod. Phys.  \textbf{A19}  (2004) 2561. 

\bibitem{MSV}G.  Marmo, A.  Simoni, F.  Ventriglia,\textit{Geometrical Structures Emerging from Quantum
Mechanics}, in  J. C.  Gallardo, E. Martinez (Eds.), \textit{Groups,
Geometry and Physics},  Monografias de la Real Academia de
Ciencias,  Zaragoza, 2006. 

\bibitem{Mar17}G.  Marmo, G.  Vilasi,.   \textit{When do Recursion Operators Generate New Conservation
Laws?}, Phys. Lett.  \textbf{B277}  (1992) 137. 

\bibitem{Mar18}G.  Marmo, G.  Vilasi,  \textit{Symplectic Structures and Quantum
Mechanics}, Mod. Phys. Letters \textbf{B10} (1996) 545. 

\bibitem{MarSchwi}P. C.  Martin, J.  Schwinger,  \textit{Theory of Many-Particle Systems. 
I}, Phys. Rev.  \textbf{115} (1959)  1342. 

\bibitem{Mau}G.  Mauceri, \textit{The Weyl Transform and Bounded Operators in
 $\mathcal{L}^{p}\left(  \mathbb{R}^{n}\right)
 $}, J. Funct. Anal.  \textbf{39}  (1980) 408. 

\bibitem{Mes}A.  Messiah,  \textit{Mecanique Quantique}, Vol. I,
Dunod, Paris, 1958. 

\bibitem{Mor}G.  Morandi, \textit{Quantum Hall Effect}, Bibliopolis, Naples,
1988. 

\bibitem{Mor2}G.  Morandi, \textit{The Role of Topology in Classical and Quantum Physics},
 Springer-Verlag, Berlin and New York, 1992. 

\bibitem{MFVMR}G.  Morandi, C.  Ferrario, G.  LoVecchio,G.  Marmo, C.  Rubano, 
\textit{The Inverse Problem in the Calculus of Variations and the
Geometry of the Tangent Bundle}, Phys. Repts.  \textbf{188}
(1990) 147. 

\bibitem{MNE}G.  Morandi, F.  Napoli, E.  Ercolessi,  \textit{Statistical Mechanics.  An Intermediate
Course}, World Scientific, Singapore, 2001. 

\bibitem{Mosta}A.  Mostafazadeh,  \textit{Pseudo-Hermitian Quantum
Mechanics}, arXiv 0810.5643 (2008). 

\bibitem{Moy}J. E.  Moyal, \textit{Quantum Mechanics as a Statistical Theory}, Proc.  Cambdridge Phil.  Soc.  \textbf{45} (1940) 90. 


\bibitem{Muk1}N.  Mukunda, \textit{Algebraic Aspects of the Wigner Distribution in Quantum
Mechanics}, Pramana, \textbf{11} (1978) 1. 

\bibitem{Muk} N.  Mukunda,G.  Marmo, A.  Zampini, S.  Chaturvedi, R.  Simon,
\textit{Wigner-Weyl Isomorphism for Quantum Mechanics on
Lie Groups}, J. Math. Phys. \textbf{46} (2005) 012106. 

\bibitem{Na}M. A.  Naimark,  \textit{Normed Rings}, Wolters-Noordhoff
Publishing, Groningen, 1970. 

\bibitem{NT}N.  Narhofer, W.  Thirring, \textit{KMS States for the Weyl
Algebra}, Lett. Math. Phys.  \textbf{27} (1993) 133. 

\bibitem{Ni}A.  Nijenhuis,  \textit{Jacobi-Type Identities for Bilinear Differential
Concomitants of Certain Tensor Fields. I and II},  Indag. 
Math. \textbf{17} (1955) 390 and  398. 

\bibitem{NN}L.  Nirenberg, A. Newlander, \textit{Complex Analytic Coordinates in Almost Complex
Manifolds}, Ann. Math.  \textbf{65}  (1957) 391. 

\bibitem{OW}Y.  Ohnuki, S.  Watanabe,  \textit{Self-Adjointness of Operators in Wigner's
Commutation Relations}, J. Math. Phys.  \textbf{33} (1992) 3653. 

\bibitem{Pan}S.  Pancharatnam,  \textit{Generalized Theory of Interference and its Applications},
in  \textit{Collected Works of S.  Pancharatnam}, Oxford Univ.  Press, Oxford, 1975. 

\bibitem{Poo}J. C. T.  Pool, \textit{Mathematical Aspects of the Weyl
Correspondence}, J. Math. Phys.  \textbf{7} (1966) 66. 

\bibitem{Pu}C. R.  Putnam,  \textit{The Quantum-Mechanical Equations of Motion and Commutation Relations}, Phys. Rev. \textbf{83} (1951) 1047. 

\bibitem{Ra}M. F.  Ra$\widetilde{n}$ada,
\textit{Dynamical Symmetries, Bi-Hamiltonian Structures and
 Superintegrability of  n=2 Systems}, J. Math. Phys. \textbf{41} (2000) 2121. 

\bibitem{RS} M.  Reed, B. Simon, \textit{Methods of Modern
Mathematical Physics vol. I.  Functional Analysis}, Ac.  Press, New York and London, 1980. 

\bibitem{Re}H.  Reichenbach, \textit{Philosophical Foundations of Quantum Mechanics}, Univ.  of California Press, 1944. 

\bibitem{Ric}R. D.  Richtmyer,  \textit{Principles of Advanced Mathematical Physics.  Vol. I}, Springer-Verlag, Berlin, 1978. 

\bibitem{RN}F.  Riesz, B.  Nagy,  \textit{Lecons d'Analyse Functionnelle}, Akademiai Kado', Budapest, 1952. 

\bibitem{Rosen}G.  Rosen,  \textit{Formulations of Classical and Quantum Dynamical Theory}, Ac. Press, New-York-London, 1969. 

\bibitem{Ru}R.  Rubio,  \textit{Alg\`{e}bres Associatives Locales sur l'Espace des Sections
d'un Fibr\'{e} \`{a} Droites}, C. R.  Acad. Sc. Paris, S\'{e}rie $I$,
n. $14$, \textbf{699}  (1984) 821. 

\bibitem{Sam}J.  Samuel, \textit{The Geometric Phase and Ray Space
Isometries}, Pramana J. Phys.  \textbf{48} (1997) 959. 

\bibitem{Sant}M. R.  Santilli,  \textit{Foundations of Theoretical
Mechanics}, Springer, Berlin and New York, 1983. 

\bibitem{Schouten}J. A.  Schouten,  \textit{On the Differential Operators of First Order in Tensor Calculus}, in  \textit{Conv. Int. Geom. Diff. Italia}, Cremonese, Rome, 1954. 

\bibitem{SCH}L.  Schwartz,  \textit{Lectures on Complex Analytic
Manifolds}, Narosa, New Dehli, 1986. 

\bibitem{S1}S. S.  Schweber, \textit{A Note on Commutators in Quantized Field
Theories}, Phys. Rev.  \textbf{78}  (1950)  613. 

\bibitem{S2}S. S.  Schweber, \textit{On Feynman Quantization}, J. Math. Phys. 
\textbf{3}  (1962) 831. 

\bibitem{S3}S. S.  Schweber,  \textit{Relativistic Quantum Field Theory}, Harper and Row, New York, 1964. 

\bibitem{So}J. M.  Souriau,  \textit{Structure des Systemes
Dynamiques}, Dunod, Paris, 1970. 

\bibitem{Ste}N.  Steenrod, \textit{The Topology of Fibre Bundles},
Princeton Univ. Press, Princeton, 1951. 

\bibitem{SW}R. F.  Streater, A. S.  Wightman,  \textit{PCT, Spind and Statistics and All That}, Benjamin Inc, New York, 1964. 

\bibitem{Stu}E. C. G.  Stueckelberg,  \textit{Quantum Theory in Real Hilbert Space}, Helv. Physica Acta \textbf{33}  (1960) 727 and \textbf{34}  (1961) 621. 

\bibitem{Sud}E. C. G.  Sudarshan,  \textit{Structure of Dynamical
Theories}, in  \textit{Brandeis Lectures in Theoretical Physics
1961}, Benjamin, New York, 1961. 

\bibitem{Va}V. S.  Varadarajan, \textit{Variations on a Theme by Schwinger and
Weyl}, Lett. Math. Phys.  \textbf{34}  (1995) 319. 

\bibitem{Ve}F.  Ventriglia,
\textit{Alternative Hamiltonian Descriptions for Quantum Systems and
non-Hermitian Operators with Real Spectra}, Mod. Phys. Lett. 
\textbf{A17}  (2002) 1589. 

\bibitem{Vi}G.  Vilasi,   \textit{Hamiltonian Dynamics}, World
Scientific, Singapore, 2001. 

\bibitem{Neu1}J.  von Neumann,  \textit{Warscheinlichtkeitstheoretische Aufbau der Quantenmechanik}, Goettingenische Nachrichten
\textbf{10} (1927) 245, in  A.H.  Taub (Ed.), \textit{J.  von Neumann
Collected Papers}, vol. $I$,  Pergamon Press, Oxford, 1961. 


\bibitem{Neu2}J.  von Neumann,  \textit{Die Mathematische Grundlagen der
Quantenmechanik}, Springer, Berlin and New York, 1932 (English translation:
\textit{Mathematical Foundations of Quantum Mechanics}, Princeton Univ. Press,
Princeton, 1955. 

\bibitem{Weil}A.  Weil,  \textit{Introduction \`a  l'\'{E}tude des Variet\'{e}s
K\"{a}hleriennes}, Hermann, Paris, 1958. 

\bibitem{We}H.  Weyl,  \textit{The Theory of Groups and Quantum
Mechanics}, Dover, N. Y., 1950, Ch. $IV$ Sect. $D$. 

\bibitem{WZ}P. B.  Wiegmann, A. V.  Zabrodin, \textit{Bethe-Ansatz for the Bloch Electrons in a Magnetic
Field}, Phys. Rev. Letters \textbf{72} (1994) 1890. 

\bibitem{Wig} E. P.  Wigner, \textit{\"{U}ber die Operation der Zeitumkehr
in der Quantenmechanik}, Gott.  Nachr. \textbf{31} (1932) 546. 

\bibitem{Wig4}E. P.  Wigner, \textit{On the Quantum Correction for Thermodynamic Equilibrium},
 Phys. Rev. \textbf{40} (1932) 749. 

\bibitem{Wig3}E. P.  Wigner, \textit{Do the Equations of Motion Determine the Quantum Mechanical Commutation Relations?},  Phys Rev. \textbf{77} (1950) 711. 

\bibitem{Wig2}E. P.  Wigner, \textit{Group Theory and its Applications to the Quantum
Mechanics of Atomic Spectra}, Ac. Press, New York-London, 1959. 

\bibitem{Wil}T. J.  Willmore,\textit{The Definition of the Lie
Derivative}, Proc. Edinburgh. Math. Soc.  \textbf{12(2)}  (1960) 27. 

\bibitem{Win}A.  Wintner,  \textit{The Unboundedness of Quantum-Mechanical
Matrices}, Phys. Rev. \textbf{71} (1947) 738. 

\bibitem{Woo1}W. K.  Wootters,  \textit{Quantum Mechanics Without Probability
Amplitudes}, Found.  of Phys. \textbf{16} (1986) 391. 

\bibitem{Woo2}W. K.  Wootters,  \textit{A Wigner-Function Formulation of Finite-State Quantum
Mechanics}, Ann. Phys. (New York) \textbf{176} (1987) 1. 

\bibitem{Ya}L. M.  Yang, \textit{A Note on the Quantum Rule of the Harmonic Oscillator}, Phys. Rev. \textbf{84}
(1951) 788. 

\bibitem{Zak1}J.  Zak,  \textit{Magnetic Translation Groups},
Phys. Rev.  \textbf{A134} (1964) 1602. 

\bibitem{Zak2}J.  Zak, \textit{Dynamics of Electrons in External
Fields}, Phys. Rev.  \textbf{168}  (1968) 686. 

\bibitem{Zak3}J.  Zak,  \textit{Weyl-Heisenberg Group and Magnetic Translations in a Finite Phase
Space}, Phys. Rev.  \textbf{B39}  (1989) 694. 

\bibitem{ZK}V. E.  Zakharov, B. G.  Konopelchenko,  \textit{On the Theory of Recursion
Operators}, Comm. Math. Phys.  \textbf{94} (1984) 483. 

\bibitem{Zam}A.  Zampini,
\textit{Il Limite Classico della Meccanica Quantistica nella
Formulazione \`a la Weyl-Wigner}, Thesis, Napoli, 2001 (Unpublished). 




\end{thebibliography}
